\documentclass[14pt,twoside,floatfix]{mbook}
\usepackage{graphicx}
\usepackage{amssymb}
\usepackage{amsmath}
\usepackage{mathtools}
\usepackage[T1]{fontenc}
\usepackage[cp 1250]{inputenc}
\usepackage{textcomp}
\usepackage{amsfonts}
\usepackage{epsfig}
\usepackage{here}
\usepackage{cite}
\usepackage{esint}
\usepackage{fancyhdr}
\usepackage{color}
\title{{\bf Test of charge conjugation invariance in $\eta\to\pi^0 e^+e^-$ and $\eta\to\pi^0 e^+e^-$ decays}}
\author{Marcin Zieli\'{n}ski}
\date{January 25th, 2012}
\begin{document}

\thispagestyle{empty}
\newpage
\thispagestyle{empty}
\begin{center}
{\Large INSTITUTE OF PHYSICS  }\\
{\Large FACULTY OF PHYSICS, ASTRONOMY  }\\
{\Large AND APPLIED COMPUTER SCIENCE}\\
{\Large JAGIELLONIAN UNIVERSITY}\\
\end{center}
\begin{center}\end{center}
\begin{center}\end{center}
\begin{center}\end{center}
\begin{center}\end{center}
\begin{center}
{\LARGE{\bf Test of charge conjugation invariance}}
\end{center}
\begin{center}
{\LARGE{\bf in $\eta\to\pi^0 e^+e^-$ and $\eta\to\pi^+\pi^-\pi^0$ decays}}
\end{center}
\begin{center}
\end{center}
\begin{center}
\end{center}
\begin{center}\end{center}
\begin{center}\end{center}
\begin{center}
{\Large Marcin Zieli\'{n}ski}
\end{center}
\begin{center}\end{center}
\begin{center}\end{center}
\begin{center}
{\normalsize Doctoral Dissertation  }\\
{\normalsize prepared in the\\}
{\normalsize{\bf{Nuclear Physics Division of the Jagiellonian University\\}}}
{\normalsize and\\}
{\normalsize{\bf{Institute of Nuclear Physics at the Research Center J\"{u}lich\\}}}
{\normalsize supervised by } 
\end{center}
\begin{center}
{\Large{\bf{Prof. dr hab. Pawe\l{} Moskal}}}
\end{center}
\begin{center}\end{center}
\begin{center}\end{center}
\begin{center}
\begin{figure}[h]
\hspace{6.6cm}
\vspace{-3.5cm}
\parbox{0.10\textwidth}{\centerline{\epsfig{file=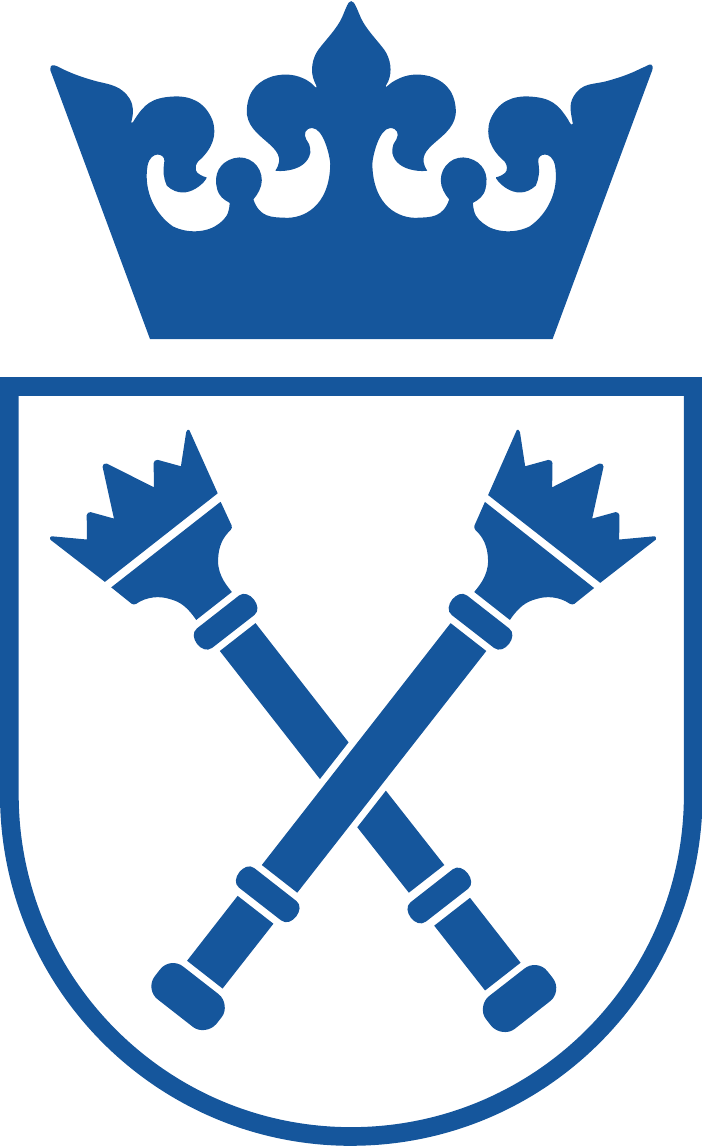,width=0.13\textwidth}}}
\end{figure}
\end{center}
\begin{center}
{\vspace{-.2cm}\normalsize }
\end{center}
\begin{picture}(0,0)
\put(0,580){\line(1,0){440}} 
\put(0,-35){\line(1,0){440}} 
\put(175,-50){CRACOW 2012}
\end{picture}

\newpage
\thispagestyle{empty}
\begin{center}\end{center}

\thispagestyle{empty}
\newpage
\thispagestyle{empty}
\begin{center}
{\Large INSTYTUT FIZYKI  }\\
{\Large WYDZIA\L{} FIZYKI, ASTRONOMII  }\\
{\Large I INFORMATYKI STOSOWANEJ}\\
{\Large UNIWERSYTET JAGIELLO\'{N}SKI}\\
\end{center}
\begin{center}\end{center}
\begin{center}\end{center}
\begin{center}\end{center}
\begin{center}\end{center}
\begin{center}
{\LARGE{\bf Test symetrii sprz\k{e}\.{z}enia \l{}adunkowego}}
\end{center}
\begin{center}
{\LARGE{\bf w rozpadach $\eta\to\pi^0 e^+e^-$ i $\eta\to\pi^+\pi^-\pi^0$}}
\end{center}
\begin{center}
\end{center}
\begin{center}
\end{center}
\begin{center}\end{center}
\begin{center}\end{center}
\begin{center}
{\Large Marcin Zieli\'{n}ski}
\end{center}
\begin{center}\end{center}
\begin{center}\end{center}
\begin{center}
{\normalsize Rozprawa doktorska  }\\
{\normalsize przygotowana\\}
{\normalsize{\bf{w Zak\l{}adzie Fizyki J\k{a}drowej Uniwersytetu Jagiello\'{n}skiego\\}}}
{\normalsize oraz\\}
{\normalsize{\bf{Instytucie Fizyki J\k{a}drowej Centrum Badawczego J\"{u}lich\\}}}
{\normalsize pod kierunkiem } 
\end{center}
\begin{center}
{\Large{\bf{Prof. dr hab. Paw\l{}a Moskala}}}
\end{center}
\begin{center}\end{center}
\begin{center}\end{center}
\begin{center}
\begin{figure}[h]
\hspace{6.6cm}
\vspace{-3.5cm}
\parbox{0.10\textwidth}{\centerline{\epsfig{file=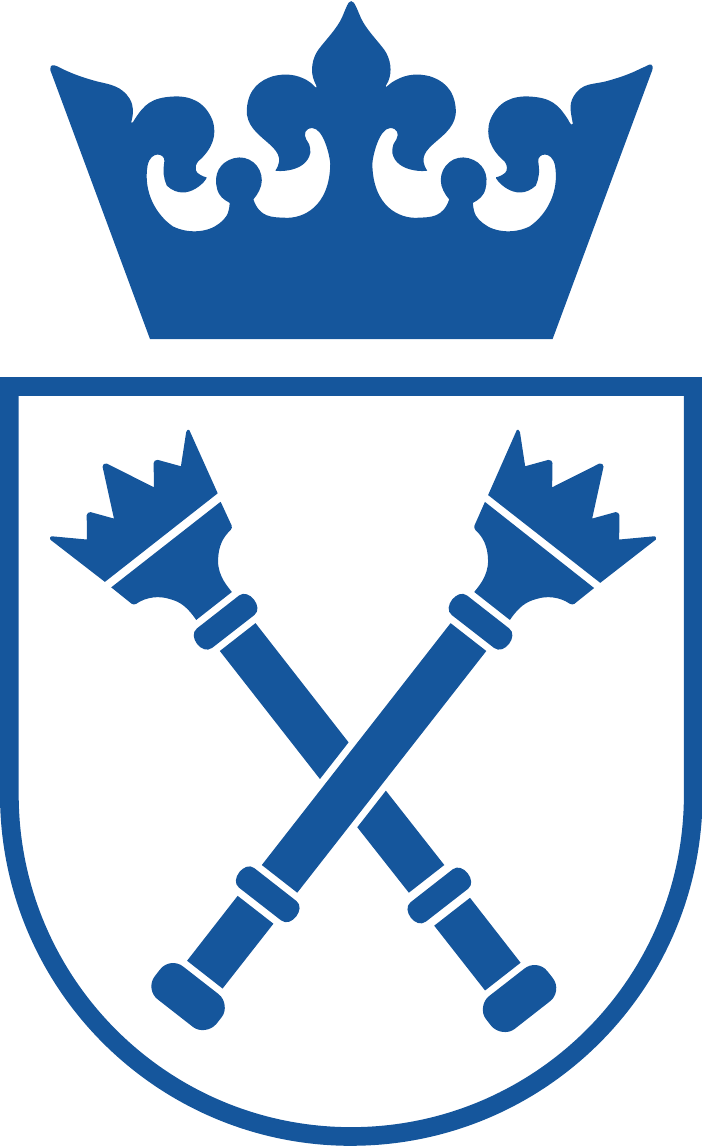,width=0.13\textwidth}}}
\end{figure}
\end{center}
\begin{center}
{\vspace{-.2cm}\normalsize }
\end{center}
\begin{picture}(0,0)
\put(0,580){\line(1,0){440}} 
\put(0,-35){\line(1,0){440}} 
\put(175,-50){KRAK\'{O}W 2012}
\end{picture}

\newpage
\thispagestyle{empty}
\begin{center}\end{center}

\newpage
\thispagestyle{empty}
\begin{center}\end{center}
\begin{center}\end{center}\begin{center}\end{center}
\begin{center}\end{center}\begin{center}\end{center}
\begin{center}\end{center}\begin{center}\end{center}
\begin{center}\end{center}\begin{center}\end{center}
\begin{center}\end{center}
\begin{center}\end{center}
\begin{center}\end{center}
\begin{center}\end{center}
\begin{center}\end{center}
\begin{center}\end{center}
\begin{center}\end{center}
\begin{center}\end{center}
\begin{center}\end{center}
\begin{center}\end{center}
\begin{center}\end{center}
\begin{flushright}
{\bf{"Imagination is more important than knowledge."}}\\
(Albert Einstein)
\end{flushright}

\newpage
\thispagestyle{empty}
\begin{center}\end{center}

\newpage
\thispagestyle{empty}
\begin{center} 
\vspace{2.cm}
\Large{\bf Abstract} 
\end{center}

\vspace{2.cm}
Charge conjugation C is one of the fundamental symmetries in nature which 
transforms particles into antiparticles. This symmetry was studied in weak interaction 
where it is fully violated, but it is poorly known in the strong and electromagnetic interactions. 
Therefore, it is important to test this symmetry accurately for a better understanding of the nature 
of the strong interaction and for the understanding of the significantly larger abundance of matter 
over antimatter in the Universe.
 
To this end, in this thesis we investigated $\eta\rightarrow\pi^{+}\pi^{-}\pi^{0}$ and $\eta\rightarrow\pi^{0}e^{+}e^{-}$ decays, which might violate charge conjugation symmetry. 
The violation of C symmetry in $\eta\rightarrow\pi^{+}\pi^{-}\pi^{0}$  process could manifest 
itself as an asymmetry between energy spectra of charged pions, and can be studied using event density 
distribution on Dalitz plot.
The $\eta\rightarrow\pi^{0}e^{+}e^{-}$ decay is forbidden by C symmetry in the first order of electromagnetic 
interaction, and can only proceed by emission of two virtual photons with the branching ratio 
on a level of $10^{-8}$, as predicted in the framework of the Standard Model. 
Therefore, observation of a larger branching ratio could indicate a 
mechanism involving first order electromagnetic interaction violating charge conjugation. 

Both decays were investigated by means of the WASA-at-COSY detector operating at the COSY synchrotron 
at the Forchungszentrum J\"{u}lich in Germany. The $\eta$ meson was produced via $pp\to pp\eta$ 
reaction at the proton beam momentum of 2.14~GeV/c which corresponds to kinetic energy of 1.4~GeV. 
The measurement was done at the turn of October and November in the year 2008. In total around 
$5\cdot10^7$ $\eta$ mesons were collected during the two weeks of data taking.
The tagging of the $\eta$ meson was done by means of the missing mass technique
and the decay products were identified by the invariant mass reconstruction. 

As a result of the analysis conducted in the framework of this thesis a Dalitz Plot distribution 
for the $\eta\rightarrow\pi^{+}\pi^{-}\pi^{0}$ decay was obtained. From this distribution we extracted 
asymmetry parameters sensitive to C symmetry violation for different isospin values of the final state 
and we have established that all are consistent with zero within the obtained accuracy.

For the $\eta\to\pi^0 e^+e^-$ decay we have not observe signal and thus we estimated an upper limit 
for the branching ratio. The established upper limit amounts to $BR_{\eta\to\pi^0 e^+e^-} < 3.7\cdot 10^{-5}$
at the 90\% confidence level. This result is more precise than previously obtained in other experiments.
We intend to continue the research, and thanks to the 20 times higher statistics of already collected data by WASA-at-COSY, the upper limit will be improved significantly.

\newpage
\thispagestyle{empty}
\begin{center}\end{center}
\newpage
\thispagestyle{empty}

\begin{center} 
\vspace{2.cm}
\Large{\bf Streszczenie} 
\end{center}

\vspace{0.5cm}
Sprz\k{e}\.{z}enie \l{}adunkowe C jest jedn\k{a} z podstawowych symetrii w przyrodzie, kt\'{o}ra zamienia 
cz\k{a}stki na antycz\k{a}stki. Symetria ta by\l{}a badana w oddzia\l{}ywaniach s\l{}abych w kt\'{o}rych 
jest ca\l{}kowicie \l{}amana, natomiast do tej pory jest s\l{}abo poznana z punktu widzenia oddzia\l{}ywa\'{n}
silnych i elektromagnetycznych. Dlatego wa\.{z}nym aspektem jest badanie stopnia zachowania tej symetrii 
dla lepszego zrozumienia natury oddzia\l{}ywania silnego oraz wyja\'{s}nienia wi\k{e}kszej abundancji materii 
ni\.{z} antymaterii we Wszech\'{s}wiecie. 

W tym celu zbadali\'{s}my dwa procesy $\eta\to\pi^+\pi^-\pi^0$ i $\eta\to\pi^0 e^+ e^-$, kt\'{o}re mog\k{a}
\l{}ama\'{c} symetri\k{e} sprz\k{e}\.{z}enia \l{}adunkowego. 
Niezachowanie sprz\k{e}\.{z}enia \l{}adunkowego w procesie $\eta\to\pi^+\pi^-\pi^0$, mo\.{z}e ujawni\'{c}
si\k{e} jako asymetria pomi\k{e}dzy rozk\l{}adami energii pion\'{o}w na\l{}adowanych i mo\.{z}e zosta\'{c}
zaobserwowana za pomoc\k{a} badania g\k{e}sto\'{s}ci obsadze\'{n} na wykresie Dalitza.
Natomiast rozpad $\eta\to\pi^0 e^+ e^-$ jest zabroniony przez symetri\k{e} \l{}adunkow\k{a} w pierwszym 
rz\k{e}dzie oddzia\l{}ywa\'{n} elektromagnetycznych i mo\.{z}e zachodzi\'{c} tylko przez emisje 
dw\'{o}ch wirtualnych foton\'{o}w ze stosunkiem rozga\l{}\k{e}zie\'{n} na poziomie $10^{-8}$, wed\l{}ug 
przewidywa\'{n} w ramach Modelu Standardowego. Jednka\.{z}e, zaobserwowanie wi\k{e}kszego stosunku 
rozga\l{}\k{e}zie\'{n}  \'{s}wiadczy\l{}oby
o innym mechani\'{z}mie reakcji ni\.{z} ten przewidywany na gruncie Modelu Standardowego, kt\'{o}ry 
m\'{o}g\l{}by niezachowywa\'{c} symetrii sprz\k{e}\.{z}enia \l{}adunkowego. 

Oba rozpady by\l{}y badane detektorem WASA-at-COSY zainstalowanym na sychrotronie COSY znajduj\k{a}cym 
si\k{e} w Centrum Badawczym J\"{u}lich w Niemczech. Mezon $\eta$ by\l{} produkowany w reakcji 
$ pp\to pp\eta$ przy p\k{e}dzie wi\k{a}zki protonowej 2.14~GeV/c, co odpowiada\l{}o energii kinetycznej 1.4~GeV. Pomiar zosta\l{} wykonany na prze\l{}omie pa\'{z}dziernika i listopada 2008 roku.
W trakcie dw\'{o}ch tygodni zebrano pr\'{o}bk\k{e} danych zawieraj\k{a}c\k{a} oko\l{}o $5\cdot10^7$ 
mezon\'{o}w $\eta$. Mezon $\eta$ by\l{} identifikowany za pomoc\k{a} widma
masy brakuj\k{a}cej, natomiast produkty jego rozpadu by\l{}y zidentyfikowane za pomoc\k{a} rekonstrukcji 
masy niezmienniczej. 

W oparciu o przeprowadzon\k{a} w ramach pracy doktorskiej analiz\k{e} danych do\'{s}wiadczalnych 
zrekonstruowano wykres Dalitza. Na jego podstawie oszacowano warto\'{s}ci parametr\'{o}w 
asymetrii czu\l{}ych na niezachowanie symetrii sprz\k{e}\.{z}enia \l{}adunkowego C dla r\'{o}\.{z}nych 
warto\'{s}ci izospinu cz\k{a}stek w stanie ko\'{n}cowym. Stwierdzono, \.{z}e wszystkie otrzymane warto\'{s}ci
asymetrii s\k{a} zgodne z zerem w granicach oszacowanych niepewno\'{s}ci. 
 
Dla rozpadu $\eta\to\pi^0 e^+e^-$ w wyniku przeprowadzonej analizy nie zaobserwowano sygna\l{}u
i dlatego oszacowano g\'{o}rn\k{a} granic\k{e} stosunku rozga\l{}\k{e}zie\'{n} na ten rozpad. Obliczona 
warto\'{s}\'{c} g\'{o}rnej granicy wynosi $BR_{\eta\to\pi^0 e^+e^-} < 3.7\cdot 10^{-5}$ na poziomie 
ufno\'{s}ci 90\%. Wynik ten jest bardziej dok\l{}adny ni\.{z} uzyskany w poprzednich eksperymentach.
W najbli\.{z}szej przysz\l{}o\'{s}ci dzi\k{e}ki zebranej do tej pory 20 krotnie wi\k{e}kszej pr\'{o}bce danych przez grup\k{e} WASA-at-COSY wynik ten mo\.{z}e zosta\'{c} znacz\k{a}co poprawiony. 

\newpage
\thispagestyle{empty}
\begin{center}\end{center}

\newpage
\pagestyle{fancy}
\fancyhead{}
\fancyfoot{}
\renewcommand{\headrulewidth}{0.8pt}
\fancyhead[RO]{\textbf{\sffamily{{{\thepage}}~}}}
\fancyhead[RE]{\bf\footnotesize{\nouppercase{\leftmark~}}}
\fancyhead[LE]{\textbf{\sffamily{{{\thepage}}}}}
\fancyhead[LO]{\bf\footnotesize{{\nouppercase{\rightmark~}}}}
\advance\headheight by 5.3mm
\advance\headsep by 0mm
\tableofcontents
\markboth{Contents}{Contents}

\chapter{Introduction}
\hspace{\parindent}
Mesons -- the states of the quark\footnote{Quarks are elementary particles with 
electric charge $q = \pm\frac{1}{3}$ or $q = \pm\frac{2}{3}$.} and anti-quark ($q\bar{q}$) -- for more then 60 years play
an important role in experimental and theoretical physics. Previously and nowadays  
physicists use different mesons to study limits of applicability of the 
Standard Model~\cite{Glashow:1961tr,Weinberg:1967tq,Englert:1964et,Higgs:1964ia,Higgs:1964pj,Guralnik:1964eu},
which is well established theory describing the strong, electromagnetic and weak 
interactions between elementary particles. 
The examination of mesons production and their decay modes give a possibility to probe fundamental 
symmetries such as: charge conjugation (C), space reflection (P), time reversal (T) and 
their combinations: CP~\cite{Kobayashi:1973fv} and CPT~\cite{Luders:1957zz,Lueders:1992dq}. 
Moreover  investigation of such processes can be used to determine the parameters of the Standard Model.

One of the particle used for these studies is the $\eta$ meson discovered in Berkeley Bevatron 
Laboratory in 1961~\cite{Pevsner:1961pa}. It is one of the Goldstone bosons in the quantum 
chromodynamics (QCD)~\cite{Greiner:1989aa} with no electric charge, flavorless and the mass of
$m_{\eta} = 547.853 \pm 0.024$~MeV~\cite{Nakamura:2010zzi}.
From the theoretical point of view it is a superposition of the SU(3) octet $\eta_8$ and singlet $\eta_1$  
states, with the wave function:
\begin{equation}
\vert\eta\rangle = \cos\Theta\vert\eta_8\rangle - \sin\Theta\vert\eta_1\rangle,
\label{stan_ety}
\end{equation}
where $\Theta$ denotes the pseudo-scalar mixing angle between singlet and 
octet state\footnote{The pseudo-scalar mixing angle was established in~\cite{Bramon:1997va} 
and amounts to $\Theta\approx -15.5^o\pm1.3^o$.}, and:
\begin{equation}
\vert\eta_8\rangle = \frac{1}{\sqrt{6}}\vert \bar{u}u + \bar{d}d - 2\bar{s}s\rangle,
\label{eta8}
\end{equation}
\begin{equation}
\vert\eta_1\rangle = \frac{1}{\sqrt{3}}\vert  \bar{u}u + \bar{d}d + \bar{s}s\rangle.
\label{eta1}
\end{equation}

The $\eta$ meson belongs to the pseudo-scalar family with isospin and angular momentum equal to zero,  
negative parity and charge conjugation equal to +1 ($I^G(J^{PC}) = 0^+(0^{-+}$)), 
along with the $\eta'$, $\pi^0$, $\pi^+$, $\pi^-$ and $K^+$, $K^-$, $K^0$, $\bar{K}^0$ mesons.  
It is an eigenstate of the charge conjugation (C) and parity (P) operators, and thus it constitutes 
an important experimental tool for investigations of the degree of conservation of these symmetries 
in strong and electromagnetic interactions. Moreover, the total width of the $\eta$ meson is equal to 
$\Gamma_{\eta} = 1.30 \pm 0.07$~keV~\cite{Nakamura:2010zzi} which is five orders of magnitude smaller 
than the typical width of neutral particles which may decay due to the strong interaction. 
Therefore, the decays of the $\eta$ meson are very sensitive to C and P violation. 

The main decay modes of the $\eta$ meson can be divided into two groups: hadronic decays and radiative decays. 
All of these strong and electromagnetic processes are forbidden in the first order~\cite{Nefkens:2002sa}. 
The most energetically favorable strong decays into $\pi^+\pi^-$ and $\pi^0\pi^0$ 
are P and CP violating, with predicted very low branching ratios. Moreover, strong decay into $4\pi$
is also forbidden because of the P and CP invariance and a small available phase space. 
Therefore, the $\eta$ meson decays predominantly into $\pi\pi\pi$ although this decay violates isospin
symmetry and G-parity~\cite{Zielinski:2008gi}. Historically the $\eta$ meson decays into $\pi\pi\pi$ were 
considered as electromagnetic
process but it was shown that these effects are small~\cite{Sutherland:1966zz},
and instead it is expected that the decay occurs only due to the difference in the mass of the 
$u$ and $d$ quarks. This fact permits to study this mass difference by comparing the measured 
branching ratios with the calculations based 
on the Chiral Perturbation Theory (ChPT)~\cite{Bijnens:2002qy}. 
Furthermore, the first order electromagnetic decays like $\pi^0\gamma$, $\pi^0\pi^0\gamma$ and 
$\pi^+\pi^-\gamma$ are also forbidden and they can occur only due
to QCD anomalies found in current algebra~\cite{Adler:1969gk,Bardeen:1969md,Wess:1971yu}.
In the massless quark limit   
the radiative decays are driven by the QCD box anomaly~\cite{Redmer:2010phd}. The second order 
electromagnetic decay $\eta\to\gamma\gamma$ is also forbidden, and occurs due to the QCD triangle 
anomaly~\cite{Sutherland:1967vf}. The above highlighted properties makes the $\eta$ meson 
specially suitable for tests of the discrete symmetries. In this work we use the $\eta$ meson for
the studies of the C symmetry.

The charge conjugation invariance was studied in weak interaction, and already in 
1957 it was discovered that breaking of this symmetry occurs in decays of 
$\pi^+$ and $\mu^+$~\cite{Garwin:1957hc}. Furthermore, it was also realized that the C operator should 
turn left-handed neutrinos into left-handed anti-neutrinos, but the experimental studies show that
all neutrinos are left-handed and anti-neutrinos right-handed. This implies that C symmetry should be fully violated in weak interactions. It is also interesting to notice that the Big-Bang model predicts the same amount of matter and antimatter 
in the Universe but the experimental observations show that there is significantly larger 
abundance of matter over antimatter~\cite{Cohen:1993nk,Sakharov:1967dj}.
The known CP breaking effect~\cite{Kobayashi:1973fv} is insufficient to explain this phenomenon, 
but it is hoped that investigations of the charge conjugation may help in clarification of this problem.  

One of the main purposes of this thesis is to study the charge conjugation invariance in strong interactions
by means of the Dalitz plot density population for the $\eta\to\pi^+\pi^-\pi^0$ decay. 
The $\eta\to\pi^+\pi^-\pi^0$ hadronic mode is one of the most frequently occurring decay of the 
$\eta$ meson with the branching ratio equal to $22.74\pm 0.28\%$~\cite{Nakamura:2010zzi}. 
The C invariance in this decay can manifest itself as an asymmetry between the 
energy distribution of the $\pi^+$ and $\pi^-$ mesons in the rest frame of the $\eta$ meson. 
The studies of the density population of the Dalitz plot can also reveal details of contribution to 
possible C violation from various isospin states of the final particles. 
Such effects can be investigated by means of three asymmetry parameters: 
(i) $A_{LR}$ -- left-right asymmetry sensitive to violation averaged over all isospin states, 
(ii) $A_{Q}$ -- quadrant asymmetry sensitive for the $I=2$, and (iii) $A_{S}$ -- sextant asymmetry which 
can test the C violation in $I=1$ state~\cite{Layter:1973ti}. 

Furthermore, we intend to extract the branching ratio or estimate an upper limit for the rare 
$\eta\to\pi^0 e^+e^-$ process which might not conserve charge conjugation. 
In the framework of the Standard Model, the decay $\eta\to\pi^0 e^+e^-$ may only proceed via C-conserving exchange of two virtual 
photons with the branching ratio of about $10^{-8}$~\cite{Smith:1968aa}.
But in principle it may also be realized by one photon intermediate state, 
forbidden by  the C invariance and increasing the branching ratio. At present  only an upper 
limit is set for this branching ratio at the level of $4\times 10^{-5}$~\cite{Nakamura:2010zzi}.
Thus, there is still more than three orders of magnitude difference between Standard Model 
predictions and experimentally measured upper limit, and therefore increase of the experimental 
sensitivity gives a chance to observe a signal which would indicate violation of C symmetry.
The possible charge conjugation breaking could be indicated if the branching ratio would be larger 
than $10^{-8}$. 

The measurement aming at the charge conjugation studies described in this dissertation was carried 
out in the Research Center J\"{u}lich in Germany, by means of the WASA-at-COSY detector. 
The $\eta$ meson was produced in proton--proton collisions at beam momentum of 2.14~GeV/c. 
Identification of the investigated reactions was based on the selection of events corresponding to the 
$\pi^+\pi^-\pi^0$ and $\pi^0 e^+e^-$ final state. The $\eta$ meson signal was extracted using 
missing mass spectrum of two outgoing protons registered in the Forward Detector, and the 
decay products were identified based on invariant mass distribution reconstructed from signals 
detected in the Central Detector of the WASA-at-COSY system. 

In the next chapter the main theoretical motivation for conducted investigations is outlined.

The WASA-at-COSY detector facility and the measurement methods is described in Chapter 3.

The Chapter 4 is devoted to the description of the analysis methods and simulation of the detector 
response.  

In Chapter 5 the track selection methods and reconstruction algorithms are discussed. 

Chapter 6 is committed to the identification method of the $pp\to pp\eta$ reaction.

Chapter 7 comprises the description of the $\eta\rightarrow\pi^{+}\pi^{-}\pi^{0}$ decay signal extraction,
and the discussion of multi-pion background reduction methods. Moreover, the kinematic fit procedure is explained in this chapter.

The final results concerning the  Dalitz plot and the asymmetry parameters studies for the 
$\eta\rightarrow\pi^{+}\pi^{-}\pi^{0}$ decay are presented in Chapter 8. 
In addition to that, in Chapter 8 the physical background subtraction, the acceptance and efficiency corrections are discussed.
Finally, the achieved experimental results are compared to theoretical predictions.

Further on, the procedure of the $\eta\rightarrow\pi^{0}e^{+}e^{-}$ decay identification
will be presented in Chapter 9.

In Chapter 10, the branching ratio results for the $\eta\rightarrow\pi^{0}e^{+}e^{-}$
decay are discussed.

The summary and final conclusions followed by perspectives are presented in Chapter~11. 

\chapter{Charge conjugation invariance tests}
\hspace{\parindent}
In the Standard Model of particles and fields, the charge conjugation C, along with the spatial 
parity P and the time reversal T, is one of the most fundamental symmetries. 
The C operator in quantum field theory applied to a particle state $\vert\psi\rangle$, 
changes all additive quantum numbers of this particle to opposite sign, leaving the mass, momentum and spin  
unchanged, and making it an antiparticle state:
\begin{equation}
C \vert\psi\rangle = \vert\bar{\psi}\rangle.
\end{equation}
In the Quantum Electrodynamics (QED) and Quantum Chromodynamics (QCD) it is postulated that C 
holds in all electromagnetic and strong interactions on the level smaller than $10^{-8}$.
Therefore, the C-invariance should imply the balance between 
matter and antimatter, which is not the case in the observed Universe. 
The Standard Model of the weak interactions allows for full C and P violation, as well as 
for small CP violation. However, the model does not explain why and how the violation occurs.
Also the discovered small CP breaking does not explain the larger abundance of matter over antimatter.
Thus, the investigation of the charge conjugation symmetry is one of the most interesting, valuable
and significant field in modern experimental nuclear and particle physics.

Difficulties in studies of the charge conjugation arise from the fact that
there are only very few known particles in nature which are the eigenstates of the C operator. 
The most suitable candidates are neutral and flavorless mesons and the particle-antiparticle systems.
The particularly interesting appears the $\eta$ meson, which plays a crucial 
role for understanding of the low energy Quantum Chromodynamics, and can be also used to tests of the 
fundamental symmetries. 

In this thesis study of the charge conjugation invariance C is presented, by conducting the analysis
of the $\eta\to\pi^+\pi^-\pi^0$ and $\eta\to\pi^0 e^+e^-$ decay modes measured in the 
experiment where the $\eta$ meson was produced in the proton-proton interactions.

\section{Decay of the $\eta$ meson into $\pi^+\pi^-\pi^0$}
\hspace{\parindent}
In view of the tests of charge conjugation invariance C, the strong hadronic decay
$\eta\to\pi^+\pi^-\pi^0$ which does not conserve isospin since the Bose symmetry forbids the three 
pions with the $J^{P} = 0^{-}$, is particularly interesting. However this decay has to be treated 
in special way because in QCD, at low energies, the strong coupling constant $\alpha_s$ is large, 
and the perturbative approach is not valid any more. Therefore, one applies the Chiral Perturbation 
Theory (ChPT) as an effective field theory specially suited for low energies regime. This theory is based on the 
approximate chiral symmetry and expansion in external momenta and quark masses. In this approach the 
role of dynamic degrees of freedom of strong interaction are given to hadrons composed of confined 
quarks and gluons~\cite{Goldstone:1962es}. One can identify them with eight Goldstone bosons members of the  pseudoscalar meson nonet, which are the result of 
the spontaneous chiral symmetry breaking~\cite{Koch:1995vp}. The effective Lagrangian is expanded in 
definite number of derivatives or powers of quark masses given as:
\begin{equation}
L_{ChPT}^{eff} = L_2 + L_4 + L_6 + ..., 
\end{equation} 
where the subscripts stands for the chiral order. Furthermore, effective Lagrangian shares the same 
symmetries with QCD, namely: C, P, T, Lorentz invariance, and the chiral $SU(3)_L \times SU(3)_R$ symmetry.
One can see that only even chiral powers arise since the Lagrangian is Lorentz scalar which implies 
that indices of derivatives appear in pairs. The lowest order of chiral Lagrangian has only two 
constants: $B_0$ which is the quark condensate parameter, and $F_\pi$ which is the pion decay constant, 
and the decay mechanism is given by Current Algebra\cite{Witten:1979vv,Witten:1983tx}. 

Historically the $\eta\to\pi^+\pi^-\pi^0$ decay was treated 
as an electromagnetic process with partial width smaller than second order electromagnetic decay. 
But as it was shown the electromagnetic contributions are small~\cite{Sutherland:1966zz,Bell:1996mi}
and instead the process is dominated by the isospin violating term in the strong 
interaction\cite{Leutwyler:1996qg}. Therefore, it is very interesting to concern hadronic decays 
of the $\eta$ meson into three pion system in terms of the different isospin states. The wave 
function for $I=0$ state can be written in the following form~\cite{Zielinski:2008gi}:
\begin{equation}
(3\pi)_{I=0} = \sqrt{\frac{1}{3}}\left[ (\pi^{+}\pi^{0})_{I=1}\vert\pi^{-}\rangle - 
                                        (\pi^{+}\pi^{-})_{I=1}\vert\pi^{0}\rangle +
                                        (\pi^{-}\pi^{0})_{I=1}\vert\pi^{+}\rangle \right].
\end{equation} 
where final 3$\pi$ can be in the isospin zero state only if the $\pi\pi$ subsystem is in $I=1$ state.
For the two pion subsystems with $I=1$ one can write the wave functions as:
\begin{equation}
\nonumber
(\pi^{+}\pi^{0})_{I=1} = \sqrt{\frac{1}{2}}\left[ 
                  \vert\pi^{+}\rangle\vert\pi^{0}\rangle - \vert\pi^{0}\rangle\vert\pi^{+}\rangle\right]  ,
\end{equation}
\begin{equation}
\nonumber
(\pi^{+}\pi^{-})_{I=1} = \sqrt{\frac{1}{2}}\left[ 
                  \vert\pi^{+}\rangle\vert\pi^{-}\rangle - \vert\pi^{-}\rangle\vert\pi^{+}\rangle\right] ,
\end{equation}
\begin{equation}
\nonumber
(\pi^{-}\pi^{0})_{I=1} = \sqrt{\frac{1}{2}}\left[ 
                  -\vert\pi^{-}\rangle\vert\pi^{0}\rangle + \vert\pi^{0}\rangle\vert\pi^{-}\rangle\right]  .
\end{equation}
Thus the full wave function for the 3$\pi$ system in $I = 0$ state reads:
\begin{eqnarray}
\nonumber
(3\pi)_{I=0} = \sqrt{\frac{1}{6}}\left[ \vert\pi^{+}\rangle\vert\pi^{0}\rangle\vert\pi^{-}\rangle -
                                        \vert\pi^{0}\rangle\vert\pi^{+}\rangle\vert\pi^{-}\rangle -
                                        \vert\pi^{+}\rangle\vert\pi^{-}\rangle\vert\pi^{0}\rangle +\right.\\
                                        \left.\vert\pi^{-}\rangle\vert\pi^{+}\rangle\vert\pi^{0}\rangle -
                                        \vert\pi^{-}\rangle\vert\pi^{0}\rangle\vert\pi^{+}\rangle + 
                                        \vert\pi^{0}\rangle\vert\pi^{-}\rangle\vert\pi^{+}\rangle \right].
\end{eqnarray}
This wave function given above is antisymmetric under any exchange of pions: 
$\pi^{0}\leftrightarrow\pi^{+}$, $\pi^{-}\leftrightarrow\pi^{+}$ and $\pi^{0}\leftrightarrow\pi^{-}$. 
In particular, by applying charge conjugation (equivalent to $\pi^+\leftrightarrow\pi^-$ exchange) one gets:
\begin{equation}
C(3\pi)_{I=0} = - (3\pi)_{I=0}.
\end{equation}
\begin{figure}[t]
\hspace{3.0cm}
\parbox{0.6\textwidth}{\centerline{\epsfig{file=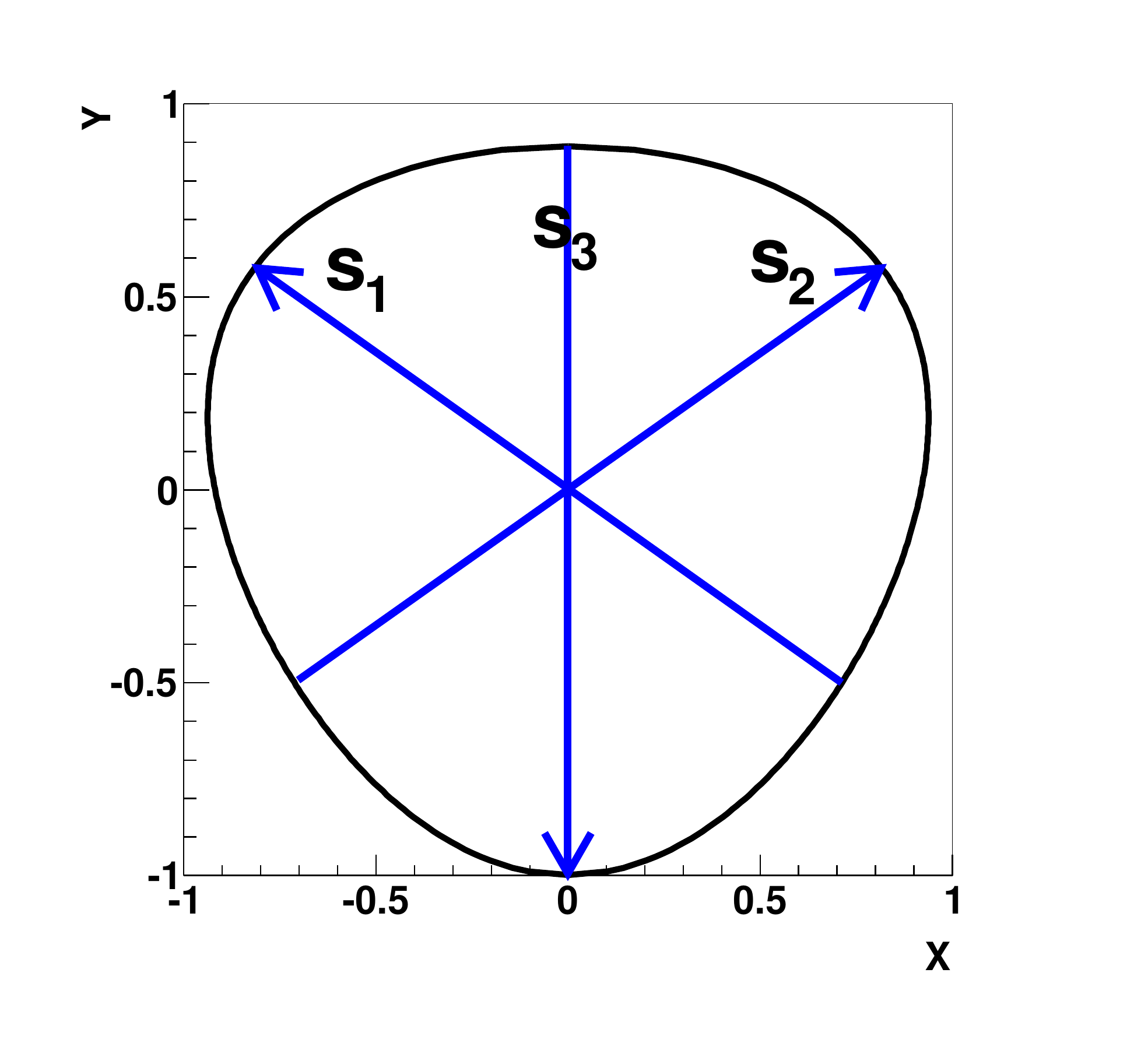,width=0.65\textwidth}}}
\caption{
Kinematical boundaries of the Dalitz plot for the $\eta\to\pi^+\pi^-\pi^0$ decay. 
}
\label{DalitzGranice}
\end{figure}
This is in contradiction with $\lambda_C$ = +1 of the $\eta$ meson. Therefore, the decay
$\eta\to\pi^{+}\pi^{-}\pi^{0}$ in the isospin state $I=0$ violates C symmetry.
One can show that in this case also G symmetry is broken. The G operator is given by:
\begin{equation}
G = C e^{i\pi I_2},
\end{equation}
where $C$ stands for charge parity, and $I_2$ denote the operator of the second component of the isospin.
The eigenvalue of this operator is given by $\lambda_G = (-1)^I\lambda_C$, thus $\lambda_G = -1$ for pions 
and $\lambda_G = +1$ for the $\eta$ mesons. Hence, the $\eta\to\pi^+\pi^-\pi^0$ in the $I=0$ does not 
conserve $G$ symmetry. Therefore, the decay $\eta\to\pi^+\pi^-\pi^0$ may occur if the isospin 
is conserved but then it has to violate C and G symmetry, or it may decay conserving C symmetry but then  
isospin is violated.

As it was mentioned before this decay has a strong isospin breaking part which is driven by the
term of the QCD Lagrangian proportional to the $m_{d} - m_{u}$. This isospin non-conserving interaction 
result in the final isospin state of three pions with $I=1$ and $\lambda_C=+1$. However, the interference between
conserving and not-conserving charge symmetry production amplitudes of the $\lambda_C=+1$ and 
$\lambda_C=-1$ states are also possible. This can result in asymmetry between kinetic energy 
distribution of charged pions $\pi^+$ and $\pi^-$.

\begin{table}[t]
\centering
\begin{tabular}{|c|c|c|c|c|c|}
\hline
Theory/Experiment  & $A_0$                            &   a    & b      & d      & f\\\hline\hline
 ChPT LO~\cite{Bijnens:2007pr,Bijnens:2007jk}  &  120 & -1.039 & 0.27   & 0.0    & 0.0\\
 ChPT NLO~\cite{Bijnens:2007pr,Bijnens:2007jk} &  314 & -1.371 & 0.452  & 0.053  & 0.027\\
 ChPT NNLO~\cite{Bijnens:2007pr,Bijnens:2007jk}&  538 & -1.271 & 0.394  & 0.055  & 0.025\\\hline
 Gromley~\cite{Gormley:1970qz}                 &      & -1.18$\pm$0.02  & 0.20$\pm$0.03 & 0.04$\pm$0.04 & --\\
 Layter~\cite{Layter:1973ti}                   &      & -1.080$\pm$0.014& 0.034$\pm$0.027 & 0.046$\pm$0.031  & --\\
 Amsler~\cite{Amsler:1995sy}                   &      & -0.94$\pm$0.15  & 0.11$\pm$0.27 & --  & --\\
 Abele~\cite{Abele:1998yj}                     &      & -1.22$\pm$0.07  & 0.22$\pm$0.11 & 0.06(fixed) & --\\
 Ambrosini~\cite{Ambrosino:2008ht}             &      & $-1.090_{-0.024}^{+0.013}$ & $0.124\pm 0.016$  & $0.057_{-0.022}^{+0.013}$  & $0.14\pm0.03$\\\hline 
\end{tabular}
\caption{
Dalitz plot parameters obtained from the theoretical predictions of the ChPT (first three rows), and 
the same parameters obtained from various experimental measurements of the $\eta\to\pi^+\pi^-\pi^0$ decay.
}
\label{tab:parametry}
\end{table}
A convenient way to study the $\eta\to\pi^+\pi^-\pi^0$ decay in view of the possible 
C violation is to use the Dalitz plot.  
For that purpose one can use the Mandelstam variables defined as:
\begin{equation}
s_i = (p_{\eta} - p_{i})^2 = (m_\eta - m_i)^2 - 2\cdot m_\eta T_i,
\end{equation}
where $p_i$ and $m_i$ denote the four-momentum vectors and masses of final state particles, and $T_i$ stands
for the kinetic energy in the rest frame of the $\eta$ meson. However, in case of the $\pi^{+}\pi^{-}\pi^{0}$ final state where 
$m_{\pi^{+}} = m_{\pi^{-}}$, one can use the symmetrized and dimensionless variables 
defined as:
\begin{equation}
X = \sqrt{3}\left(\frac{T_{+} - T_{-}}{Q}\right), 
\label{dX}
\end{equation}
\begin{equation}
Y = \frac{3T_{0}}{Q} - 1, 
\label{dY}
\end{equation}
where $Q = T_{+} + T_{-} + T_{0} =  m_{\eta} - 2m_{\pi^{\pm}} - m_{\pi^0}$ is the excess energy.
The kinematical boundary of the Dalitz plot in the X,Y plane for the $\eta\to\pi^+\pi^-\pi^0$ decay 
is shown in Fig.~\ref{DalitzGranice}.

The distribution inside the boundaries is symmetric and flat when the transition matrix element is constant. 
In general the density distribution is given by the matrix element squared which can be described 
by expanding the amplitude in the powers of X and Y:
\begin{equation}
\vert M \vert^2 = A^2_0 (1 + aY + bY^2 + cX + dX^2 + fY^3 + ...),
\label{amplituda}
\end{equation}
where $a, b, c, d, f, ...$ are the parameters which can be obtained phenomenologically or on the ground of theory, and $A_0$ stands for the normalization factor. By extracting the parameters from the experimental data 
and comparing them to the theoretical predictions one can test the assumptions of the Chiral 
Perturbation Theory. Furthermore, $c$ coefficient, and other parameters standing in the odd-powers 
of X are sensitive to charge conjugation violation. The values of the parameters obtained from 
previous measurements and from calculations calculated based on the ChPT are collected in Tab.~\ref{tab:parametry}. In all previous experiments $c$ parameter was found to be consistent with zero, but was not explicitly given in the publications due to large errors. Furthermore, only KLOE~\cite{Ambrosino:2008ht} 
experiment collected enough statistics to establish value of the f parameter.  

\begin{figure}[t]
\hspace{-0.2cm}
\parbox{0.31\textwidth}{\centerline{\epsfig{file=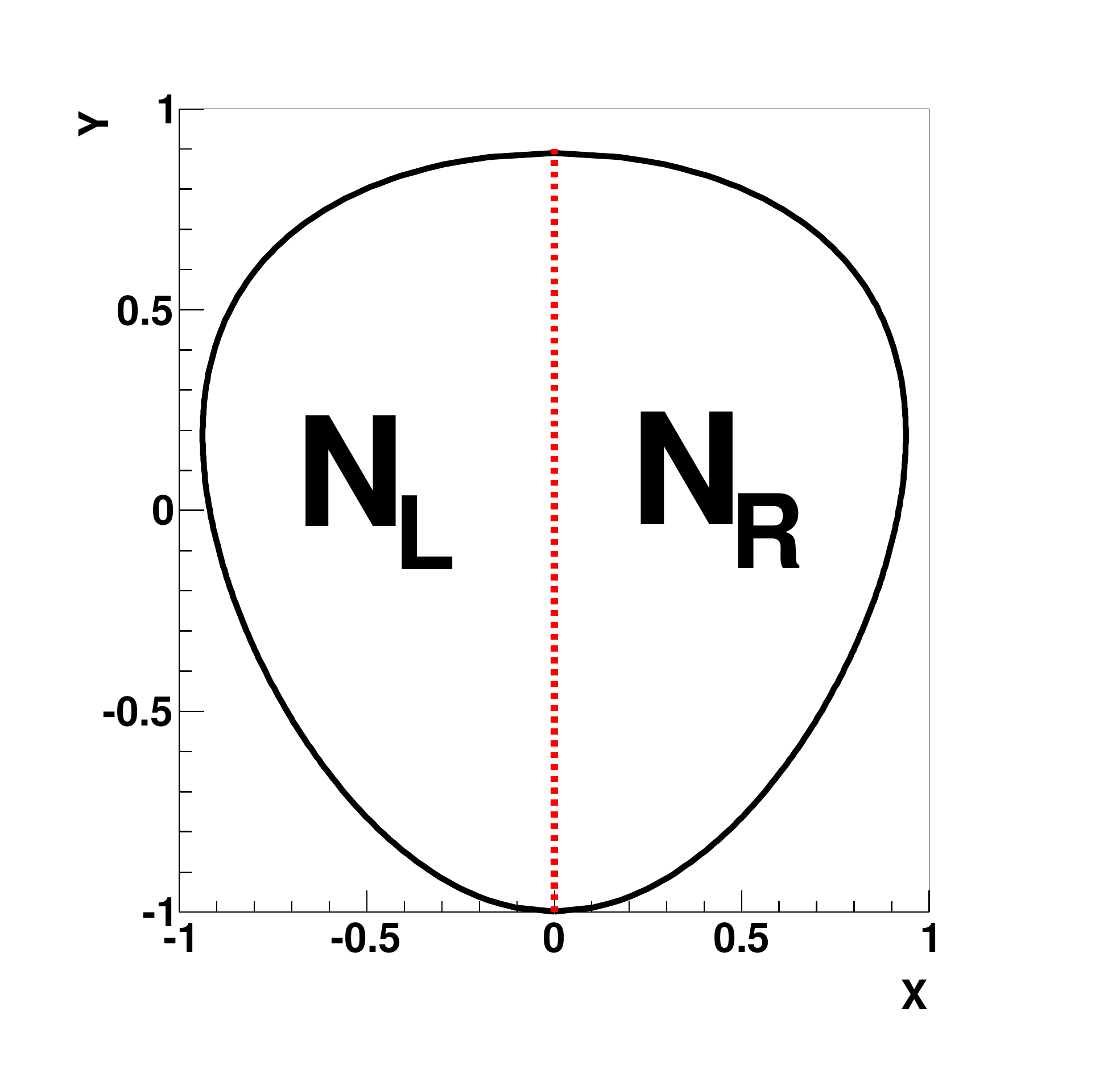,width=0.37\textwidth}}}
\hspace{0.4cm}
\parbox{0.31\textwidth}{\centerline{\epsfig{file=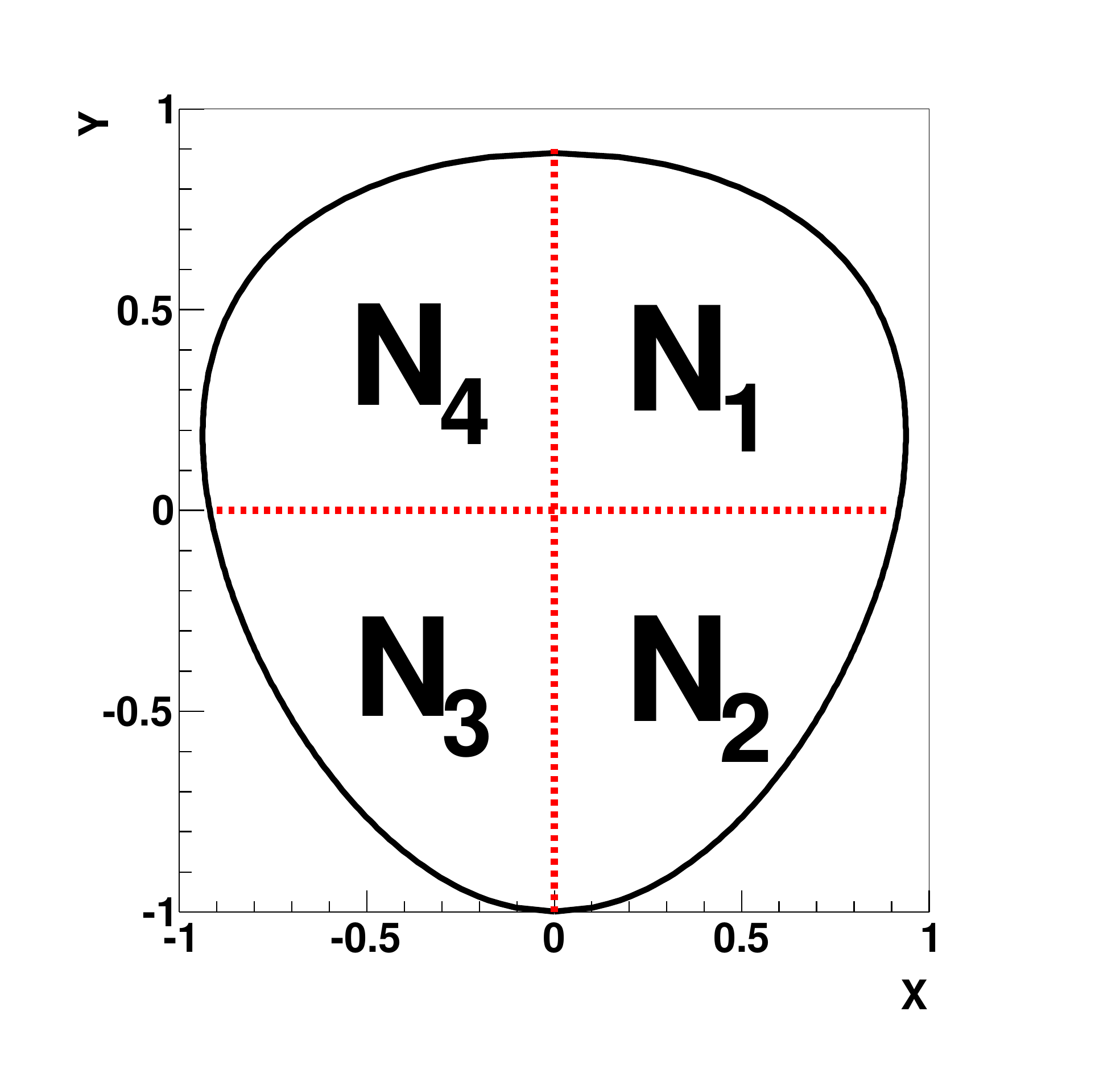,width=0.37\textwidth}}}
\hspace{0.0cm}
\parbox{0.31\textwidth}{\centerline{\epsfig{file=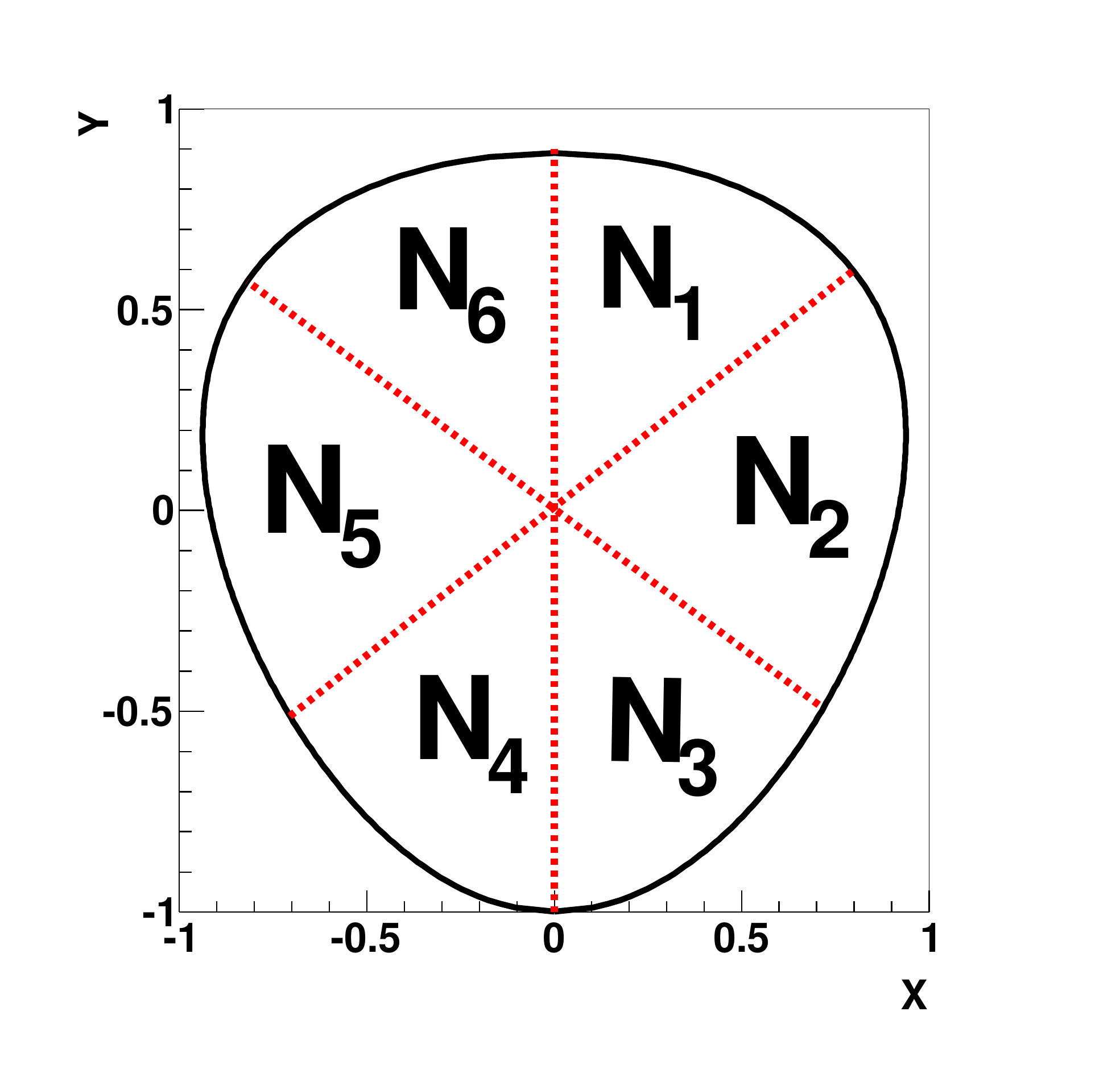,width=0.37\textwidth}}}
\caption{
The partition of the Dalitz plot into sectors in order to define the asymmetries:
{\bf{(left)}} left-right ($A_{LR}$), 
{\bf{(middle)}} quadrant ($A_{Q}$),
{\bf{(right)}} sextant ($A_{S}$).
}
\label{DXY-asymetrie}
\end{figure}

The amplitude mixing between $\lambda_C=-1$ and $\lambda_C=+1$, describing the transition into isospin 
state $I=1$ 
and $I=0,2$, respectively, can be investigated by studying of the symmetries of population in different 
parts of the 
Dalitz plot. In particular the possible presence of C violation could be observed in three parameters:
(i) left-right asymmetry -- $A_{LR}$, (ii) quadrant asymmetry -- $A_{Q}$, 
and (iii) sextant asymmetry -- $A_{S}$. Each of these parameters depends on 
different isospin states of the final three pions. The asymmetries are defined as number of events 
observed in different sectors of the Dalitz plot divided as it is shown in Fig.~\ref{DXY-asymetrie}.

The left-right asymmetry is defined as: 
\begin{equation}
A_{LR} = \frac{N_{R} - N_{L}}{N_{R} + N_{L}},
\label{ALR}
\end{equation}
where the $N_{L}$ stands for the number of events where $\pi^-$ has a larger energy than 
$\pi^+$ and and $N_{R}$ denotes the number of events where the $\pi^+$ has greater energy than $\pi^-$.
It is sensitive to C violation averaged over all isospin states. However, it is possible to test the 
charge conjugation invariance in given $I$ state. For this one uses the quadrant and sextant 
asymmetries which are defined as:
\begin{equation}
A_{Q} = \frac{N_{1} + N_{3} - N_{2} - N_{4}}{N_{1} + N_{2} + N_{3} + N_{4}},
\label{AQ}
\end{equation}
\begin{equation}
A_{S} = \frac{N_{1} + N_{3} + N_{5} - N_{2} - N_{4} - N_{6}}{N_{1} + N_{2} + N_{3} + N_{4} + N_{5} + N_{6}},
\label{AS}
\end{equation}
where $N_i$ denotes the number of observed events in $i$-th sector of the Dalitz plot. 
The quadrant asymmetry tests the C invariance in transition into the $3\pi$ final state 
with $I=2$, and the sextant asymmetry is sensitive to the $I=1$~\cite{Jarlskog:2002zz}.  
Table~\ref{tab:asymetrie} summarizes all experimentally measured values of three asymmetries. 
The previously measured values of the asymmetry parameters indicate no C violation with respect to calculated uncertainties. 
\begin{table}[!h]
\centering
\begin{tabular}{|c|c|c|c|}
\hline
 Experiment        & $A_{LR}\times 10^{-2}$   &   $A_{Q}\times 10^{-2}$  & $A_{S}\times 10^{-2}$\\\hline\hline
 Layter~\cite{Layter:1973ti}         & $-0.05 \pm 0.22$ & $-0.07 \pm 0.22$ & $ 0.10 \pm 0.22$\\
 Jane~\cite{Jane:1974mk}             & $ 0.28 \pm 0.26$ & $-0.30 \pm 0.25$ & $ 0.20 \pm 0.25$\\
 Ambrosino~\cite{Ambrosino:2008ht}   & $ 0.09 \pm 0.10$ & $-0.05 \pm 0.10$ & $ 0.08 \pm 0.10$\\
 PDG average~\cite{Nakamura:2010zzi} & $-0.09_{-0.12}^{+0.11}$ & $-0.09 \pm 0.09$ & $0.12_{-0.11}^{+0.10}$\\\hline
\end{tabular}
\caption{
The values of Dalitz plot asymmetries obtained experimentally together with the average value from the PDG
for the $\eta\to\pi^+\pi^-\pi^0$ decay.
}
\label{tab:asymetrie}
\end{table}

\section{Decay of the $\eta$ meson into $\pi^0 e^+e^-$}
\begin{figure}[!h]
\hspace{3.3cm}
\parbox{0.40\textwidth}{\centerline{\epsfig{file=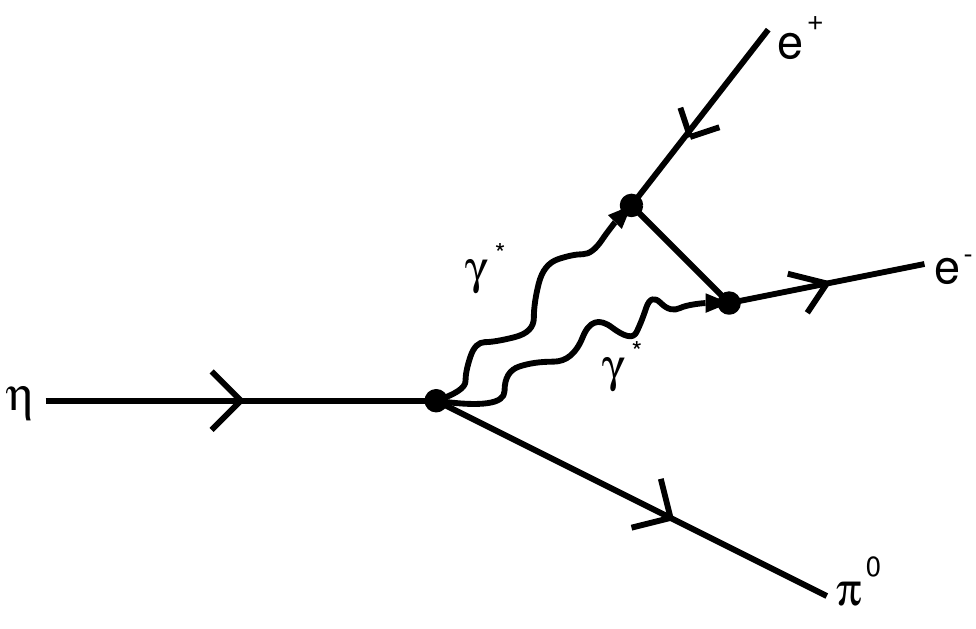,width=0.45\textwidth}}}
\caption{
Decay mode $\eta\to\pi^0\gamma^*\gamma^*\to\pi^0 e^+ e^-$ occurring by the C-conserving 
second order electromagnetic process.
}
\label{grafyETAEEPI-1}
\end{figure}

\hspace{\parindent}
The investigation of the charge conjugation invariance C in the electromagnetic  
interactions can be done by studying the $\eta\to\pi^0 e^+e^-$ decay. 
In the framework of the Standard Model and the QED the matrix element for this process 
should involve the two virtual photon exchange~\cite{Smith:1968ab} as it is presented in 
Fig.~\ref{grafyETAEEPI-1} with the transition according to the reaction:
\begin{equation}
\eta \to \pi^0 + \gamma^* + \gamma^* \to \pi^0 + e^+ + e^- .
\end{equation}
Therefore, the wave function of the $\pi^0 \gamma^* \gamma^*$ system transforms with the C operator
as follows:
\begin{equation}
C(\pi^0\gamma^*\gamma^*) = \lambda^{\pi^0}_C\lambda^{\gamma^*}_C\lambda^{\gamma^*}_C(\pi^0\gamma^*\gamma^*)
 = (+1)\cdot(-1)\cdot(-1)(\pi^0\gamma^*\gamma^*) = +1(\pi^0\gamma^*\gamma^*).
\end{equation}
The eigenvalue of the charge parity for the $\eta$ meson is $\lambda_C = +1$ which is in agreement with above 
shown charge parity of the decay system, thus this process hold the C invariance. 
The decay rate of this C-conserving process, predicted theoretically ranges from $10^{-11}$ to $10^{-8}$ depending on the undertaken assumptions:
\begin{equation}
BR(\eta  \to \pi^0 e^+ e^-  ) \approx ( 1.5 \pm 0.4 )\cdot 10^{-11}~\text{\cite{Llewellyn:1967tt}},
\end{equation}
\begin{equation}
BR(\eta  \to \pi^0 e^+ e^-  ) \approx 1.1 \cdot 10^{-8}~\text{\cite{Cheng:1967zza}},
\end{equation} 
\begin{equation}
BR(\eta  \to \pi^0 e^+ e^-  )  \approx (1 - 6) \cdot 10^{-9}~\text{\cite{Ng:1993sc}}.
\end{equation} 
It is worth to mention that the second and third predictions are based on the approach 
of the Vector Dominance Model (VMD)~\cite{Sakurai:1960ju}. In the framework of this model it is assumed
that the decay is dominated by the virtual transition $\eta\to V\gamma^*$ followed by the $V\to\pi^0\gamma^*$
and $\gamma^*\gamma^*\to e^+e^-$, where the $V$ denotes all the neutral vector mesons of zero 
strangeness: $\omega,\rho,\phi$. 

\begin{figure}[t]
\hspace{3.3cm}
\parbox{0.40\textwidth}{\centerline{\epsfig{file=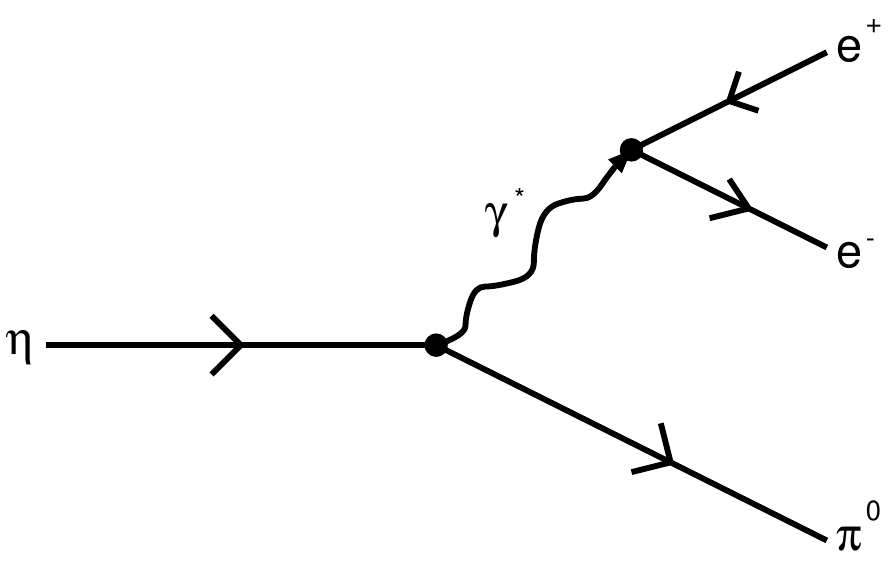,width=0.45\textwidth}}}
\caption{
Diagram for the C invariance violating transition $\eta\to\pi^0\gamma^*\to\pi^0 e^+ e^-$
 occurring by the first order electromagnetic process. 
}
\label{grafyETAEEPI-2}
\end{figure}

However, in principle the decay $\eta\to\pi^0 e^+e^-$ may also be realized in first order of the 
electromagnetic interaction with only single $\gamma$ quantum in the intermediate state (see Fig.~\ref{grafyETAEEPI-2}) via transition:
\begin{equation}
\eta \to \pi^0 + \gamma^* \to \pi^0 + e^+ + e^- .
\end{equation}
In this case the C operator acting on the wave function of the intermediate state leads to:
\begin{equation}
C(\pi^0\gamma^*) = \lambda^{\pi^0}_C\lambda^{\gamma^*}_C(\pi^0\gamma^*)
 = (+1)\cdot(-1)(\pi^0\gamma^*) = -1(\pi^0\gamma^*),
\end{equation}
which is in contradiction with the charge parity of the $\eta$ meson ($\lambda^{\eta}_C=+1$). Thus, this process 
introduces the violation of charge conjugation. The decay width for the first order electromagnetic processes are larger than for second order mechanism. Therefore, experimentally the C-invariance breaking 
would then  manifest itself with increasing the branching ratio with respect to the predictions listed in equations 2.16, 2.17 and 2.18.
At present only an experimental upper limit for the rate of the branching ratio $BR(\eta\to\pi^0 e^+e^-)$
was determined~\cite{Jane:1975nt,Bernstein:1965hj,Billing:1967aa,Bazin:1968zz,Price:1965zza,Rittenberg:1965zz}, and it amounts to $4\cdot 10^{-5}$~\cite{Nakamura:2010zzi}. Therefore, still at least 
three orders of magnitude remains to be experimentally investigated until value predicted based on the 
Standard Model will be reached. The observation of higher branching ratio than one calculated in the framework of the Standard Model could provide the evidence that the decay $\eta\to\pi^0 e^+e^-$ is not conserving 
C-invariance.

One has to stress that C-violating first order electromagnetic process is  the most 
probable possibility, but not the only one. Another conceivable process which may lead to the increase of 
the branching ratio could be e.g. $\eta\to\chi^0\pi^0\to\gamma^*\gamma^*\pi^0$, where $\chi^0$ is an 
unknown particle not included in the Standard Model~\cite{Kupsc:priv2010}.

The aim of this thesis is to contribute in searches of the C violation by determining the 
$BR(\eta\to\pi^0 e^+e^-)$ or lowering the present upper limit for this branching ratio. 

\chapter{Experimental methods}
\hspace{\parindent}
The measurement described in this thesis was carried out in October and November 2008 by means of the 
Wide Angle Shower Apparatus (WASA)~\cite{Adam:2004ch} and  
the Cooler Synchrotron (COSY)~\cite{Maier:1997zj} operating at the Research Center J\"{u}lich, Germany.
The $\eta$ meson was created in proton-proton collisions via $pp\to pp\eta$ reaction with 
the beam momentum of $p_{beam}$~=~2.142~GeV/c which corresponds to the kinetic beam energy of 
$T_{beam}$~=~1.4~GeV. The experiment was based on measurement of four-momentum vectors of outgoing 
nucleons and of decay products of unregistered short lived $\eta$ meson which was 
identified using the missing and invariant mass techniques.

\section{Cooler Synchrotron COSY}
\begin{figure}[H]
\hspace{.1cm}
\vspace{2.0cm}
\parbox{0.99\textwidth}{\centerline{\epsfig{file=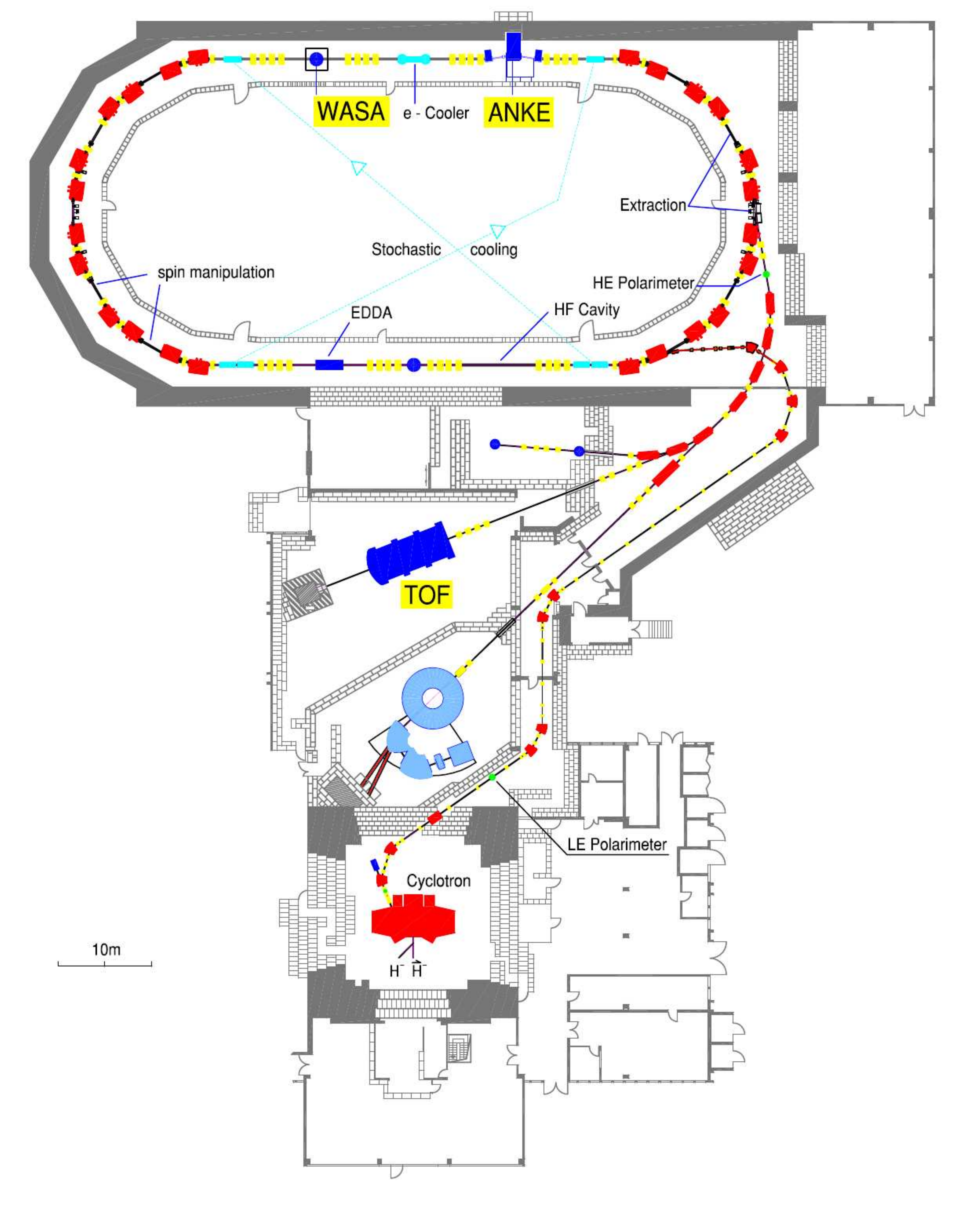,width=0.99\textwidth}}}
\caption{
Schematic view of the COSY Cooler Synchrotron~\cite{Maier:1997zj} storage ring at the Research 
Centre J\"{u}lich. Presently used detector systems: WASA~\cite{Adam:2004ch}, COSY-TOF 
(on external beam line)~\cite{Bohm:2000rx} and ANKE~\cite{Barsov:2001xj}, are marked in yellow.
}
\label{cosy2d}
\end{figure}

\hspace{\parindent}
The COoler SYnchrotron ''COSY'' is a storage ring which can deliver unpolarized and 
polarized proton and deuteron beams in momentum range between 300~MeV/c and 3700~MeV/c. 
The ring consists of 24 dipoles and 56 quadrapole magnets which are used to keep and focus particle 
trajectories during the acceleration process, and also sextupole magnets which are used to 
deflect beam what results in achieving better beam optics.  
The acceleration process takes place in two steps: (i) first the ions ($H^-$ or $D^-$)
are accelerated in the isochronous cyclotron (JULIC) and next (ii) the beam is stripped of electrons and finally
injected into the 184~m long COSY ring (see Fig.~\ref{cosy2d}),
where particles are stored and accelerated up to the demanded momentum. 
The beam is then directed into internal or external 
experimental targets. The beam energy range allows for production of all basic pseudoscalar and vector mesons 
up to mass of $\phi(1019)$ particle.

Additionally COSY accelerator is equipped with two types of beam cooling systems: (i) an electron and 
(ii) stochastic, used for low and high energies, respectively~\cite{Prasuhn:2000eu}.
Both cooling methods allow to reduce the momentum and spatial spread of the beam. 
The beam momentum spread $\frac{\Delta p}{p}$ after applying both types of cooling method can be reduced 
to around $10^{-3}$~\cite{Moskal:2000em}.

The whole process of acceleration together with the beam cooling phase takes a few seconds. 
COSY storage ring can be filled with up to $10^{11}$ particles, and the life time of the 
circulating beam varies from minutes to hours depending on the thickness of used target. 
Presently in COSY accelerator facility two internal experiments: WASA~\cite{Adam:2004ch} and ANKE~\cite{Barsov:2001xj}  
are in operation, and one external experiment: COSY-TOF~\cite{Bohm:2000rx}. 
\section{WASA-at-COSY apparatus}
\begin{figure}[H]
\hspace{.8cm}
\parbox{0.90\textwidth}{\centerline{\epsfig{file=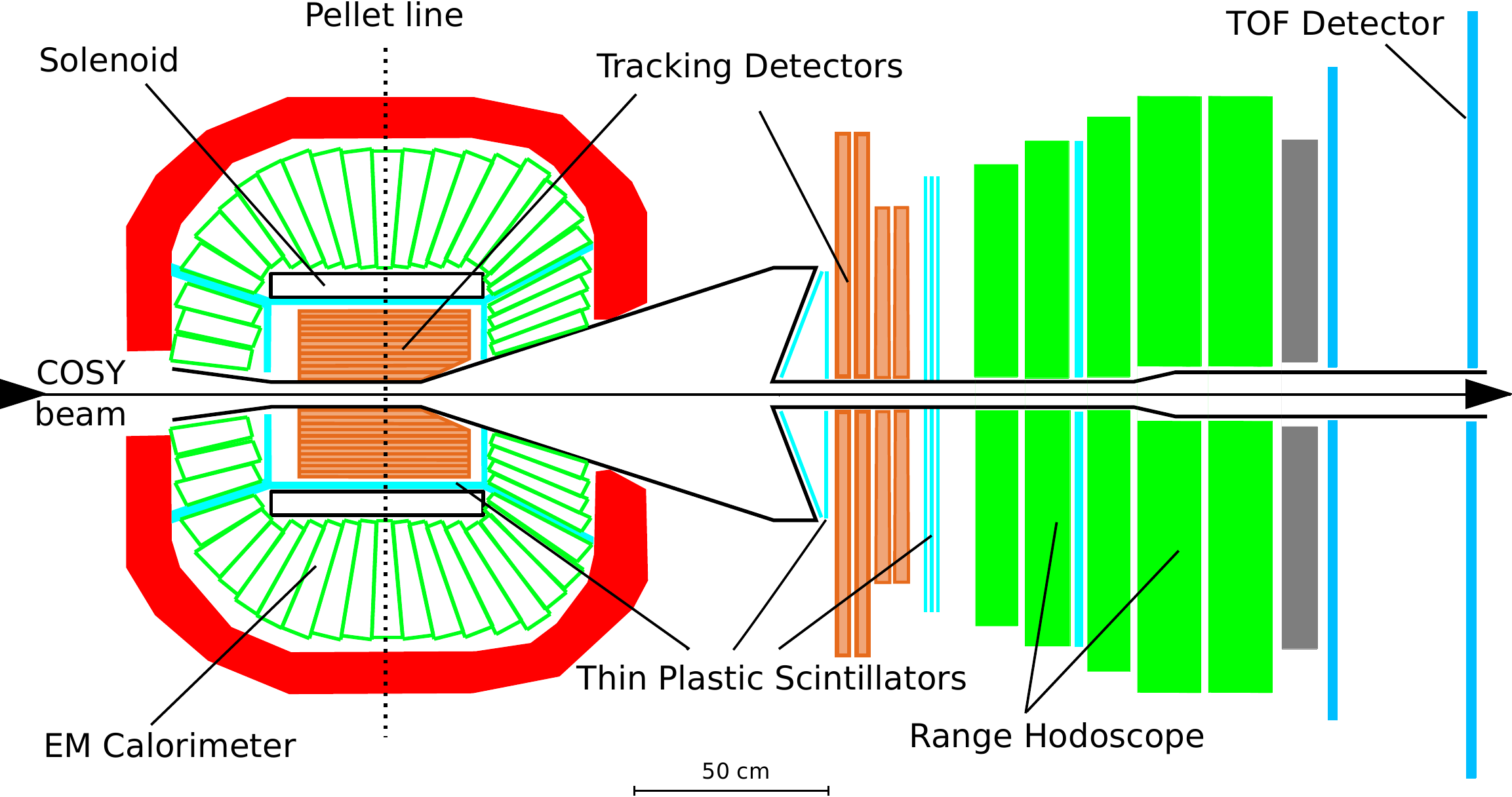,width=0.95\textwidth}}}
\caption{
Schematic cross view of the WASA-at-COSY apparatus. 
The detector components are described in the text. 
}
\label{wasa2d}
\end{figure}

\hspace{\parindent}
The Wide Angle Shower Apparatus -- WASA detector -- originally operating at the 
CELSIUS~\cite{Bargholtz:2008ze,Holm:1986jq} facility in Uppsala, 
Sweden was transferred to COSY accelerator facility in 2006~\cite{CernCourier:2005}.
The new WASA-at-COSY~\cite{Adam:2004ch} detector, shown schematically 
in Fig~\ref{wasa2d}, is a large acceptance detector consisting of three main parts: the Central Detector (CD),
the Forward Detector (FD) and the Pellet Target system. 
WASA-at-COSY detector system is capable to register neutral and charged particles emerged in the collision of beam and target and also 
particles originating from the decays of short lived mesons.
It was developed mainly for production and detection of the $\pi^0$ and $\eta$ meson decay products in order 
to study the fundamental symmetries and to test the Standard Model. 

\subsection{Central Detector (CD)}
\hspace{\parindent}
The WASA Central Detector (CD)~\cite{Schubert:1995,IKoch:2004} is positioned around the 
beam and target interaction point. It is used to detect and identify light neutral and charged particles 
like: $\gamma, e^{+}, e^{-}, \pi^{+},\pi^{-}$ which originate from decay of the short lived mesons and 
direct collision of nucleons. 
The most inner part of CD is the Mini Drift Chamber (MDC) surrounded by the Plastic Scintillator Barrel (PSB)
and the yoke of the super-conducting solenoid~\cite{ruber:1999sc} which together enables to measure 
momenta of charge particles. The outer part constitutes the Scintillating Electromagnetic Calorimeter (SEC) which is used to measure particles energy.  

\vspace{1.0cm}
{\bf{Mini Drift Chamber -- MDC}}

\begin{figure}[t!]
\hspace{.5cm}
\parbox{0.45\textwidth}{\centerline{\epsfig{file=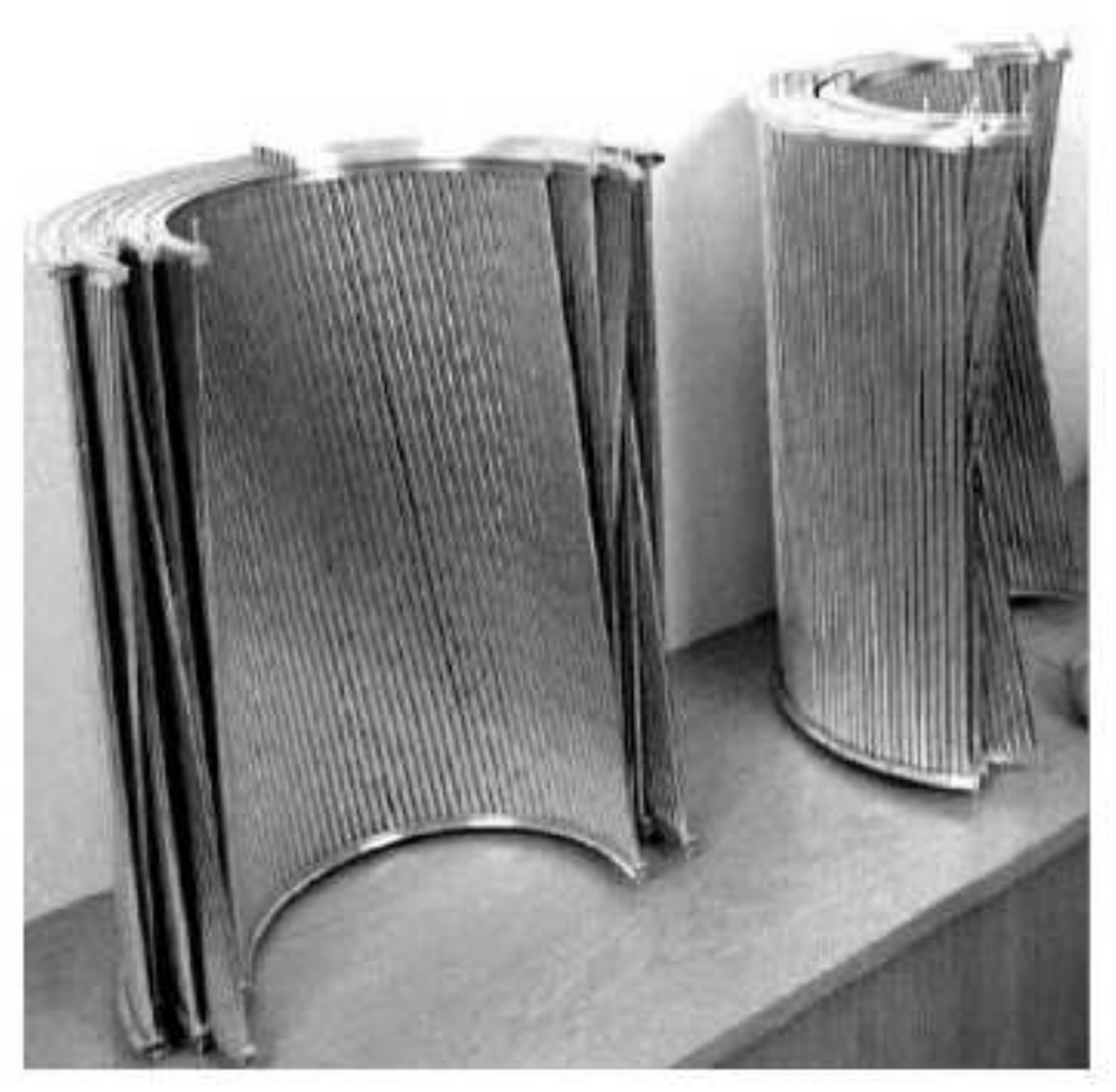,width=0.50\textwidth}}}
\hspace{.5cm}
\parbox{0.50\textwidth}{\centerline{\epsfig{file=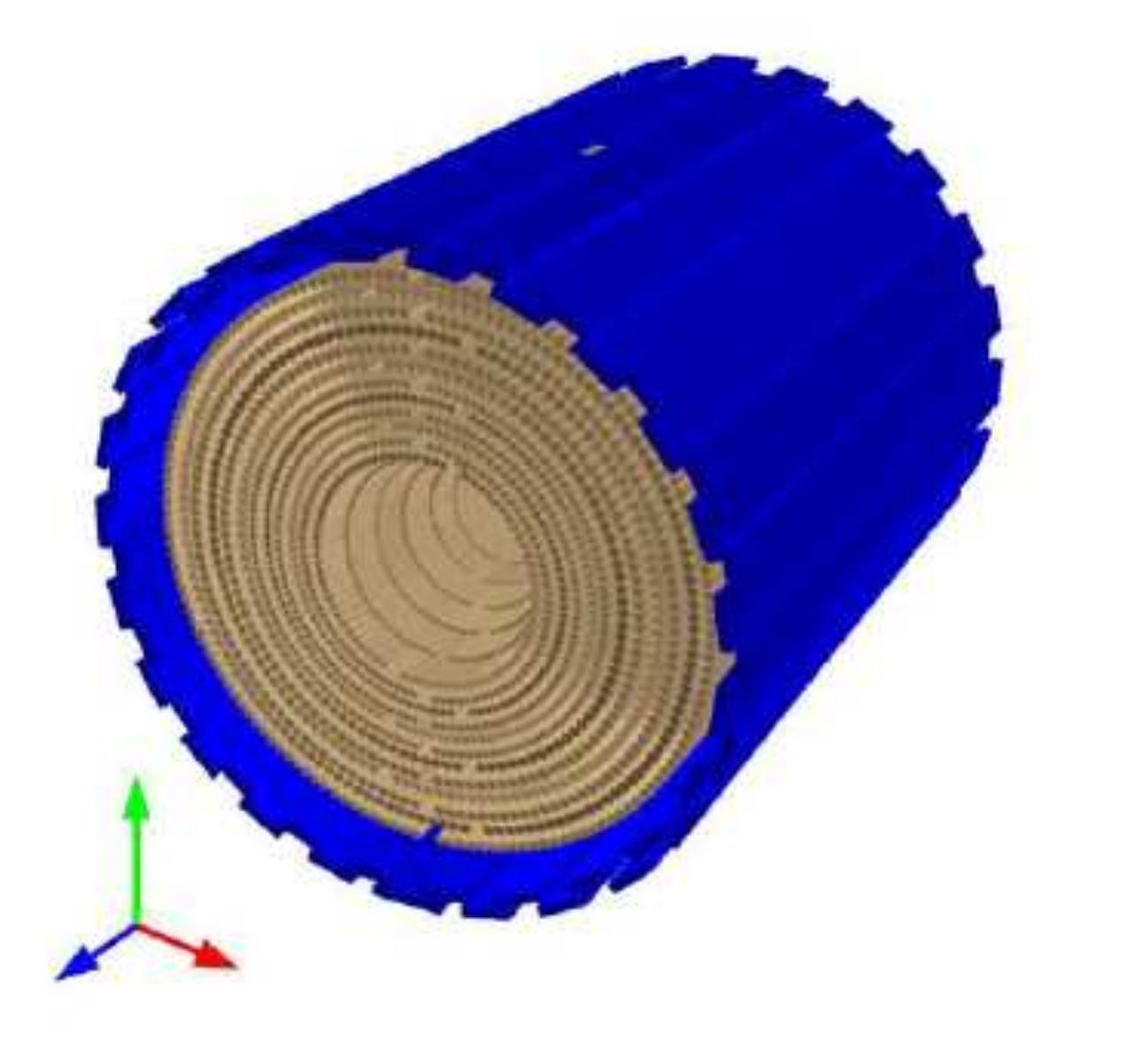,width=0.50\textwidth}}}
\caption{ 
{\bf{(left)}}~Photo of the MDC detector during the construction phase. 
{\bf{(right)}}~3D view of the MDC detector surrounded with the Plastic Scintillator Barrel detector.
}
\label{mdc}
\end{figure}

\indent
The MDC detector is mounted around the beam pipe inside the superconducting solenoid which provides 
an axial magnetic field up to B~=~1~T. It is arranged in a cylindrical shape consisting 
of 1738 straws stacked in 17 layers (see Fig~\ref{mdc}). Each straw tube is made out of thin 
($24\mu m$) mylar foil, aluminized from inside, and with a gold plated sense wire in its center 
with diameter of $20\mu m$\cite{Jacewicz:2004phd}.
The MDC has nine layers with straws parallel to the beam direction and eight with a small skew angle	
with respect to the beam line. The tubes are filled with gas mixture: 80\% of argon and 20\% of ethane 
to ensure that each charged particle which pass through a single tube will cause an ionization. 
The MDC vertex detector enables to determine the momenta of charged particles~\cite{Yurev:FZJ08} 
with accuracy of $\Delta p/p < 1\%$ for electrons and positrons, $\Delta p/p < 4\%$ for charged pions, and 
$\Delta p/p < 5\%$ for protons~\cite{Yurev:201phd}. 

\vspace{.5cm}
{\bf{Plastic Scintillator Barrel -- PSB}}

\indent
The PSB detector is used to determine the energy loss 
of charged particles and particle identification by $\Delta E - \vert\vec{p}\vert$ method.
The 52 scintillator bars of PSB detector are arranged in cylindrical 
shape~\cite{Podkopal:FZJ07,Podkopal:FZJ08} 
around the straw drift chamber where each bar is overlapping with next one with 7$^o$ to assure the 
coverage of a full geometrical acceptance (see Fig~\ref{mdc} (right)).
Additionally the forward and backward 
part is equipped with ''end-cups'' made out of trapezoidal elements arranged around the beam pipe. 

\vspace{.5cm}
{\bf{Scintillating Electromagnetic Calorimeter -- SEC}}

\begin{figure}[t!]
\hspace{.5cm}
\parbox{0.45\textwidth}{\centerline{\epsfig{file=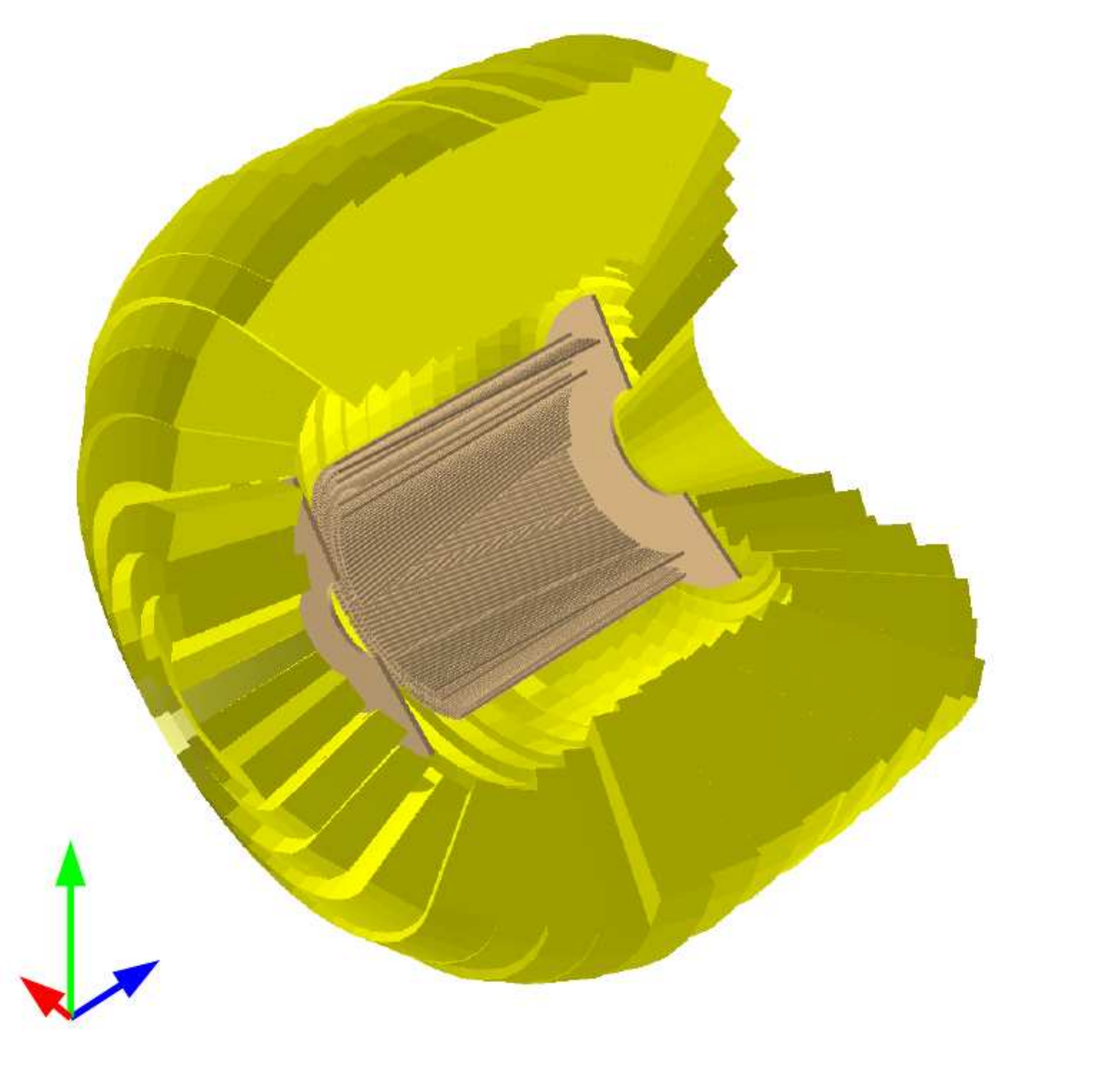,width=0.50\textwidth}}}
\parbox{0.50\textwidth}{\centerline{\epsfig{file=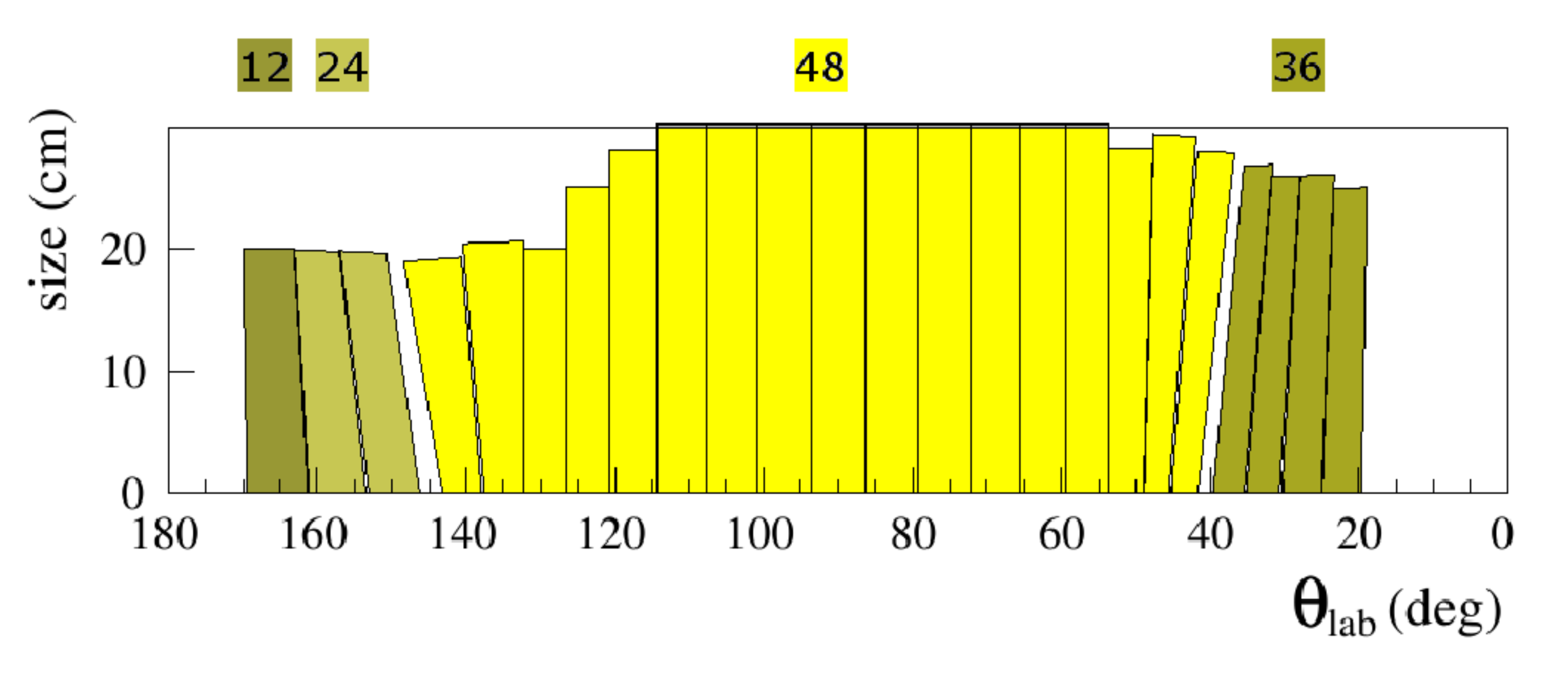,width=0.50\textwidth,angle=-90}}}
\caption{
{\bf{(left)}}~Schematic 3D view of the Scintillating Electromagnetic Calorimeter (SEC) consisting of 
              1024 scintillating modules arranged in 24 layers.
{\bf{(right)}}~Angular coverage of the Scintillating Electromagnetic Calorimeter (SEC).
}
\label{sec}
\end{figure}

\indent
The Scintillating Electromagnetic Calorimeter consists of 1012 sodium doped 
cesium iodide CsI(Na) scintillating crystals
in a shape of the pyramids in order to be arranged spherically around the interaction 
point (see Fig.~\ref{sec} (left)).
The CsI(Na) scintillating material provides a large light yield and short radiation length
making them a very good material for measuring the energy and scattering angles of neutral and 
charged particles such as gamma quanta, electrons, positrons and pions~\cite{IKoch:2004,BRJany:2006dip}. 
The crystals are grouped in 24 layers covering almost the $4\pi$ acceptance (in polar 
angle: $20^{o} \leq \theta \leq 169^{o}$, and azimuthal angle: $0^{o} \leq \phi \leq 360^{o}$).
Depending on the layer the length of the crystals varies from 30~cm in central part, to 20~cm in backward and
25~cm in forward part (see Fig.~\ref{sec} (right)).
The overall energy resolution of the SEC for photons can be described by the relation: 
$\frac{\sigma(E)}{E} = \frac{5\%}{\sqrt{E}}$ and the angular resolution for scattering angle 
is equal to $5^{o}$~(FWHM)~\cite{IKoch:2004}.
The "punch-through" kinetic energy for pions is 190~MeV and for protons 400~MeV. The electrons, 
positrons and photons are stopped in the calorimeter depositing all its energy.

\subsection{Forward Detector (FD)}
\hspace{\parindent}
The Forward Detector (FD) of the WASA apparatus consists of fourteen scintillating layers 
and a proportional straw drift chamber, and its geometrical acceptance covers in laboratory frame 
a range in polar angle from 3$^o$ to 18$^o$. 
Such a setup enables to measure the energy loss and trajectories
of recoil particles, mainly protons, deuterons and $^3$He nuclei. The  
particle identification in FD is based on the $\Delta E - E$ method
which enables to reconstruct proton energy with overall resolution of about 10\%. 
To further improve energy resolution and particle identification a new reconstruction technique based 
on the Time-Of-Flight measurement is under 
development~\cite{Moskal:WACnote2006,Zielinski:FZJ07,Zielinski:FZJ11,Zielinski:FZJ12,Zielinski:2008gi}. 
Another possible upgrade which is in progress is a Cherenkov DIRC detector~\cite{Vlasov:2007mv,Adolph:FZJ09,Fohl:FZJ09}.

\vspace{.5cm}
{\bf{Forward Window Counter -- FWC}}

\begin{figure}[t!]
\hspace{.1cm}
\parbox{0.35\textwidth}{\centerline{\epsfig{file=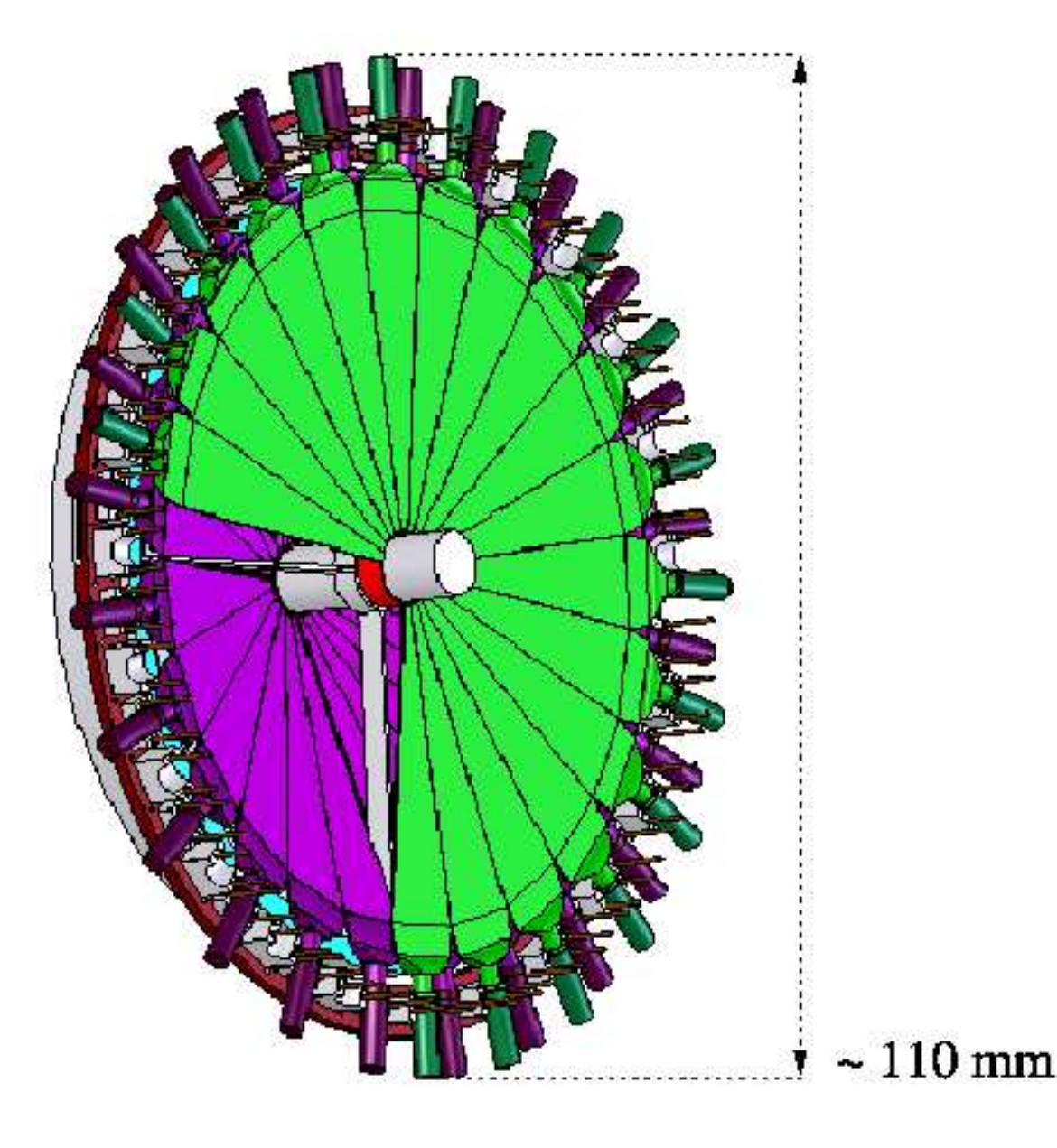,width=0.40\textwidth}}}
\hspace{.6cm}
\parbox{0.55\textwidth}{\centerline{\epsfig{file=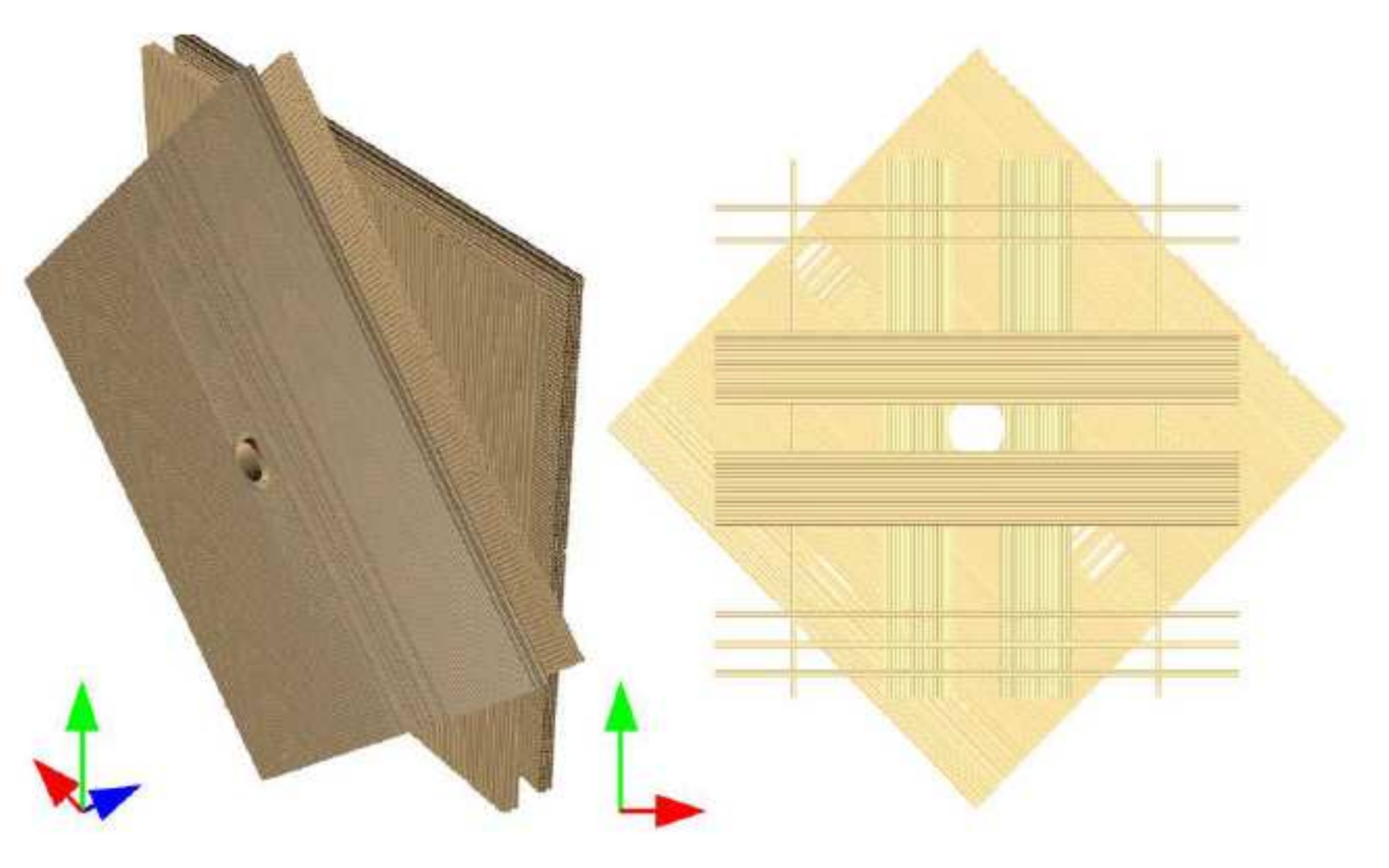,width=0.60\textwidth}}}
\caption{
{\bf{(left)}}~Forward Window Counter (FWC) build out of two layers each 
              consisting of 24 cake-piece modules.  
{\bf{(middle and right)}}~Forward Proportional Chamber (FPC) build out of 4~layers of straw tubes.    
}
\label{fwc_fpc}
\end{figure}

\indent
The Forward Window Counter (FWC) is a closest detector to the conical 
exit window of the axially symmetric scattering chamber. 
The detector is 48-fold segmented and it is composed of two layers \'{a} 24 
cake-pieced elements made out of 3~mm thick plastic scintillator~\cite{Pricking:2007FZJ,Pricking:2010phd}. 
The first layer is arranged in a conical shape whereas the elements of the second
layer are assembled in a vertical plane (see Fig.~\ref{fwc_fpc} (left)). 
The elements of the second plane are rotated by one 
half of module -- 7.5$^o$ -- with respect to the first layer. Such an arrangement 
ensures a complete coverage of the forward scattering area. The light collection 
in the hodoscope is optimized to keep the detection efficiency as homogeneous as 
possible over the full detector. 
It is worth to mention that this detector will serve as a ''start'' detector 
for the Time-of-Flight method.

\vspace{.5cm}
{\bf{Forward Proportional Chamber -- FPC}}

\indent
The Forward Proportional Chamber (FPC) is a detector located directly after the FWC and it is used 
for trajectory reconstruction purpose. It can measure particles scattering angles with a 
precision better than 0.2$^o$~\cite{Janusz:FZJ06,Janusz:FZJ07}. It consists of 1952 thin straws stacked 
by 122 in sixteen layers and grouped in four detection modules. Each module is turned with respect to 
each other by 45$^o$~\cite{Dyring:1997phd}. With respect to the x-axis the 
angles position of subsequent layers are: 315, 45, 0 and 90 degrees (see Fig.~\ref{fwc_fpc} (right)).
Each straw has a diameter of 8~mm and it is made out of mylar foil aluminized from inside with 
a 20~$\mu m$ stainless steel sense wire in its center. All straws are filled with a gas mixture: 
80~\% of argon and 20~\% of ethane to ensure an efficient ionization.  
 
\vspace{1.cm}
{\bf{Forward Trigger Hodoscope -- FTH}}

\begin{figure}[t!] 
\hspace{.5cm}
\parbox{0.90\textwidth}{\centerline{\epsfig{file=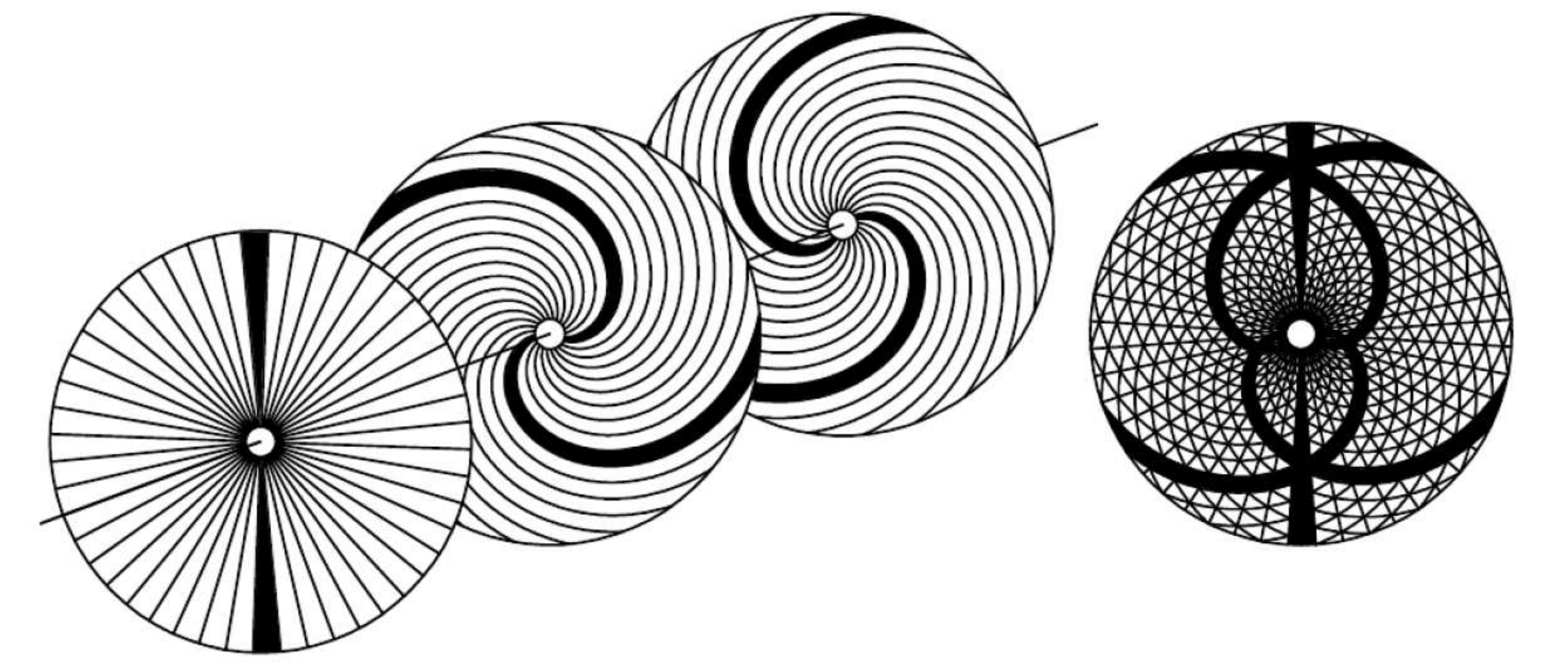,width=0.95\textwidth}}}
\caption{
Forward Trigger Hodoscope (FTH) detector arranged in three layers: one with cake-piece shaped modules,
and two with modules in form of Archimedean spirals. 
}
\label{fth}
\end{figure}

\indent
The Forward Trigger Hodoscope (FTH) consists out of 96 individual plastic scintillator modules arranged in 
three layers: (i) two layers with a 24 elements each, in a form of Archimedean spiral rotated 
clockwise and counterclockwise, (ii) and one layer with 48 cake-piece 
shaped elements~\cite{Waters:1994phd,Redmer:2006dip,Goldenbaum:FZJ06}. 
Overlap of these three layers gives 24~x~24~x~24 pixel map (see Fig~\ref{fth}). 
Whole detector setup of the FTH has highly homogeneous detection efficiency and shows 
a fairly uniform behavior~\cite{Pauly:2007FZJ}.
The FTH is used as a first level trigger and gives information about 
particles multiplicities. It is also possible to use FTH to real time scattering 
angle reconstruction of individual tracks and combine it with the information about 
deposited energy in successive layers 
and to use it to determine the missing mass of forward going particles on line on the 
trigger level~\cite{Pauly:2007FZJ}.
FTH can be also used to determine the energy losses and thus can be used as $\Delta E$ 
detector for identification of recoil particles via $\Delta E - E$ method. 

\vspace{.5cm}
{\bf{Forward Range Hodoscope -- FRH}}

\begin{figure}[h!]
\hspace{.3cm}
\parbox{0.50\textwidth}{\centerline{\epsfig{file=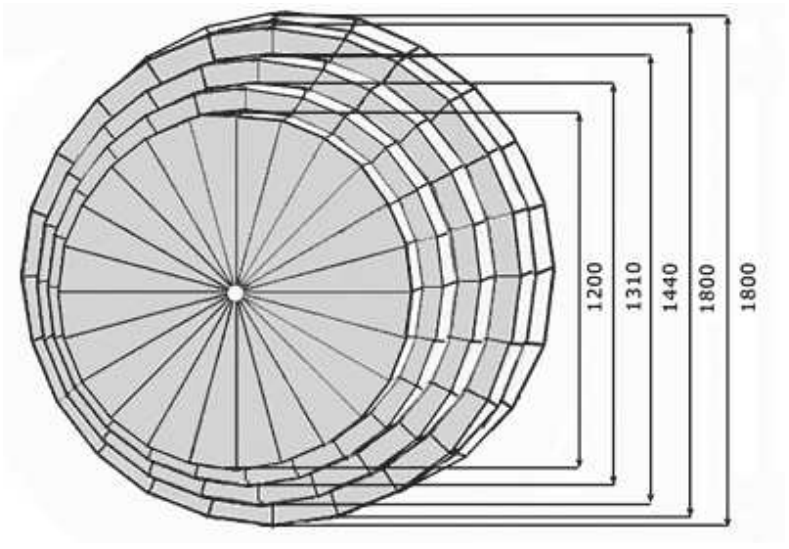,width=0.55\textwidth}}}
\hspace{.5cm}
\parbox{0.40\textwidth}{\centerline{\epsfig{file=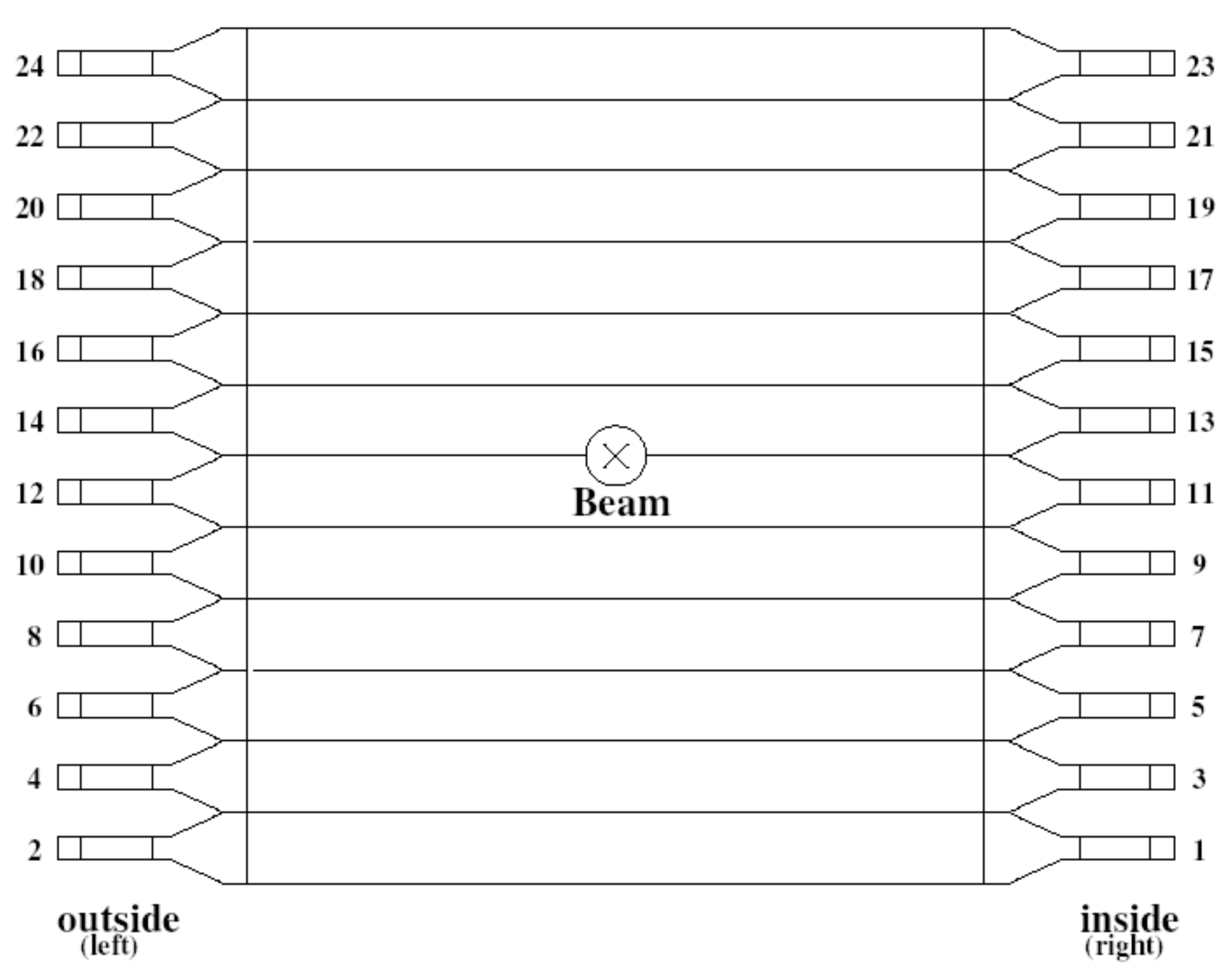,width=0.45\textwidth}}}
\caption{
{\bf{(left)}}~View of the Forward Range Hodoscope (FRH) build out of the cake-pieced scintillating modules
              arranged in five layers.
{\bf{(right)}}~Front view of the Forward Veto Hodoscope (FVH)  with 12 horizontal 
             scintillation modules readed from both sides. 
}
\label{frh_fvh}
\end{figure}

\indent
The Forward Range Hodoscope (FRH) is build out of five thick plastic scintillator layers, each cut into 
24 cake-piece elements~\cite{Calen:1996ft}. 
The first three layers have thickness of 11~cm while layers 4 and 5 have 15~cm 
(see Fig.~\ref{frh_fvh} (left))~\cite{Calen2007:FZJ}.
The FRH enables to reconstruct kinetic energy of charged particles and to 
use the $\Delta E - E$ method for particle identification. The relative kinetic energy
resolution for protons up to 360~MeV which are stopped in FRH, is almost constant and is equal to about 3~\%. 
For more energetic protons resolution worsens linearly to be about 10~\% for protons with energies about 600~MeV.
FRH provides also information for the trigger matching algorithm to verify alignment of each track 
in the azimuthal plane.  

\vspace{1.cm}
{\bf{Forward Veto Hodoscope -- FVH}}

\indent
The Forward Veto Hodoscope (FVH) consists of 12 horizontally and 12 vertically placed plastic scintillator 
bars equipped with photomultipliers on both sides~\cite{Brodowski:1995phd,Pricking:2010phd} 
(see Fig.~\ref{frh_fvh} (right)). 
This enables to reconstruct particle hit position from time signals 
registered on both sides of the module. 
The bars are arranged in two layers with the relative distance of 77~cm. 
The FVH is mainly used to detect particles punching through the 
FRH and to reject them as too energetic for the $pp \to pp\eta$ reaction. 
Depending on the measured reaction, a passive iron absorber can be placed between FRH and FVH. 
The thickness of the absorber can be chosen from 5~mm up to 100~mm. Usage of the absorber enables to 
disentangle between slow and fast recoil protons coming from meson production reaction  and elastic scattering. 

\subsection{Pellet Target system}
\begin{figure}[h!]
\hspace{0.2cm}
\parbox{0.5\textwidth}{\centerline{\epsfig{file=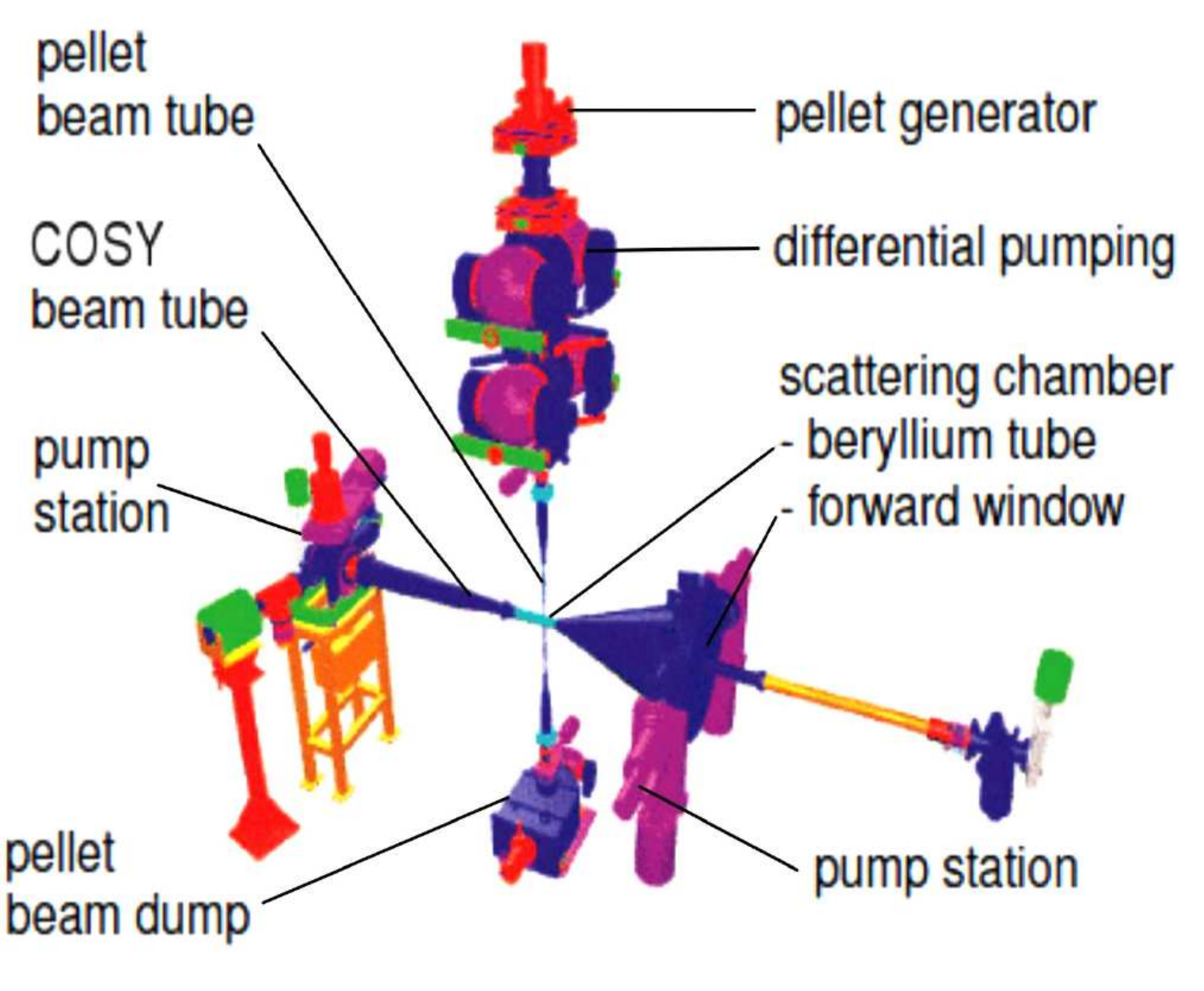,width=0.55\textwidth}}}
\hspace{0.5cm}
\parbox{0.4\textwidth}{\centerline{\epsfig{file=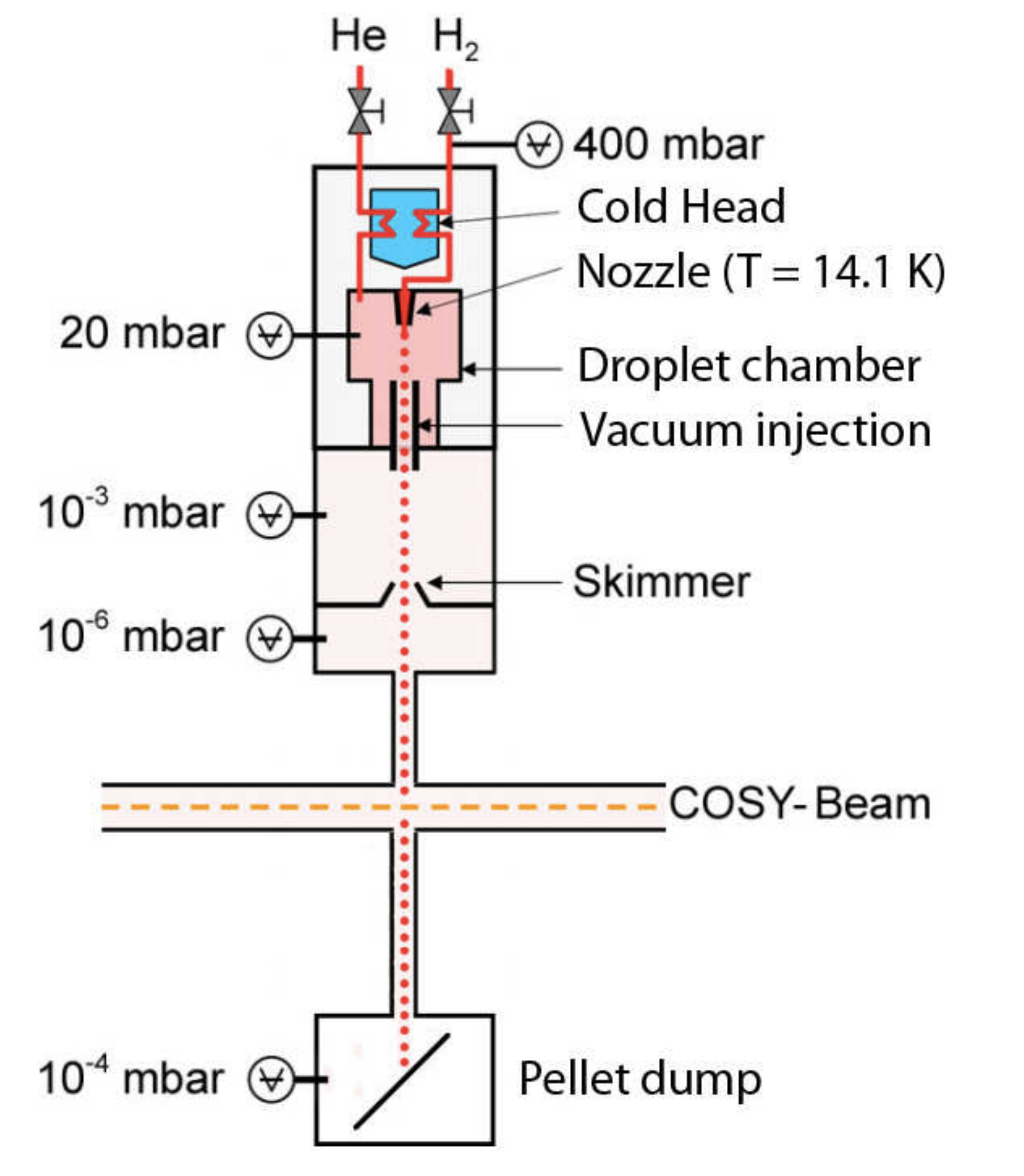,width=0.45\textwidth}}}
\caption{
View of the WASA-at-COSY Pellet Target 
System~\cite{Tros:1995nm,Ekstrom:1996po,Winnemoeller:2007FZJ,Bergmann:2008FZJ}.
}
\label{pellet}
\end{figure}

\hspace{\parindent}
The WASA-at-COSY detector is equipped with a specially designed 
target system~\cite{Tros:1995nm,Ekstrom:1996po,Winnemoeller:2007FZJ,Bergmann:2008FZJ}, providing 
high density frozen droplets of hydrogen or deuterium, called {\it{pellets}}. It is located on a platform 
above the Central Detector, delivering pellets of an average size of 35~$\mu m$, with frequency rate of about
10~kHz, to the interaction region by a thin 2~m long pipe. 
The droplets are produced by a piezoelectric transducer, which 
induce vibration of the nozzle and brakes liquid stream into pieces.
After passing through the scattering chamber, pellets are captured in a cryogenic dump.
Fig.~\ref{pellet} shows schematically the target-beam arrangement (left) and the 
scheme of the pellet generation process.  
Such a construction satisfies requirements for achieving 
high densities of particles up to $10^{15}$~atoms/cm$^{2}$ resulting in luminosities 
up to 10$^{32}$~cm$^{-2}$~s$^{-1}$ with combination of the 10$^{11}$~particles of the COSY beam. 
Thus, WASA-at-COSY is capable to 
carry out high statistics experiments, needed to study rare and very rare decays of mesons.  

\section{Data Acquisition System and trigger logic}
\begin{figure}[t!]
\hspace{1.cm}
\parbox{0.85\textwidth}{\centerline{\epsfig{file=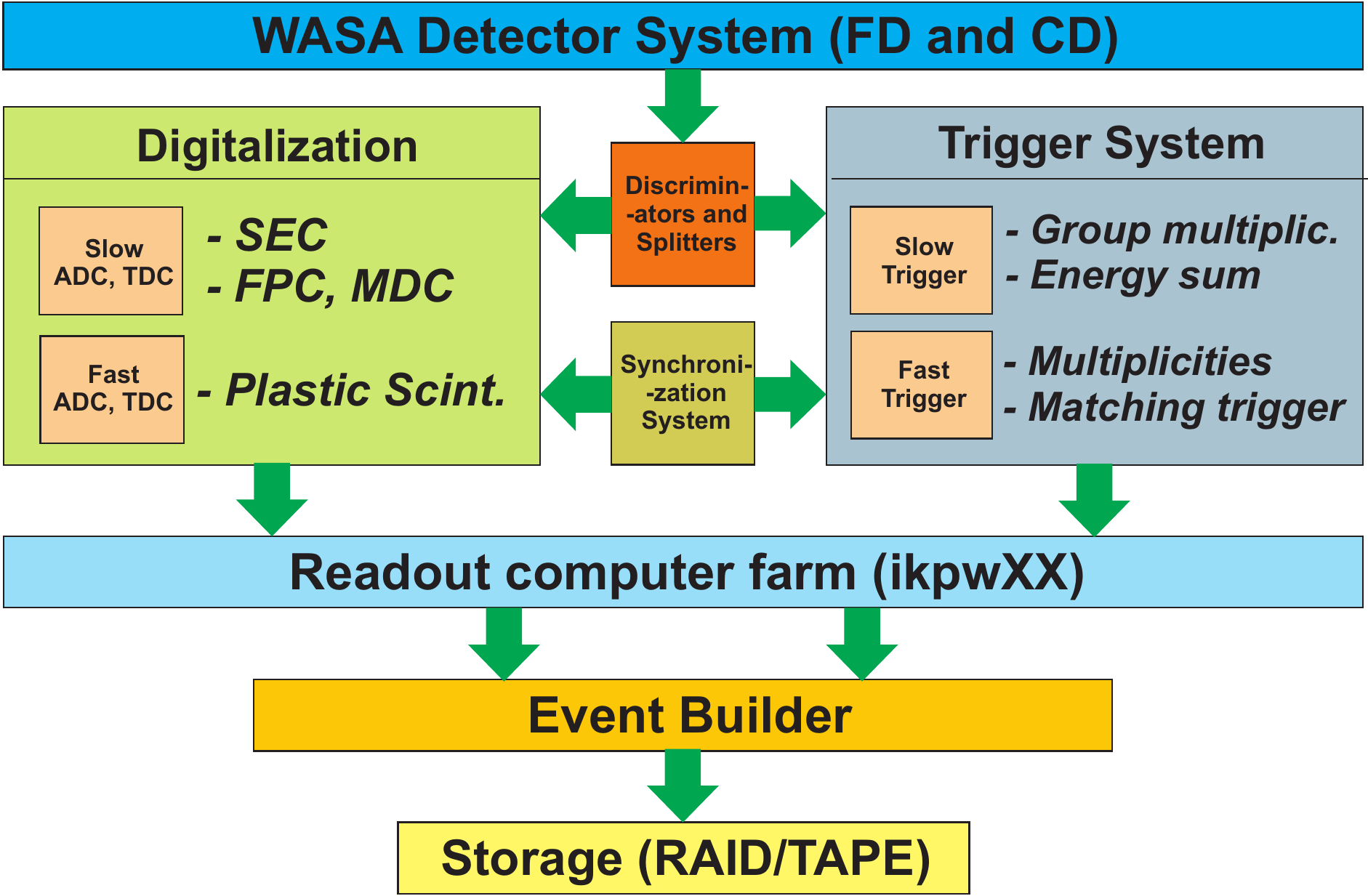,width=0.90\textwidth}}}
\caption{
Block scheme of the Data Acquisition System (DAQ) of the WASA-at-COSY experiment.
}
\label{daq}
\end{figure}

\hspace{\parindent}
In WASA-at-COSY experiment, to handle a high event rates which are typically larger than 10~kHz, an efficient 
Data Acquisition System (DAQ) was developed to digitalized and store data for further off-line 
analysis~\cite{Kleines:2006cy}. The schematic view of the system is show in Fig~\ref{daq}. 

Variety of detectors used for registering desired processes, forces to use different digitalization modules. 
The plastic scintillator elements (FWC, FTH, FRH, FVH and PS) are read out by photomultipliers, 
from where the signal is directed to the splitters which divide it (i) and directs to the leading edge discriminators 
and then to the trigger system, and (ii) to the shaper which stretches the signal to about 100~ns 
and then to the Fast-Charge-to-Digital-Converters (F-QDC) which converts 
analog signals to digits (numbers) which can be further processed by computers\cite{Hejny:2007sv}.
Also the same signal is used to digitize time information by Fast-Time-to-Digital-Converters 
(F-TDC-GPX) which has a time resolution of about 85~ps~\cite{Kleines:2006zz}.

The signals from the Electromagnetic Calorimeter crystals are splitted into two branches. One goes to readout 
system built out of Slow-QDC's (Flash-ADC chips) and then to the Field-Programmable-Gate-Array (FPGA) 
to be integrated for charge measurement. The Slow-QDC's can run in ''floating gate'' and ''fixed gate'' mode
depending on the experimental demands. 
Second signal is sent to the discriminators and used for summing different groups of signals together 
and then to be applied as a logic signals which are sent directly to the trigger system. 

The straw chambers (FPC and MDC) signals are first amplified and then sent to the discriminators and then
they are digitalized by the F1-TDC's (Slow-TDC), with the time resolution of around 120~ps~\cite{Kleines:2007zz}.

The whole system consists of 14 crates equipped with QDC's and TDC's digitalization modules, which continuously 
samples data streams and stores signals from each module in 2~$\mu s$ FIFO\footnote{An acronym for {\it{First In,
First Out}}, an abstraction related to ways of organizing and manipulation of data relative to time and
prioritization.} queue. This type of buffer allows to run the measurement without a trigger delay. 
Each one of the digitalized signal is marked with a time stamp relative to the trigger, which is 
broadcasted by a Synchronization System connected to each crate by Low-Voltage-Differential-Signaling (LVDS) bus. 
After the digitalization, signals are matched by the time stamp and marked with the same event number. 

Before event can be saved it has to be checked to fulfill the trigger logic~\cite{Fransson:2002ak}. 
The conditions of the trigger 
are based on the time and geometrical coincidences and hit multiplicities in different detection modules. 
In the Forward Detector it is also possible to apply a simple track finding algorithm on a trigger level.
The so called ''Matching Trigger'', compares hit position in consecutive layers of FWC, FTH, FRH and 
decides if they are coming from the same particle~\cite{Zheng:FZJ08}. 
This technique allows to select interesting 
events during the measurement and reduce the data rate to be later stored on the disk. 
Depending on the studied physical process several trigger conditions can be imposed simultaneously 
on the event by applying logical ''and'' operation. Also a multiple trigger conditions can be applied to 
the same data stream at the same time. 

Further on, after passing through the trigger level, events are sent to Event Builder which stores them 
in files of 20~GB size on a RAID\footnote{An acronym for {\it{Redundant Array of Independent Disks}} which is a 
storage technology that combines multiple disk drive components into a logical unit.} system. 
Each file is marked by a unique run number and after some time it is transferred from RAID to the tape 
archive for permanent storage.  

\section{Missing and invariant mass techniques}
\hspace{\parindent}
In order to evaluate physical observable like branching ratios and asymmetry parameters,
one has to extract clean signal of desired reaction from the measured data sample. 
In this thesis two complete reaction chains have to be identified:
\begin{equation}
 pp \to pp \eta \to pp \pi^+ \pi^- \pi^0 \to pp \pi^+ \pi^- \gamma \gamma,
\end{equation}
\begin{equation}
 pp \to pp \eta \to pp \pi^0 e^+ e^- \to pp e^+ e^- \gamma \gamma.
\end{equation}
In both cases reaction identification will relay on determining the four-momentum vectors of all 
particles in final state. The reconstruction of protons in FRH will be based on measurement of energy loss 
and the direction vector $\vec{r}$ in the FPC. 
The charged pions and electrons will be registered in CD, where MDC will provide the information about the
momentum vector $\vec{p}$, and energy losses will be measured in PS and SEC. 
The gamma quanta originating from the decay of neutral pion will be registered in SEC and their, 
four-momentum vectors will be reconstructed based on the energy losses and the positions of the hits. 

The procedure of extraction of signal from interesting reaction is divided in several analysis steps, where 
conditions for the minimal thresholds of energy are set and time coincidences between different tracks 
are checked. In general the signal identification will be based on the reconstruction of missing and 
invariant masses. The missing mass is defined as follows:
\begin{equation}
m_x = \sqrt{(E_x^2 - \vec{p_x}^2)} 
    = \sqrt{(E_{b} + m_{t} - E_{p_1} - E_{p_2})^2 - (\vec{p}_{b} - \vec{p_1} - \vec{p_2})^2},
\label{defMM}
\end{equation}
where the $m_x$ denotes the mass of unregistered short lived meson (in our case $\eta$ meson), 
$E_{b}$, $\vec{p}_{b}$ corresponds to the energy and momentum vector of the beam, respectively, and 
$E_{p_1}$, $E_{p_2}$, $\vec{p_1}$, $\vec{p_2}$ represent energies and momenta of recoil protons, and the $m_t$ stands for the mass of the target. 
While the invariant mass reads:
\begin{equation}
m_x = \sqrt{\left(\sum_i E_i\right)^2 - \left(\sum_i \vec{p_i}\right)^2},
\label{defIM}
\end{equation}
where $E_i$ and $p_i$ corresponds to the energies and momenta of decay products of short lived meson. 

As a first stage of reaction identification a missing mass of the system $p_{b} p_{t}\to p_1 p_2X$ 
will be determined according to the equation~\ref{defMM}. Next charge particles ($\pi^{\pm}$,$e^{\pm}$) 
will be identified via invariant mass:
\begin{equation}
m_{\pi^{\pm}(e^{\pm})} = \sqrt{E_{\pi^{\pm}(e^{\pm})}^2 - \vec{p}_{\pi^{\pm}(e^{\pm})}^2}.
\end{equation}
Finally the signal from the $\pi^0$ meson will be reconstructed as the invariant mass of two gamma quanta
according to the equation~\ref{defIM}. 

\chapter{Reaction kinematics, analysis methods and detector simulation}
\hspace{\parindent}
In order to understand properly the physical processes, the computer simulations of
the detector response for both investigated reactions were carried out.
For that purpose we used the hadronic event generator Pluto++~\cite{Frohlich:2007bi} 
and a WASA Monte Carlo Software which is 
based on the GEANT package~\cite{geant:1994} used to generate the response of whole detector setup. 
Based on this we were able to reproduce the kinematics of the investigated reactions and to
calculate the geometrical acceptance of the detector.

\section{Event generator: Pluto++}
\hspace{\parindent}
The simulations were performed using the Pluto++ event generator, which 
enables to simulate kinematics of reactions at beam momentum given by the user. 
As an output the four-momentum vectors of all particles in the final state are returned. 
In most of the cases, 
four-momentum vectors of all particles in the exit channel are weighted
according to the Phase Space\footnote{In this context the {\it{Phase Space}} 
term refers to an isotropically and uniformly distributed values of the four-momentum 
vectors of particles produced in the {\it{'s'}} wave with the relative angular momentum J=0.} 
using GENBOD routines~\cite{James:1968}.
For a few reactions and decay channels a phenomenological properties of a given process like 
angular distributions and transition matrix elements are implemented into the code.

\section{WASA Monte Carlo package}
\hspace{\parindent} 
To simulate the response of the detector for the generated data, a virtual model of the WASA detector was 
built based on a GEANT~3.21 framework~\cite{geant:1994}. 
The description of the detector elements and support structures were fully modelled 
using an abstract objects called ''{\it{volumes}}'' which are filled with appropriate materials
and embedded in a 3D coordinate system with an interaction point in its center. 
Each volume is described by the geometrical dimensions and positions relative to the center of the 
coordinate system. The implementation includes the sensitive, as well the passive materials which 
creates the whole apparatus.

The interactions of the particles with the detector is done by propagating each of them 
through the volumes and simulating according to the known cross sections 
the physical processes like: photon conversion, 
production of secondary particles, quenching effects, multiple Coulomb scattering, hadronic interactions,
energy losses. However, the package do not include the light propagation 
processes in the scintillator, electronic noise, and the response of the photomultipliers. 
But still, the package provides a very precise comparison to the measured experimental data. 

The Monte Carlo simulation starts with the input of four-momentum vectors from Pluto++ event 
generator, and is continued with the stepping action of each particle through the volumes. Generated  
responses for each event include the map of activated detector elements,
values of deposited energy and times. Further, matching of actual experimental conditions 
with the simulations, an additional smearing of the observable is possible via separate 
input filter files. 

\section{Root-Sorter work flow scheme}
\hspace{\parindent}
The main analysis software used for off-line data processing is the Root-Sorter framework~\cite{Hejny:FZJ03}
which bases on the object oriented C++ language~\cite{Eckle:2000aw} and CERN-ROOT package~\cite{Brun:1997pa}.
The software is organized in a modular way, such that, each class is responsible 
for different tasks like: tracks or cluster finding, calibration, and reconstruction. 
\begin{figure}[t!]
\hspace{1.cm}
\parbox{0.85\textwidth}{\centerline{\epsfig{file=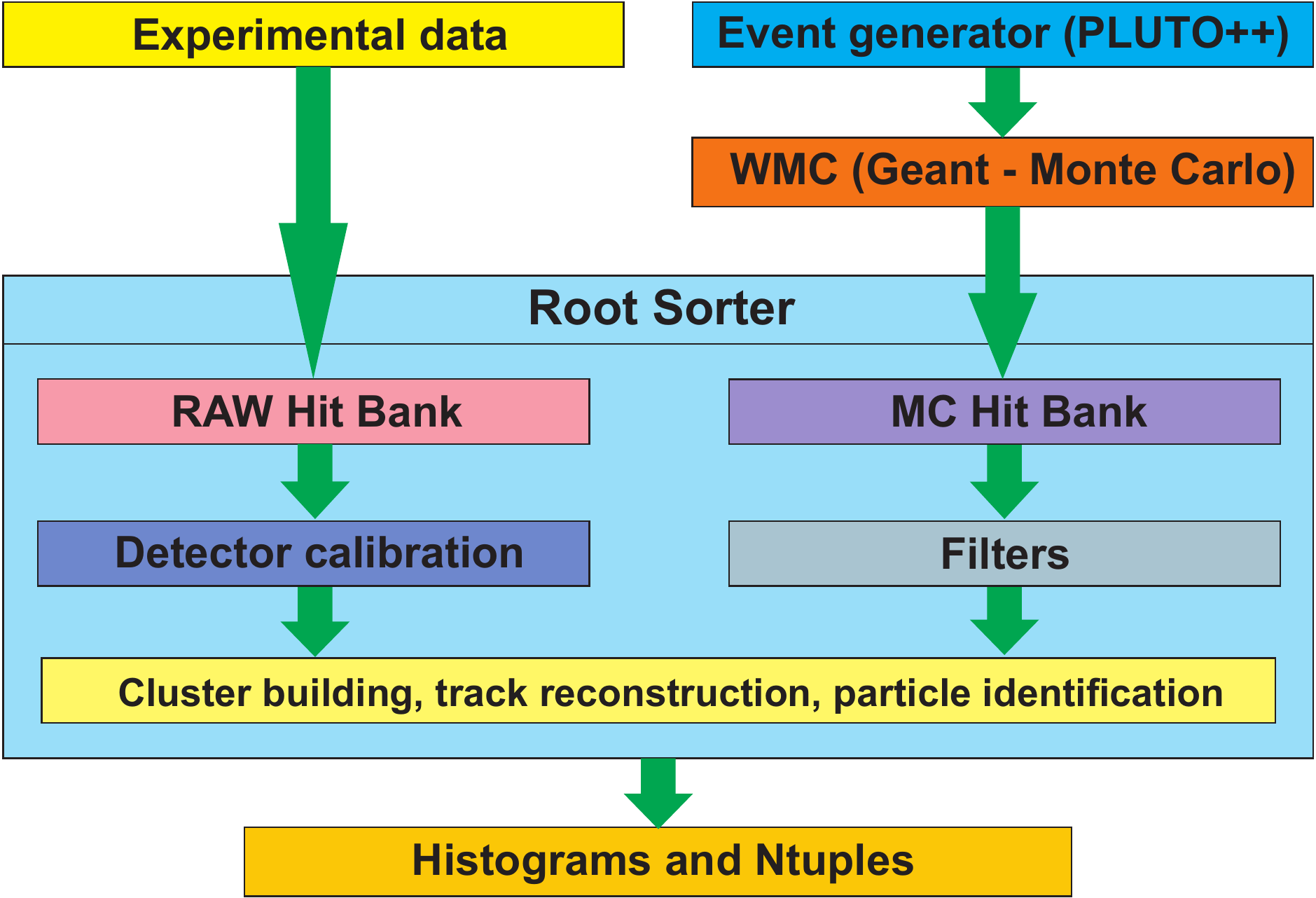,width=0.9\textwidth}}}
\caption{
The schematic diagram of the Root-Sorter analysis workflow.
}
\label{rootsorter}
\end{figure} 
The ROOT software delivers very rich library of standard mathematical functions, 
histogramming and fitting procedures, and also provides the management of the input/output streams. 
It provides very useful classes for the vector and four-momentum vectors manipulations as well
most of the non-standard and dedicated methods of reconstruction, calibration and track finding
which were prepared by the members of the WASA-at-COSY and CELSIUS/WASA collaborations.

The event processing is divided into two stages: (i) low level, and (ii) high level analysis
(see Fig.~\ref{rootsorter}).
In the low level analysis, the raw signals from stored data files are loaded into the computer 
memory and then calibrated from QDC and TDC units, to the units of energy 
represented in gigaelectronvolts (GeV) and time expressed in nanoseconds (ns).
Also, the offset due to the electronic delays and non-uniformity corrections
are applied on this stage.
For the Monte Carlo data, instead of the calibration procedures, the filters are applied,
which stores the smearing values of energy and times 
in order to account for the experimental resolution of the detection system.  

The data flow for experiment after decoding and calibration, and for simulations 
after filtering is organized in the same way. At the beginning the 
signals from individual detection modules are composed into hits, 
and then they are stored in a HitBanks separately allocated for each detector. 
Then the cluster algorithm groups the adjacent hits coming from one detector 
into clusters which are stored in Cluster-Banks. 
Subsequently, the track finding algorithm is applied to the cluster banks and creates from 
the clusters belonging to different detectors an object, which is representing a particle track.

The high level analysis refers in most of the cases to the stage, in which user 
by himself writes an analysis procedures to extract signal of desired reaction, using 
reconstructed tracks and constructing missing an invariant mass spectra, and also 
applying cuts. 
On this stage user loads the low level analysis modules and, the event processing is then 
automatically performed, so that the user can access 
information about registered particles in the form of the deposited energy and track direction $\vec{r}$.
Furthermore, for backward consistency, the track-, cluster- and hit- banks have an 
ability to inherit properties from its predecessors, so that it is possible for the 
user at any stage of the analysis to access to low level information from top-level data structures. 

\section{Kinematics of the $pp\to pp\eta\to pp\pi^+\pi^-\pi^0(\gamma\gamma)$ 
         and $pp\to pp\eta\to pp\pi^0(\gamma\gamma) e^+ e^-$ reactions}\label{sec:kinematyka}
\hspace{\parindent}
In the experiment the production of the $\eta$ meson was performed by colliding the proton beam 
of 1.4~GeV energy with a stationary proton pellet target.   
For the simulation the Pluto++ event generator was used which includes properties of the $\eta$ meson 
production mechanism. One of them is the anisotropy parameter of the $\eta$ polar angle 
in the center of mass frame $\theta_{\eta}^{cm}$. This anisotropy was observed by the DISTO 
Collaboration~\cite{Balestra:2004kg} at the beam kinetic energies of 2.15, 2.5 and 2.85~GeV. 
In case of the investigated reaction the beam energy corresponds to the excess energy of Q = 55~MeV. 
As it was shown by the COSY-11 and COSY-TOF collaborations~\cite{Moskal:2000pu,Smyrski:1999jc,Moskal:2003gt,AbdelBary:2002sx},
in reactions closer to the kinematic threshold for the $\eta$ meson production this anisotropy vanishes. 
In addition, the proton angular distribution, which with the increasing beam momentum shows tendency
to be aligned stronger to forward and backward directions was included.
The decay products of the $\eta$ meson were distributed homogeneously in the whole phase space.  
\begin{figure}[t!]
\mbox{
\hspace{.3cm}
\parbox{0.40\textwidth}{\centerline{\epsfig{file=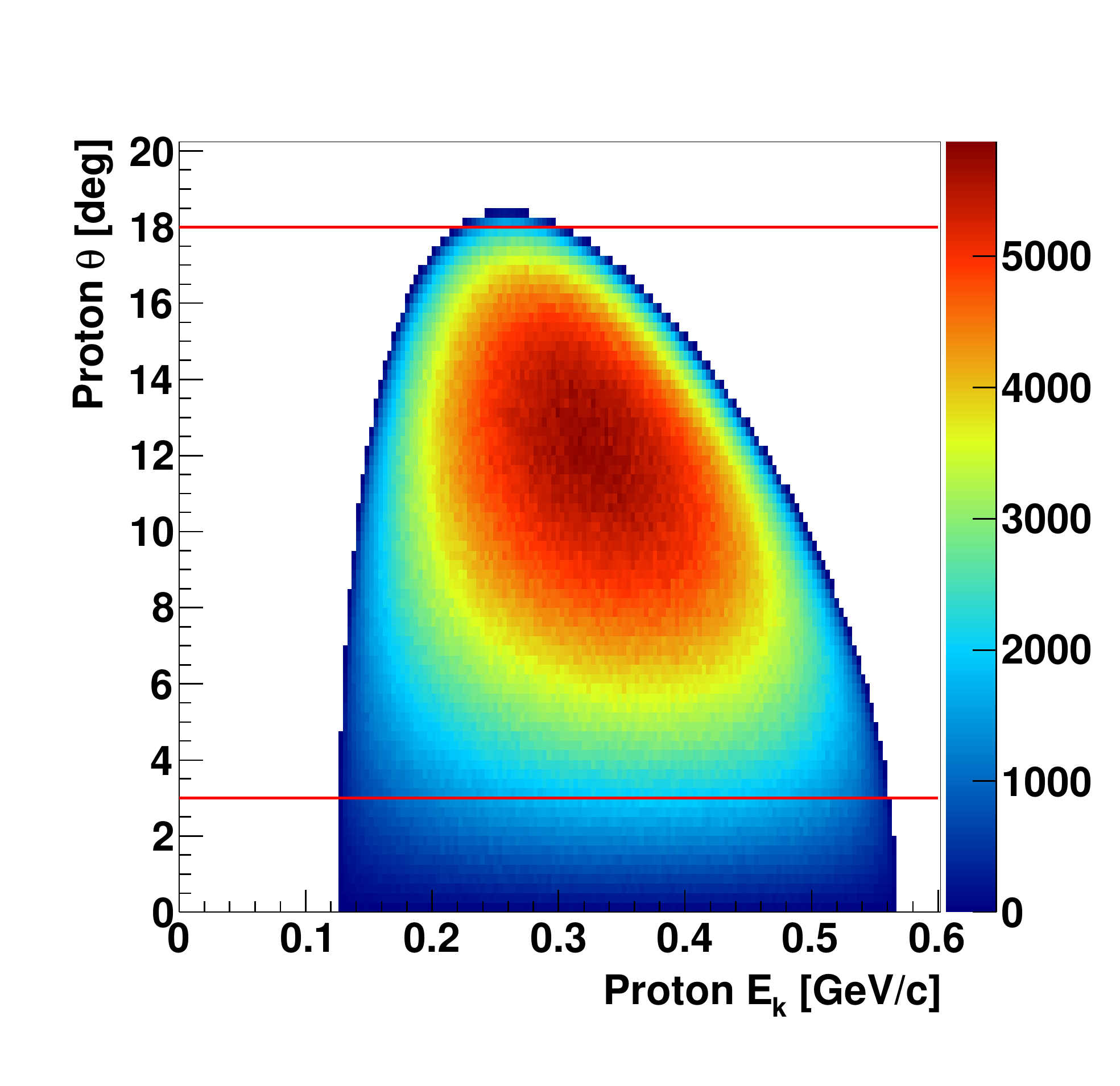,width=0.45\textwidth}}}
\hspace{1.1cm}
\parbox{0.40\textwidth}{\centerline{\epsfig{file=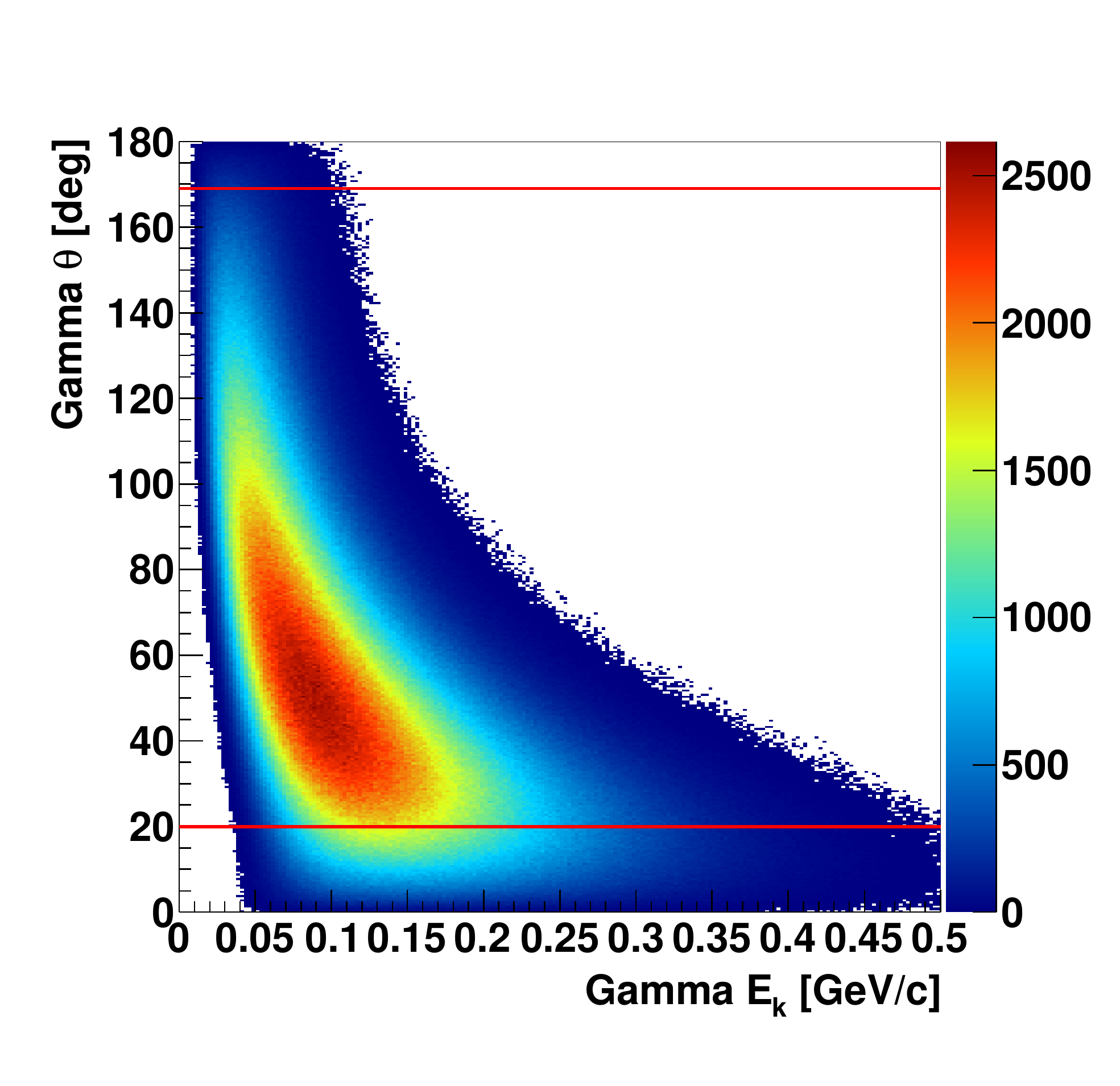,width=0.45\textwidth}}}}
\mbox{
\hspace{0.3cm}
\parbox{0.40\textwidth}{\centerline{\epsfig{file=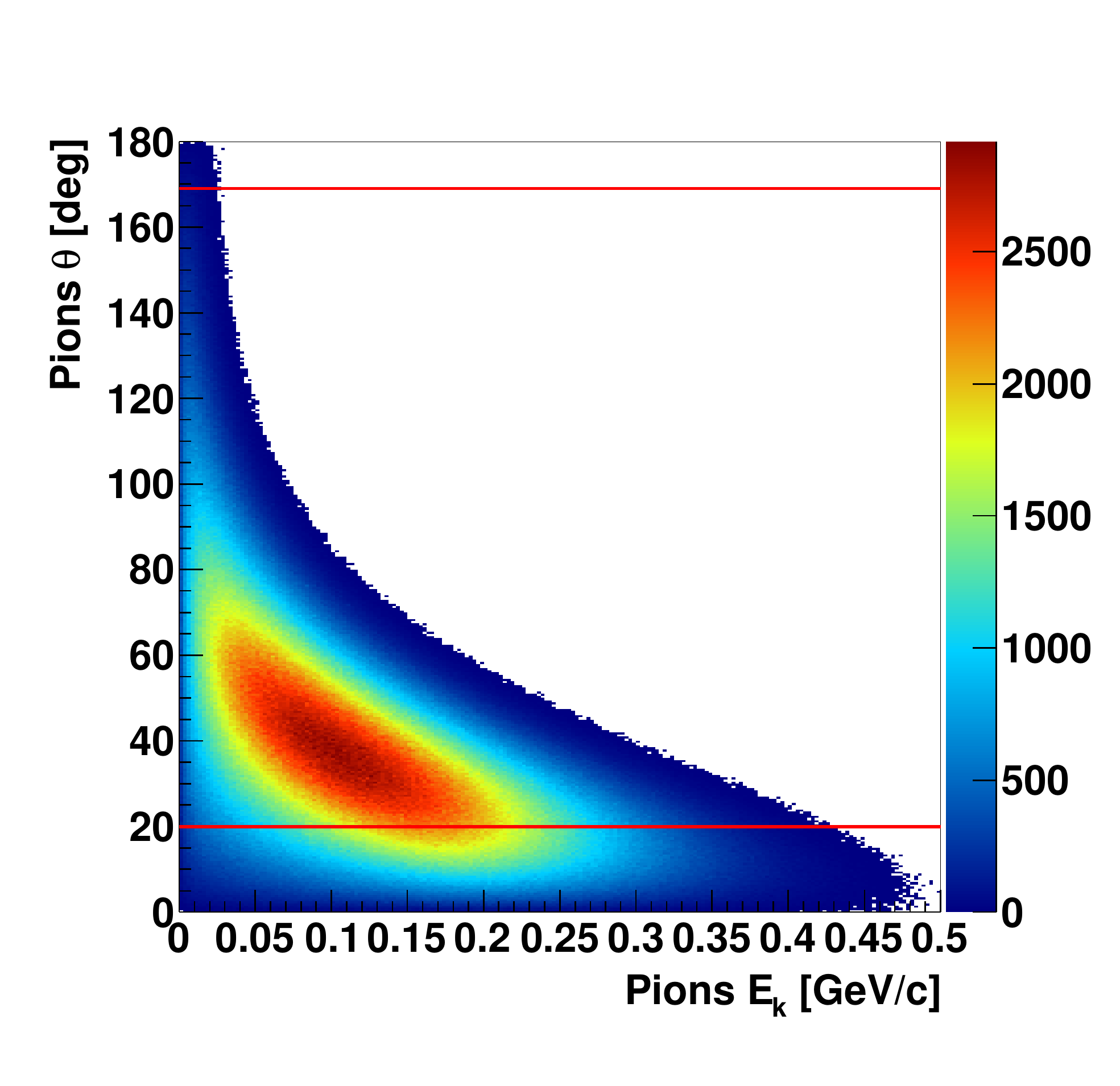,width=0.45\textwidth}}}
\hspace{1.2cm}
\parbox{0.40\textwidth}{\centerline{\epsfig{file=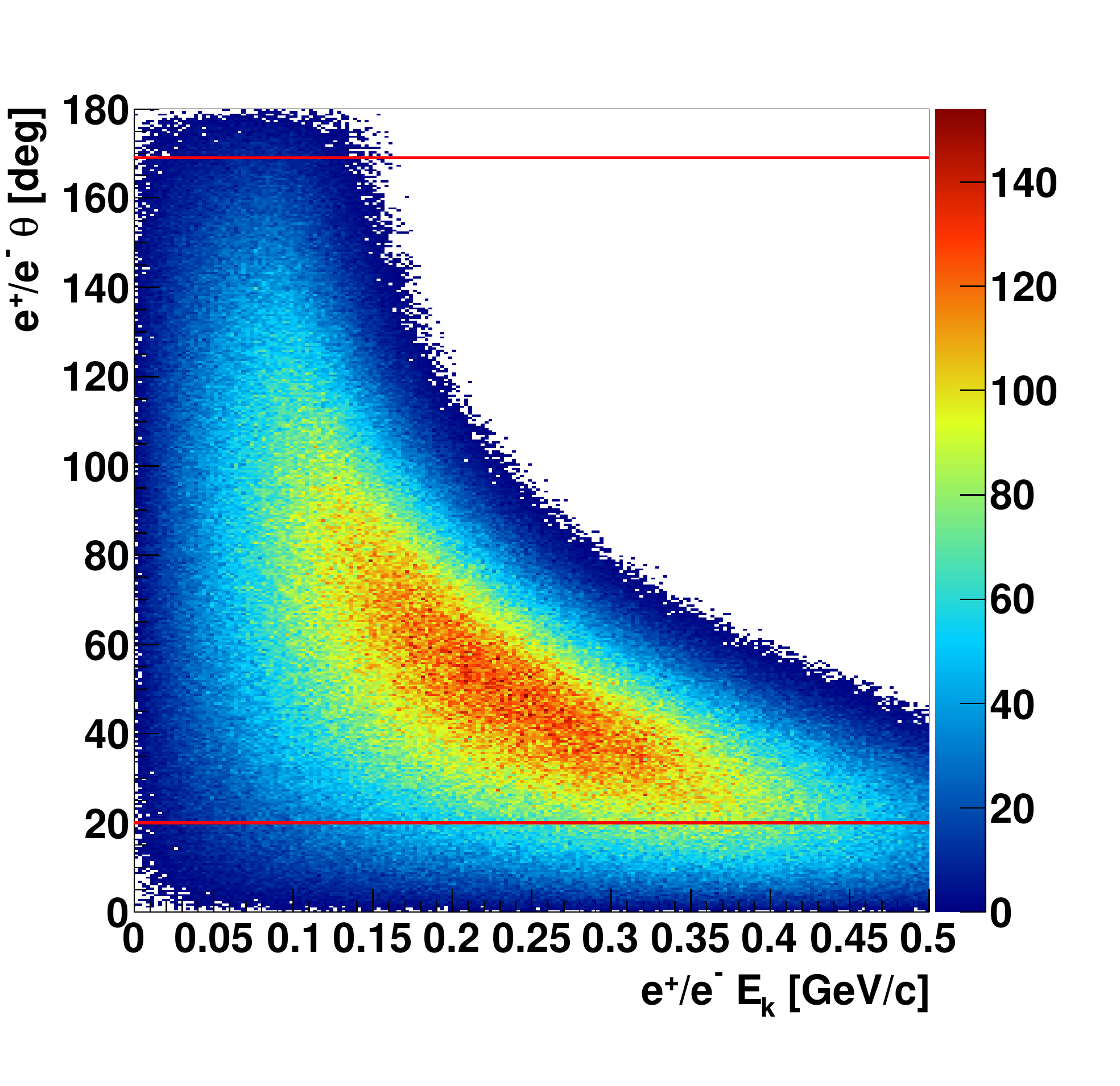,width=0.44\textwidth}}}}
\caption{
The simulated kinematics plots shows the correlation of kinetic energy with the scattering 
angles for particles in the final state: {\bf{(upper left)}} protons, {\bf{(upper right)}} gamma quanta,
{\bf{(lower left)}} pions, {\bf{(lower right)}} electrons and positrons. 
The red lines indicate the geometrical acceptance of the WASA-at-COSY detector.
}
\label{ppkin}
\end{figure} 
For both reactions the distribution of the angles and energy of recoil protons and gamma quanta 
originating from the $\pi^0$ decay, are the same. The differences appears in the spectra of the 
charged particles. As it is shown in Fig.~\ref{ppkin} (lower left) the pions $\pi^\pm$
populate more frequently the lower kinetic energy region and lower polar angles. 
While, the electrons and positrons have more broader distribution of the energy and 
$\theta$ (Fig.~\ref{ppkin} (lower right)). 
This difference is due to around 300 times lower mass of the electrons than pions. 

In order to estimate the acceptance, the geometrical coverage of the CD and FD was taken into account.
For detecting protons in forward direction the acceptance is equivalent to the angular range of $3^o \leq \theta \leq 18^o$ and 
for pions and gamma quanta in central part is $20^o \leq \theta \leq 169^o$ (see Fig.~\ref{ppkin}).
For the reaction $pp\to pp\eta\to\pi^+\pi^-\pi^0(\gamma\gamma)$ the fraction of events
which were found in the sensitive range of the detector (red lines) is equal to 35.3\%. 
While, for the $pp\to pp\eta\to\pi^0(\gamma\gamma) e^+ e^-$ process the geometrical acceptance 
is equal to 46.9\%. 

To check the event generator properties the Dalitz plot expressed in
variables defined by equations (\ref{dX}) and (\ref{dY}) was plotted, 
using the simulated four-momentum vectors. 
\begin{figure}[h!]
\mbox{
\hspace{.1cm}
\parbox{0.30\textwidth}{\centerline{\epsfig{file=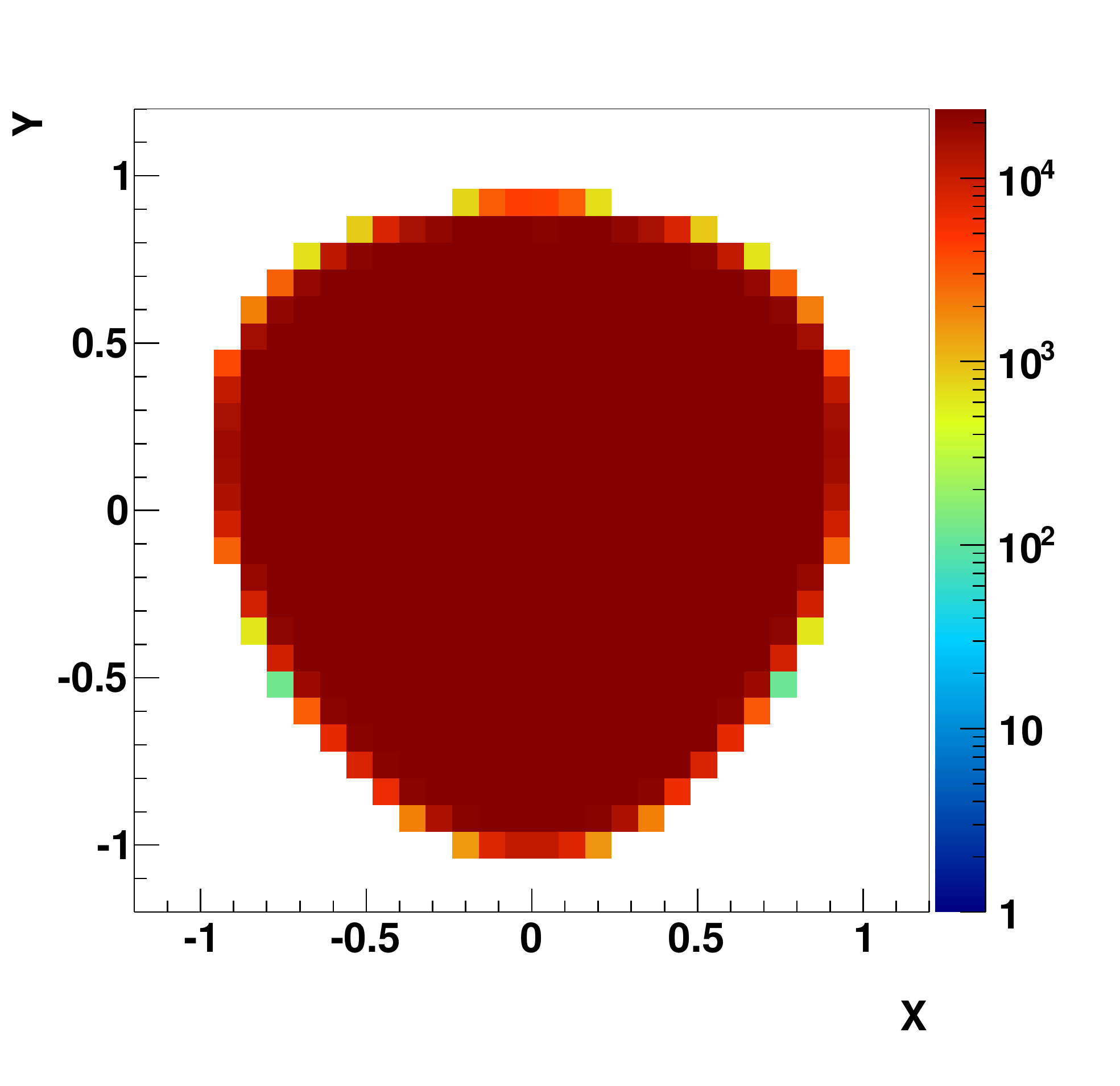,width=0.35\textwidth}}}
\hspace{0.6cm}
\parbox{0.30\textwidth}{\centerline{\epsfig{file=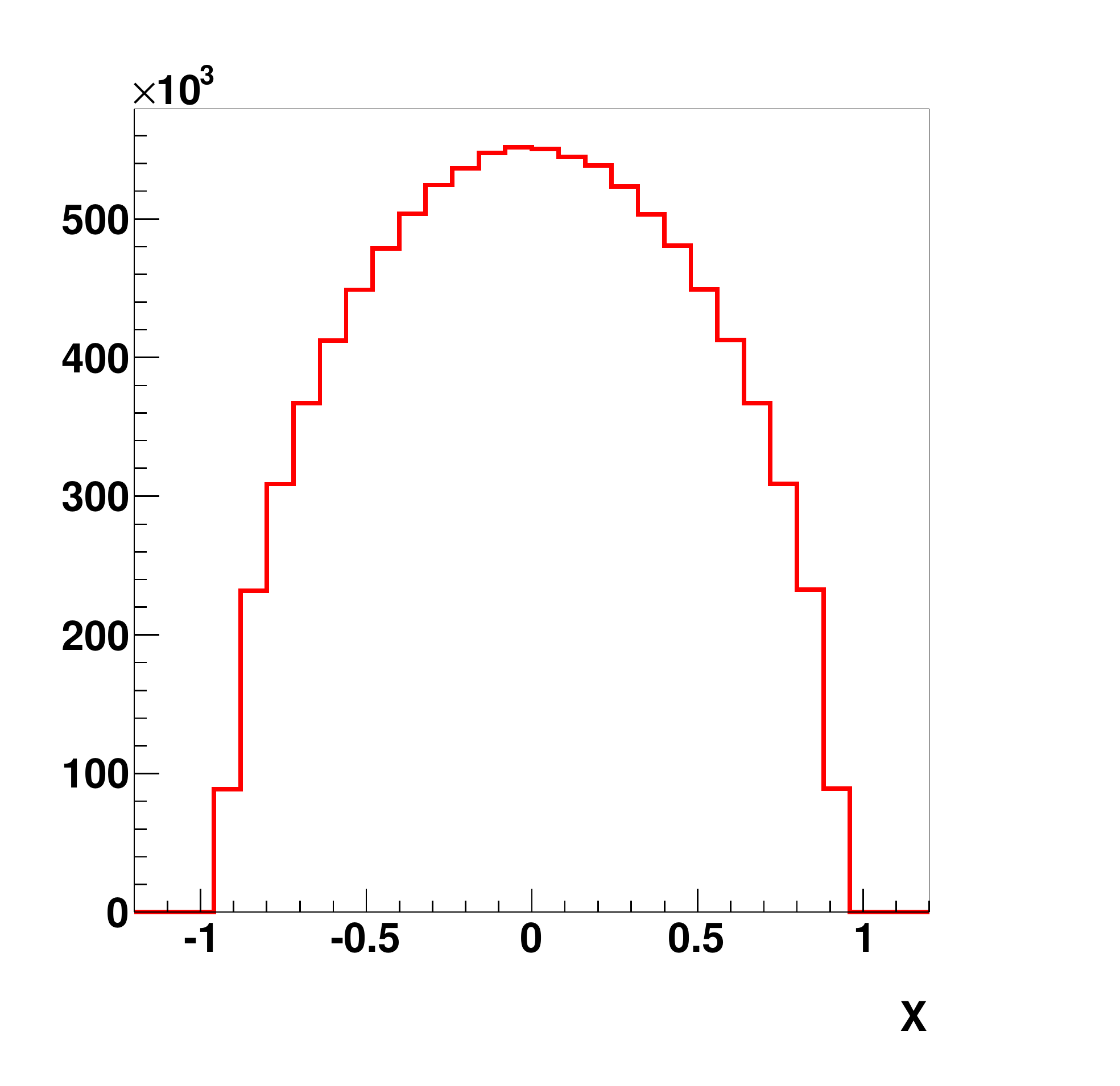,width=0.35\textwidth}}}
\hspace{0.3cm}
\parbox{0.30\textwidth}{\centerline{\epsfig{file=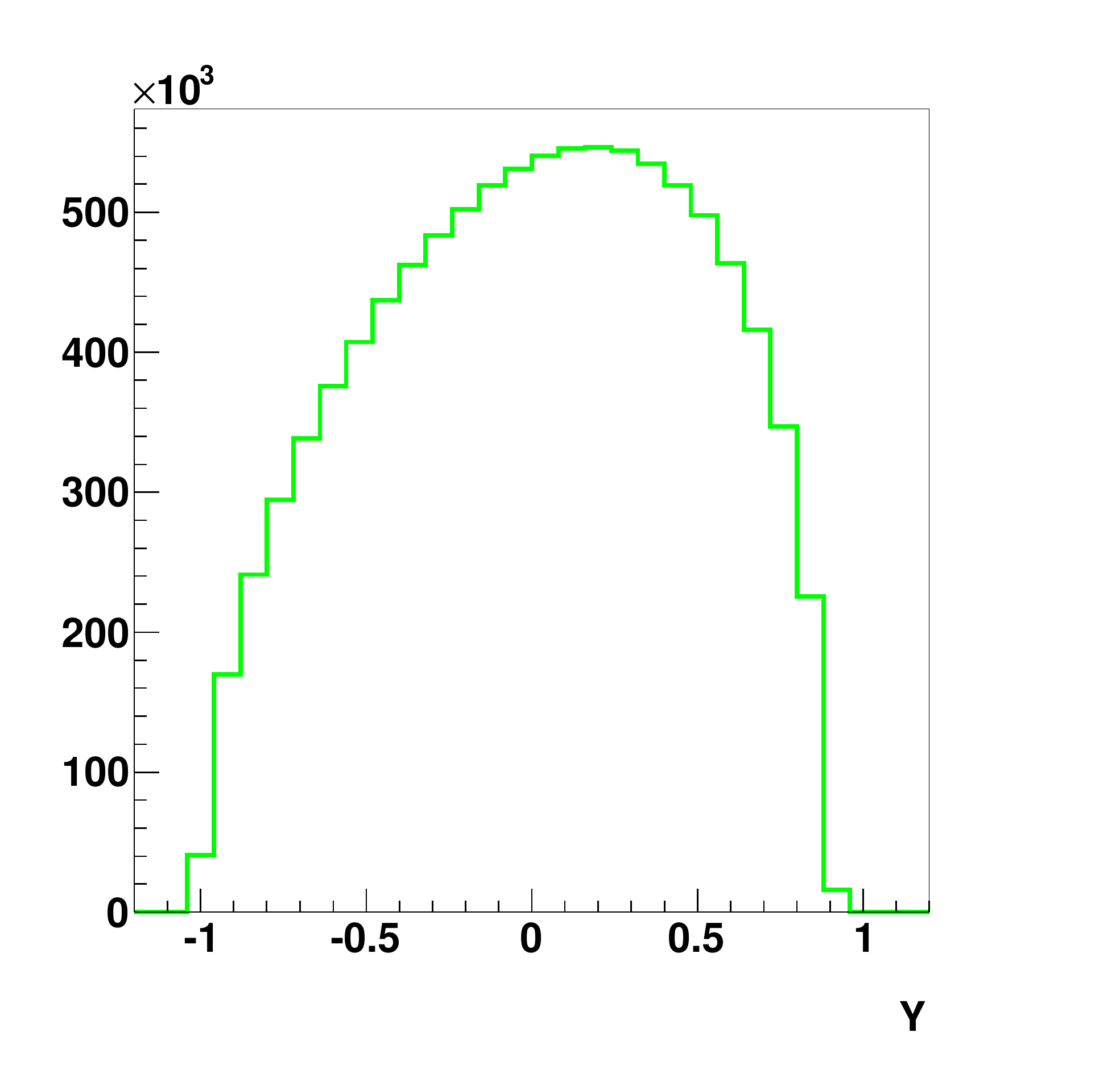,width=0.35\textwidth}}}}
\caption{
Simulations of the Dalitz plot for the $\eta\to\pi^+\pi^-\pi^0$ decay according the 
to the homogeneous phase space population. 
{\bf{(left)}} The two dimensional Dalitz plot in normalized X,Y coordinate system. 
{\bf{(middle)}} Projection of the Dalitz plot to X axis.
{\bf{(right)}} Projection of the Dalitz Plot to Y axis.
}
\label{DPps}
\end{figure} 
As it was mentioned in Chapter 2 the Dalitz plot represents a population of kinematically 
available phase space. In the case of simulation, performed under the assumption of no interactions 
between particles, the Dalitz plot should be populated homogeneously.
As it is shown in Fig.~\ref{DPps} (left), the results of the simulations 
are consistent with these expectations. Additionally in Fig.~\ref{DPps} (middle) 
and Fig.~\ref{DPps} (right) the projections of the Dalitz plot into the X and Y axes are presented.
 
\chapter{Track selection and reconstruction}
\hspace{\parindent}
The data analysis aiming at identification of $pp\to pp\eta\to pp\pi^+\pi^-\pi^0(\gamma\gamma)$
and $pp\to pp\eta\to pp\pi^0(\gamma\gamma) e^+ e^-$ reaction chains, at the first stage is 
the same, thus the procedures for both reactions, will be discussed simultaneously. 
The track reconstruction is based on signals registered in Forward and Central Detector 
of WASA setup. At the beginning of the reconstruction process hits in detector elements
which belong to the same particle are merged into clusters. Next, the clusters reconstructed in different 
detectors are combined into single particle track. 
For both investigated processes the trigger conditions and track reconstruction procedures are the same. 

\section{Data set and trigger conditions}\label{sec:Tryger}
\hspace{\parindent}
The data analyzed in this thesis were collected during the two week run in the year 2008.
The $\eta$ meson was produced in a reaction of proton beam with 
momentum of $p_{b} = 2.142$~GeV/c and a stationary proton target. 
The excess energy for the $pp\eta$ system was equal to Q~=~60~MeV, for which the production cross 
section of the $\eta$ meson amounts to $\sigma = 9.8 \pm 1.0~\mu b$~\cite{Chiavassa:1994ru}.
This cross section allowed to measure large event rates enabling to study rare and very rare 
decay processes. 
However, in the proton-proton interactions one has to deal with a large background 
originating from the direct multi-meson production. For the interesting $\eta\to\pi^+\pi^-\pi^0$ 
decay the physical background comes from direct production of three pions via the 
$pp \to pp\pi^+\pi^-\pi^0$ reaction channel. The total cross section for this reaction is in the same 
order of magnitude as the one for the $\eta$ production process. 
Therefore, part of this background with an invariant mass of $\pi^+\pi^-\pi^0$ system 
close to the mass of the $\eta$ meson will contribute to the signal range. 
However, expected signal to background ratio should be about ten in case of tagging by means of the 
missing mass of two protons, with resolution of a few MeV~\cite{Zielinski:2011dt}. 
Yet, the situation for a two pions direct production ($pp \to pp\pi^+\pi^-$) is worser,
because the total cross section is hundred times larger~\cite{Skorodko:2009ys} than for the $\eta$ 
meson production. But in this case only events with misidentified $\pi^0$ can be mistakenly 
taken as the signal and therefore one can suppress the background due to different final state than signal 
reaction.

In case of the $\eta\to\pi^0 e^+e^-$ final state, the background comes mostly from the reaction 
$\eta\to e^+e^-\gamma$ where the additional gamma quantum is reconstructed due to the splitting of signals in the calorimeter, and reactions with  
final state of $\pi^+\pi^-\pi^0$ for which the electrons and positrons can be misidentified with 
charged pions. Another source of background constitutes the conversion of gamma quanta on the beam pipe,
where the interaction of photons in the beryllium can cause emission of the electron and positron pair. 
To suppress this type of background the studies of the conversion process has to be performed.  
Moreover, the direct production of two neutral pion with subsequent decays, into $\pi^0\to\gamma\gamma$ and 
$\pi^0\to e^+e^-\gamma$ ($pp\to \pi^0\pi^0\to e^+e^-\gamma\gamma\gamma$) can obscure the searched signal. 
Thus the experimental conditions and large hadronic background make the investigation 
of the $\eta$ meson decays in proton-proton collisions experimentally challenging, 
and demands a very selective trigger. 

In order to reconstruct $2p$, $2\pi$ (or $2e$) and $2\gamma$ one has to construct 
a trigger, which basing on the simultaneous fulfilment of specific conditions in FTH, FRH, FVH, PSB 
and SEC, will select two charged particles in the Forward Detector and two charged and two neutral 
particles in the Central Detector. 
The first requirement was that at least two charged particles in second layer of the FRH will be 
detected $T^{FRH2}_{\mu\geqslant 2}$ (where $\mu$ indicates the hit multiplicity in the detectors), 
together with the Matching Trigger condition $T^{MT}_{\mu\geqslant 2}$ requiring that at least two groups of
clusters from different detectors corresponds to the same azimuthal angle.
Furthermore we set a rejection '{\it{veto}}' condition for events when  
at least one charged particle was registered in FVH detector ($T^{FVH}_{\mu = 0}$). 
This requirement was used to discard events with too energetic protons coming most probably from the 
elastic scattering reaction. 
\begin{figure}[!t]
\hspace{3.1cm}
\parbox{0.6\textwidth}{\centerline{\epsfig{file=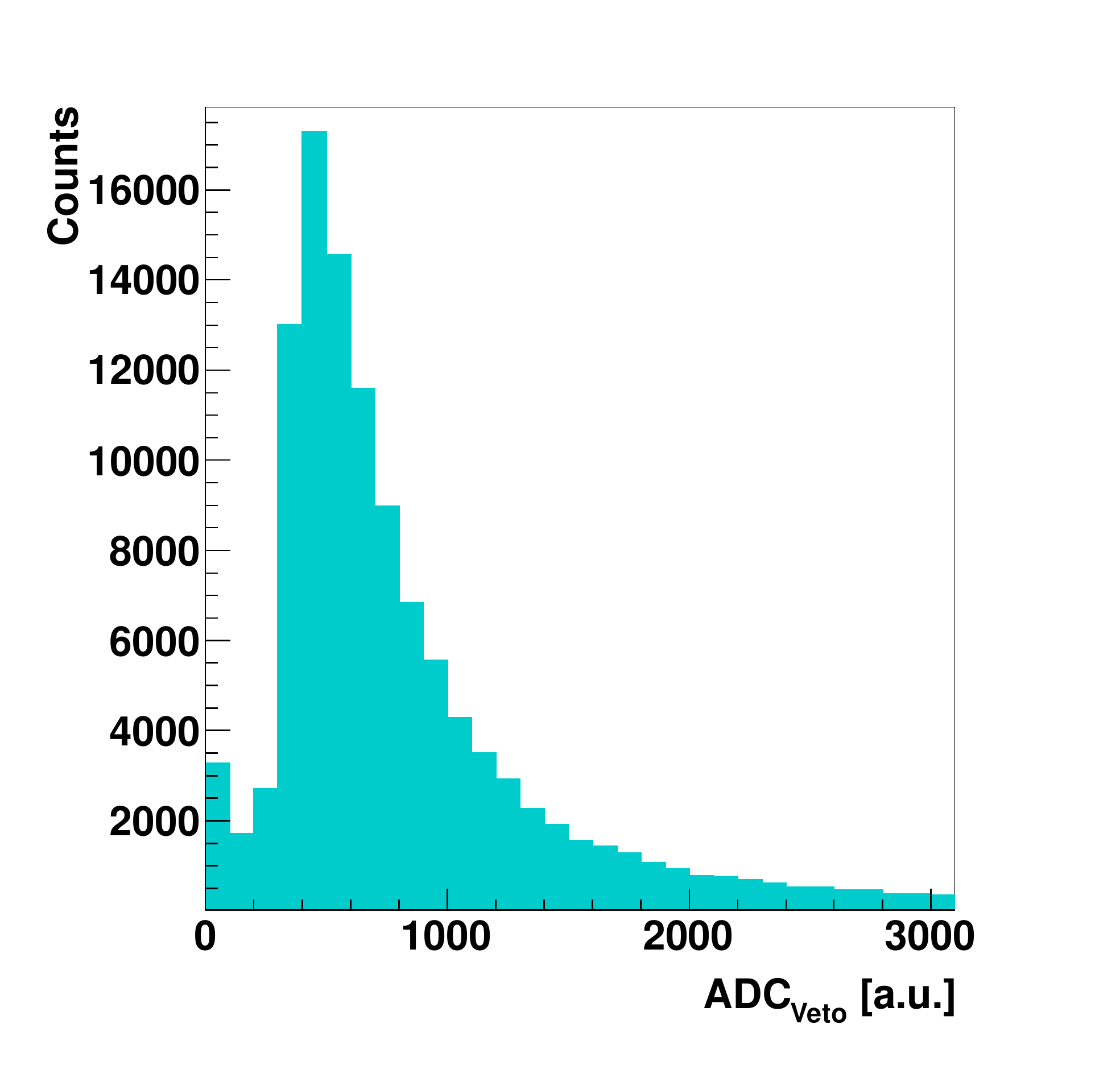,width=0.65\textwidth}}}
\caption{
The ADC distribution for the Forward Veto Hodoscope detector after pedestal subtraction. 
The threshold on the discriminator used for the $T^{FVH}_{\mu = 0}$ was set to 22~mV which 
corresponds effectively to the value of 300 in the units of the ADC channels. 
}
\label{vetoADC}
\end{figure}
The corresponding ADC spectrum of the FVH detector is shown in Fig.~\ref{vetoADC}.
In the Central Detector at least two charged particles in Plastic Scintillating Barrel 
$T^{PSB}_{\mu\geqslant 2}$ were required and at least one particle in the Scintillating 
Electromagnetic Calorimeter $T^{SEC}_{\mu\geqslant 1}$.
Thus summarizing, the main trigger for detection of the interesting reactions, reads:
\begin{equation}
T_{\eta} = T^{FRH2}_{\mu\geqslant 2} \wedge T^{MT}_{\mu\geqslant 2} \wedge T^{FVH}_{\mu = 0} 
           \wedge T^{PSB}_{\mu\geqslant 2} \wedge T^{SEC}_{\mu\geqslant 1}.
\label{trigger}
\end{equation} 
All the events which fulfilled  this requirements were saved to the disk.  

\section{Selection of tracks in the Forward Detector}
\hspace{\parindent}
Finding and reconstruction of particle track in the Forward Detector relay on geometrical 
overlap of clusters, which are formed based on hits in different detectors. 
The place where a particle crosses the FTH, is determined as an overlapping region of three elements
from each detection layer. 
This allows to determine a line from an assumed 
interaction point of the beam and target to the center of the pixel. Also a time information is determined
for tracks as an average time of signals in each layer of FTH. Then, the angular information is 
improved by reconstruction of tracks in FPC, where the polar ($\theta$) and azimuthal ($\phi$) 
angles of the track are reconstructed based on position of sense wires and the drift times. 
For the reconstruction procedure a line parameters determined in FTH are used as an initial values.  
This procedure typically improves the angular resolution by a factor of two and more.
\begin{figure}[h!]
\hspace{0.5cm}
\parbox{0.35\textwidth}{\centerline{\epsfig{file=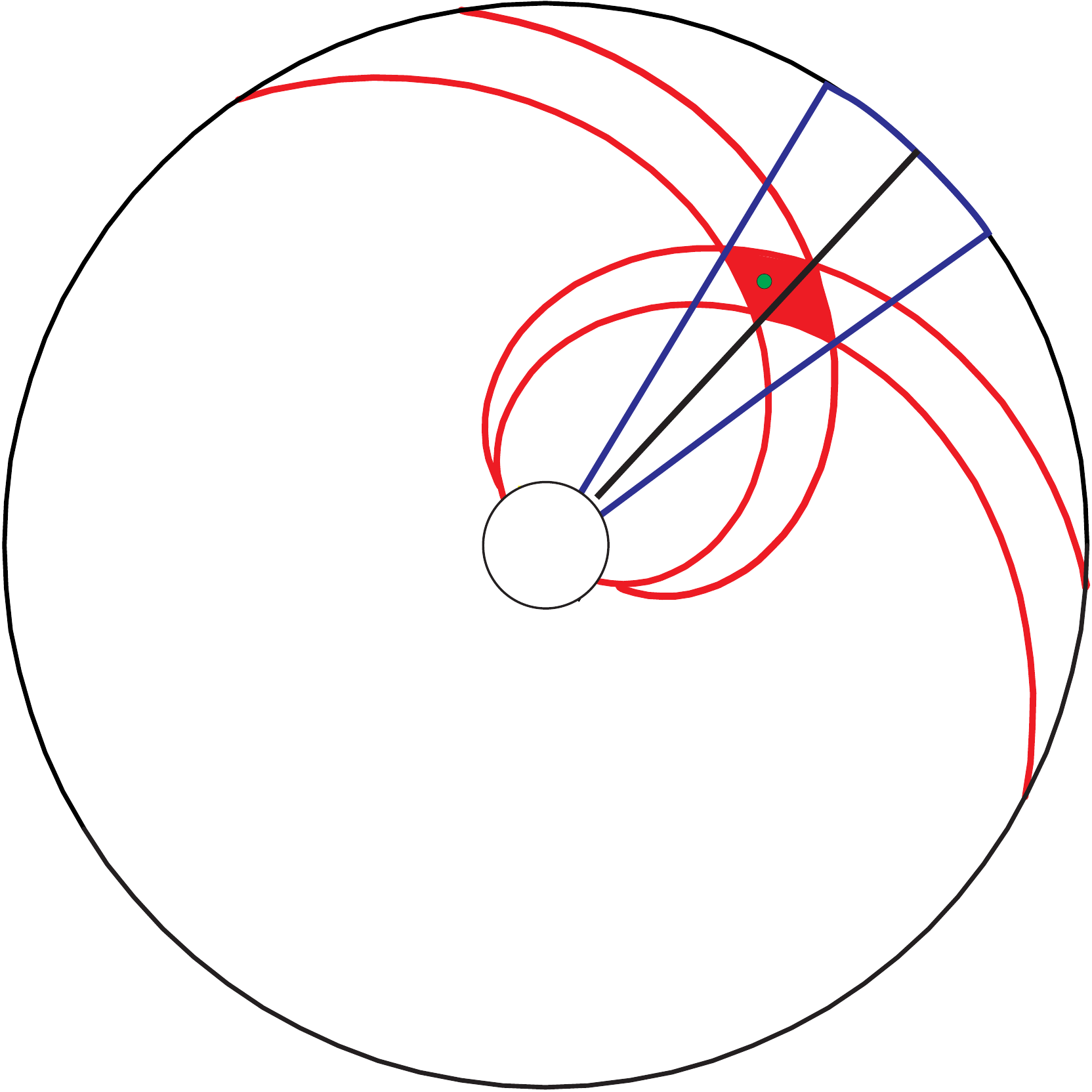,width=0.40\textwidth}}}
\hspace{1.0cm}
\parbox{0.5\textwidth}{\centerline{\epsfig{file=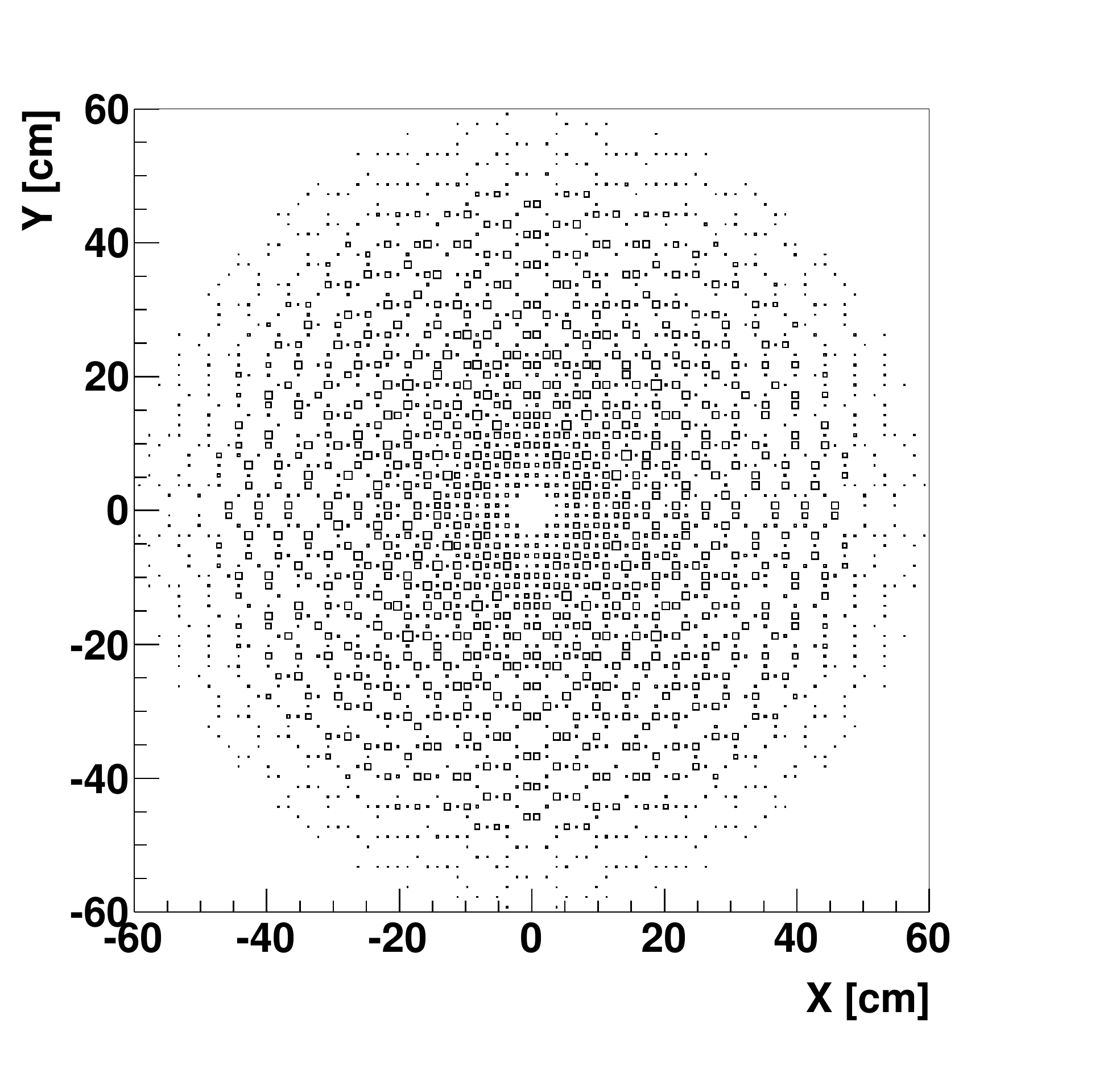,width=0.55\textwidth}}}
\caption{
{\bf{(left)}} Schematic view of the FTH pixel formation in the XY plane perpendicular to the beam line.
              The green point denotes the center of the pixel which is taken as the hit position.
{\bf{(right)}} Example of the hit distribution in the FTH. The spectrum shows the hits in the 
               center of each pixel.
}
\label{CDdiff1}
\end{figure}
After the track formation, the information from clusters about azimuthal overlapping, energy deposits,
and time differences from the FWC, FRH and FVH are taken into account. 

\section{Selection of tracks in the Central Detector}
\label{sec:cdtracks}
\hspace{\parindent}
The Central Detector allows to register charged and neutral particles coming 
from the $\eta$ meson decays. 
As a first step, signals registered in each of the detection elements are formed into a cluster, for each 
detector separately. Next, clusters from different detectors are matched in order to 
reconstruct a track of particle which passes through the detection system. 
The track reconstruction algorithm, based on the signals registered in MDC and PSB searches for all charged 
particle tracks. Afterwords, remaining signals in SEC not associated with a charged particle, 
are treated as originating from neutral particles. 
\begin{figure}[b!]
\hspace{0.2cm}
\parbox{0.5\textwidth}{\centerline{\epsfig{file=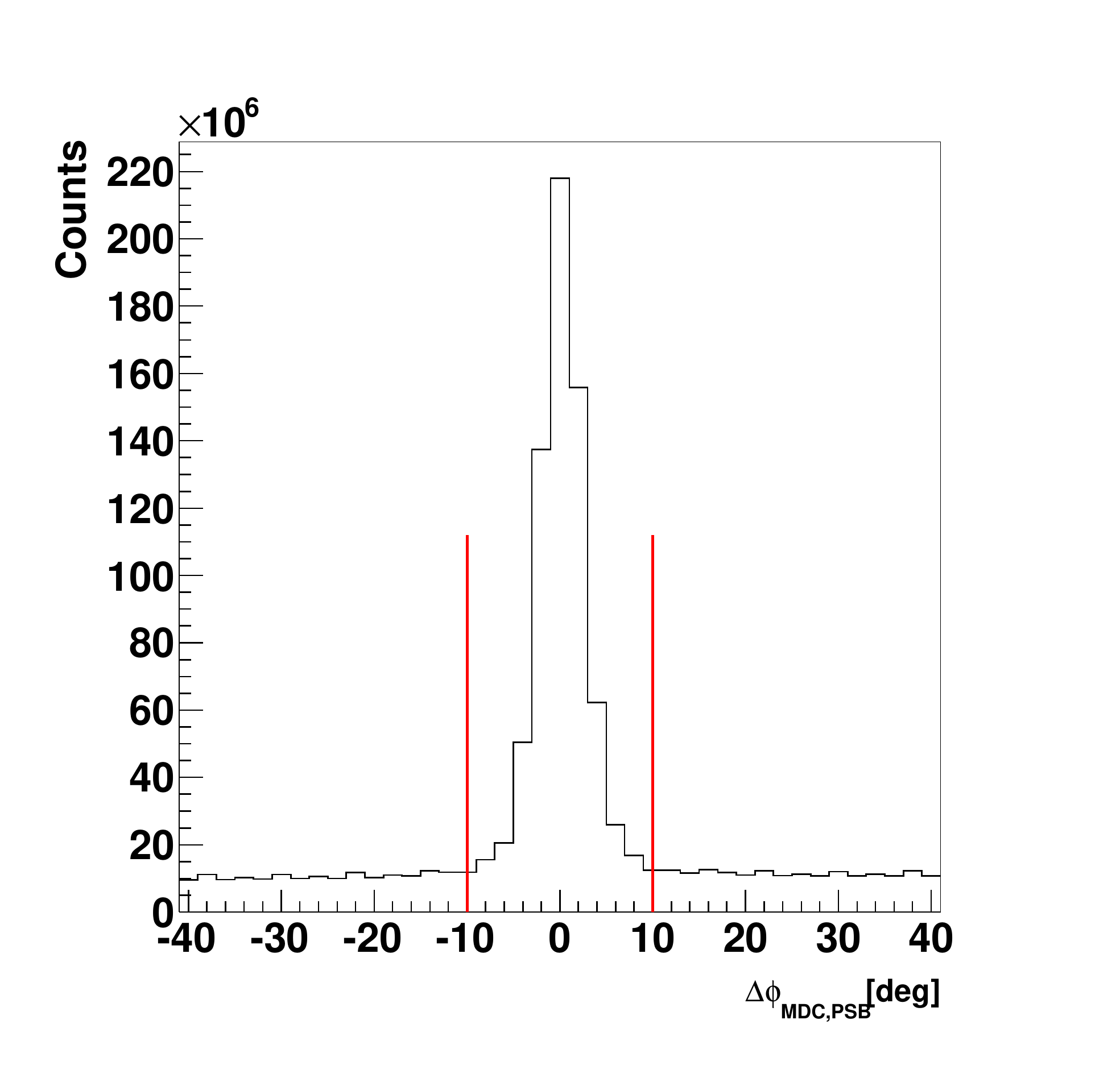,width=0.55\textwidth}}}
\hspace{0.3cm}
\parbox{0.5\textwidth}{\centerline{\epsfig{file=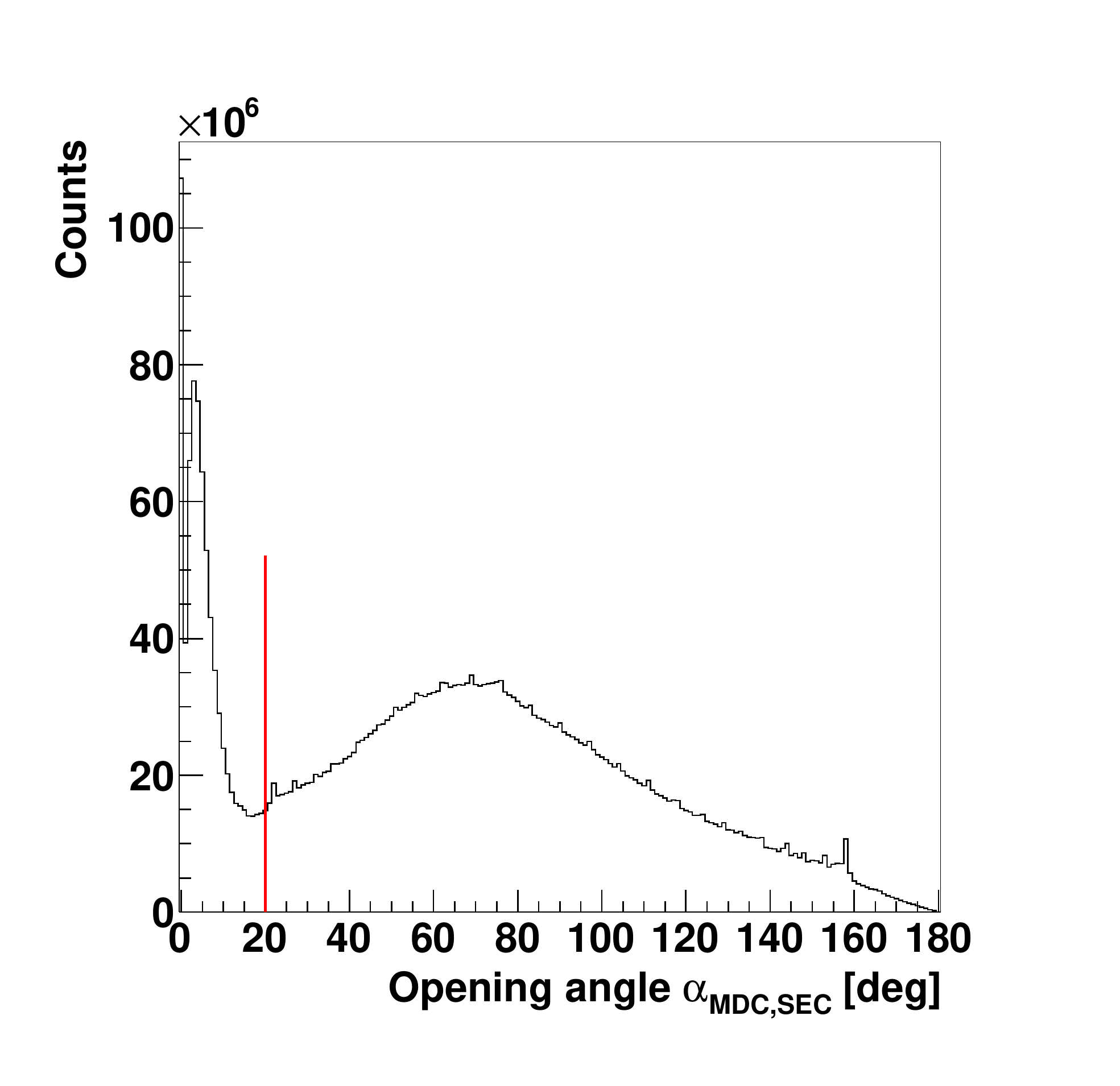,width=0.55\textwidth}}}
\caption{
{\bf{(left)}} Difference in the azimuthal angle $\Delta\phi_{MDC,PSB}$  of the exit coordinate of the 
MDC and the PSB detectors.  
{\bf{(right)}} Distribution of the opening angle $\alpha_{MDC,SEC}$ between 
cluster position in SEC calculated from the extrapolation of the helix from the MDC and the 
real cluster position measured in the SEC. The small spike structures seen in the spectrum are due to the 
granularity of the detector. In both figures red lines indicates the conditions 
under which clusters were assigned to one track.}
\label{CDdiff1}
\end{figure}
\begin{figure}
\hspace{0.2cm}
\parbox{0.5\textwidth}{\centerline{\epsfig{file=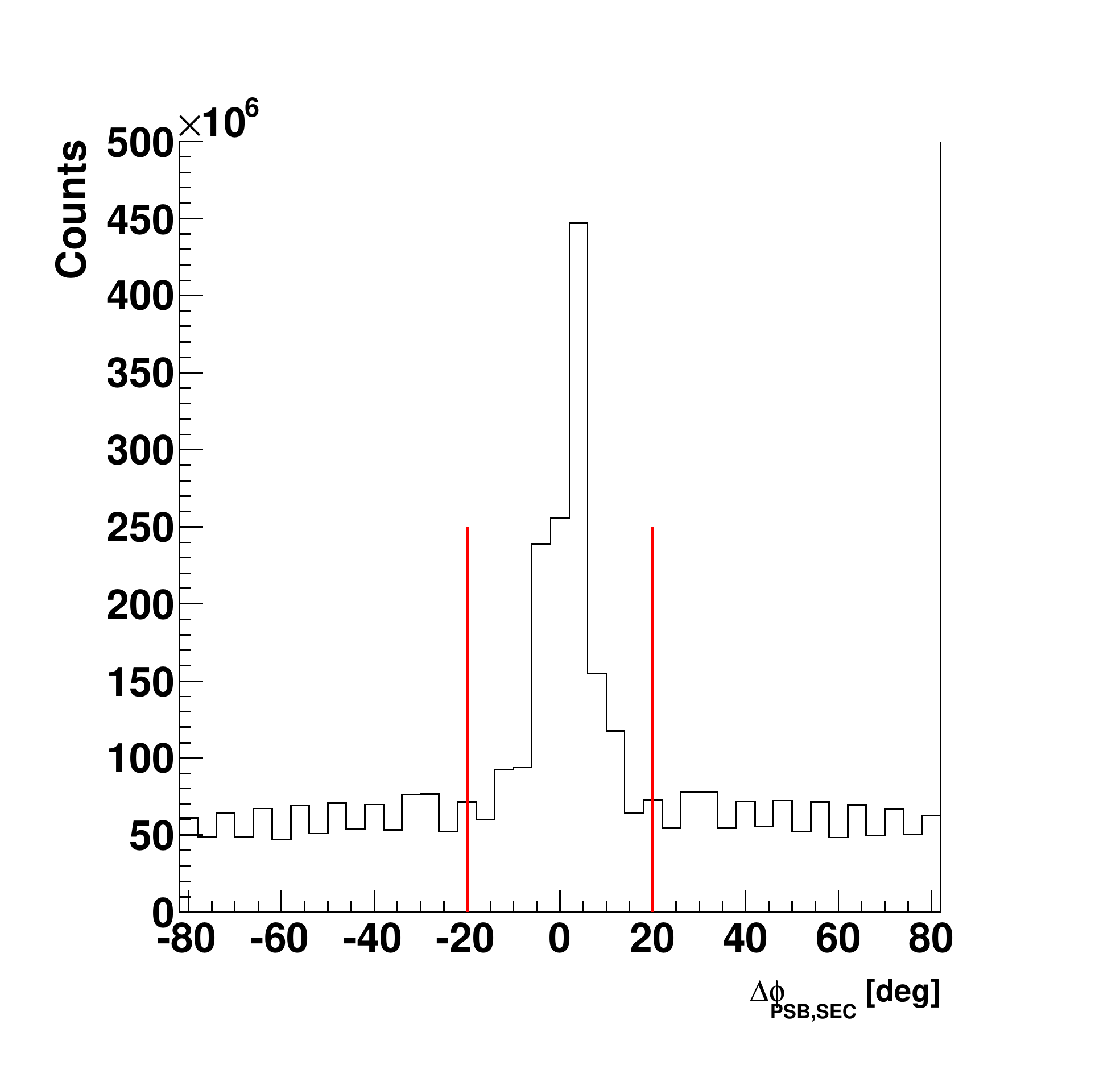,width=0.55\textwidth}}}
\hspace{0.3cm}
\parbox{0.5\textwidth}{\centerline{\epsfig{file=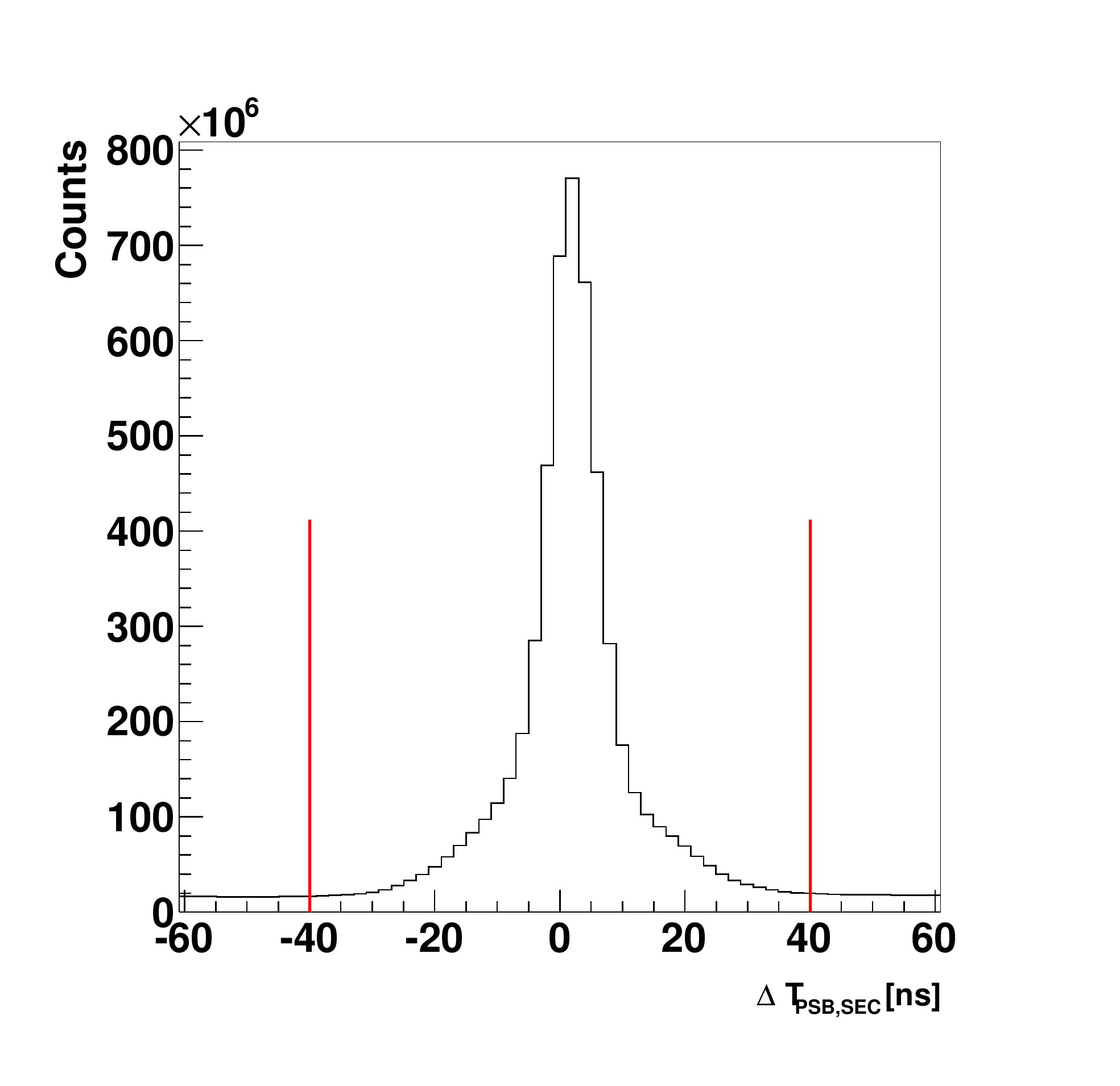,width=0.55\textwidth}}}
\caption{
{\bf{(left)}} The difference in the azimuthal angle $\Delta\phi_{PSB,SEC}$ of the cluster in the PSB and 
cluster in the SEC detectors. The bumpy structures on the spectrum are due to the granularity of the detector.
{\bf{(right)}} Difference between the time of the cluster in PSB and time of the cluster in SEC.
In both figures red lines indicates the conditions under which clusters were assigned to one track.}
\label{CDdiff2}
\end{figure}

In SEC when a particle hits the crystal an electromagnetic shower is produced,
which spreads over nearest elements. The reconstruction procedure starts with a 
search for a crystal with deposited energy greater than 5~MeV. When such a crystal is found, 
the routine puts it in a center of a pre-cluster which is a square of $3\times3$ elements.
Next, the routine checks energy in the adjacent elements. If the energy is greater than 2~MeV
the element is added to the cluster. These elements also become a reference in the next step of 
the procedure. Finally, the algorithm terminates if there is no more crystals with the energy greater
than 2~MeV. In addition only elements which have a time signals within 
50~ns window are included into a cluster. Further on, only clusters with sum of energy of individual 
crystals greater than 20~MeV are considered for further analysis. 
This condition allows to remove low energy noise signals
and as a result suppresses the background. 

The PSB clusters produced by one particle are build from neighboring detection 
elements which are in a time window of 10~ns. The cluster finding routine 
allows maximally for three detection elements to build one cluster. 
Similarly as for the SEC the procedure is searching for most energetic hits, and then checks 
for hits in closest detection elements. The minimal deposited energy of a hit which is 
taken into account is 0.5~MeV. 

The MDC clustering and reconstruction algorithms rely on finding the hits in 17 layers of straw
tube detectors. The procedure is realized in two steps. First the routine uses the pattern 
recognition where hits in each tube are projected into plane perpendicular to the beam 
axis~\cite{komogorov:1997jh}. Then the procedure fits a circle, and minimizes the weighted sum of distances 
between the hit position and the center of the circle. 
In the second stage, algorithm fits a straight line to hits 
in the 'Rz' plane creating a three dimensional helix. Reconstruction of particle momentum vector 
components is done from the curvature of the helix assuming the homogeneous magnetic field 
inside the detector.

After obtaining cluster information from all central detectors the track finding algorithm 
is used. This procedure matches clusters from MDC, PSB and SEC detectors and assigns them
to one particle. First the helix from the drift chamber is extrapolated to pass through 
PSB and SEC. Next, the procedure matches exit position of the helix from the MDC with the 
cluster position in PSB, by checking the difference in the azimuthal angle $\phi$. 
The clusters are grouped into one track if the $\vert\Delta\phi_{MDC,PSB}\vert = \vert\phi_{MDC} - \phi_{PSB}\vert$ is less than $10^{o}$. The experimental spectrum of the $\Delta\phi_{MDC,PSB}$ is shown in 
Fig.~\ref{CDdiff1} (left).

Furthermore, the matching between the MDC and SEC is done by checking the opening angle 
$\alpha_{MDC,SEC}$ between calculated position of the hit in the calorimeter from the extra\-polation 
of the helix to the SEC, and measured cluster position. 
In this case, the procedure matches the clusters into one track if the opening angle 
$\alpha_{MDC,SEC}$ is less or equal to $20^{o}$. 
The experimental distribution of the $\alpha_{MDC,SEC}$ parameter is shown in Fig.~\ref{CDdiff1} (right).

Also, an additional matching criteria are checked between clusters in PSB and SEC detectors. 
In this case angular and time information for clusters are available. The absolute value of  
difference in azimuthal angle $\Delta\phi_{PSB,SEC}$ between clusters in PSB and SEC has 
to be less than $20^{o}$, to be accepted as originating from one particle. 
Also difference between time of the cluster registered in PSB and time of 
the cluster in SEC, has to be in  time coincidence window of 80~ns ($\pm 40$~ns). The experimental 
distribution of the azimuthal angle between PSB and SEC is shown in Fig.~\ref{CDdiff2} (left),
and the time difference distribution is shown in Fig.~\ref{CDdiff2} (right).

\chapter{Identification of the $pp\to pp\eta$ reaction}
\hspace{\parindent}
After tracks reconstruction described in the previous Section as a next 
step of analysis the identification of protons registered in the Forward 
Detector and reconstruction of their four-momentum vectors was done. Events corresponding to
the $pp\to pp\eta$ reaction were identified based on the missing mass technique. 
The main trigger in the experiment aiming at selection of the $pp\to ppX$ reaction 
required two hits in second layer of the Range Hodoscope detector,
azimuthal angle agreement between FRH, FTH and FWC, and no hits in Veto Hodoscope.
The trigger conditions were described in details in Section~\ref{sec:Tryger}.

\section{Identification of recoil protons}
\hspace{\parindent}
In order to select protons from $pp\to pp\eta$ reaction one has to deal with large 
amount of background reactions and random coincidences. 
First the reconstruction algorithm recognizes tracks based on signals registered in different detectors
as it was described in Section~5.2. Then tracks which corresponds to a signals in the FPC detector 
are marked as coming from charged particle and considered for further analysis. 
Next, the conditions to reject noise signals were applied.
Signals assigned to tracks are checked to satisfy the 25~MeV minimal energy deposit condition in the whole 
forward detector. Further on, tracks are verified if they are inside geometrical acceptance of the Forward Detector which can detect particles only in the polar angle range of $3^o - 18^o$. 

The particle identification is done by means of the $\Delta E - E$ method. 
In this technique the energy deposited by charged particle in the first layer of the 
Forward Range Hodoscope is plotted as a function of the energy deposited in whole FRH.    
\begin{figure}[!h]
\hspace{0.2cm}
\parbox{0.5\textwidth}{\centerline{\epsfig{file=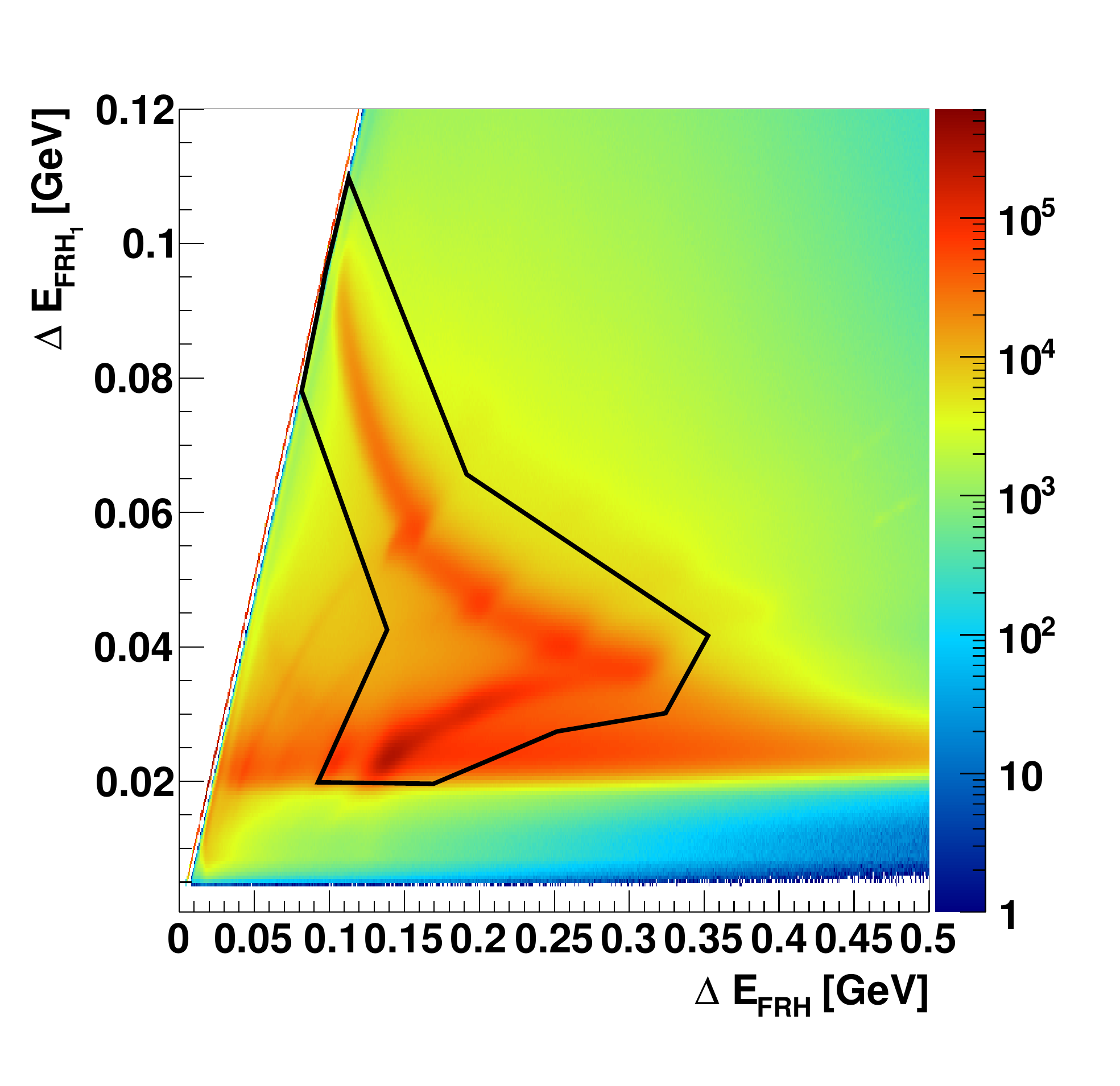,width=0.55\textwidth}}}
\hspace{0.2cm}
\parbox{0.5\textwidth}{\centerline{\epsfig{file=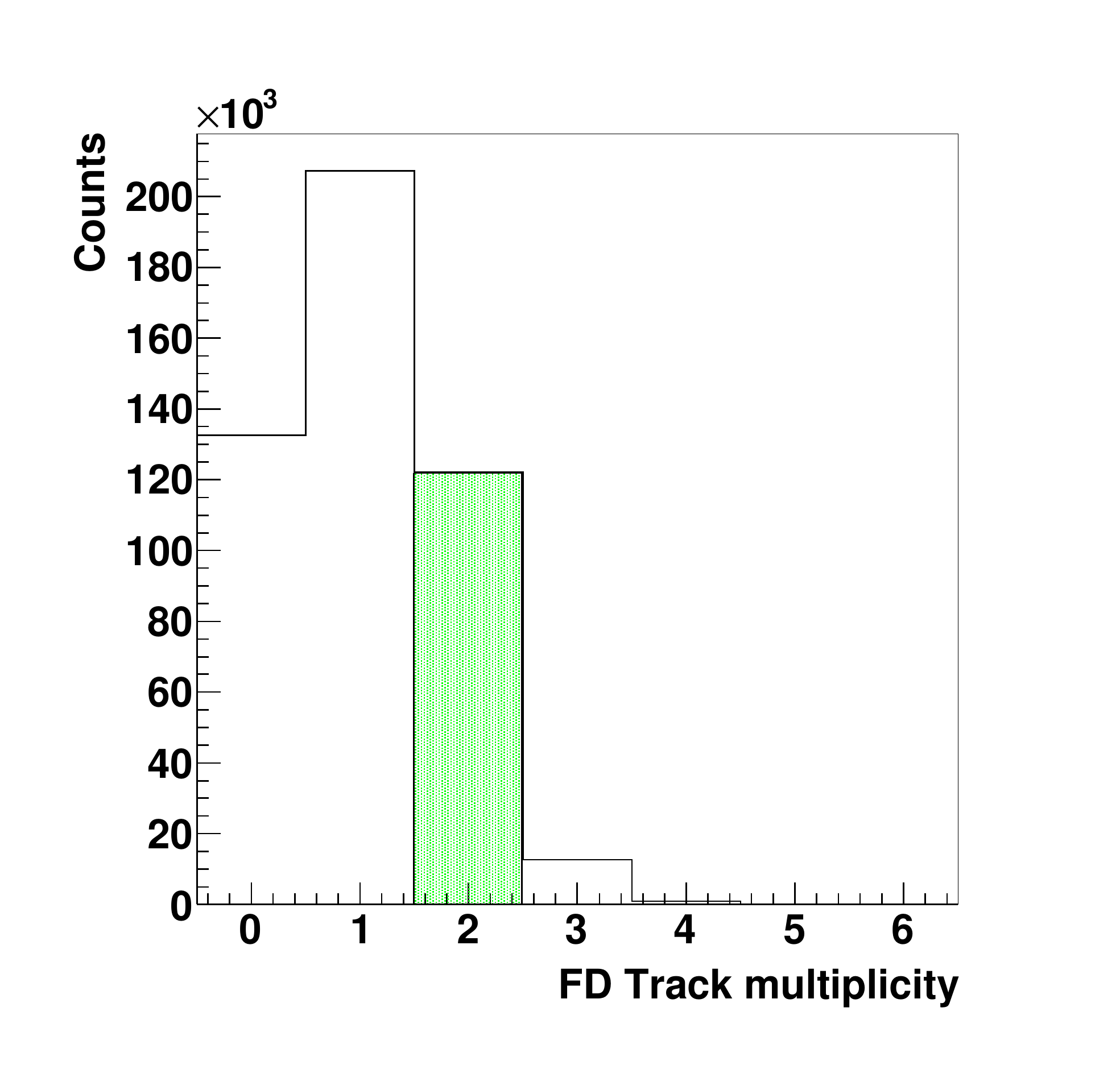,width=0.55\textwidth}}}
\caption{
{\bf{(left)}} The particle identification plot for Forward Detector.  
The correlation of energy deposited by charged particles in the first layer of the FRH detector 
as a function of energy deposited in whole Range Hodoscope. The superimposed black line indicates 
cut region from which event were taken for further analysis.  
{\bf{(right)}} Charge particle multiplicity in Forward Detector after particle identification.
The green shaded area indicate that for the further analysis only events with exactly two protons 
reconstructed were chosen.
}
\label{FD}
\end{figure}
\begin{figure}[!h]
\hspace{3.5cm}
\parbox{0.5\textwidth}{\centerline{\epsfig{file=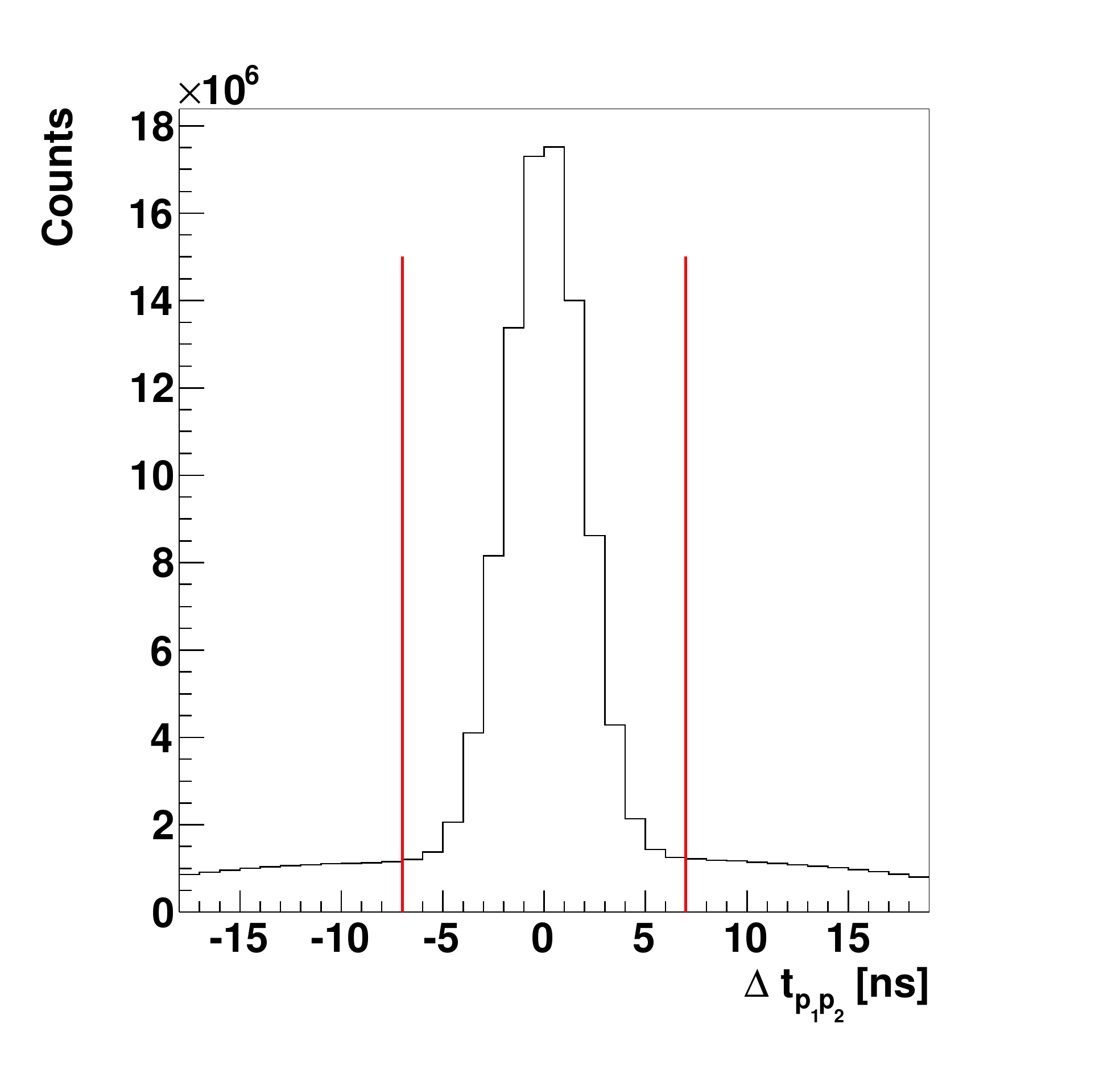,width=0.55\textwidth}}}
\caption{
The time difference between signals from protons measured in FTH. In order to 
minimize the random coincidence, the proton candidate pair has to have time 
difference within window of $\pm 7$~ns.
}
\label{pptime}
\end{figure}
The protons coming from the $pp\to ppX$ reaction are seen in Fig.~\ref{FD} (left) as two-arm band.
The condition for proton identification is indicated on the plot as a region within 
the borders of the black lines. After these cuts, only events, with two properly identified protons
were chosen for for further analysis (see Fig.~\ref{FD} (right)).

The several discontinuity in the upper proton band seen in Fig.~\ref{FD} (left) are due to 
passive material (8~mm plexi) 
which is holding elements of each layer together. When a particle is passing through 
region between layers it losses some part of its energy in plexi material. 
The detection of this loss is impossible, which is revealed as a discontinuity on the 
$\Delta E - E$ spectrum. One can also observe a structure at 
low energy deposition region, where the most of minimum ionizing pions are giving signals. 
The long tail which is localized behind the lower band originates from 
nuclear interactions of protons with the material of the detector.  
Also the continuous background is due to the nuclear interactions. 
The fraction of events where two protons were identified in the Forward Detector after applying 
above conditions is equal to 26\% (Fig.~\ref{FD} (right)). 

Further, in order to properly select two protons 
the time difference between two subsequent signals in the Forward Detector 
was checked. The pair with the time difference $\Delta t_{p_{1}p_{2}}$ 
within time window of 14~ns ($\pm 7$~ns) are selected for further analysis. 
The corresponding distribution of $\Delta t_{p_{1}p_{2}}$ is shown in Fig.~\ref{pptime}. 
One can see a very steep peak around 0~ns which is coming from protons
and almost flat background originating from random coincidences.

\section{Identification of the $\eta$ meson}\label{sec:ppeta}
\begin{figure}[!h]
\hspace{3.2cm}
\parbox{0.6\textwidth}{\centerline{\epsfig{file=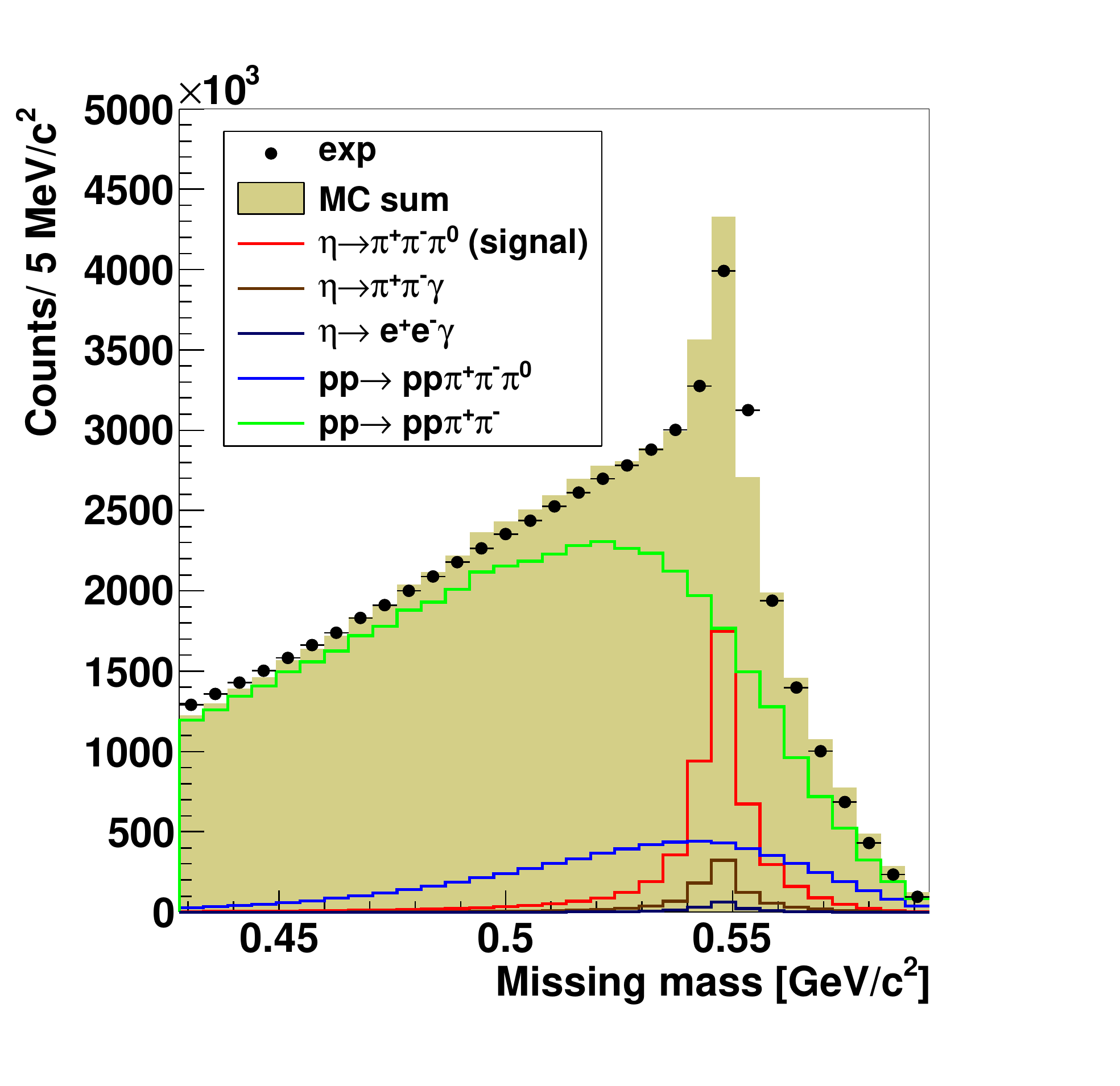,width=0.65\textwidth}}}
\caption{
Missing mass distribution for the $pp\to ppX$ reaction measured at beam momentum of $p_b = 2.14~GeV/c$
after the preselection described in this Section and identification of two protons in 
Forward Detector (black rectangles).
The solid lines indicate results of the simulations of signal and background channels as given in the figure.
The shaded area indicates the fitted sum off all simulated reaction channels.  
}
\label{ppX}
\end{figure}

\hspace{\parindent}
After reconstruction and identification of two protons in the Forward Detector originating 
from $pp\to ppX$ reaction one can plot a missing mass distribution according to 
formula~\ref{defMM} in order to identify the production of the $\eta$ meson. 
The resulting spectrum of the missing mass is shown in Fig.~\ref{ppX}. 
A peak originating from the production of the $\eta$ meson is clearly seen on the top
of a continues and broad background. The shape of the distribution outside the $\eta$ peak region
originates form the direct pions production.    
The peak seen in Fig.~\ref{ppX} contains not only the events corresponding to the final state 
of the decay $\eta\to \pi^+\pi^-\pi^0$ but also other charged decays of the $\eta$ meson which fulfil the condition of the trigger and preselection in the Central Detector. Mainly it may be the 
$\eta\to\pi^+\pi^-\gamma$ decay which has a branching ratio of 4.60~\%. 

In order to describe the shape of the background and estimate the signal to background ratio on 
this stage of the analysis several reaction channels were simulated. The main background 
constitute the $pp\to pp\pi^+\pi^-$  and $pp\to pp\pi^+\pi^-\pi^0$ processes.
For both reactions the cross section for beam energy of $E_b = 1.4~GeV$ are unknown.
However, from the shape of the excitation function for these reactions one can roughly predict that 
the ratio of the cross sections is equal to about 
$\frac{\sigma_{pp\to pp\pi^+\pi^-}}{\sigma_{pp\to pp\pi^+\pi^-\pi^0}} \approx 100$.
Moreover the simulations of the signal decays $\eta\to\pi^+\pi^-\pi^0$ and background processes like:
$\eta\to\pi^+\pi^-\gamma$ and $\eta\to e^+ e^-\gamma$ were performed.
Simulated data samples were analyzed in the same way as it was done for the experimental data. 
The results of the simulations in comparison to the experimental data are presented in Fig.~\ref{ppX}.

In order to estimate signal to background ratio simulated missing mass spectra of background reactions
were fitted to the data, excluding the signal region from 0.52 to 0.57~GeV, according to the formula:
\begin{equation}
B(mm,\alpha,\beta) = \alpha\cdot f_{pp\to pp\pi^+\pi^-}(mm) + \beta\cdot f_{pp\to pp\pi^+\pi^-\pi^0}(mm),
\end{equation}
where the $\alpha$ and $\beta$ denotes the free parameters varied during the fit, and functions
$f_{pp\to pp\pi^+\pi^-}(mm)$ and $f_{pp\to pp\pi^+\pi^-\pi^0}(mm)$  indicates the missing mass 
spectra of simulated background reactions. One can see that the simulations are in a good agreement with the 
experimental data. Obtained signal to background ratio on this stage of the analysis is equal to 0.6.

The method to further improve the signal to background ratio and select investigated decay channel 
will be presented in Section~\ref{sec:7}.

\chapter{Extraction of signal for the $\eta\to\pi^0\pi^+\pi^-$ channel}\label{sec:7}
\hspace{\parindent}
The $\eta$ meson is a particle with an average life time of $10^{-21}$~s,
thus it decays almost immediately after being produced into lighter mesons. 
In the WASA facility the decay products are detected and identified in the 
Central Detector (the detail description how the track reconstruction and clustering algorithm 
work was given in section~\ref{sec:cdtracks}).

Initially the tracks originating from the decay of the $\eta$ meson registered in Central 
Detector have to be correlated in time with two protons identified in the Forward Detector.
This is done by checking the time coincidences between protons and particles in CD.
The time of two protons in Forward Detector is determined based on signals from the FTH detector 
which is positioned 1.5~m downstream from the interaction point. 
Thus this time has to be corrected for the time-of-flight according to the equation:
\begin{equation}
t_{p} = t_{FTH} - \frac{l}{\beta\cdot c},
\end{equation}
where $l$ denotes the distance from the interaction point to the hit position in the FTH detector, 
$\beta$ indicates the proton velocity and $t_p$ stands for time when proton left the interaction region.
Finally the time of particles registered in CD is compared to the average time of both identified protons.

In the Central Detector the timing for the charged particles is provided by the PSB detector,
and for the neutral particle it is readed from the SEC. 
Both types of particles are treated separately due to different distance to the interaction point
and different time resolution of the PSB and SEC. 
The PSB is a plastic scintillating detector giving a very fast time signals with a very 
good accuracy. While, the organic crystals building calorimeter are ''slower'' and are characterized by
worser time resolutions. 
\begin{figure}[!h]
\hspace{0.1cm}
\parbox{0.5\textwidth}{\centerline{\epsfig{file=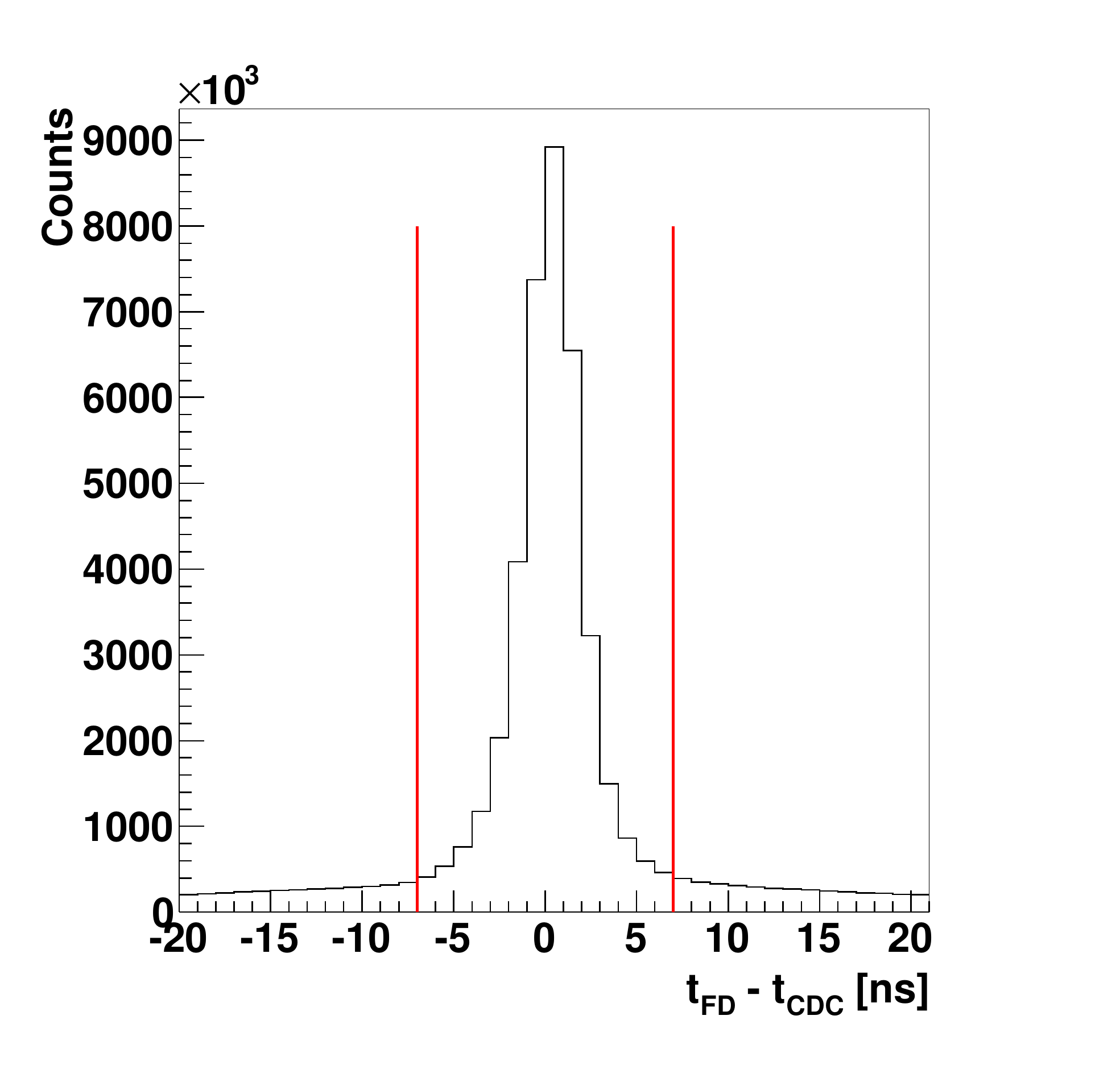,width=0.55\textwidth}}}
\hspace{0.1cm}
\parbox{0.5\textwidth}{\centerline{\epsfig{file=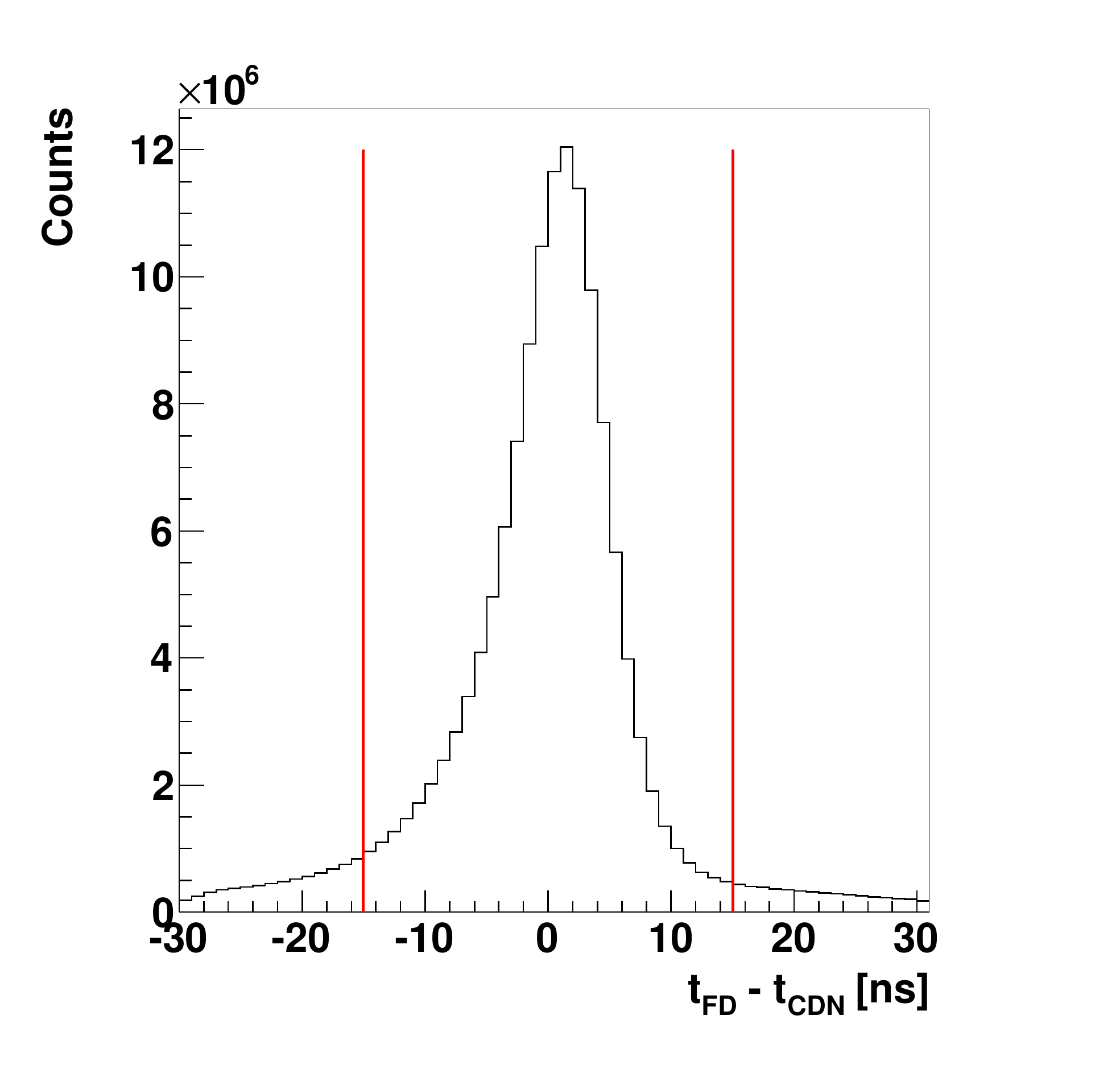,width=0.55\textwidth}}}
\caption{
Distribution of differences between time of particles registered in CD and average time of protons in FD corrected for the 
time-of-flight: 
{\bf{(left)}} for charged particles registered in PSB,
{\bf{(right)}} for neutral clusters detected in SEC.
The superimposed red lines indicate the time window of the event acceptance, where for charge 
particles it is $-7~ns \leq \Delta t_{CDC} \leq 7~ns$ and for neutral $-15~ns \leq \Delta t_{CDN} \leq 15~ns$
}
\label{timeCD}
\end{figure}
\begin{figure}[!h]
\hspace{-0.1cm}
\parbox{0.5\textwidth}{\centerline{\epsfig{file=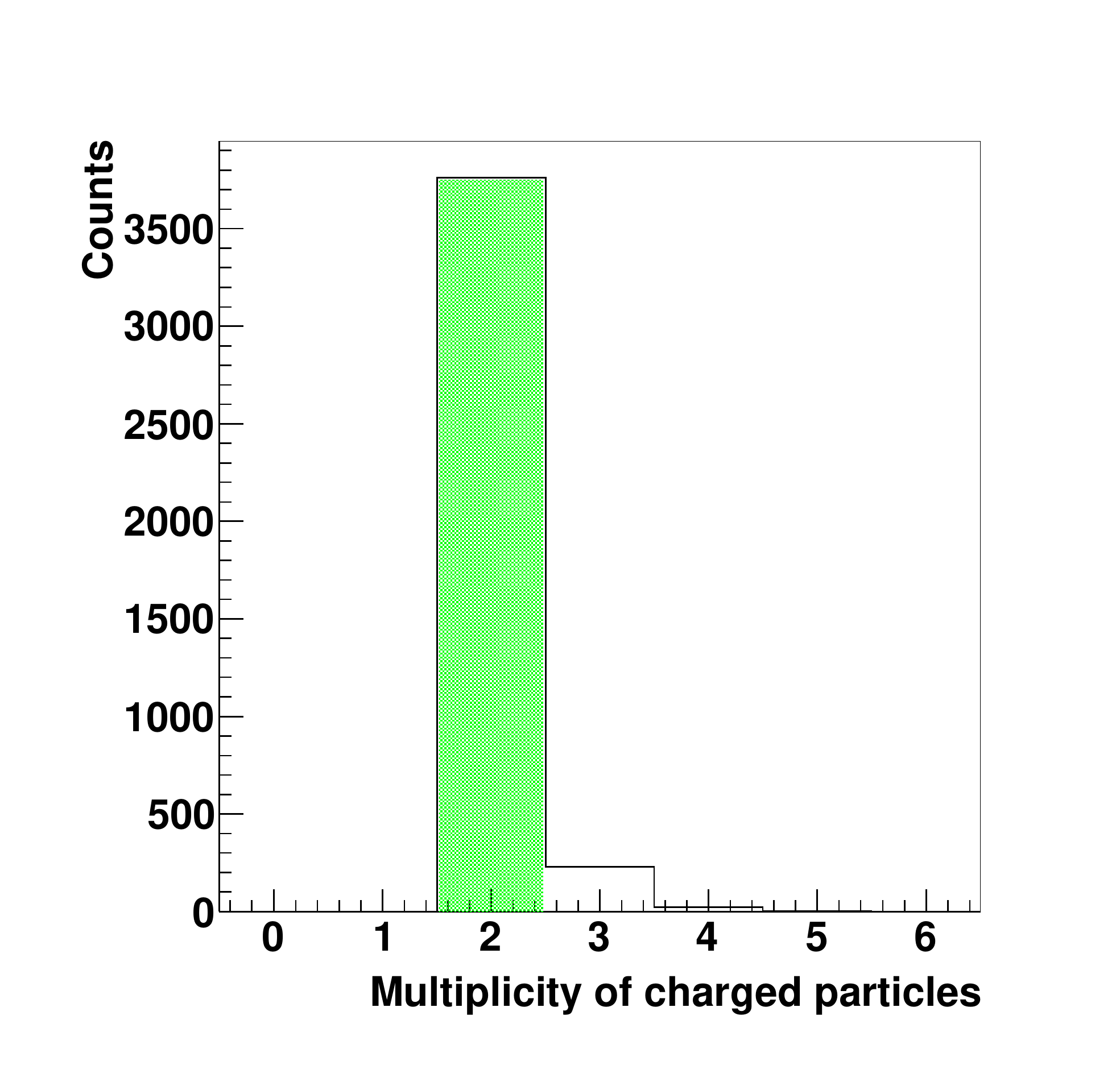,width=0.55\textwidth}}}
\hspace{0.1cm}
\parbox{0.5\textwidth}{\centerline{\epsfig{file=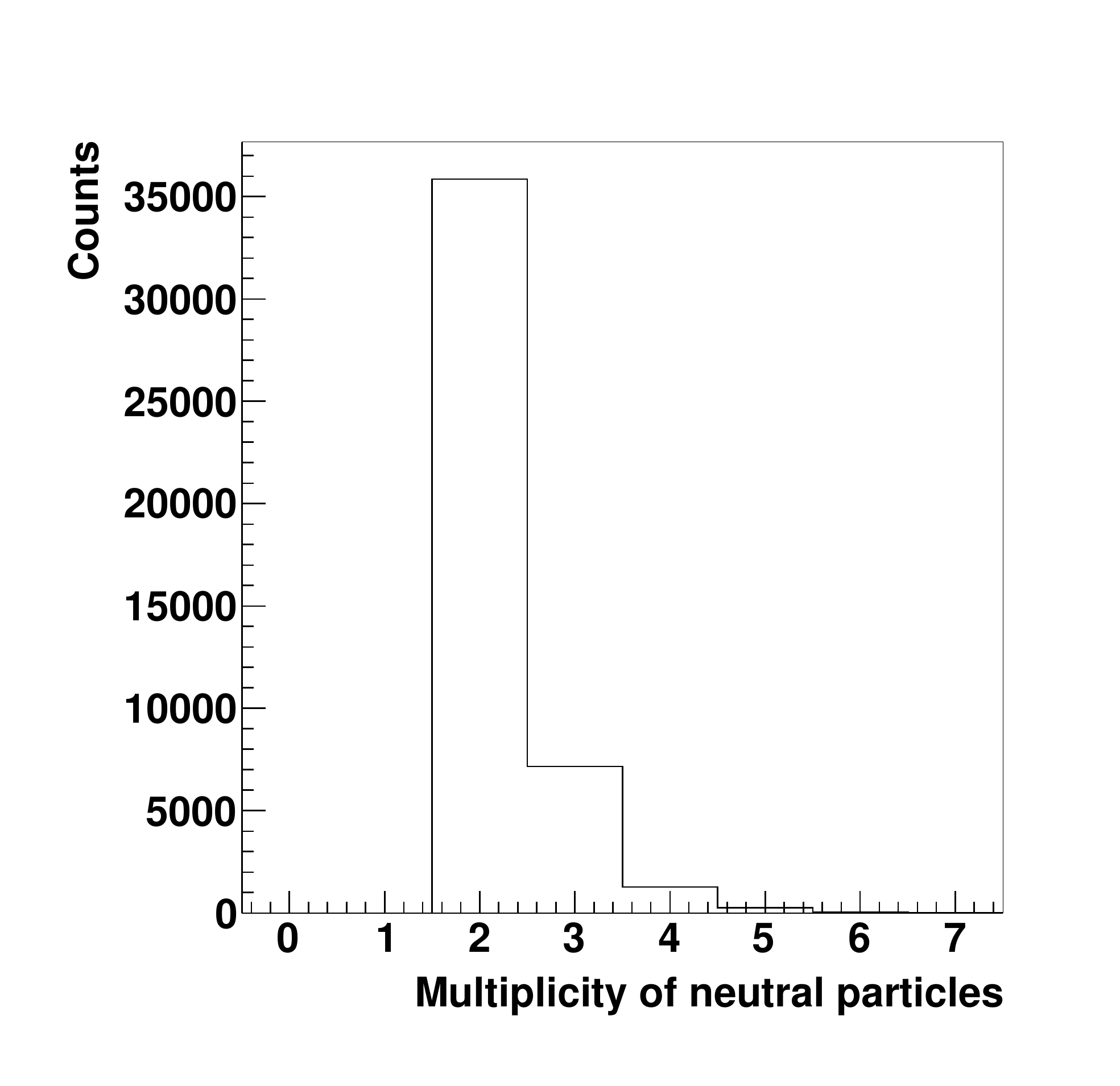,width=0.55\textwidth}}}
\caption{
{\bf{(left)}}
Multiplicity of charged particles registered in Central Detector. 
The green area indicates multiplicities equal to 2, which is taken for further 
analysis.
{\bf{(right)}}
Experimental distribution of multiplicity for neutral particles detected in Central Detector 
after identification of two protons and selection of two oppositely charged particles.
}
\label{CDCmulti}
\end{figure}
For both charge and neutral particles the time distributions (see Fig.~\ref{timeCD})
shows a pronounced peak on an almost flat background. In case of charged particles the selection window
is 14~ns wide ($\pm$~7~ns), while for the neutral particles it is 30~ns wide 
($\pm$~15~ns). 

Furthermore, the tracks are checked to fit the geometry of the Central Detector. The charged particles 
are identified only if at least 2 axial and 3 stereo straws gives a signal and this is possible 
only if the particle passes at least five layers of the MDC. Therefore, the range of the polar 
angle for the acceptance of the charged particles is equal to $24^o$ - $159^o$. 
For the neutral particles, SEC acceptance enables to register signals in the
range of $20^o$ - $169^o$.

To select candidates for the $\eta\to\pi^+\pi^-\pi^0$ decay,
the reconstructed tracks which fulfilled time and acceptance conditions 
are taken into account. Only events with exactly two charged particles 
having opposite charges, and at least two neutral clusters are saved for further analysis. 
The corresponding graph of multiplicity for charged particles is shown in Fig.~\ref{CDCmulti} (left).
The distribution shows that only in about 5~\% of all events there are 
more than two charged particles identified. Therefore, for further analysis we 
take only these events were exactly two oppositely charged particles were reconstructed in the Central Detector.

The multiplicity of neutral particles registered in SEC, after identification 
of protons in FD and two oppositely charged particles in CD is shown in Fig.~\ref{CDCmulti} (right). 
From this distribution one can see that in most cases the two 
neutral particles were registered in the calorimeter.
The events in which number of neutral clusters is greater than two is partially due 
to splitting and hadronic interactions of charged particle with the scintillator. 
At this stage of analysis we accept all the events independently of the 
multiplicity of neutral particles candidates.

\section{Identification of charged pions $\pi^\pm$}\label{sec:pipi}
\hspace{\parindent}
The selection of charged pions bases on momentum reconstruction from the curvatures of particle 
trajectories in magnetic field in the Mini Drift Chamber 
combined with the information about energy losses in the Plastic Scintillator Barrel
and the Scintillating Electromagnetic Calorimeter ($\Delta E - \vert\vec{p}\vert$ method). 

\begin{figure}[!h]
\hspace{0.0cm}
\parbox{0.5\textwidth}{\centerline{\epsfig{file=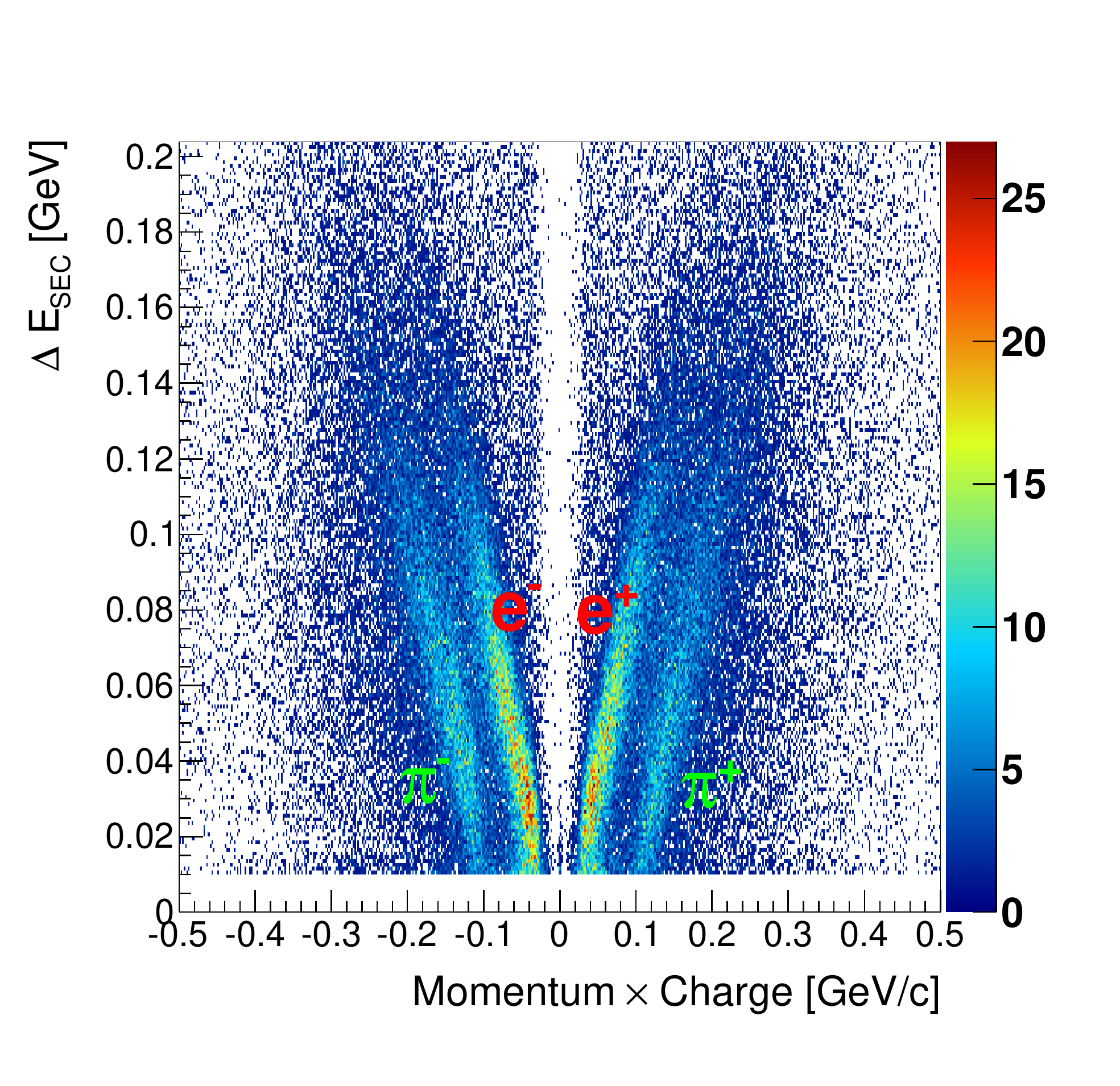,width=0.55\textwidth}}}
\hspace{0.2cm}
\parbox{0.5\textwidth}{\centerline{\epsfig{file=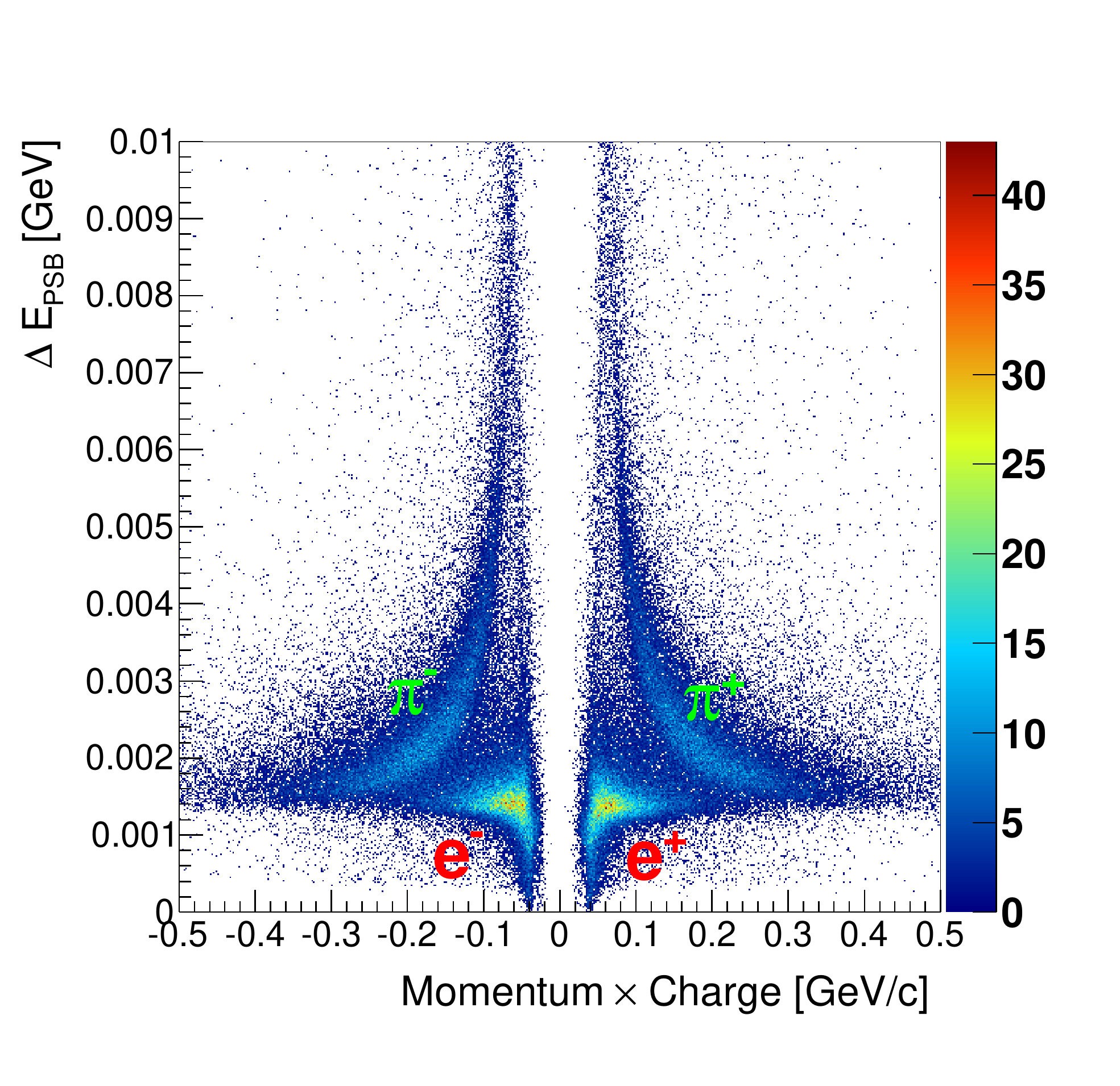,width=0.55\textwidth}}}
\caption{
Simulated distribution of the energy deposited in the calorimeter {\bf{(left)}}, 
and in plastic scintillator {\bf{(right)}} as a function of the particles momentum and charge. 
The structures on the left part of both spectra corresponds to particles 
with -1 and on the right side with +1 charge state. 
The bands corresponding to pions and electrons are marked respectively.
}
\label{MCpidCDC}
\end{figure}

For both detectors PSB and SEC the correlation between energy 
deposited in the scintillating material by charged particle and absolute value of momentum 
can be graphed. 
In this method for the different types of particles the correlation of energy loss and momentum 
differ and leave distinct bands which can be used for separation and identification purpose.
Figure~\ref{MCpidCDC} shows corresponding identification spectra obtained from the 
Monte Carlo simulations for the $\eta\to\pi^+\pi^-\pi^0 (\pi^0\to e^+e^- \gamma)$ decay. 
For both detectors four densely populated areas are clearly visible.
For better visualization the momenta are multiplied by sign of charge which enable 
to separate in the figure negatively and positively charged particles.

\begin{figure}[!h]
\hspace{0.0cm}
\parbox{0.5\textwidth}{\centerline{\epsfig{file=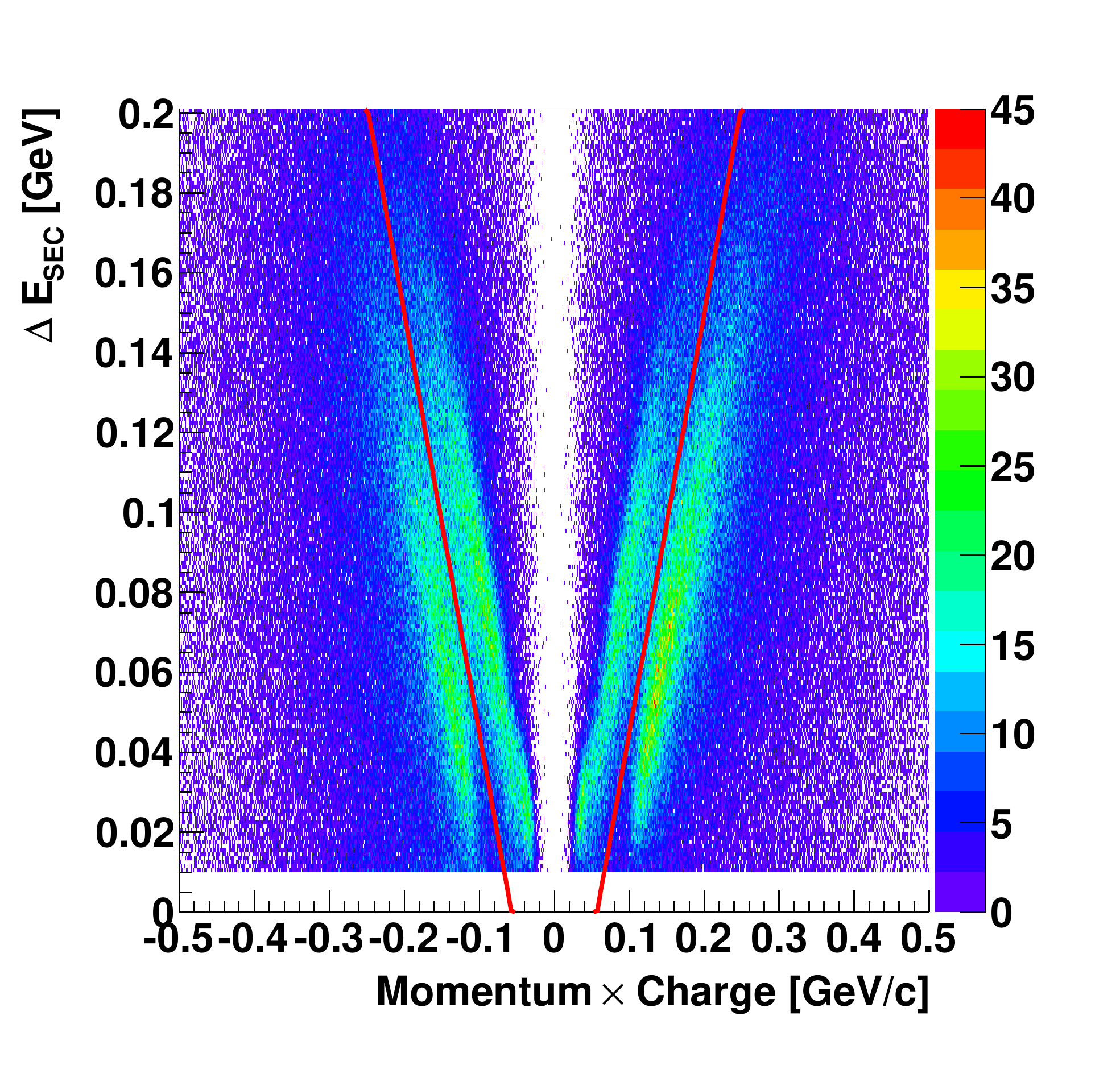,width=0.55\textwidth}}}
\hspace{0.2cm}
\parbox{0.5\textwidth}{\centerline{\epsfig{file=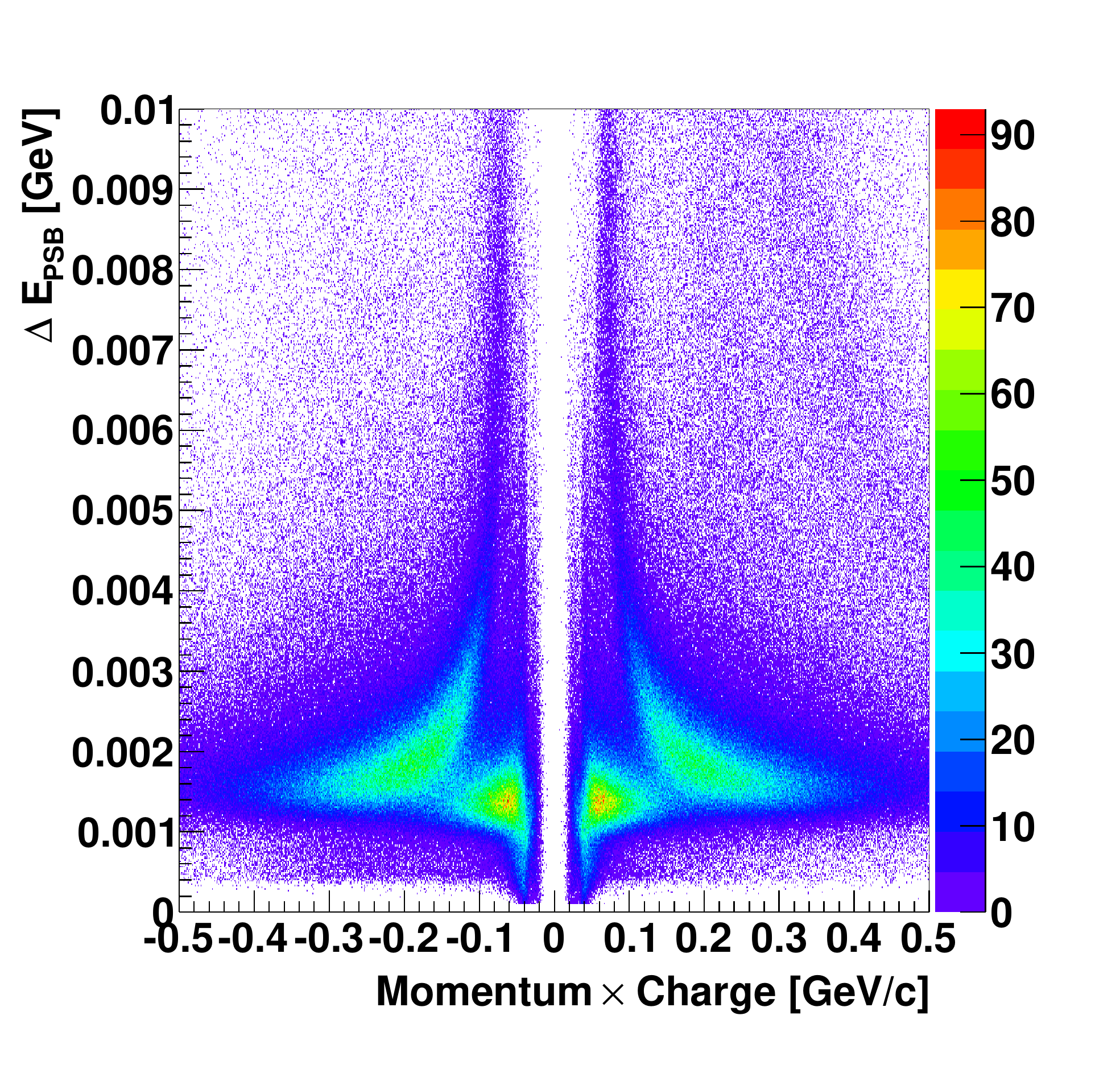,width=0.55\textwidth}}}
\caption{
Experimental $\Delta E - \vert\vec{p}\vert$ spectra used for particle 
identification in Central Detector. Energy deposit as a function on momentum and charge: 
in the SEC {\bf{(left)}}, and in PSB before applying identification condition in SEC {\bf{(right)}}.
The superimposed lines in left panel indicate the cut regions used to select signal events.
}
\label{pidCDC}
\end{figure}

In the calorimeter pions and electrons deposit their total kinetic energy, which at given 
momentum is larger for light electrons than for heavier pions.  
This allows to impose the identification conditions to distinguish between both type of particles as
can be seen in Fig.~\ref{pidCDC} (left). The red line imposed on the plot
indicates the pion identification condition given by the inequality: 
$\Delta E_{SEC} < 1.05\cdot\vert\vec{p}\vert\cdot q - 0.06$. 

To apply the same method in PSB an energy loss is normalized to the path length of the particle
when passing through the detector. 
One can also see that in case of $e^-$ and $e^+$ particles the energy loss in the whole range 
of the momentum is almost constant. In this case for a given momentum electrons deposit less
energy than pions because energy loss decreases with increase of velocity. 
Experimental spectra of pions and leptons similarly as in SEC, reveal separate 
bands as it is visible in Fig.~\ref{MCpidCDC} (right).

\begin{figure}[!h]
\hspace{3.5cm}
\parbox{0.5\textwidth}{\centerline{\epsfig{file=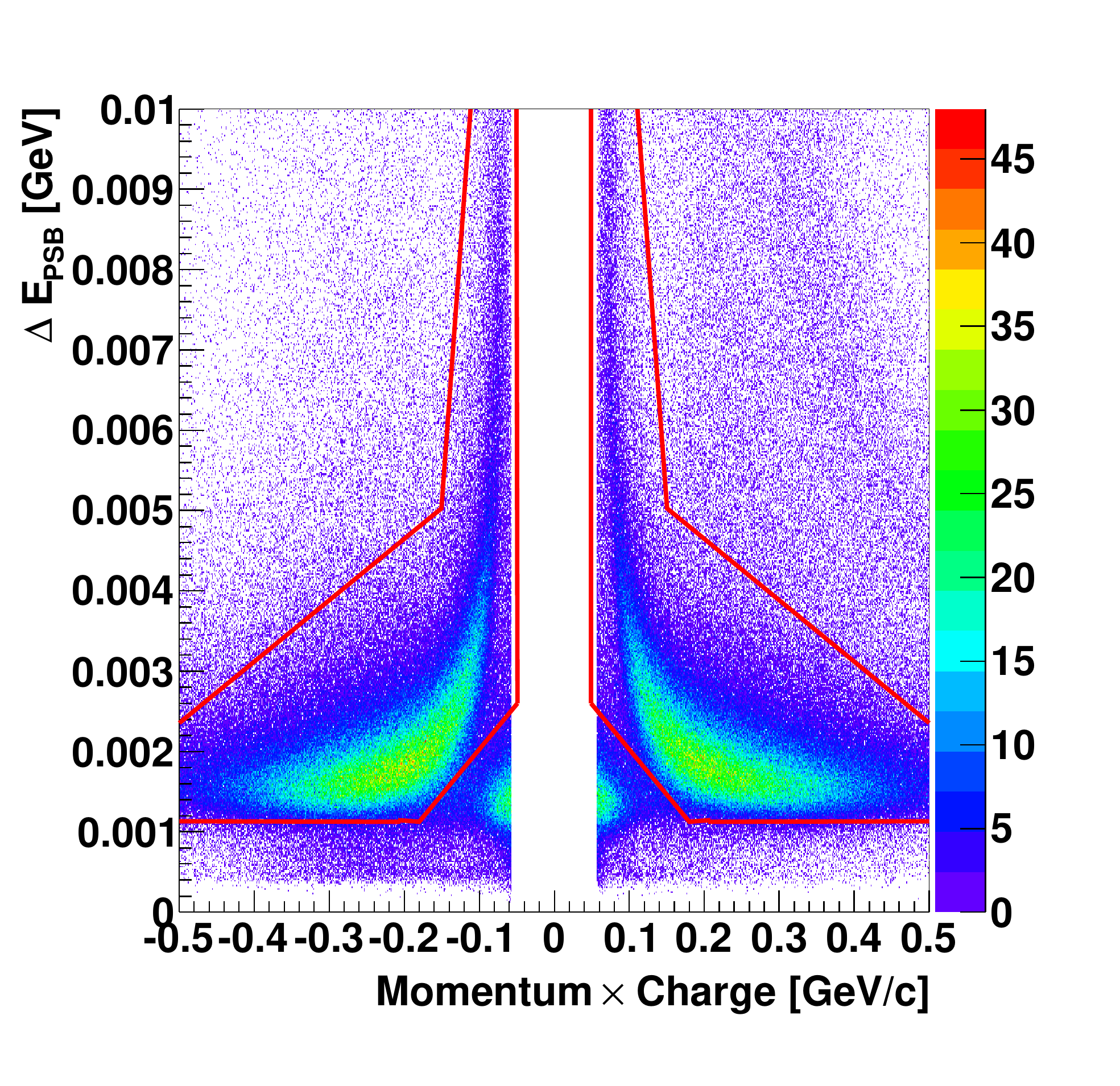,width=0.55\textwidth}}}
\caption{
Experimental $\Delta E - \vert\vec{p}\vert$ spectra used for particle 
identification in PSB after applying identification condition in SEC, with superimposed lines 
indicating the cut regions used to select signal events.
}
\label{pidPSB}
\end{figure}

In case of this analysis first selection of pions with the conditions on the correlation 
of energy and momentum in Scintillating Electromagnetic Calorimeter are applied (Fig.~\ref{pidCDC} (left)), 
and further events which met this requirement are checked in the Plastic Scintillator Barrel Fig.~\ref{pidPSB}. This enables to reject particle misidentified in the SEC. 
\begin{figure}[!h]
\hspace{0.0cm}
\parbox{0.5\textwidth}{\centerline{\epsfig{file=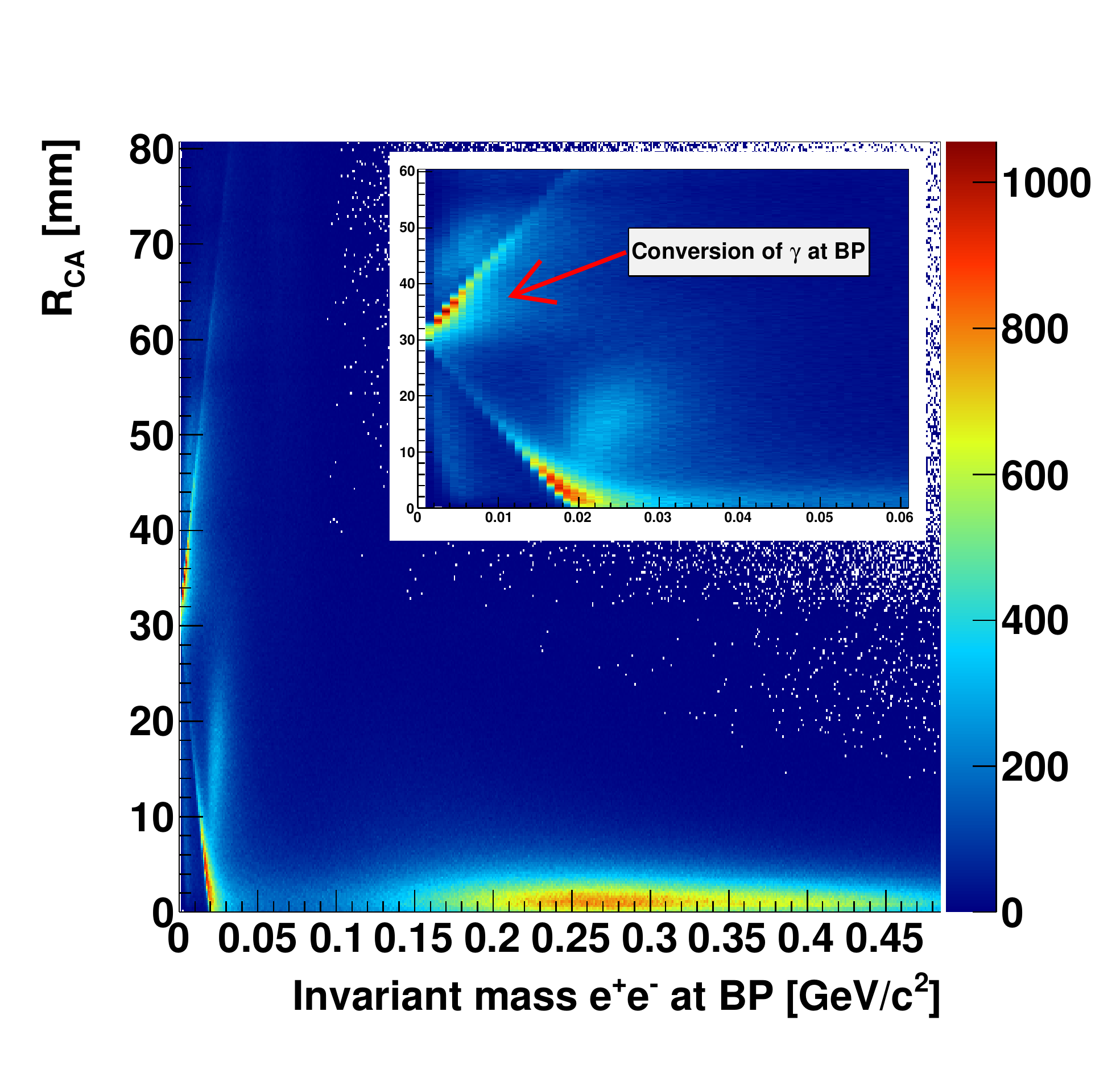,width=0.55\textwidth}}}
\hspace{0.5cm}
\parbox{0.5\textwidth}{\centerline{\epsfig{file=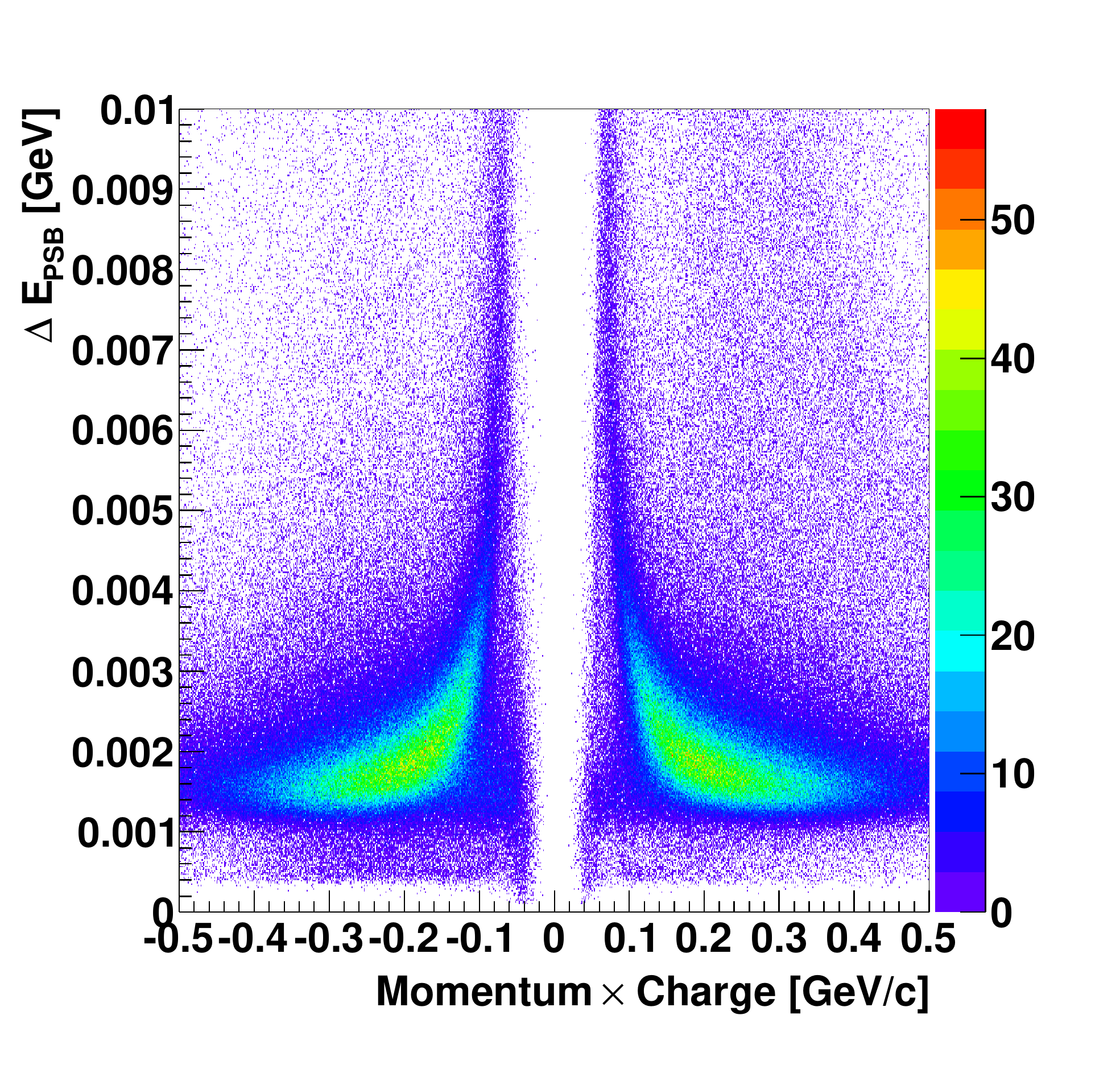,width=0.53\textwidth}}}
\caption{
{\bf{(left)}} Experimental distribution of the distance $R_{CA}$ between interaction point and the 
reconstructed $e^+e^-$ vertex versus the invariant mass of $e^+e^-$ pair calculated 
assuming that lepton pair originates from the beam pipe. 
{\bf{(right)}} Distribution of the $\Delta E - \vert\vec{p}\vert$ for PSB detector after demanding 
only events with $R_CA < 10$~mm and invariant masses of $e^+e^-$ grater than 0.07 GeV/c$^2$. 
}
\label{cabp}
\end{figure}

One can also see that in the experimental spectra for the PSB detector (Fig.~\ref{pidCDC} (right)) 
the $e^\pm$ bands are much more densely populated than $\pi^\pm$. This picture can be confusing 
because the cross section for the pion production is few orders of magnitude higher than for electrons. 
But one has to stress that the identification plots are made after demanding at least two neutral 
clusters in the electromagnetic calorimeter. Thus, the number of pions  
from the direct production via channel $pp\to pp\pi^+\pi^-$ drastically decreases and
they can only pass the requirements when two fake clusters (for e.g. due to hadronic 
split-off) are wrongly identified as photon. 
Furthermore, requirement of at least two neutral particles, enhances in the selected sample 
fractional number of events corresponding to the 
conversion process of photons on a beryllium beam pipe. One can suppress the background 
originating from the external photon conversion based on the correlation of the distance between 
the center of the interaction region and the point of 
closest approach $R_{CA}$ of two helices, and the invariant mass $IM_{BP}(e^+e^-)$ calculated 
assuming that the $e^+e^-$ pair was created in the beam pipe.
The dilepton pair originating from the conversion process creates small values of the invariant mass and 
the minimal $R_{CA}$ value should be around 30~mm (for particles flying perpendicular to the beam pipe) 
which is the radius of the beam pipe. The corresponding correlation between the radius of closest approach 
and the invariant mass calculated on the beam pipe is illustrated in Fig.~\ref{cabp} (left).
The conversion and non-conversion events are very clearly separated. One can see that 
conversion process enhances density population in the region of low invariant masses and 
$R_{CA} > 30$~mm. The electrons and pions originating from production reactions populate 
larger invariant masses and distance $R_{CA}$ close to 0. 
Furthermore, in order to check how the conversion events influence the identification distributions
we plotted the $\Delta E - \vert\vec{p}\vert$ for the PSB detector under the condition that 
$R_CA$ is smaller than 10~mm and invariant masses $e^+e^-$ are greater than 0.07 GeV/c$^2$.
The resulting spectrum can be seen in Fig.~\ref{cabp} (right). After rejecting the conversion 
event region one can see that the $e^\pm$ band is almost invisible. However, it is important to 
stress that, this restriction is not used in the further analysis of the decay $\eta\to\pi^+\pi^-\pi^0$. 

\section{Identification of $\gamma$ quanta and $\pi^0$ mesons}\label{sec:pi0}
\hspace{\parindent}
The reconstruction of the $\pi^0$ meson relies on registration and identification of 
two gamma quanta in the Scintillating Electromagnetic Calorimeter.
As it was shown in Fig.~\ref{CDCmulti} (right) for fraction of about 15\% of events,
due to splitting, more than two clusters were reconstructed. 
For these events to select pair originating from the neutral pion decay we apply the chi-square test  
according to the equation: 
\begin{equation}
\chi^2_{ij} = \frac{(IM(\gamma_i\gamma_j) - m_{\pi^0})^2}{\sigma^2_{\pi^0}},
\label{chigg}
\end{equation}
where the $m_{\pi^0} = 134.98$~MeV denotes mass of the neutral pion, $IM(\gamma_i\gamma_j)$ 
is the invariant mass of two photon candidates, and $\sigma^2_{\pi^0}$ is the experimental mass resolution. 
The invariant mass is calculated according to the formula:
\begin{equation}
IM(\gamma_i\gamma_j) = \sqrt{\vert \mathbb{P}_{\gamma} + \mathbb{P}_{\gamma}\vert^2} =  2\sqrt{E_{\gamma_i}E_{\gamma_j}}\sin{\frac{\theta_{\gamma_i\gamma_j}}{2}},
\label{Eqimgg}
\end{equation}
where $E_{\gamma}$ and $\theta_{\gamma_i\gamma_j}$ indicate the energy of photon candidates and
opening angle between them, respectively. 
\begin{figure}[!h]
\hspace{0.0cm}
\parbox{0.5\textwidth}{\centerline{\epsfig{file=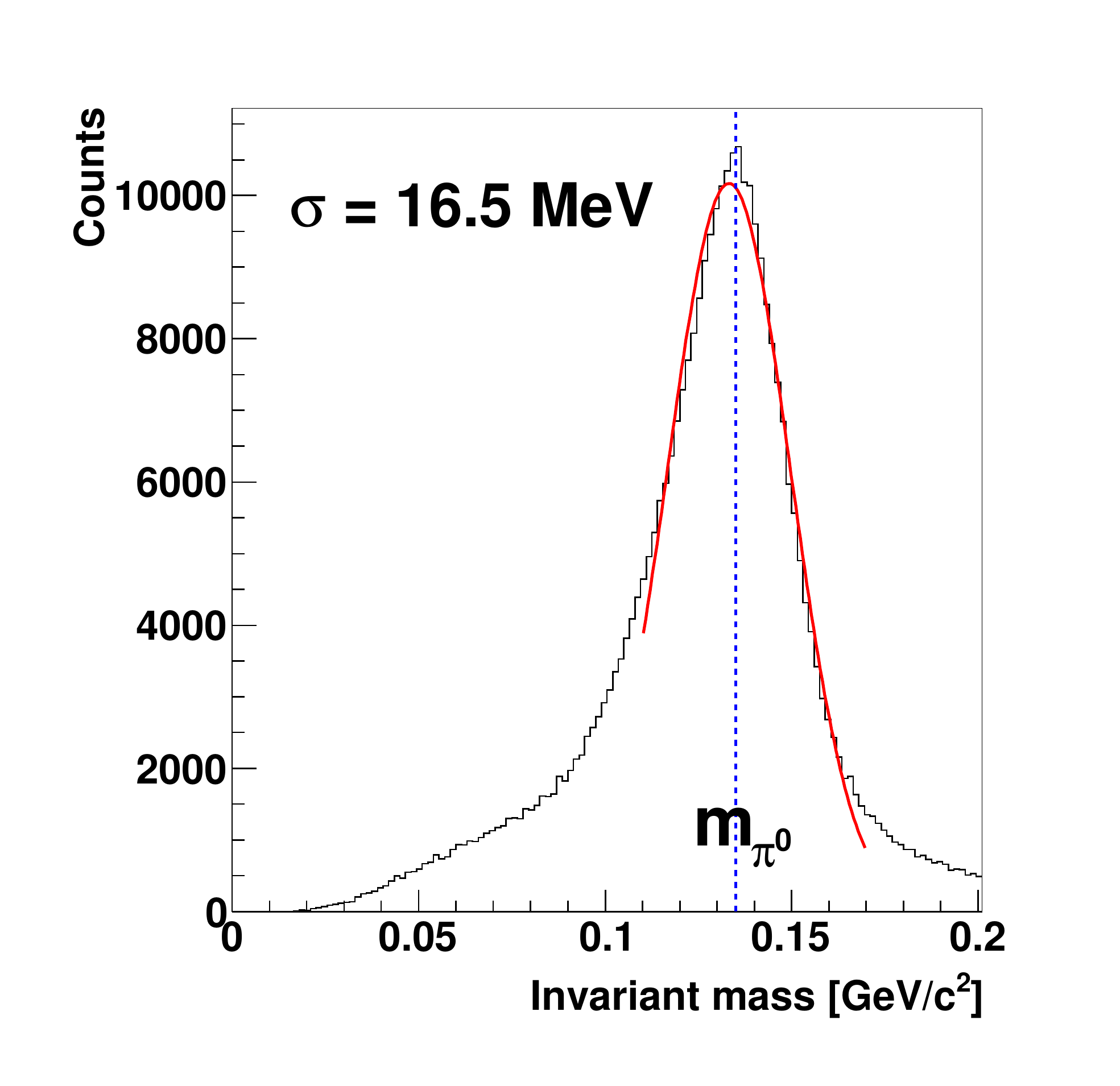,width=0.55\textwidth}}}
\hspace{0.1cm}
\parbox{0.5\textwidth}{\centerline{\epsfig{file=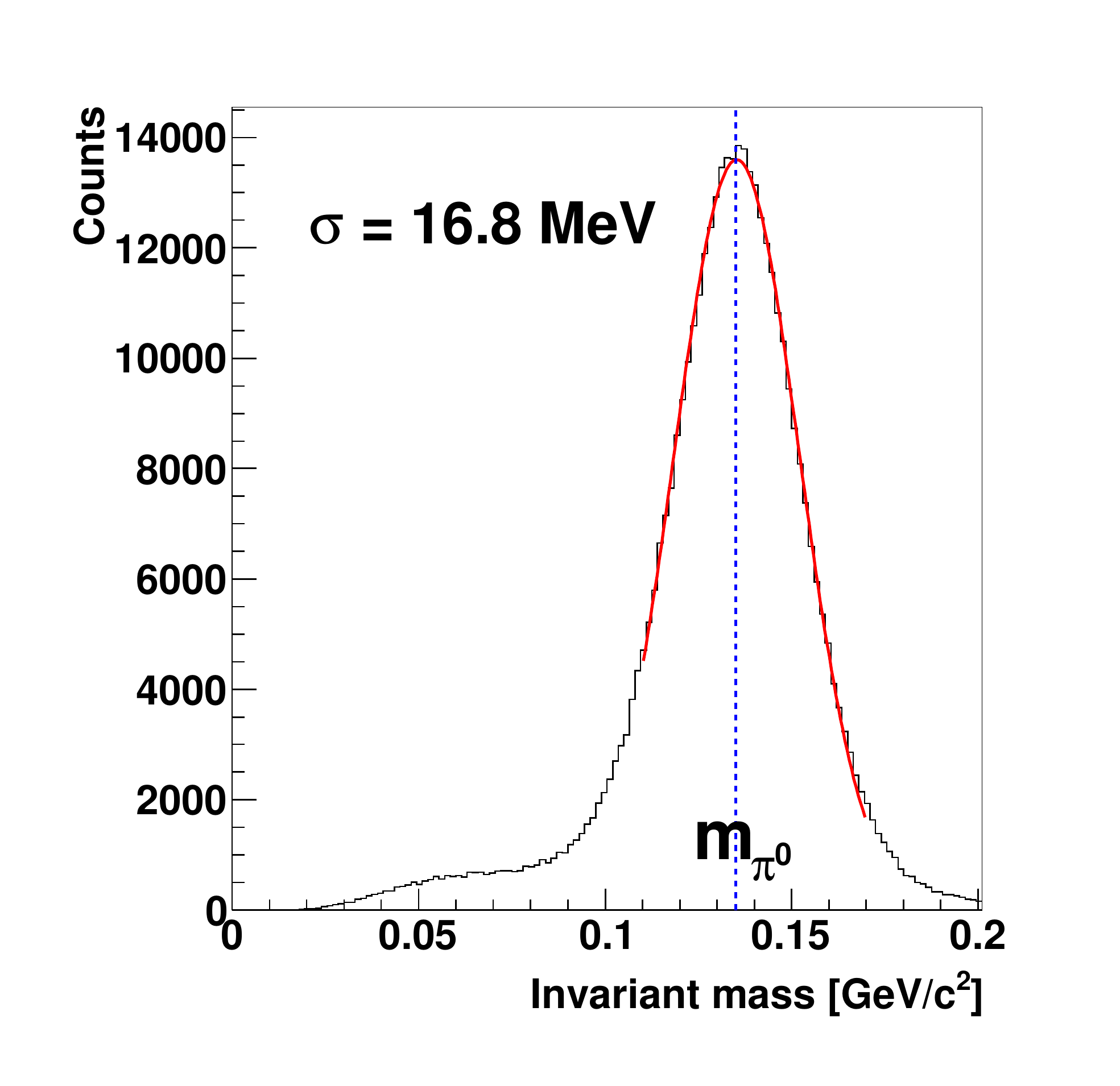,width=0.55\textwidth}}}
\caption{
The invariant mass distribution for two photons registered in the calorimeter:
{\bf{(left)}} for the experimental data, {\bf{(right)}} for the simulated sample. 
In the case of three photon candidates only these pairs were taken into account for which the 
$\chi^2$ defined in eq.~\ref{chigg} was minimal.
In both cases a clear peak is visible at the mass of the neutral pion $m_{\pi^{0}} = 134.98~MeV$. 
The superimposed solid line indicates the Gaussian fit in order to determine the resolution 
and position of the invariant mass peak. The obtained resolution for both experimental 
and simulated data equals to $\sigma \approx 17~MeV$. 
}
\label{imgg}
\end{figure}

For further analysis we accept only these two $i,j$ clusters for which the $\chi^2_{ij}$ is minimal.
Furthermore to clear the data sample and reject split-off effect of photons we have taken into account only 
pairs of clusters with opening angles greater than $40^o$.  The photon pair with lower values of 
opening angles are mainly coming from split-off processes and are contributing to the background.
The spectrum of the invariant mass of accepted pairs of photons is shown in Fig.~\ref{imgg},
where distribution on the left panel was obtained from experimental data sample and in the right panel 
from the simulated data. 

Experimentally obtained distribution of invariant mass of two photons shows a clear peak 
at the mass of the $\pi^0$ meson. To compare and tune the simulation of the detector 
response we have plotted the same spectra for the simulated sample, and fitted the 
peak region with the gauss function. Parameters  obtained for both spectra are 
in agreement. The invariant mass resolution for the experiment and simulation 
is almost the same and equals to $\sigma \approx 17 MeV$. Thus the Monte Carlo 
simulations are well tuned to the experimental conditions. 

\section{Background suppression}\label{sec:bcg}
\begin{figure}[!h]
\hspace{0.0cm}
\parbox{0.3\textwidth}{\centerline{\epsfig{file=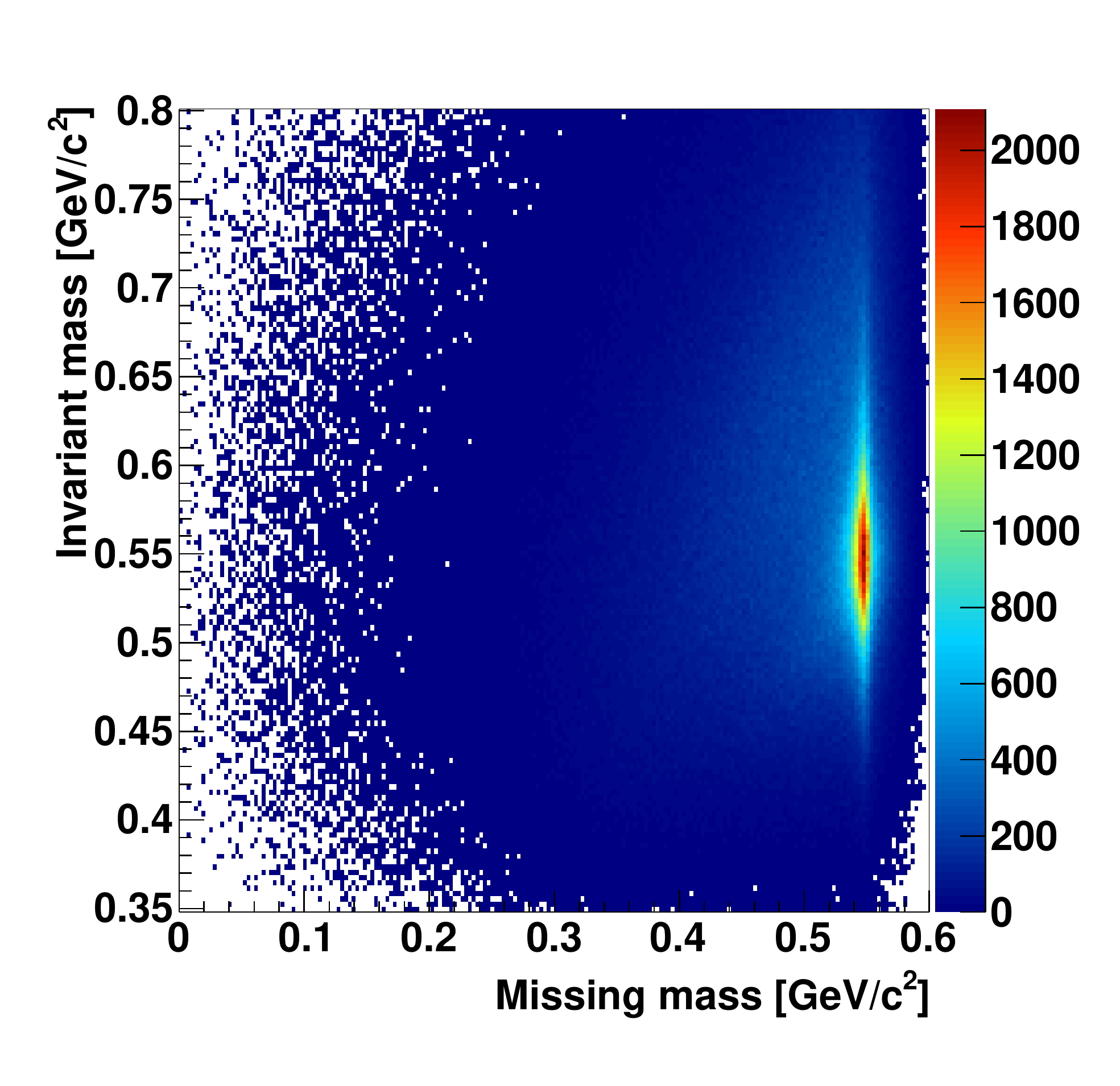,width=0.35\textwidth}}}
\hspace{.4cm}
\parbox{0.3\textwidth}{\centerline{\epsfig{file=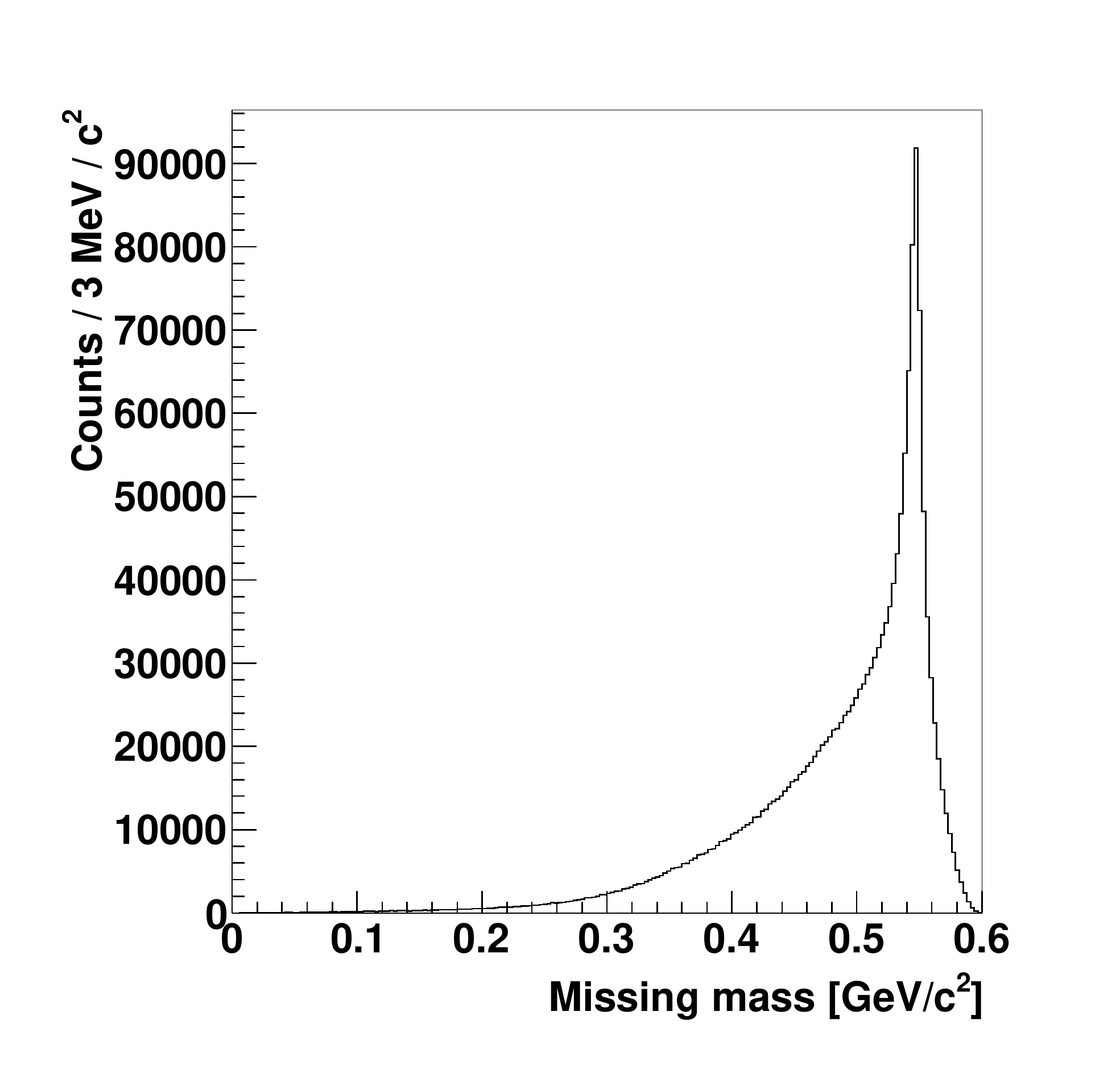,width=0.35\textwidth}}}
\hspace{.4cm}
\parbox{0.3\textwidth}{\centerline{\epsfig{file=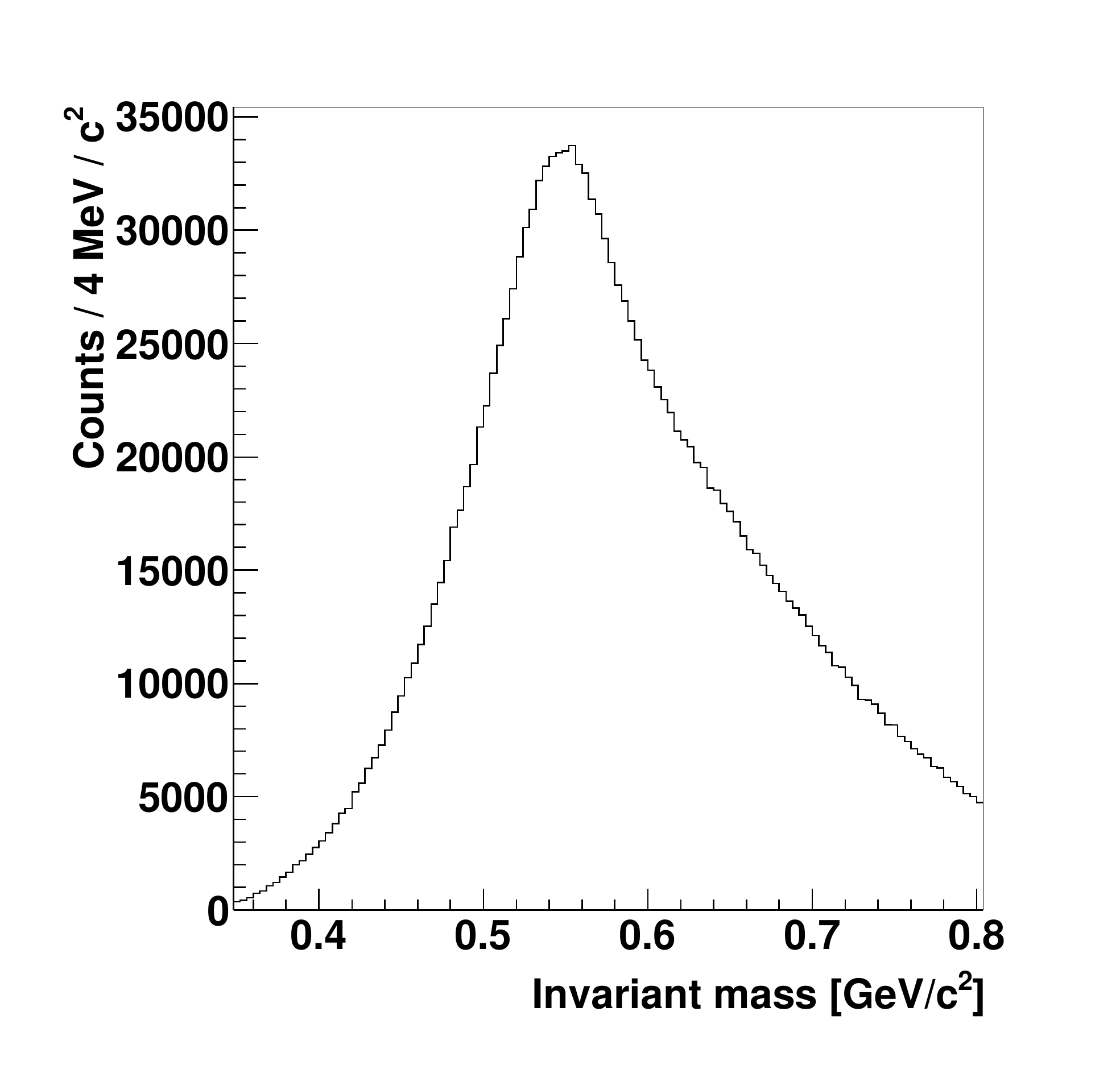,width=0.35\textwidth}}}
\caption{
{\bf{(left)}} 
The experimental distribution of the invariant mass of the $\pi^+\pi^-\pi^0(\gamma\gamma)$ 
system as a function of the missing mass for the $pp\to ppX$ reaction, after particles 
selection and identification introduced in previous sections.
{\bf{(middle)}} Experimental distribution of the missing mas for the $pp\to ppX$.
{\bf{(right)}}  Experimental distribution of the invariant mass of the $\pi^+\pi^-\pi^0$system.
}
\label{MMppIMpipipi}
\end{figure}

\hspace{\parindent}
After selecting complete reaction chain which reads $pp\to pp\eta\to pp\pi^+\pi^-\pi^0(\gamma\gamma)$
and identifying all the particles in the final state using methods described in previous sections,
one can plot distribution of the missing mass of the $pp\to ppX$ reaction as a function of the 
invariant mass of the decay products $\pi^+\pi^-\pi^0(\gamma\gamma)$. The relevant distribution is 
shown in Fig.~\ref{MMppIMpipipi}. A clear enhancement in the number of entries is visible 
at the mass of the $\eta$ meson. 
However, a large amount of background is still present in the data sample. 
The remaining background is caused mostly by the two processes: (1) direct production of 
two pions via $pp\to pp\pi^+\pi^-$ reaction, and (2) by direct production of three pions 
via $pp\to pp\pi^+\pi^-\pi^0(\gamma\gamma)$ reaction chain. For the two pion production, the previously described 
event selection and identification method turned out to be insufficient due to splitting of signal in the 
calorimeter. Nevertheless, this 
background contribution can be further suppressed using methods which will be described in this section.

\begin{figure}[!h]
\hspace{.0cm}
\parbox{0.5\textwidth}{\centerline{\epsfig{file=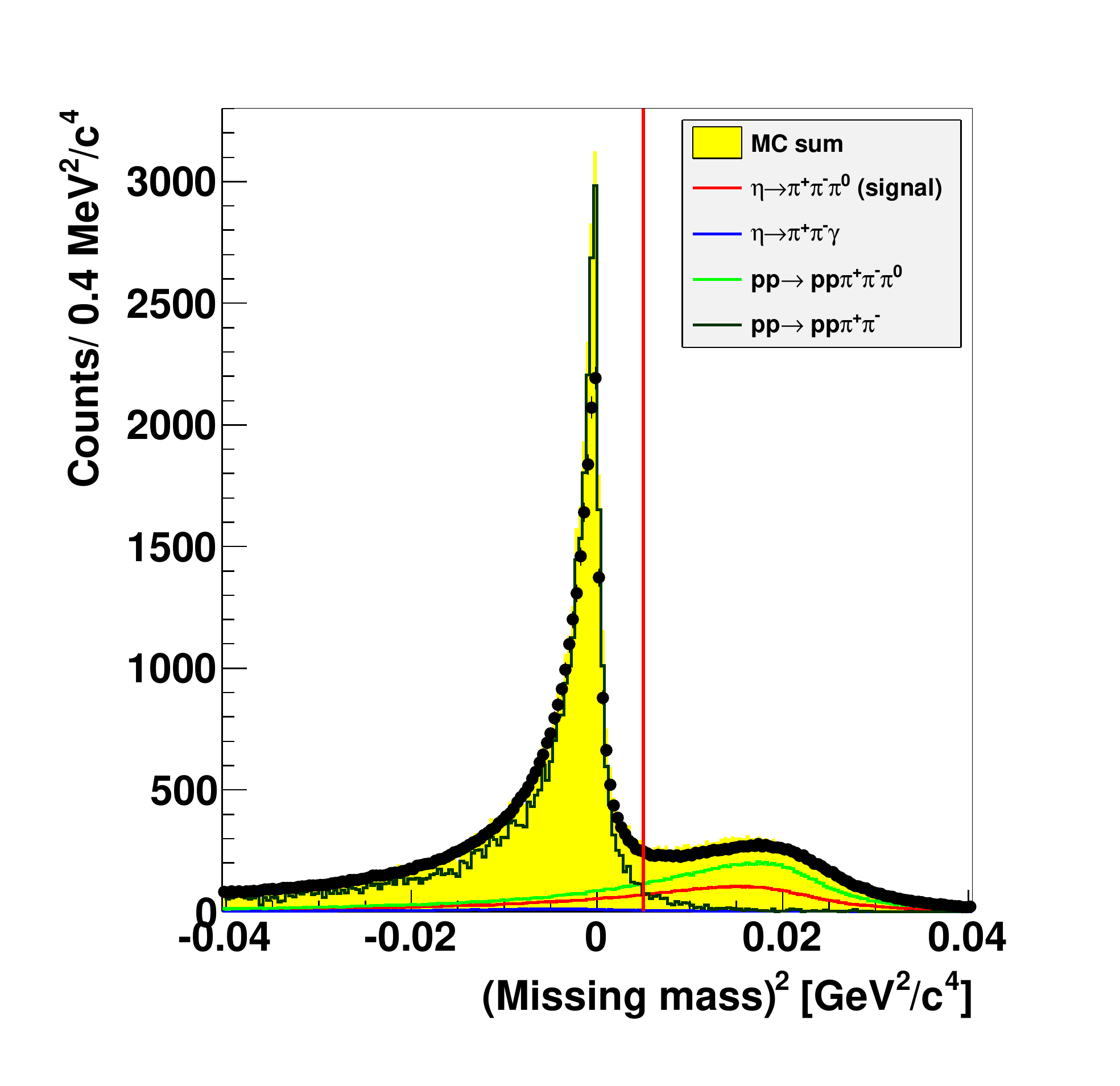,width=0.55\textwidth}}}
\hspace{.2cm}
\parbox{0.5\textwidth}{\centerline{\epsfig{file=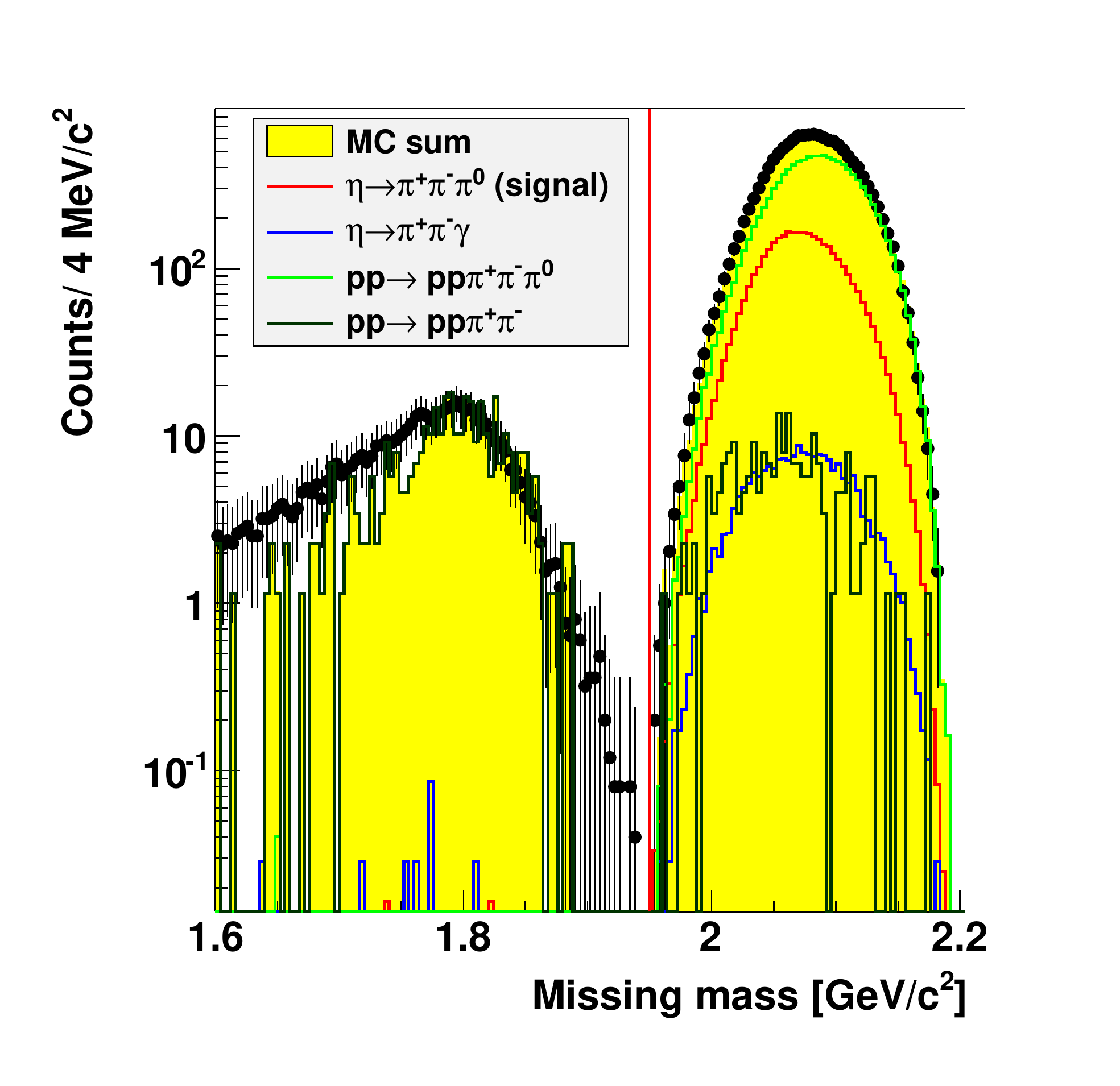,width=0.55\textwidth}}}
\caption{ 
Distribution of the missing mass squared to the system $pp \to pp\pi^+\pi^-X$ {\bf{(left)}} and 
missing mass to the $pp\to \pi^+\pi^- X$ reaction as obtained after the requirement that missing mass 
on the left figure is larger than 0.005~GeV~$^{2}/c^{2}$ {\bf{(right)}}. In both panels solid lines indicate
reconstructed events from simulations of signal reaction 
$pp\to pp\eta\to pp\pi^+\pi^-\pi^0\to pp\pi^+\pi^-\gamma\gamma$ (red histogram),
background $\eta$ decay via channel $pp\to pp\eta\to pp\pi^+\pi^-\gamma$ (blue histogram),
three pion direct production $pp\to pp\pi^+\pi^-\pi^0\to pp\pi^+\pi^-\gamma\gamma$ (green histogram) 
and two pion production $pp\to pp\pi^+\pi^-$ (dark green histogram), black points denote the 
experimentally measured distributions, and the yellow area is sum of all simulated reactions. 
The superimposed line in both plots indicates the cut chosen to suppress two pion background. 
The spectrum for  $X=pp\pi^0$ is spread below the sum of masses $2\cdot m_p + m_{\pi^0}$ because 
of the finite experimental resolution of determining the four momentum vectors.
}
\label{mmpppipiX}
\end{figure}

The contribution of the direct two pion production can be identified using the momentum 
and energy conservation. One can look at the spectrum of the square of the 
$pp\to pp \pi^+\pi^- X$ reaction, where in case of the signal reaction X denotes missing neutral pion. 
Reconstructed missing mass of desired reaction should be around the squared mass of the 
neutral pion $m_{\pi^0}^2 = 0.018~GeV^2/c^4$, and the mass for reactions where the $\pi^0$ was not
produced should be around zero. Left panel of Fig.~\ref{mmpppipiX} shows corresponding missing mass spectrum
for simulated signal and background reactions (solid lines), as well as the experimental data (black points). 
One can observe two peaks. First located around zero corresponds to the background events
where no additional particle was produced (or it was massless). The second peak corresponds 
to three pion production in the final state located around squared $\pi^0$ mass.
To further suppress the background from the two pion final state, 
all events with squared missing mass 
of the $pp \pi^+\pi^-$ system lower then 0.005 $GeV^2/c^4$ are rejected. 
The applied condition is indicated as a solid red line in  Fig.~\ref{mmpppipiX} (left). 
It reduces the 90\% of the $\eta\to\pi^+\pi^-\gamma$ background and 
97\% of the $pp\to pp\pi^+\pi^-$ decreasing the efficiency for the signal by 41\%.

The remaining background can be suppressed by looking at the missing mass for the 
$pp\to \pi^+\pi^- X$ reaction, where for the signal ($X=pp\pi^0$), $m_X$ should be larger than 
$2\cdot m_p + m_{\pi^0} = 2.01~GeV/c^2$. 
Whereas, in case of the direct production of the two 
charged pions the missing mass is only due to two protons and the distribution is shifted towards 
the lower masses. The situation is illustrated in Fig.~\ref{mmpppipiX} (right) where
the experimental and simulated data are shown in the same way as it was for the left panel. 
One can see that in the measured data the signal peaks are separated from each other. 
The cut was chosen to accept missing masses greater than $1.95~GeV/c^2$. This allows to
reduce the remaining two-pion background to negligible level, decreasing the efficiency of the signal reconstruction by 0.1\% only. 

\begin{figure}[!h]
\hspace{.0cm}
\parbox{0.3\textwidth}{\centerline{\epsfig{file=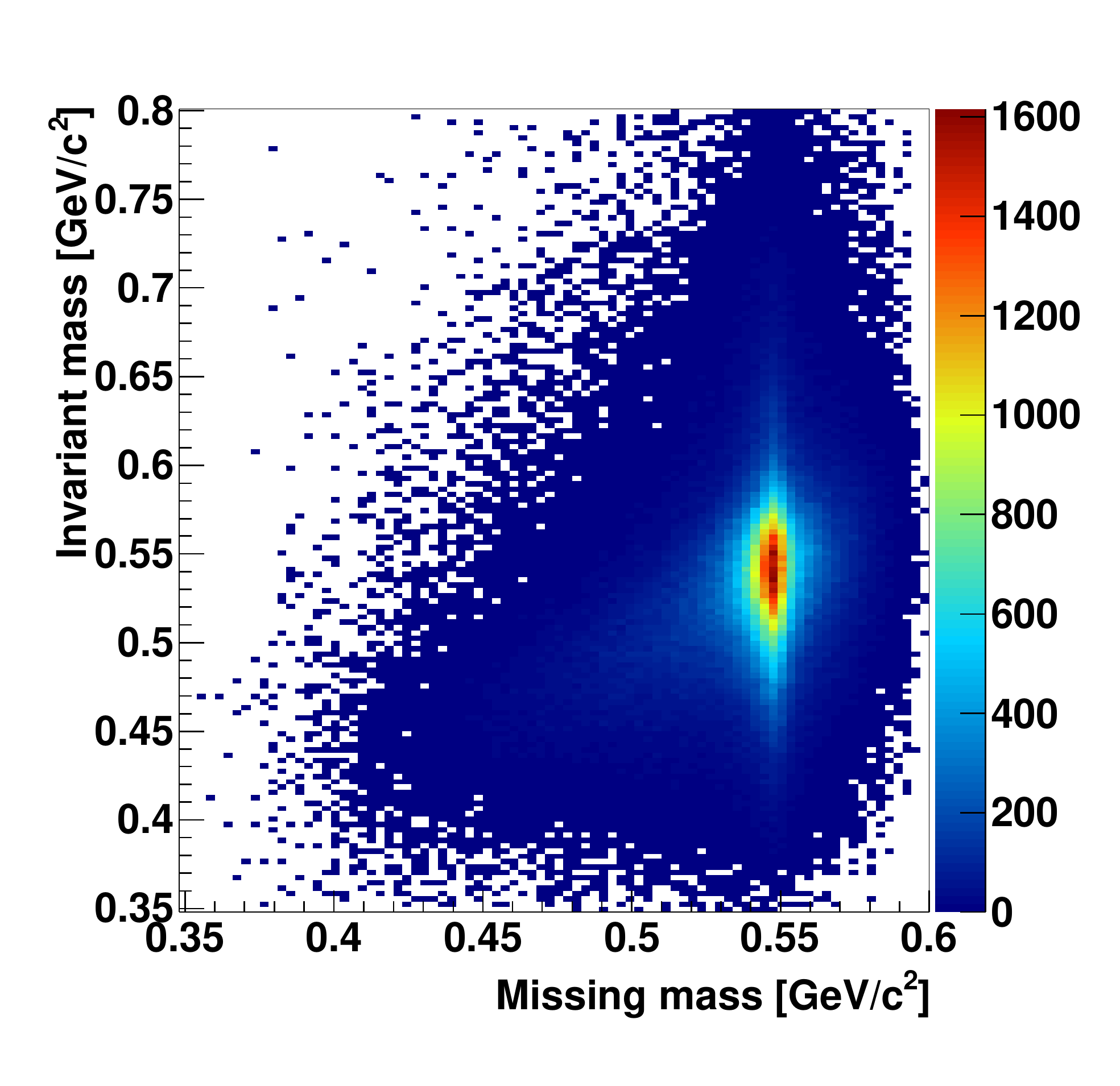,width=0.35\textwidth}}}
\hspace{.4cm}
\parbox{0.3\textwidth}{\centerline{\epsfig{file=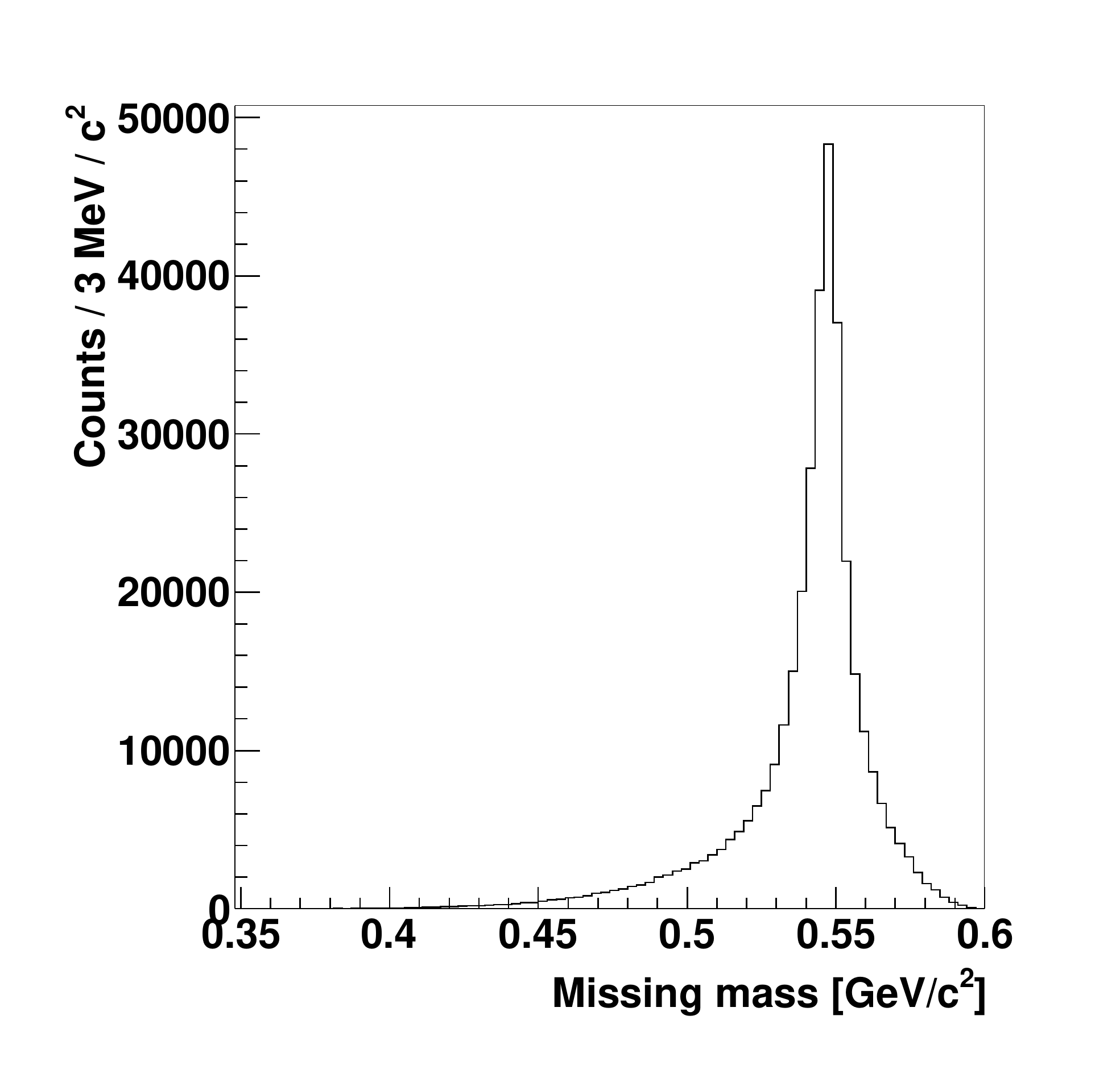,width=0.35\textwidth}}}
\hspace{.4cm}
\parbox{0.3\textwidth}{\centerline{\epsfig{file=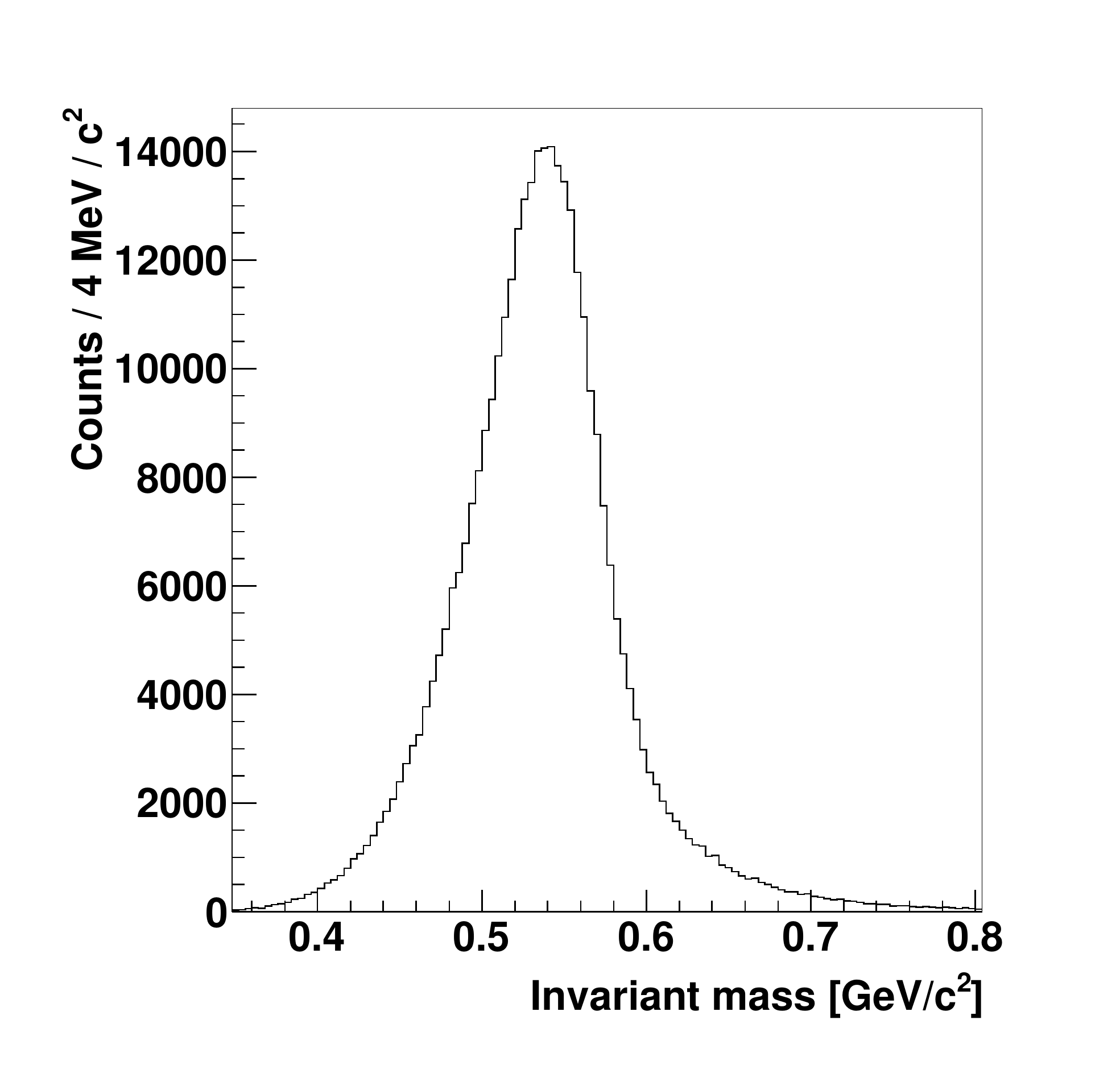,width=0.35\textwidth}}}
\caption{
{\bf{(left)}} Experimental distribution of the invariant mass for the 
              $\pi^+\pi^-\pi^0$ system as a function of the missing mass for the 
              $pp\to ppX$ reaction after applying background suppression described in the text.
{\bf{(middle)}} Experimental missing mass for the $pp\to ppX$ reaction.
{\bf{(right)}}  Invariant mass of the $\pi^+\pi^-\pi^0$ system. 
}
\label{finalcut2pi}
\end{figure}

The result of all previously applied condition is illustrated in Fig.~\ref{finalcut2pi}. 
It can be seen that the multi-meson background (mostly direct two pion production) was significantly reduced
compared to the one shown in Fig.~\ref{MMppIMpipipi}. The signal to background ratio was improved 
from 0.8 to 2.2
The spectrum of the missing mass 
(see Fig.~\ref{finalcut2pi} (middle)) shows a peak located at the mass of the $\eta$ meson 
with a tail towards the lower masses. The tail and continuous background under the 
peak is originating mostly from direct three pion production.  
The invariant mass distribution of the $\pi^+\pi^-\pi^0$ is shown in Fig.~\ref{finalcut2pi} (right).
It also shows a maximum located around the mass of the $\eta$ meson. But the invariant mass 
distribution is much broader than the missing mass spectrum. 
This is because the four momentum vector reconstruction of particles registered in Central Detector
is less accurate than in the Forward Detector.  

\section{Kinematic fit}\label{sec:kfit}
\hspace{\parindent}
The identification of the reaction $pp\to pp\eta$ and the subsequent decay $\eta\to\pi^0\pi^+\pi^-$ is
based on the measurement of momentum vectors of all final state particles 
and reconstruction of the missing and invariant masses. A finite resolution
of the four momentum vector determination is reflected in the population of kinematically 
forbidden regions of the phase space. This obstacle will be corrected by performing a kinematic 
fitting of the data.
In this experiment we have measured more variables than needed for  complete description 
of the kinematics of the final state. This redundancy will be used to improve 
the mass resolution and in consequence to improve the signal to background ratio, and to 
correct each event to follow kinematically 
allowed phase space region.  
 
The kinematic fit procedure is a least squares method which is based on variation of kinematic 
observables of all final state particles in the range of experimental resolution in such a way that after 
the change they fulfil the kinematical constraints~\cite{Kupsc:1995kf}.
A $\chi^2$ value is used as a measure of differences between values of a measured and the varied observables
which matches the assumed hypothesis. As a best solution fulfilling the required criterion one take this
minimizing the $\chi^2$ function. In our case we require that event  fulfil kinematics 
of the $pp\to pp\pi^+\pi^-\pi^0\to pp\pi^+\pi^-\gamma\gamma$ reaction chain. 
The basic conditions for the minimalization includes the check of the momentum and energy conservation.
Also, conditions such as masses of intermediate particles can be introduced additionally into the fitting procedure. 

The $\chi^2$ quantity which describes the agreement between fitted and measured variables reads:
\begin{equation}
\chi^2 = (\mathbb{P}_{F} - \mathbb{P}_{M}) S^{-1} (\mathbb{P}_F - \mathbb{P}_{M})^T 
               + \lambda_C C + \lambda_U U,
\end{equation} 
where the $\mathbb{P}_{F}$ and $\mathbb{P}_{M}$ represent vectors of fitted and measured
variables, respectively, $S$ denotes the covariance matrix of the uncertainties, 
the $\lambda_C$ and $\lambda_U$ stand for Lagrange multipliers to introduce the 
constraints for energy and momentum conservation $C$, and additional conditions $U$. 

In case of this thesis after identification of particles masses their momentum vectors are described 
by a set of three variables: kinetic energy ($E_k$), polar angle ($\theta$), and 
the azimuthal angle ($\phi$). For investigated reaction chain one has six particles 
in the final state: $p_1,~p_2,~\pi^+,~\pi^-,~\gamma_1,~\gamma_2$. 
Therefore, taking into account the four equations for the momentum and energy conservation,
the minimal number of independent variables which has to be measured to 
describe the system is equal to:
\begin{equation}
N_v^{min} = 3 \cdot 6  - 4 = 14.
\end{equation}
However, we have measured all the variables, so 
we have $N_v = 18$, and therefore for each event we have redundant information allowing us 
to perform kinematic fitting. The number of degrees of freedom can be calculated as: 
\begin{equation}
N_{ndf} = N_v - N_v^{min} + N_U, 
\label{ndf}
\end{equation}
where the $N_U$ denotes the number of additionally introduced conditions.
In case of this analysis we introduced the condition that two photons registered 
in electromagnetic calorimeter are originating from the decay of the neutral pion. 
Thus according to equation~\ref{ndf} the number of degrees of freedom reads $N_{ndf} = 5$.
Further, one can introduce the probability defining the goodness of the fit for a given $\chi^2$ 
value and a given number of degrees of freedom $N_{ndf}$, as:
\begin{equation}
P_F(\chi^2,N_{ndf}) = 1 - F(\chi^2,N_{ndf}) = \frac{1}{\sqrt{2^{\frac{N_{ndf}}{2}} \Gamma(\frac{N_{ndf}}{2})}}
\int_{\chi^2}^{\infty}e^{-\frac{t}{2}} t^{(\frac{N_{ndf}}{2} - 1)} dt,
\end{equation}
where $F(\chi^2,N_{ndf})$ is the Cumulative Distribution Function~\cite{Brandt:1999sm}, and the $\Gamma$ 
denotes the Euler special function. The distribution of the probability of the fit $P_F$ for the proper hypothesis in case of Gaussian errors should be flat~\footnote{It is because the probability density distribution of the cumulative distribution is always uniform independent of the probability density function.}

As mentioned above to use the kinematic fit procedure one has to introduce an error for each
measured observable. These inaccuracy depends on the particle type, its energy, polar and 
azimuthal angle, and the specific detection properties of different detectors. 
In order to estimate these errors using the WASA Monte Carlo package a sample of 
single particle events were simulated. The single particle events were chosen due to the technical reasons, 
because they ensure the uniform distribution in the whole acceptance of the detector in comparison to a 
specific reaction like for e.g. the $pp\to pp \eta$, where particles more often are emitted in 
forward direction. The reconstruction methods and condition for the detection of particles are identical 
as in the previously described analysis. 

The convention for error parameterization for each particle was chosen as fallows:
\begin{equation*}
\sigma_{E_k} = \frac{E_{k}^{rec} - E_{k}^{gen}}{E_{k}^{rec}},
\end{equation*}
\begin{equation}
\sigma_{\theta} = \theta^{rec} - \theta^{gen}, 
\label{kf2}
\end{equation}
\begin{equation*}
\sigma_{\phi} = \phi^{rec} - \phi^{gen}, 
\end{equation*}
where superscripts $rec$ and $gen$ denote the reconstructed and generated value
of the variable, respectively.  The kinetic energy is parameterized by a relative difference of the 
reconstructed and generated values, and for the angles the absolute values are used.  
The errors are taken as a one standard deviation calculated from 
fitting the resulting distributions given by equations~\ref{kf2} with a Gauss function.  
Finally all errors  are determined as double differential functions
which depends on the reconstructed kinetic energy and polar angle.

The bin width for the parameterization was chosen separately for different type of particles.
For protons in Forward Detector size of kinetic energy and polar angle intervals 
are equal to 50~MeV and 1$^o$, respectively. For the charged pions in Central Detector the 
bin width for energy is 50~MeV and for polar scattering angle 4$^o$.
The different approach is for parameterization of errors of energy and angle for photons registered 
in the calorimeter.
Here the energy is divided into intervals with size 50~MeV, but instead of the polar angle the 
24 bins corresponding to number of ring building the calorimeter were taken. 
The summary of the bin size chosen for the parameterization is given in Tab.~\ref{tab:kferr}.
\begin{table}[H]
\centering
\begin{tabular}{c|c|c|c}
\hline
Particle type & Detector & $E_{kin}$ & $\theta$\\\hline\hline
p        & FD     & 50 MeV &  1$^o$\\
$\gamma$ & CD/SEC & 50 MeV &  crystal size\\
$\pi^+$  & CD/MDC & 50 MeV &  4$^o$\\
$\pi^-$  & CD/MDC & 50 MeV &  4$^o$\\\hline
\end{tabular}
\caption{Summary of the bin size used in parameterization of errors in kinematic fit.}
\label{tab:kferr}
\end{table}

\begin{figure}[p]
\hspace{.0cm}
\parbox{0.3\textwidth}{\centerline{\epsfig{file=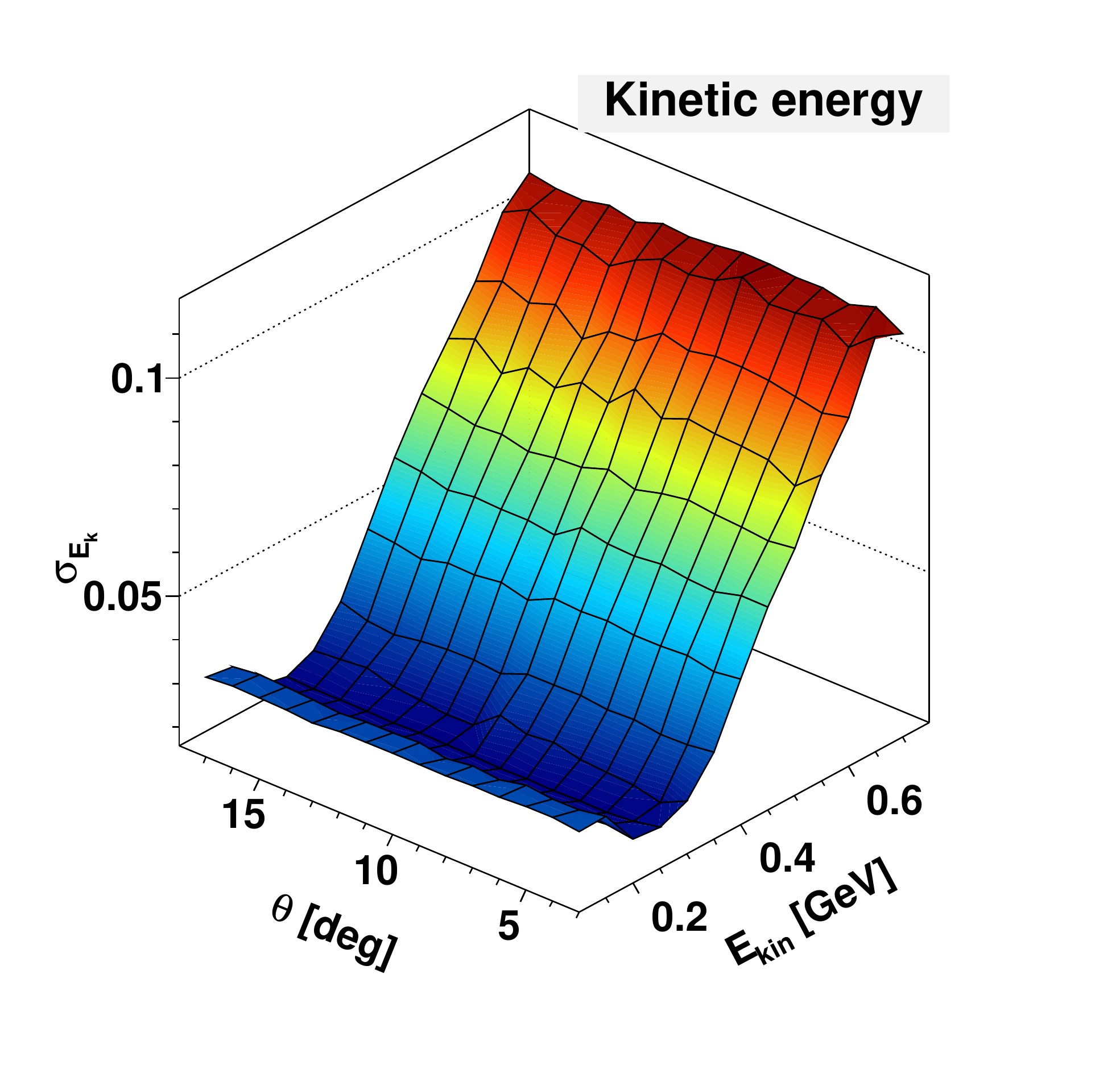,width=0.35\textwidth}}}
\hspace{.0cm}
\parbox{0.3\textwidth}{\centerline{\epsfig{file=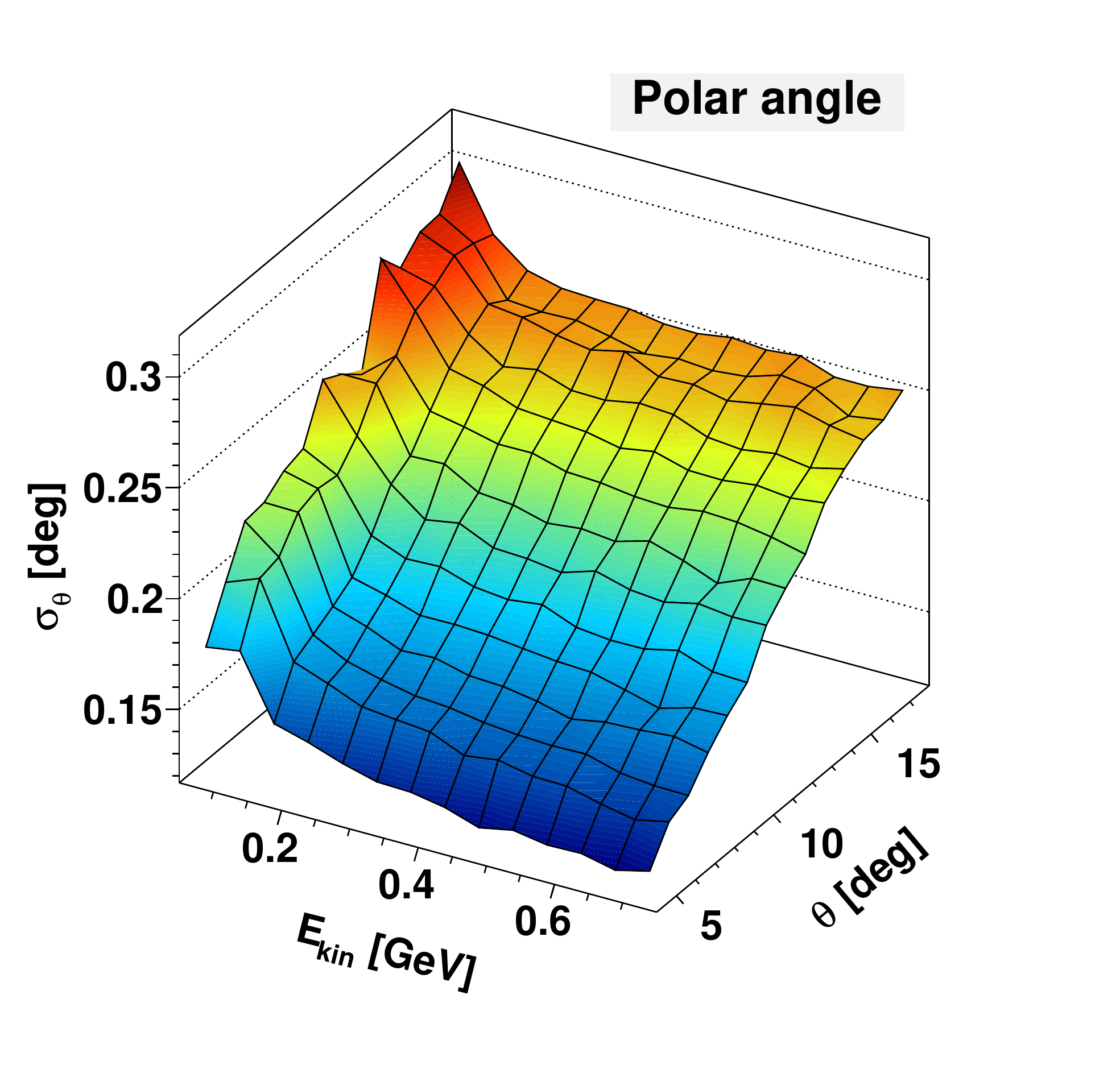,width=0.35\textwidth}}}
\hspace{.0cm}
\parbox{0.3\textwidth}{\centerline{\epsfig{file=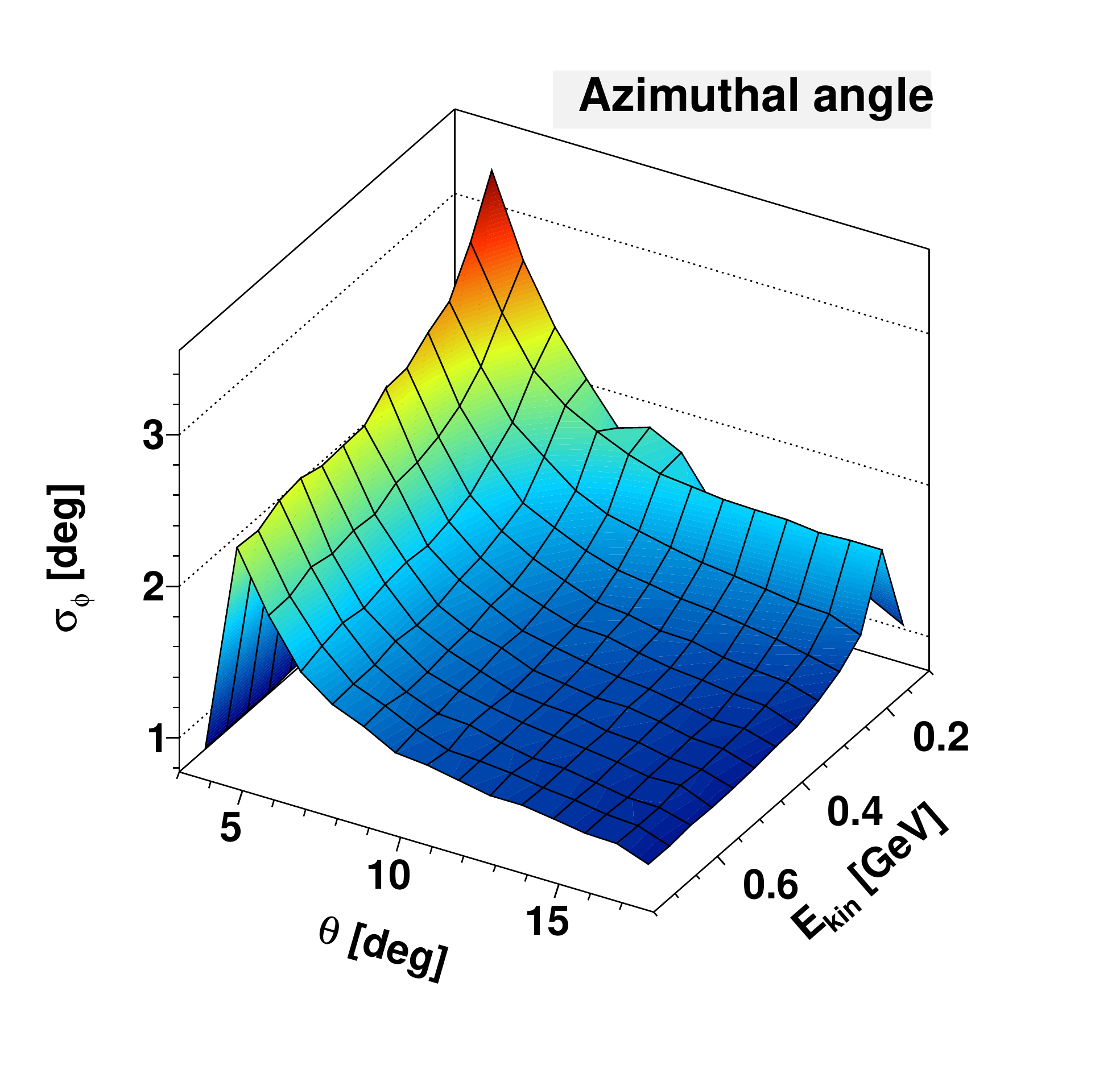,width=0.35\textwidth}}} 
\hspace{.7cm}
\parbox{0.3\textwidth}{\centerline{\epsfig{file=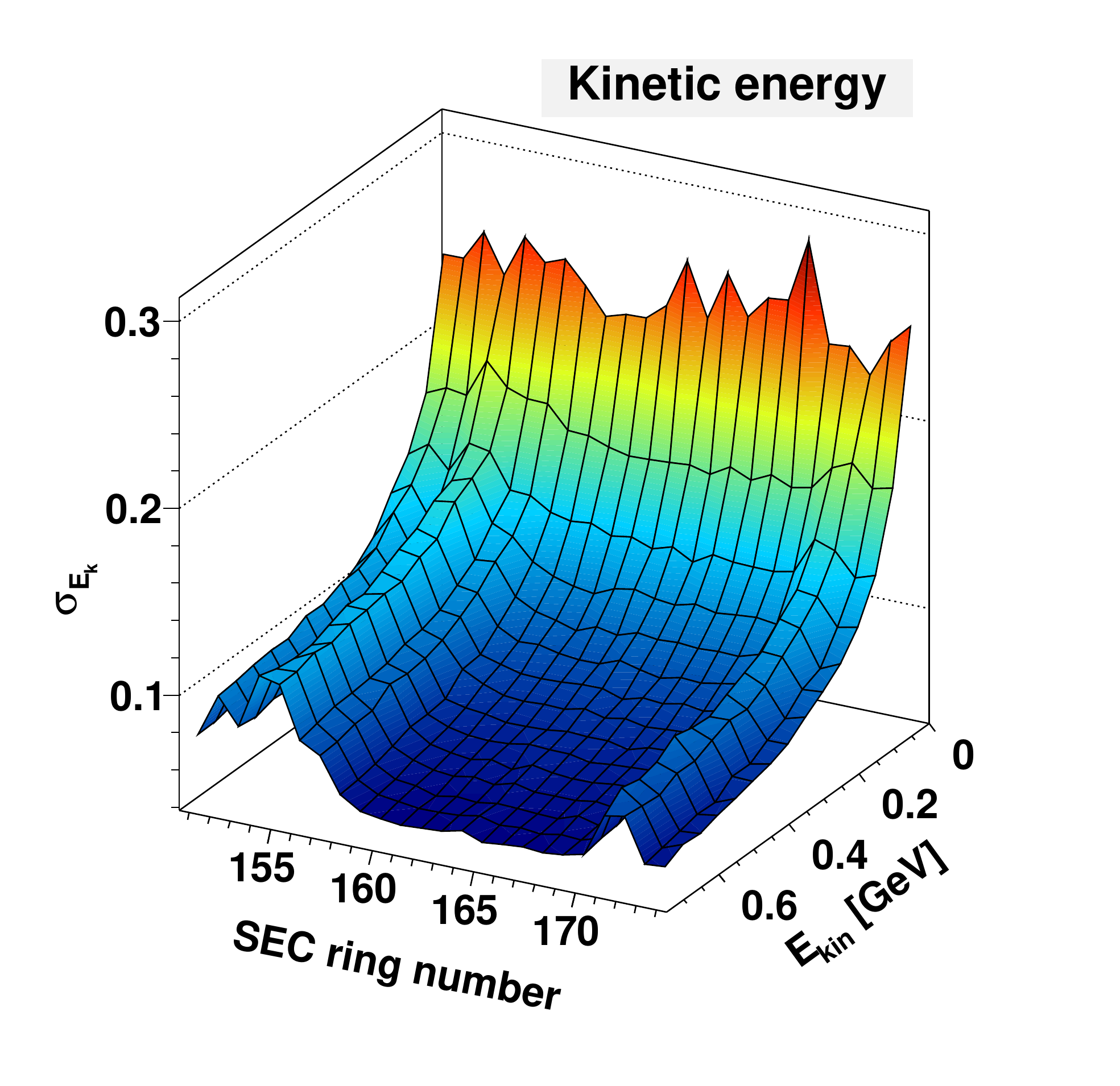,width=0.35\textwidth}}}
\hspace{.0cm}
\parbox{0.3\textwidth}{\centerline{\epsfig{file=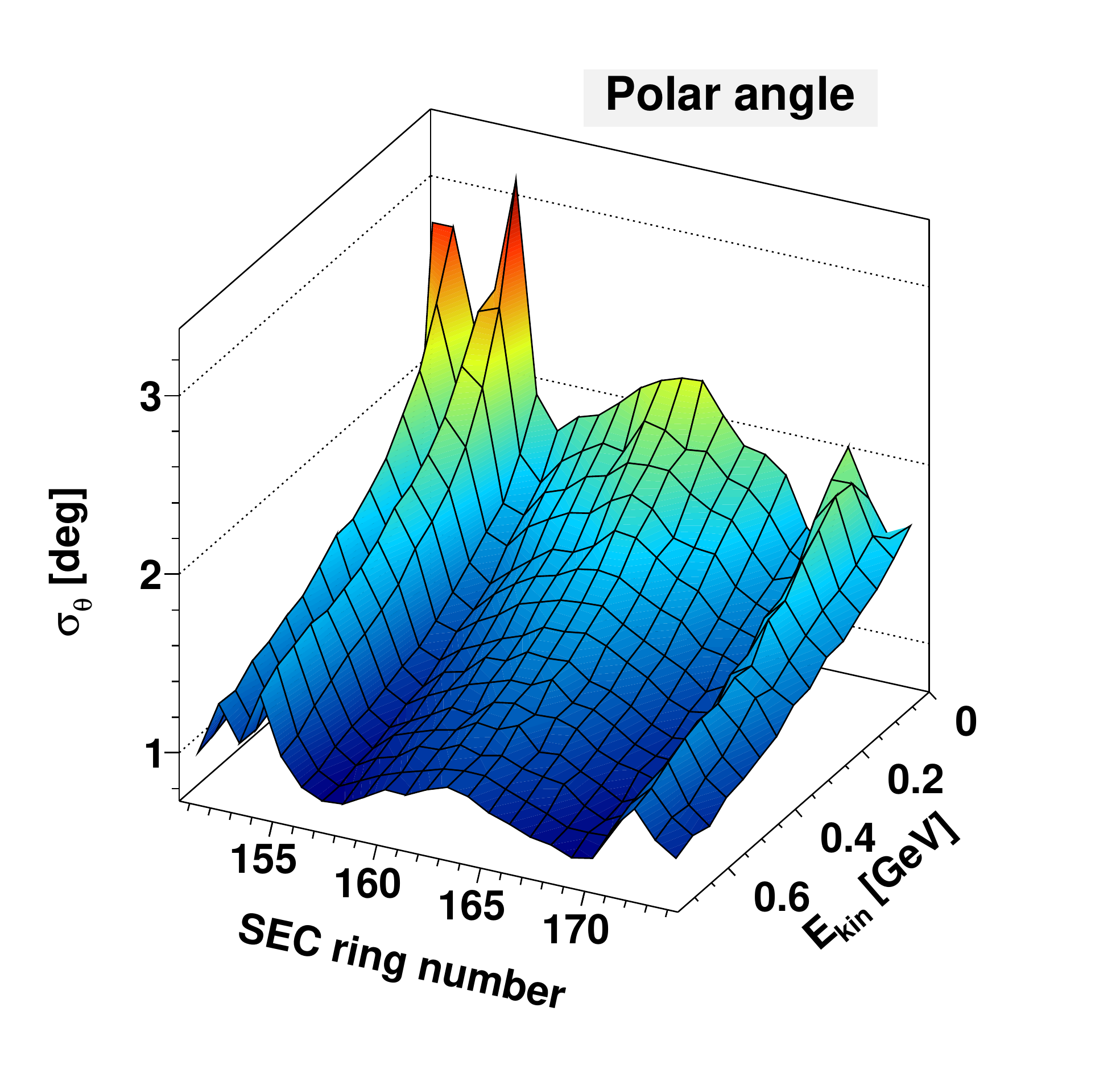,width=0.35\textwidth}}}
\hspace{.0cm}
\parbox{0.3\textwidth}{\centerline{\epsfig{file=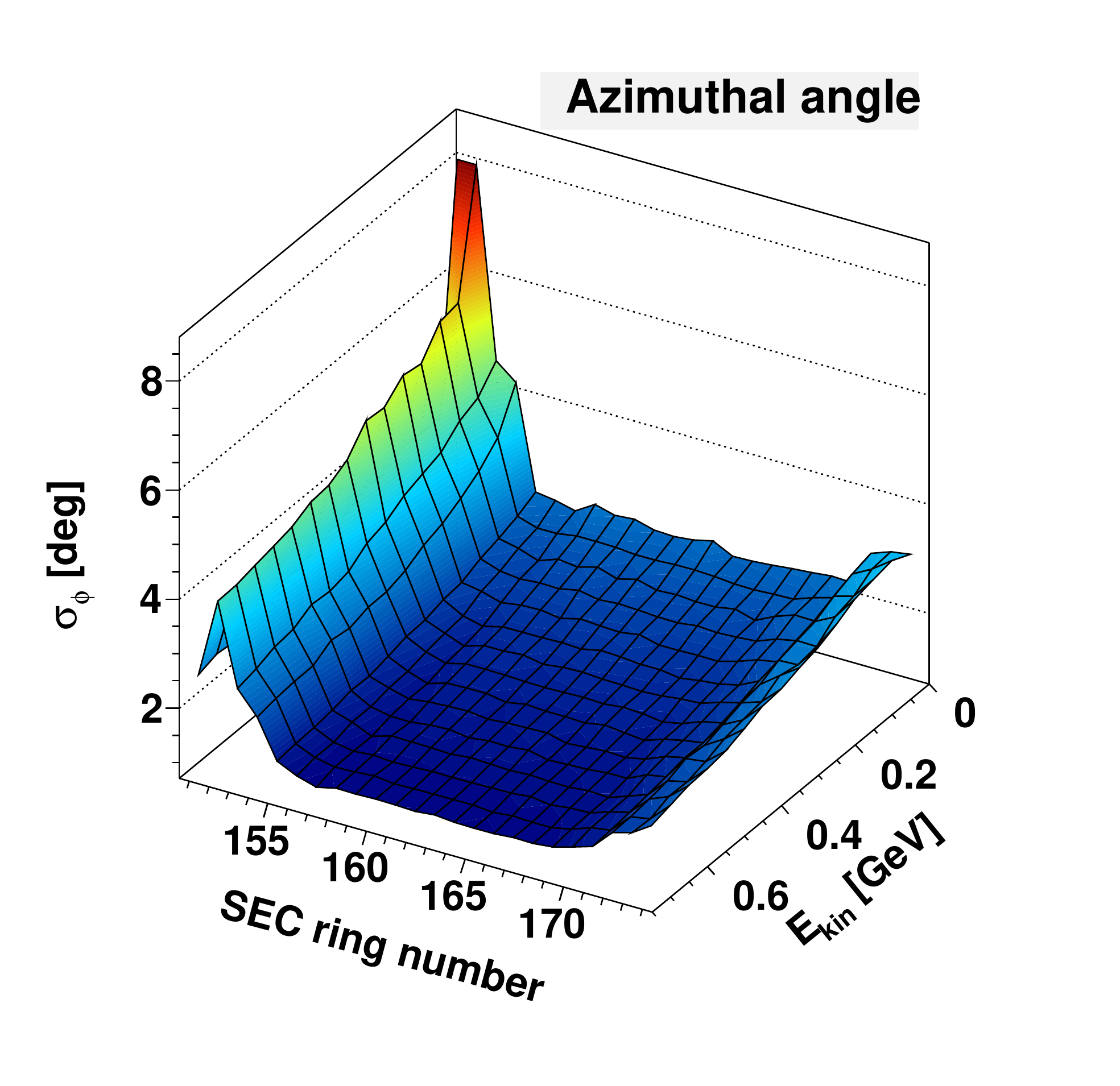,width=0.35\textwidth}}} 
\hspace{.7cm}
\parbox{0.3\textwidth}{\centerline{\epsfig{file=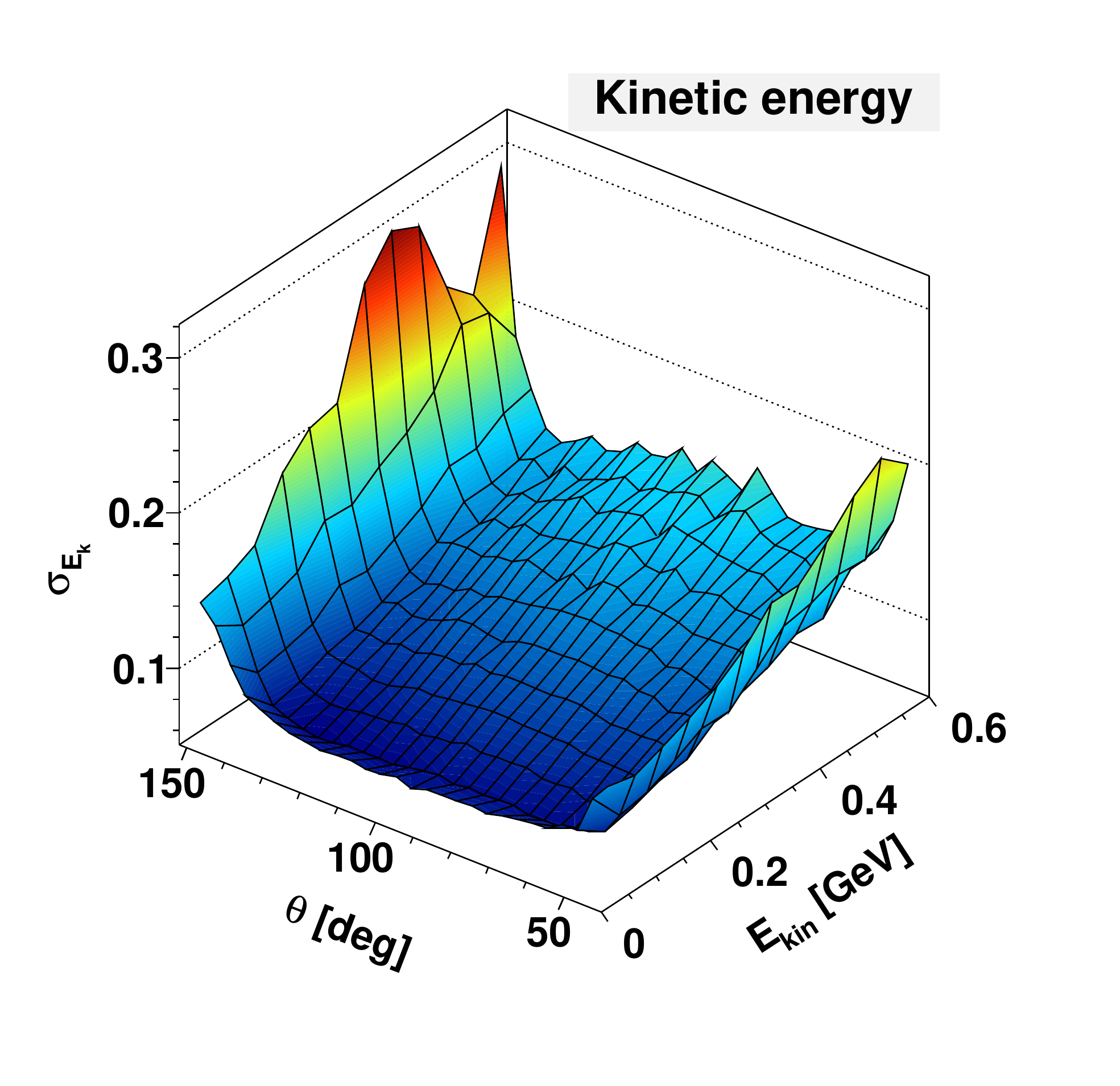,width=0.35\textwidth}}}
\hspace{.0cm}
\parbox{0.3\textwidth}{\centerline{\epsfig{file=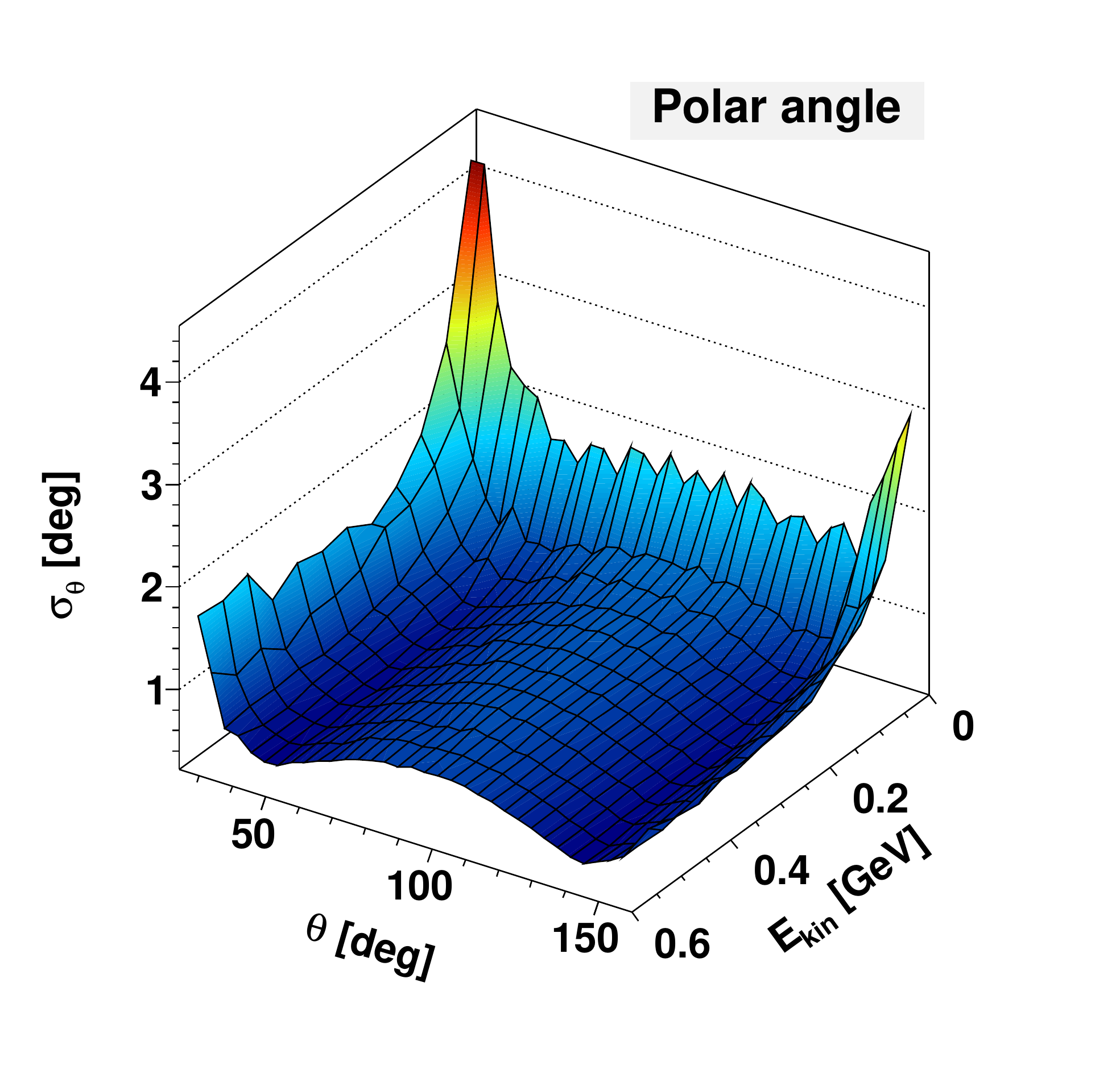,width=0.35\textwidth}}}
\hspace{.0cm}
\parbox{0.3\textwidth}{\centerline{\epsfig{file=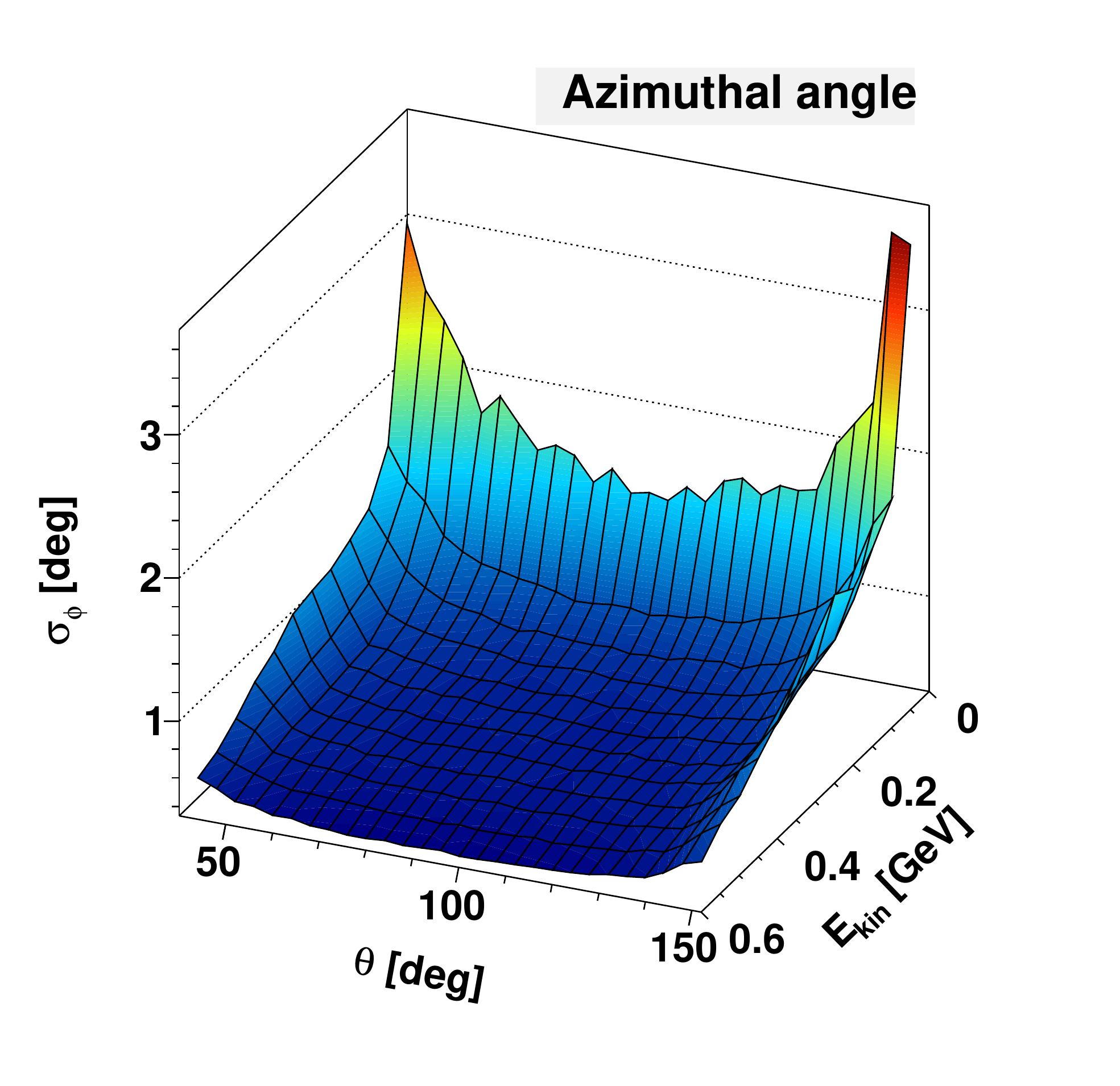,width=0.35\textwidth}}} 
\hspace{.7cm}
\parbox{0.3\textwidth}{\centerline{\epsfig{file=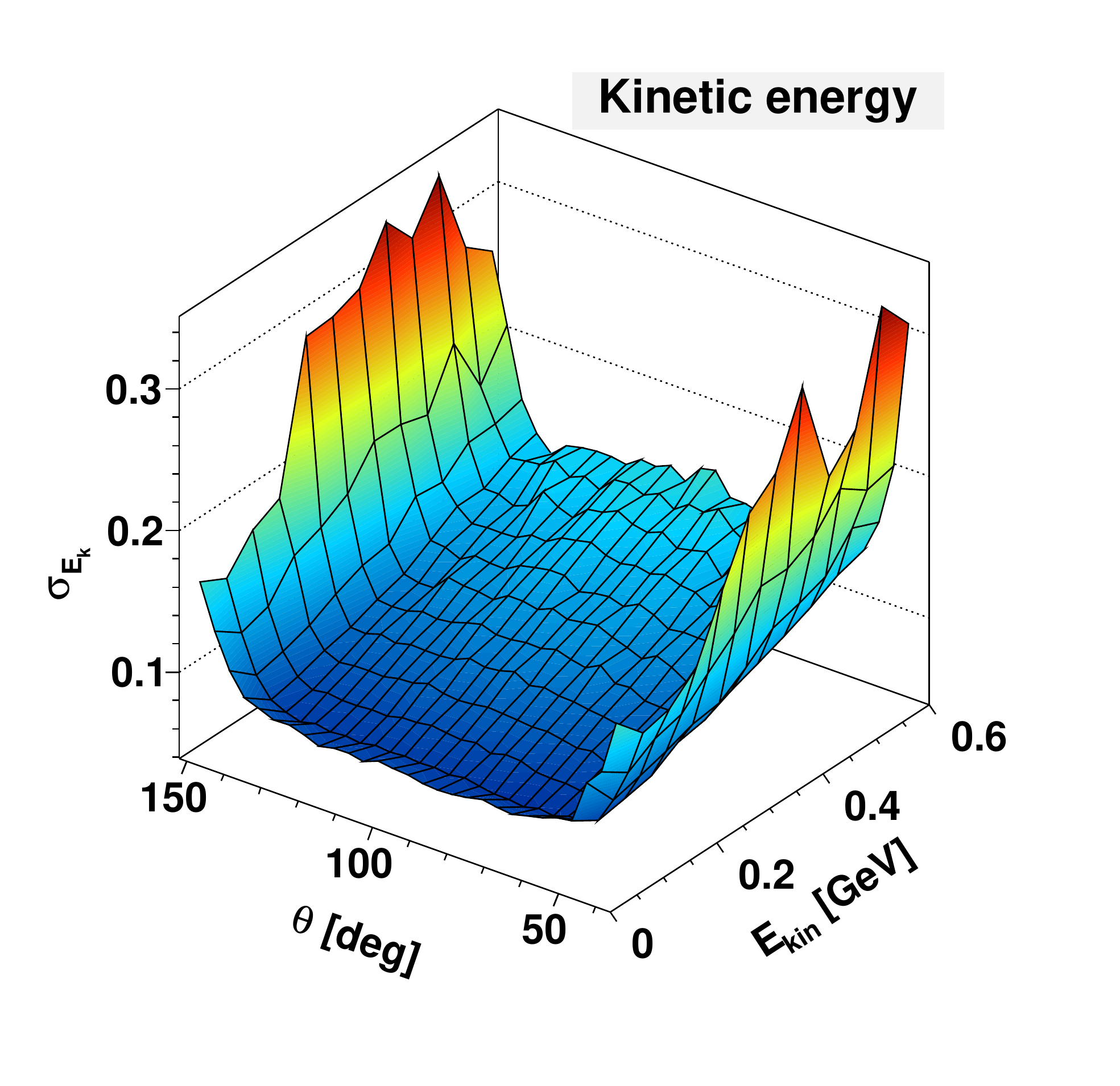,width=0.35\textwidth}}}
\hspace{.5cm}
\parbox{0.3\textwidth}{\centerline{\epsfig{file=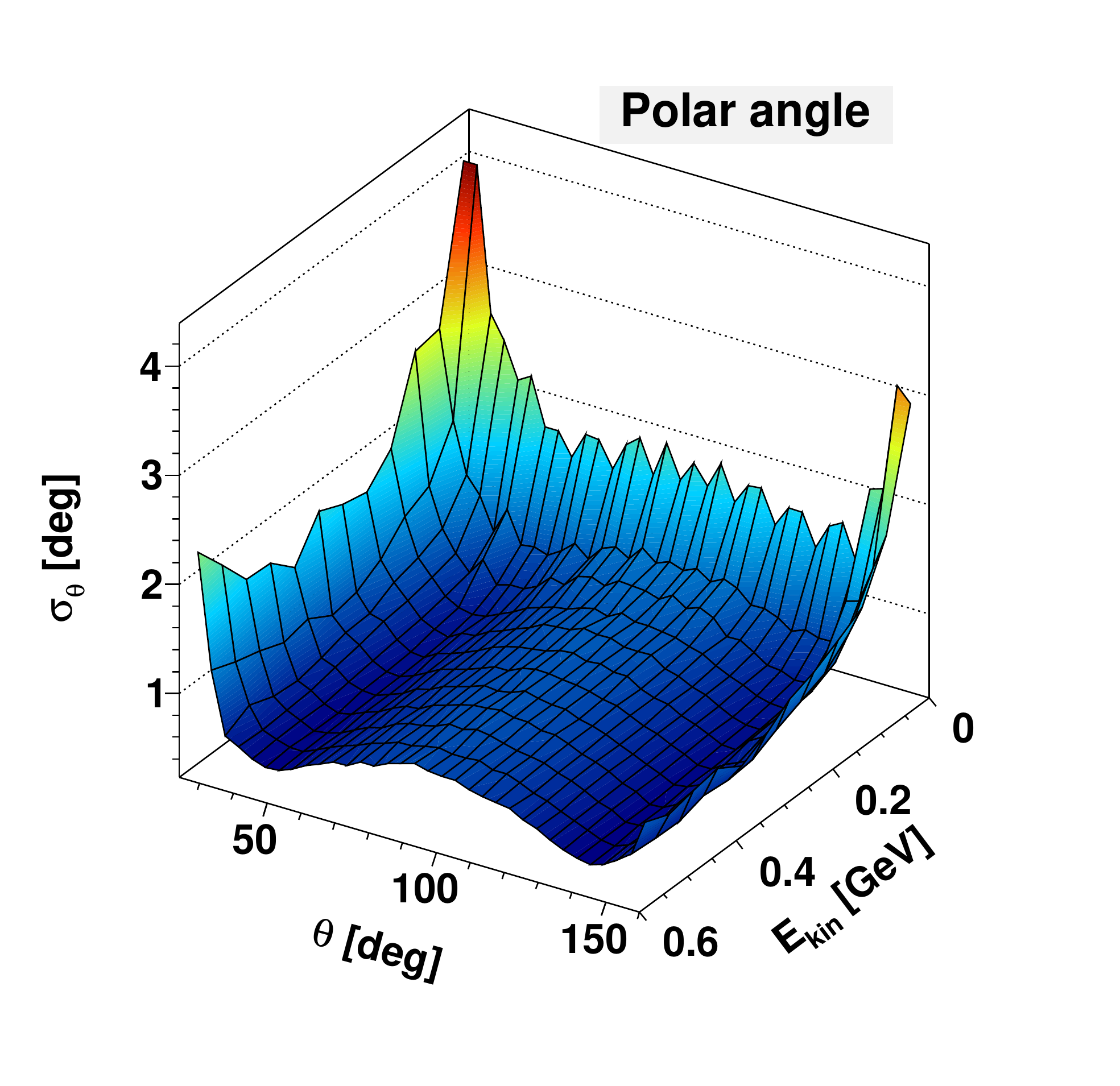,width=0.35\textwidth}}}
\hspace{.5cm}
\parbox{0.3\textwidth}{\centerline{\epsfig{file=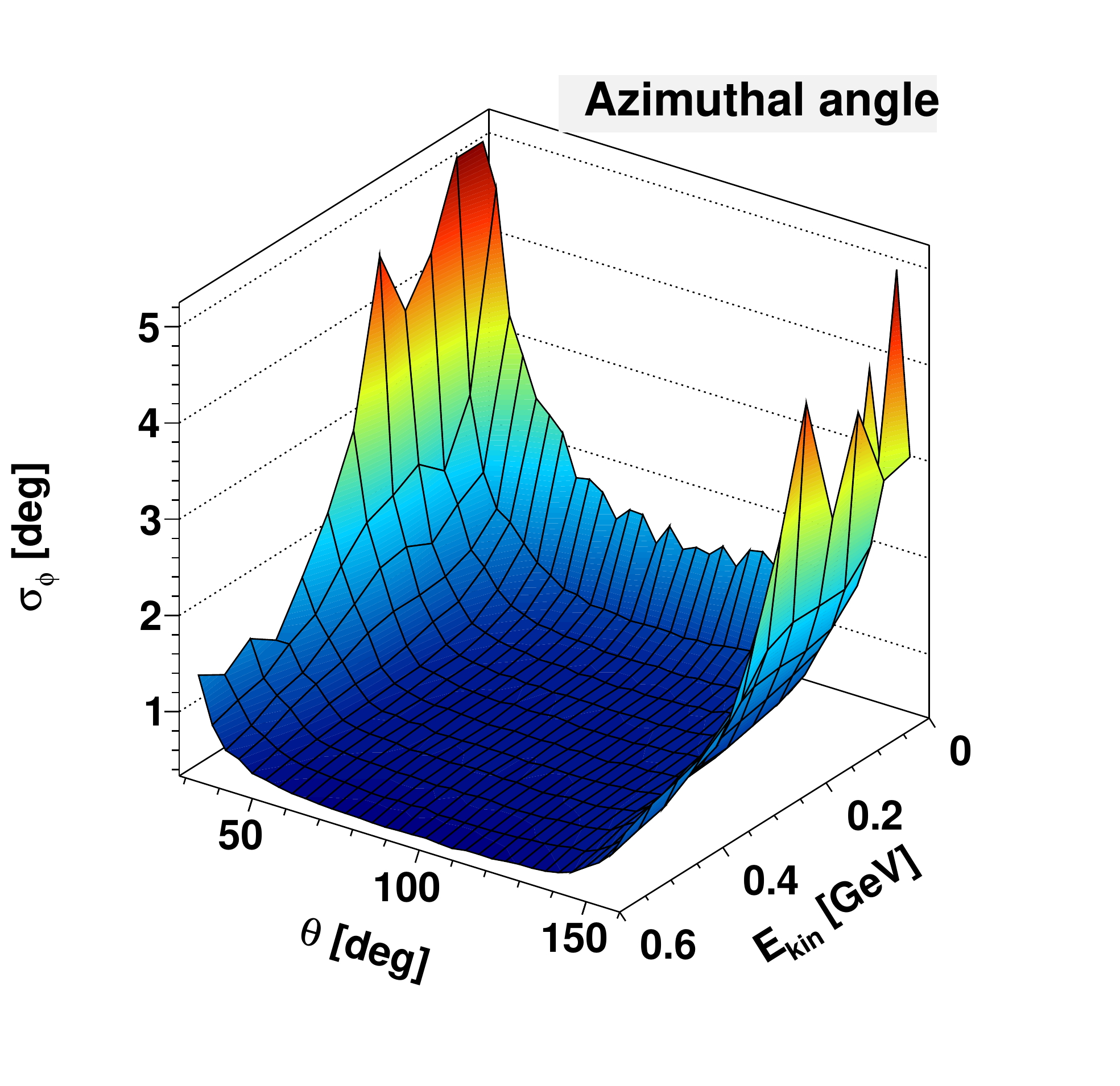,width=0.35\textwidth}}} 
\caption{
Distributions of the double differential parameterization of errors in determining the $E_{kin},~\theta$ and 
$\phi$ {\bf{(from the top in rows)}} for: protons, photons, $\pi^+$ and $\pi^-$. The columns from the left denote errors for kinetic energy, and polar and azimuthal angle, respectively. 
}
\label{errparam}
\end{figure}

The errors as a function of kinetic energy and polar angle extracted for each type of 
particle can be seen in Fig.~\ref{errparam}. The plots are aligned in rows from the 
top for: protons, photons, $\pi^+$ and $\pi^-$. The columns from the left are assigned 
accordingly to: kinetic energy, polar angle and azimuthal angle.

For protons it can be seen that relative error of kinetic energy does not depend on the scattering 
angle $\theta$ and for energies up to 360~MeV it is almost constant (around 2\%), for higher energies it
starts to increase linearly up to 11\% at 700~MeV. The increase is associated with the 
fact that protons with energy higher than 360~MeV punch through the FRH. 
The error of the polar angle is nearly independent of the energy but it increases with the 
scattering angle. Whereas, the error of the azimuthal angle decreases when polar angle increases, 
and it is almost independent of the energy. 

For photons the relative energy error decreases with increasing energy with visible 
enhancement in the regions where the central part of the calorimeter meet the backward or
forward part. In case of the azimuthal and polar angle the errors decrease with increasing 
energy of photons. For both variables errors are increasing in the end caps of the calorimeter
staying rather constant in the central part. 

\begin{figure}[!t]
\hspace{.0cm}
\parbox{0.5\textwidth}{\centerline{\epsfig{file=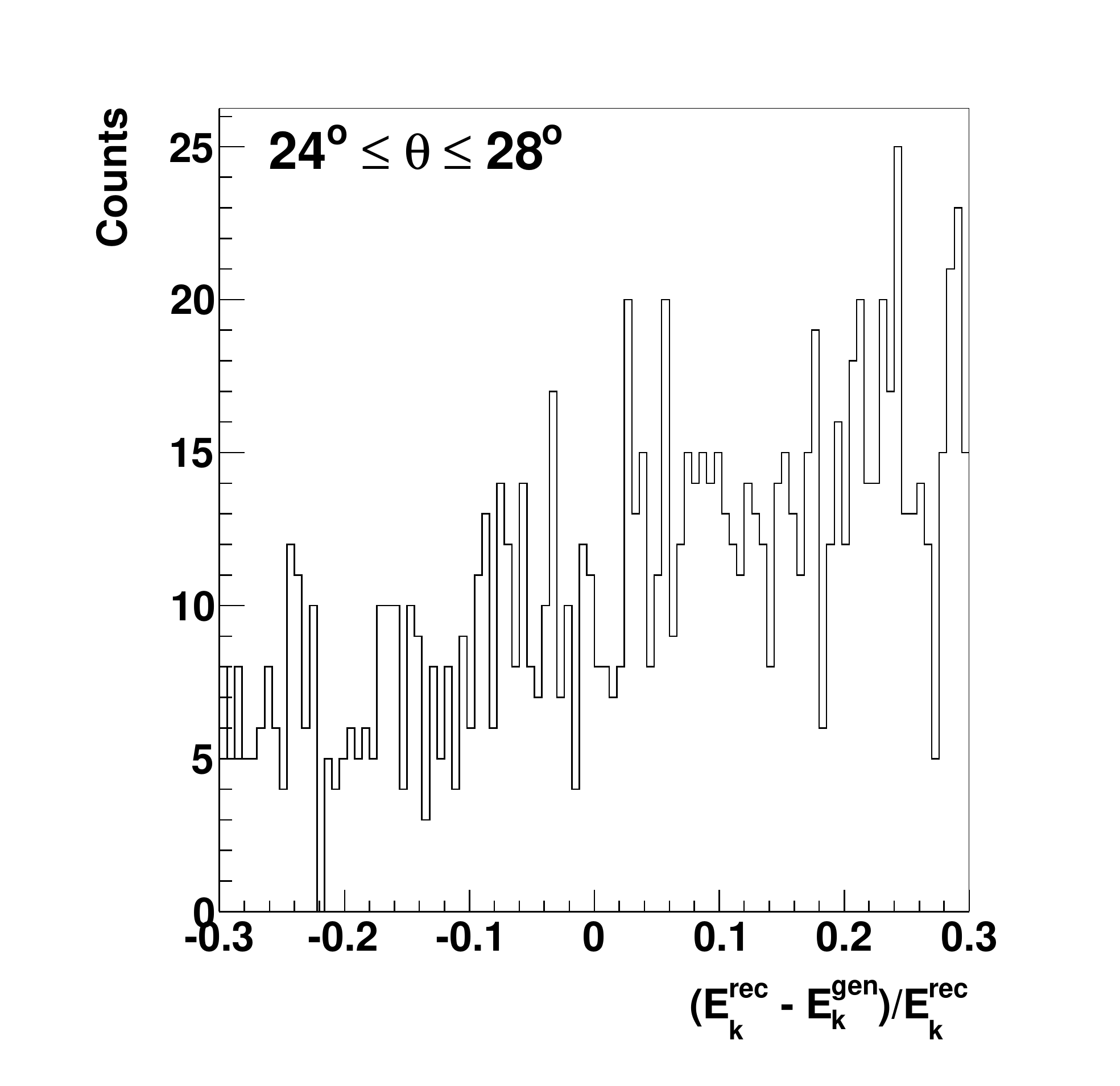,width=0.55\textwidth}}}
\hspace{.0cm}
\parbox{0.5\textwidth}{\centerline{\epsfig{file=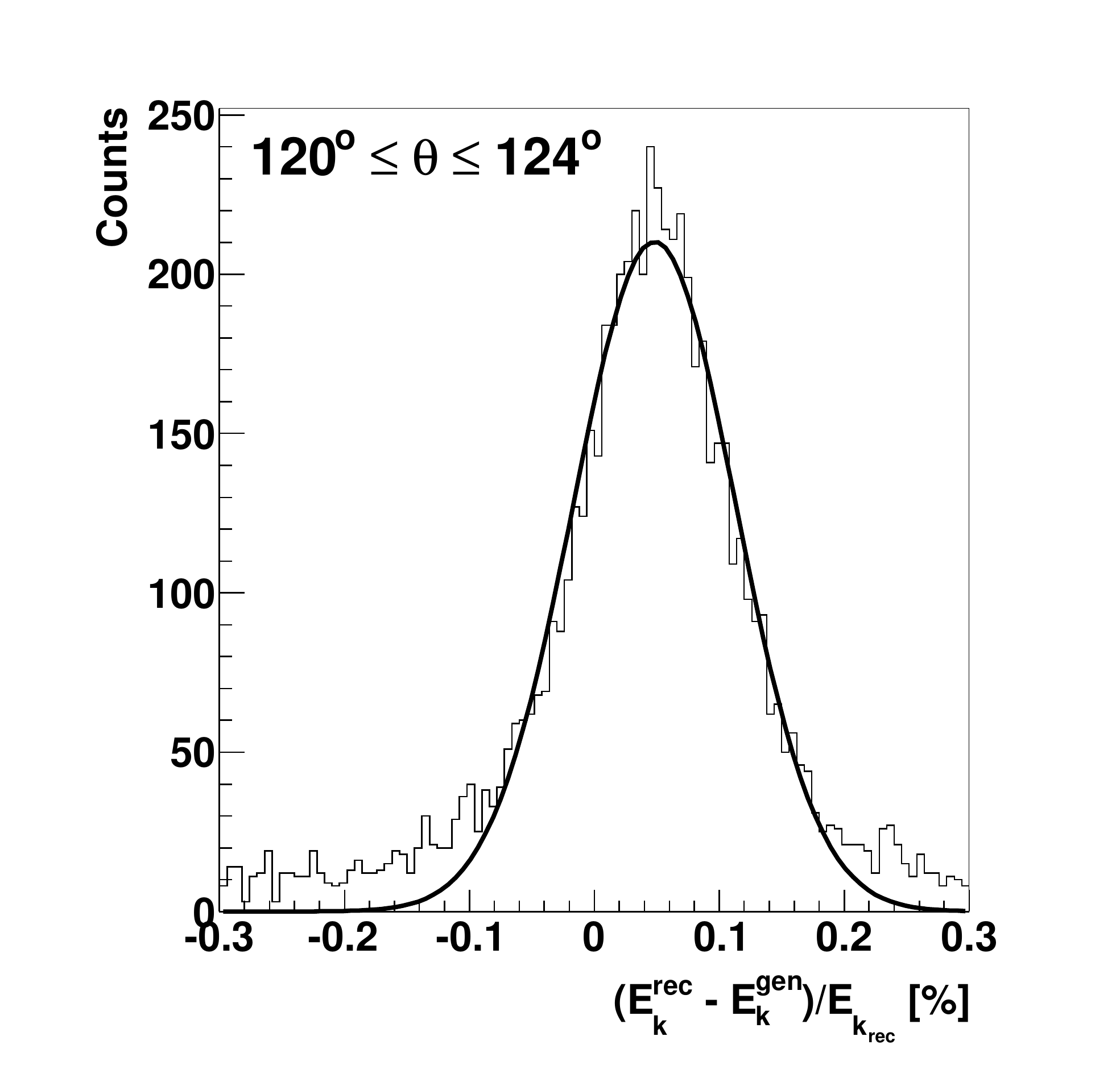,width=0.55\textwidth}}}
\caption{
Two exemplary distributions of the relative difference of the reconstructed and generated values of the 
kinetic energy of $\pi^+$ for:  
{\bf{(left)}} front region of MDC for polar angles $24^{o} \leq \theta \leq 28^{o}$  where less straws 
fires and errors are no-Gaussian, {\bf{(right)}} and middle region of MDC for 
$120^{o} \leq \theta \leq 124^{o}$ where errors are Gaussian. 
}
\label{MDCerrors}
\end{figure}

In the case of charged pions, errors for all kinematical variables in the range from 40$^o$ to 140$^o$ 
shows dependence only on the kinetic energy. For angles below 40$^o$ and 
above 140$^o$ the errors are instantly increasing. This illustrates some features of 
the reconstruction method in the MDC. These two regions reflect the front and rear
part of the drift chamber. Particles flying through these parts fire less 
straws which are contributing to the track reconstruction and thus the inaccuracy 
of the measurement increases (see Fig.~\ref{MDCerrors}). 

\begin{figure}[!b]
\mbox{
\hspace{.0cm}
\parbox{0.5\textwidth}{\centerline{\epsfig{file=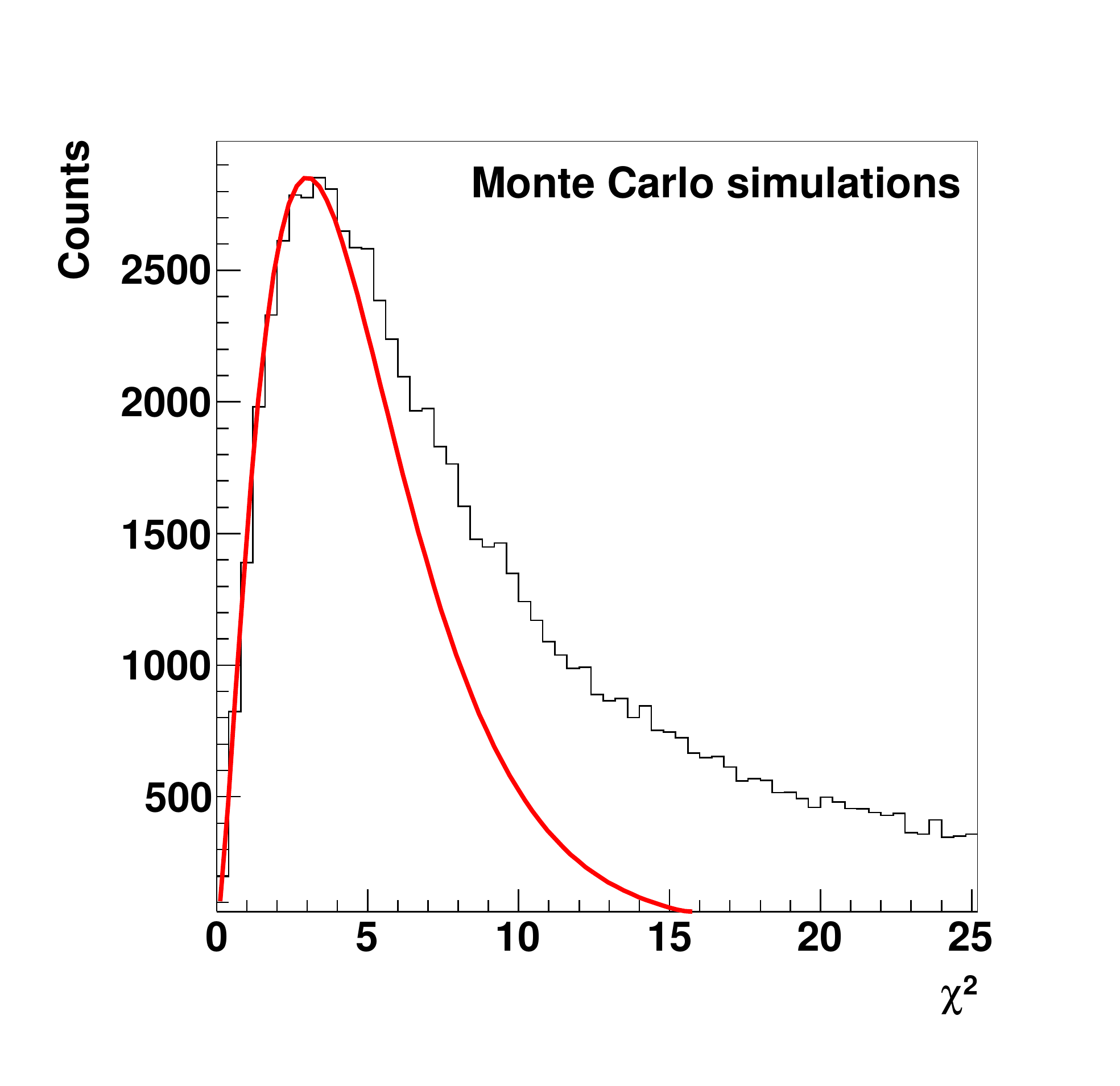,width=0.55\textwidth}}}
\hspace{.0cm}
\parbox{0.5\textwidth}{\centerline{\epsfig{file=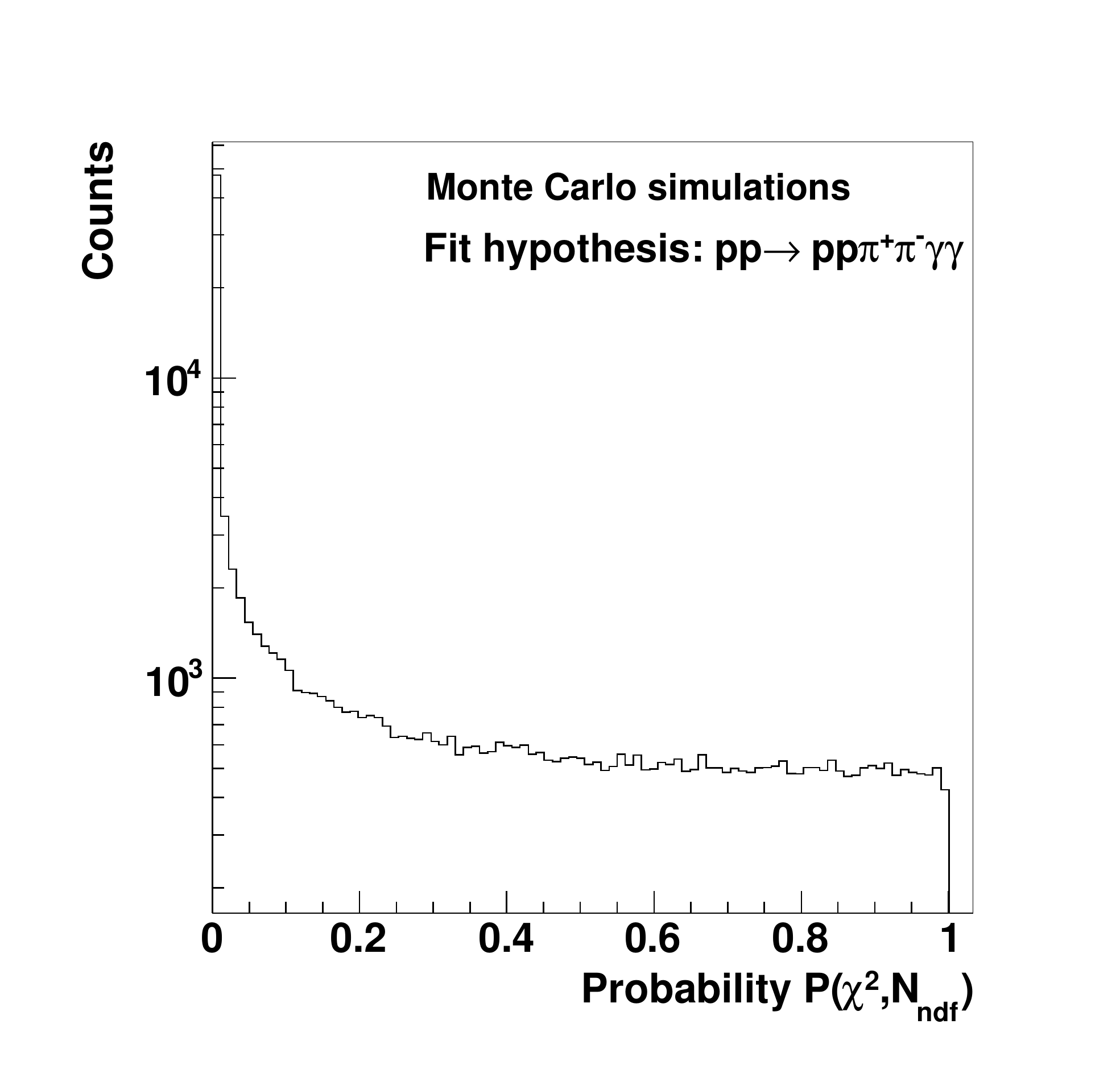,width=0.55\textwidth}}}
}
\mbox{
\hspace{.0cm}
\parbox{0.5\textwidth}{\centerline{\epsfig{file=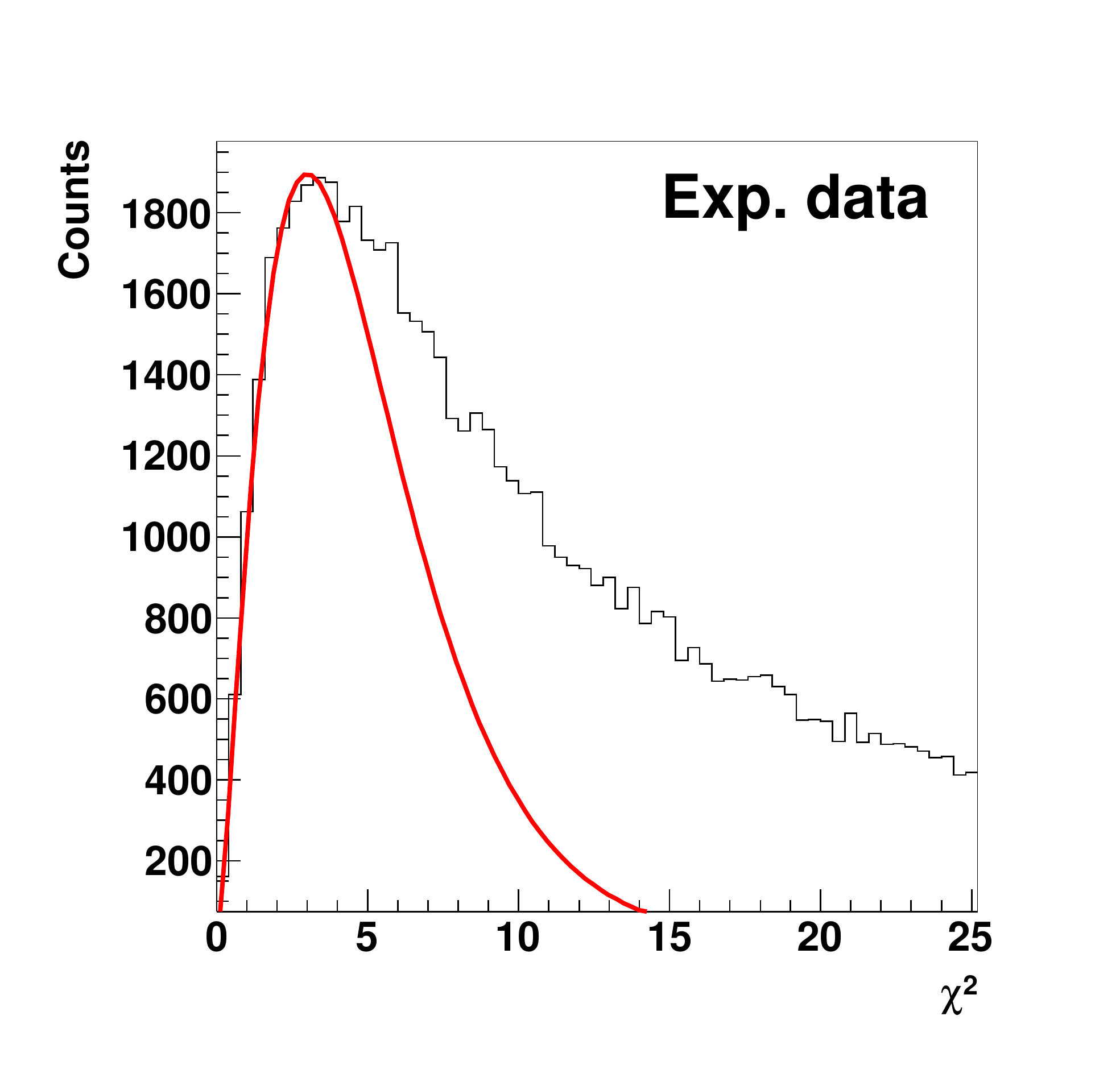,width=0.55\textwidth}}}
\hspace{.0cm}
\parbox{0.5\textwidth}{\centerline{\epsfig{file=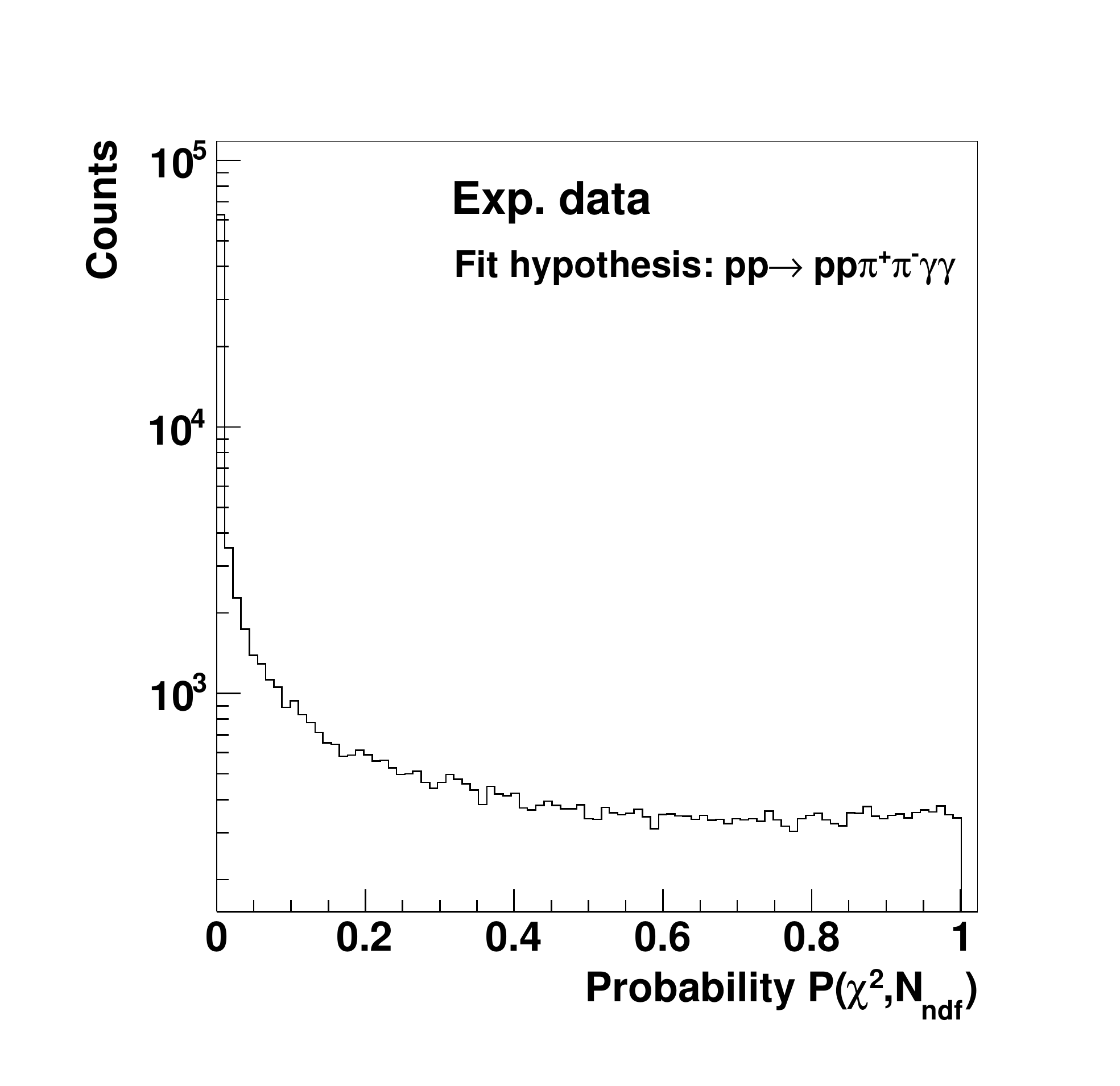,width=0.55\textwidth}}}
}
\caption{
The $\chi^2$ {\bf{(left)}} and probability $P_F(\chi^2,N_{ndf})$ {\bf{(right)}} distribution for the 
simulated {\bf{(upper row)}} and experimental data {\bf{(lower row)}} 
for the reaction $pp\to pp\eta \to pp\pi^+\pi^-\pi^0 (\gamma\gamma)$.
The superimposed red curve indicates the theoretical $\chi^2$ distribution calculated for the five degrees 
of freedom. The disagreement between theoretical and experimental $\chi^2$ distribution reflects 
the non-Gaussian errors taken into the fitting procedure and treated as Gaussian. 
}
\label{chi2}
\end{figure}

Having the errors parameterized one can execute the kinematic fitting procedure both on the 
measured and simulated data samples. Observables in each event will be varied such that the momentum and
energy conservation and additional constraint that two detected photons 
originate from the decay of the neutral pion are fulfilled.  
The obtained $\chi^2$ and $P_F(\chi^2,N_{ndf})$ spectra for simulated and experimental data are shown 
in Fig.~\ref{chi2}.

The distribution of the $\chi^2$ for the simulated data presented in the upper left 
part of Fig.~\ref{chi2} shows an significant enhancement at large values of $\chi^2$ in comparison to the 
theoretically predicted $\chi^2$ distribution for five degrees of freedom. This reflects the 
non-Gaussian error distribution which were assumed and taken into the fitting routine.
The problem were seen for charged pions flying to the forward and backward part of the 
MDC (see Fig.~\ref{MDCerrors}). The right panel of Fig.~\ref{chi2}
shows the distribution of probability $P_F(\chi^2,N_{ndf})$ which has an enhancement at zero corresponding to events seen 
in tail of the $\chi^2$ distribution. Events which are populating low probabilities are 
wrongly reconstructed. The probability distribution is relatively flat for values of probability 
larger than 0.1, and only these events will be taken in further analysis.  
\begin{figure}[!t]
\hspace{.0cm}
\parbox{0.5\textwidth}{\centerline{\epsfig{file=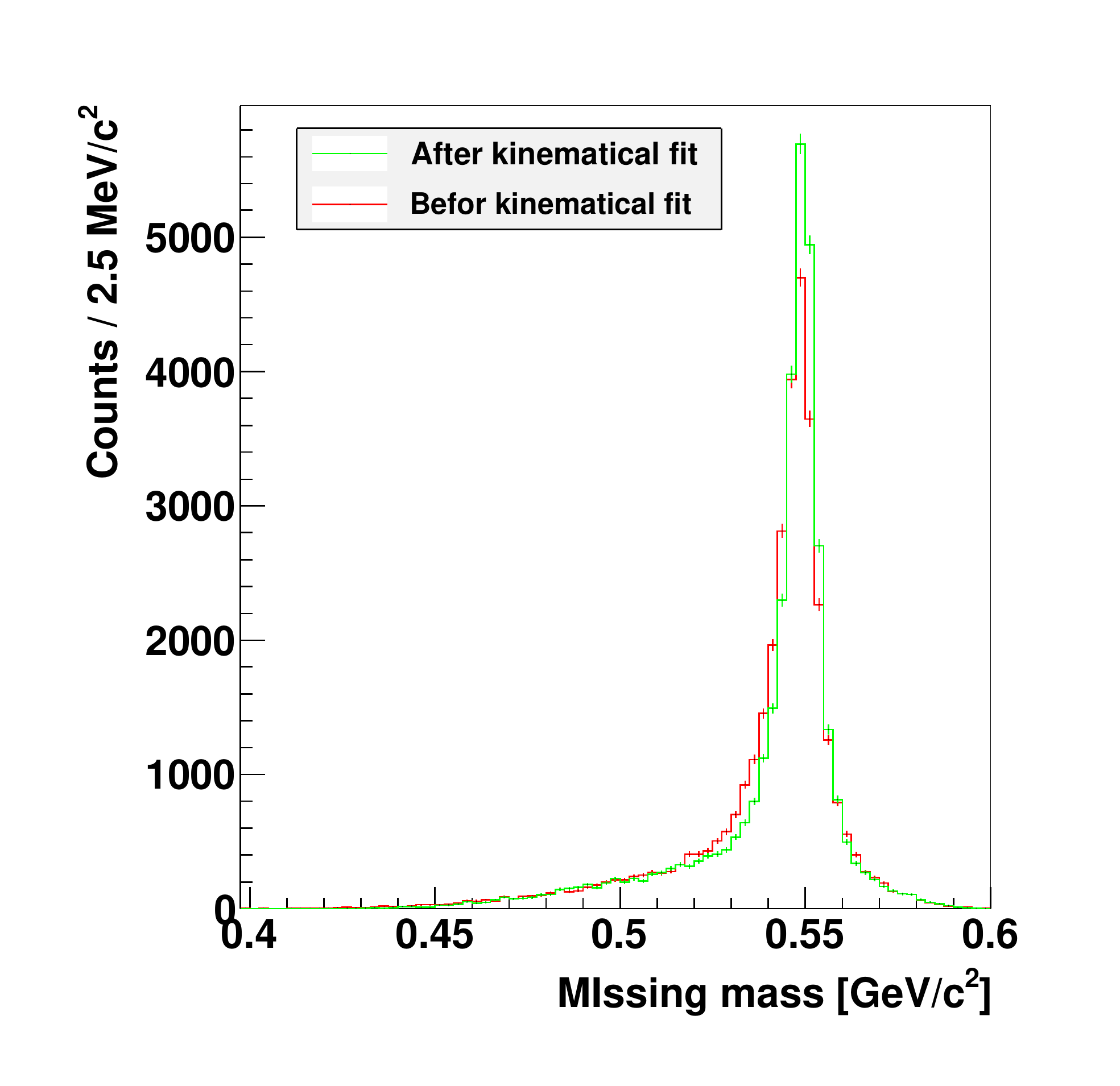,width=0.55\textwidth}}}
\hspace{.0cm}
\parbox{0.5\textwidth}{\centerline{\epsfig{file=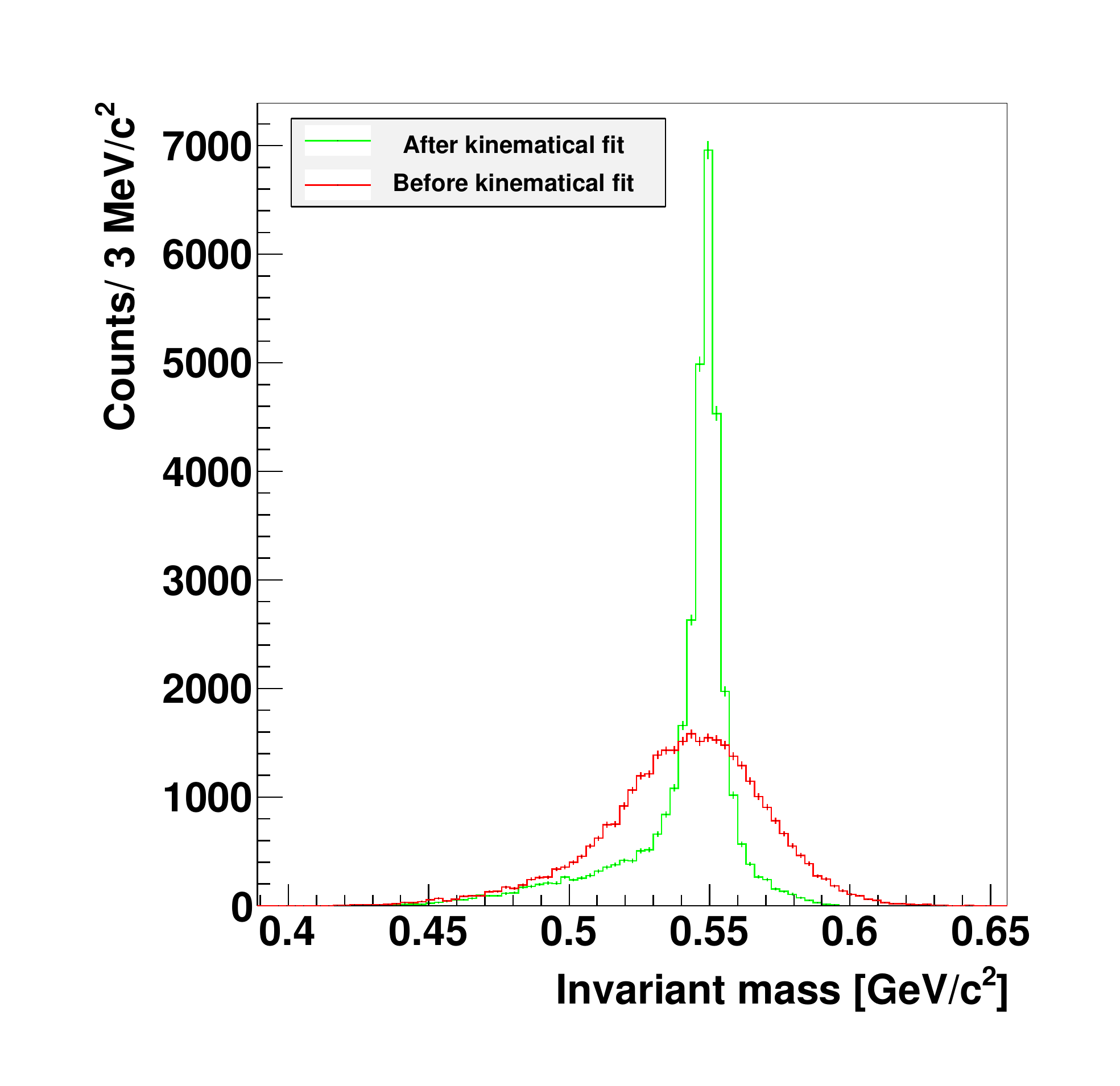,width=0.55\textwidth}}}
\caption{
Missing mass distribution {\bf{(left)}} for the $pp\to pp X$ reaction and 
invariant mass spectrum for the $\pi^+\pi^-\pi^0$ system {\bf{(right)}} before (red histogram) 
and after (green histogram) kinematic fit and rejection of events with $P_F(\chi^2,N_{ndf})$ less than 0.1.  
}
\label{mmKF}
\end{figure}

The lower two plots in Fig.~\ref{chi2} illustrate the $\chi^2$  and probability distributions 
obtained for the experimental data sample. One can see that here  
the $P_F(\chi^2,N_{ndf})$ distribution is peaked at lower probabilities even more strongly than result obtained 
from the simulations. 
This is because in the data sample still a background events are present for which the fit 
hypothesis is not justified. Also the $\chi^2$ has much more enhanced tail in comparison to 
Monte Carlo spectrum due to non-Gaussian error estimate for the charged pions which are 
coming also from the background channels. To suppress majority of the background 
and wrongly reconstructed events the region of $P_F(\chi^2,N_{ndf}) > 0.1$ was selected for further analysis.

The missing mass spectrum before and after applying the kinematic fit 
is shown in Fig.\ref{mmKF} (left). One can see a slight improvement in the resolution. 
The ratio of the signal to background increased from 
2.2 to 2.9. The invariant mass distribution of the $\pi^+\pi^-\pi^0$ system  is shown in 
Fig.~\ref{mmKF} (right). A very narrow peak is seen in the mass of the $\eta$ meson
in comparison to the distribution before the kinematical fit. The resolution of the invariant
mass after the fit is equal to $\sigma = 4~MeV/c^2$, compared $\sigma = 25~MeV/c^2$
before the fit. Thus, kinematical fit had improved significantly the resolution of invariant mass,
and as a consequence also the signal to background ratio. The remaining background is due to the direct 
three pion production. One can also see that the shape of the distribution after the fit is very similar to 
the missing mass of the two protons. This is because in the fit the information of protons four-momentum 
vectors is used, and they are determined with better precision than four-momentum vectors of other particle. 

Furthermore, it is also worth to notice that one can improve the resolution of fit parameters by 
demanding the invariant mass of the decay particles to be equal to the mass of the $\eta$ meson. 
However, this would result in, squeezing the whole distribution to the value of the mass of the $\eta$ meson, 
and thus subtraction of the background would be no longer possible. 
Therefore in this analysis we did not used this condition. 

\section{Estimation of the analysis efficiency}\label{Sec:eff}
\hspace{\parindent}
To determine the overall reconstruction efficiency  of a particular reaction channel
one has to take into account the geometrical acceptance of the detector and efficiency 
of the applied analysis chain. The geometrical acceptance can be studied using kinematic 
event generator by superimposing on the scattering angle of final state 
particles the condition that it has to be within a range of the detector. 
The geometrical acceptance was studied in Sec.~\ref{sec:kinematyka}. 

In order to study the efficiency of applied analysis, and the resulting background suppression, 
various reactions were simulated using the WASA Monte Carlo package and analyzed by the same 
software as it was used for the experimental data. The list of reactions together with the 
number of simulated events is given in the Tab.~\ref{tab:sim}.
\begin{table}[!h]
\centering
\begin{tabular}{c|c|c}
\hline
No. & Reaction & Number of generated events\\\hline\hline
1 & $pp\to pp\eta\to pp\pi^+\pi^-\pi^0\to pp\pi^+\pi^-\gamma\gamma$  (signal)   &  $20\times 10^6$ \\\hline
2 & $pp\to pp\eta\to pp\pi^+\pi^-\gamma$                               &  $5\times 10^6$ \\
3 & $pp\to pp\eta\to pp e^+ e^-\gamma$                                 &  $20\times 10^6$ \\\hline
4 & $pp\to pp\pi^+\pi^-\pi^0\to pp\pi^+\pi^-\gamma\gamma$              &  $20\times 10^6$\\
5 & $pp\to pp\pi^+\pi^-$                                               &  $58\times 10^6$\\
6 & $pp\to pp\pi^0\pi^0\to pp\gamma\gamma\gamma\gamma$                 &  $18\times 10^6$\\\hline
\end{tabular}
\caption{List of simulated reaction channels for efficiency studies. The number in right column 
indicates the number of initially generated events.}
\label{tab:sim}
\end{table}
\begin{figure}[!h]
\hspace{2.7cm}
\parbox{0.7\textwidth}{\centerline{\epsfig{file=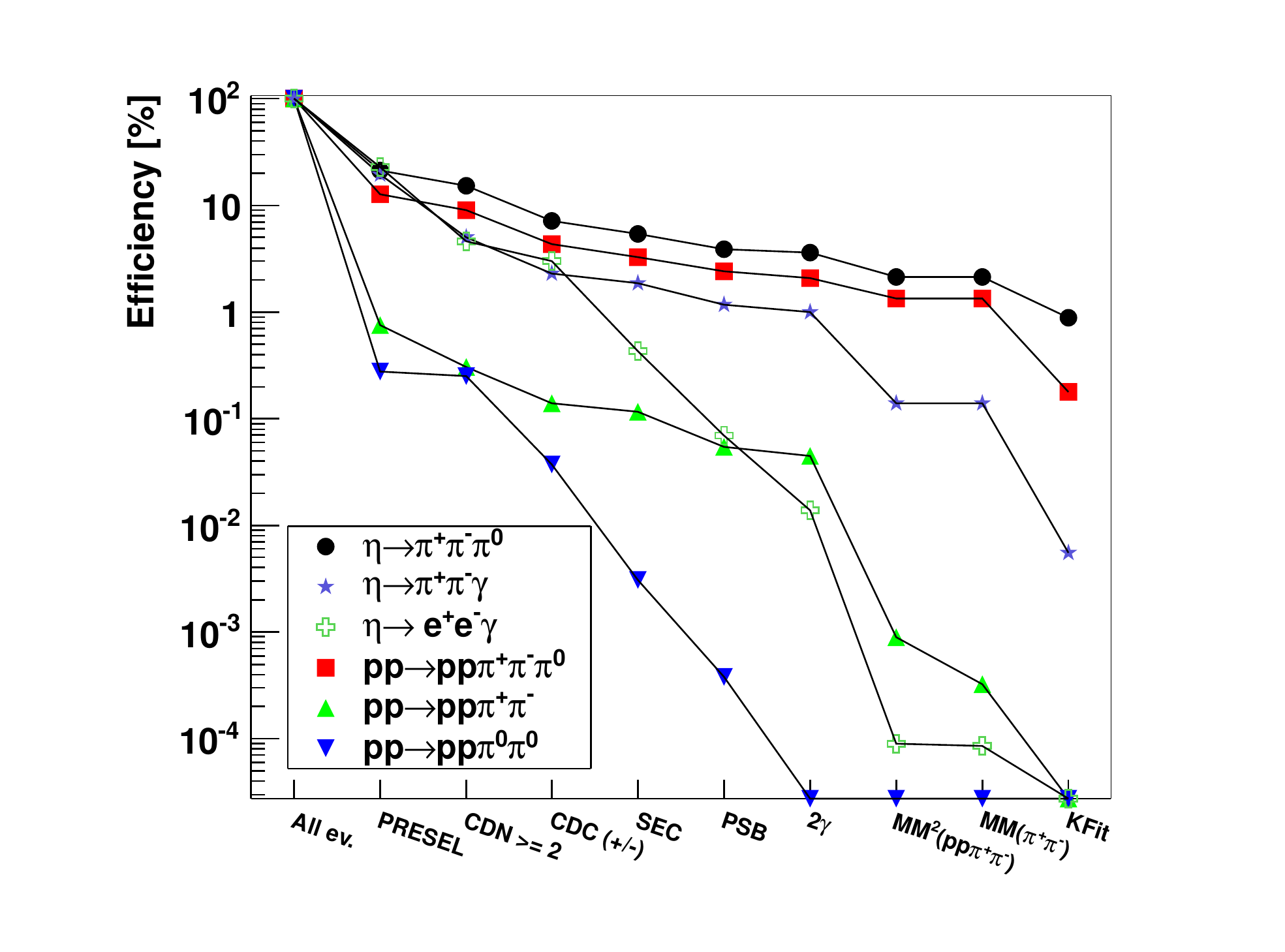,width=0.75\textwidth}}}
\caption{
Reconstruction efficiency obtained after subsequent application of conditions indicated at the bottom of the 
figure. Results are shown for simulations for reactions listed inside the figure.
}
\label{effMC}
\end{figure}
Second reaction listed, may be misidentified as a signal due to the splitting of clusters in 
the electromagnetic calorimeter. Third reaction can imitate signal due to misidentification
of $e^+$ and $e^-$ as pions and splitting of signals in the calorimeter. Fourth reaction constitute a physical background of investigated decay. The fifth process can obscure the signal channel by the bremsstrahlung 
of charged particles in the calorimeter. Sixth reaction can simulate a signal due to the external conversion 
of $\gamma$ quanta and the subsequent misidentification on the electron and positron as pion pair. 

The efficiency at each analysis stage can be studied by determining the 
number of events left after applying the particular condition. 
Figure~\ref{effMC} presents obtained reconstruction efficiency for several simulated reactions
in the subsequent steps of the analysis. The name of bins corresponds to the name of applied cut 
and they are explained in Tab.~\ref{tab:eff-1}. Furthermore in the right column of Tab.~\ref{tab:eff-1}
the efficiency of individual conditions for the signal reaction is given. 
\begin{table}[t]
\centering
\begin{tabular}{c|l|c|c}
\hline
Cut  & Cut description & Section & Efficiency\\\hline\hline
 PRESEL           & Preselection of two protons in FD, two        & Sec.~\ref{sec:ppeta}, & 21.2\%\\
                  & oppositely charged particles in CD and        &                       &\\
                  & two or more clusters in SEC.                  &                       &\\\hline
 CDN>2            & Selection of at least two clusters            & Sec.~\ref{sec:pipi},  & 71.6\%\\
                  & in CD with minimum energy of 20~MeV.          & Fig.~\ref{CDCmulti}   &\\\hline
 $CDC_{\pm}$=2    & Reconstruction of tracks corresponding to     & Sec.~\ref{sec:pipi},  & 47.2\%\\
                  & two particles with opposite charges.          & Fig.~\ref{CDCmulti}   &\\\hline
 SEC              & Particle identification in the                & Sec.~\ref{sec:pipi},  & 75.5\%\\
                  & Scintillating Electromagnetic Calorimeter.    & Fig.~\ref{pidCDC}     &\\\hline
 PSB              & Particle identification in Plastic            & Sec.~\ref{sec:pipi},  & 71.9\%\\
                  & Scintillator Barrel.                          & Fig.~\ref{pidPSB}     &\\\hline
 2$\gamma$        & Selection of two photons originating from     & Sec.~\ref{sec:pi0},   & 92.7\%\\
                  & the decay of $\pi^0$ meson.                   & Fig.~\ref{imgg}       &\\\hline
 MM$^2(pp\pi^+\pi^-)$ & Cut on the missing mass squared of the    & Sec.~\ref{sec:bcg},   & 59.3\%\\
                  & $pp\to pp\pi^+\pi^- X$ reaction.              & Fig.~\ref{mmpppipiX}  &\\\hline
 MM($\pi^+\pi^-$) & Cut on the missing mass for the reaction      & Sec.~\ref{sec:bcg},   & 99.9\%\\
                  & $pp\to\pi^+\pi^- X$.                          & Fig.~\ref{mmpppipiX}  &\\\hline
 KFit             & Kinematical fit and rejection of events with  & Sec.~\ref{sec:kfit},  & 41.8\%\\
                  & probability $P_F(\chi^2,N_{ndf})$ < 0.1.      &                       &\\\hline\hline
 Total            & Final reconstruction efficiency               &                       & 0.89\%\\\hline
\end{tabular}
\caption{Reconstruction efficiency of the individual steps of the analysis for the signal reaction 
$pp\to pp\eta\to\pi^+\pi^-\pi^0(\gamma\gamma)$ obtained from the Monte Carlo simulations.}
\label{tab:eff-1}
\end{table}

As expected first biggest drop in the efficiency comes from the preselection of candidates for the 
investigated reaction channel in Forward and Central Detectors. As it was shown in Sec.\ref{sec:kinematyka}
the geometrical acceptance of the detection apparatus causes the lost of 65\% of all events.
The rest of the decrease in the efficiency on the preselection level can be attributed to 
selective conditions in the Central Detector specially the conditions for identification of charged pions 
for which we demanded signals in all central sub-detectors. Second major decrease comes after applying 
the condition on the probability $P_F(\chi^2,N_{ndf})$ of the kinematical fit. 

The performed analysis allowed to increase significantly the signal to background ratio. The background 
from the two pion production was suppressed to the negligible level, and the signal of most dangerous decay 
($\eta\to\pi^+\pi^-\gamma$) was suppressed by more than four orders of magnitude.
The event sample resulting after applying all conditions, contains nearly in 100\% the 
$\eta\to\pi^+\pi^-\pi^0$ decay and the only background left corresponds to the same final state originating 
from direct three pion production. This background can be seen in the missing and invariant 
mass spectra shown in Fig.~\ref{mmKF} as a continues distribution spreading over much larger range than the 
$\eta$ meson peak. In the next chapter we will introduce the method how to estimate contribution of this   
background using a fit of polynomial function to the missing mass distribution. 
 
\chapter{Results for the $\eta\to\pi^0\pi^+\pi^-$ decay}
\hspace{\parindent}
Identification of all final state particles of the $\eta\to\pi^+\pi^-\pi^0$ decay enables to 
use a Dalitz plot analysis to study the dynamics of three body system which manifests itself in the population 
density distribution in this plot. The parameterization of the events is given conveniently 
in the normalized variables X and Y, dependent on the kinetic energies of 
three particles in the rest frame of the $\eta$ meson which were introduced in equations 
\ref{dX} and \ref{dY}. In this chapter we will present the final results for the Dalitz plot distributions. 

\section{Resolution of the observables determination}
\hspace{\parindent}
In order to use the Dalitz distribution one has to know the resolution for the determination of the 
$X$ and $Y$ observable and decide about the bin width. The interval size depends on the collected 
statistics and the accuracy in determination of four momentum vectors of particles in the final state. 

\begin{figure}[!h]
\hspace{.0cm}
\parbox{0.5\textwidth}{\centerline{\epsfig{file=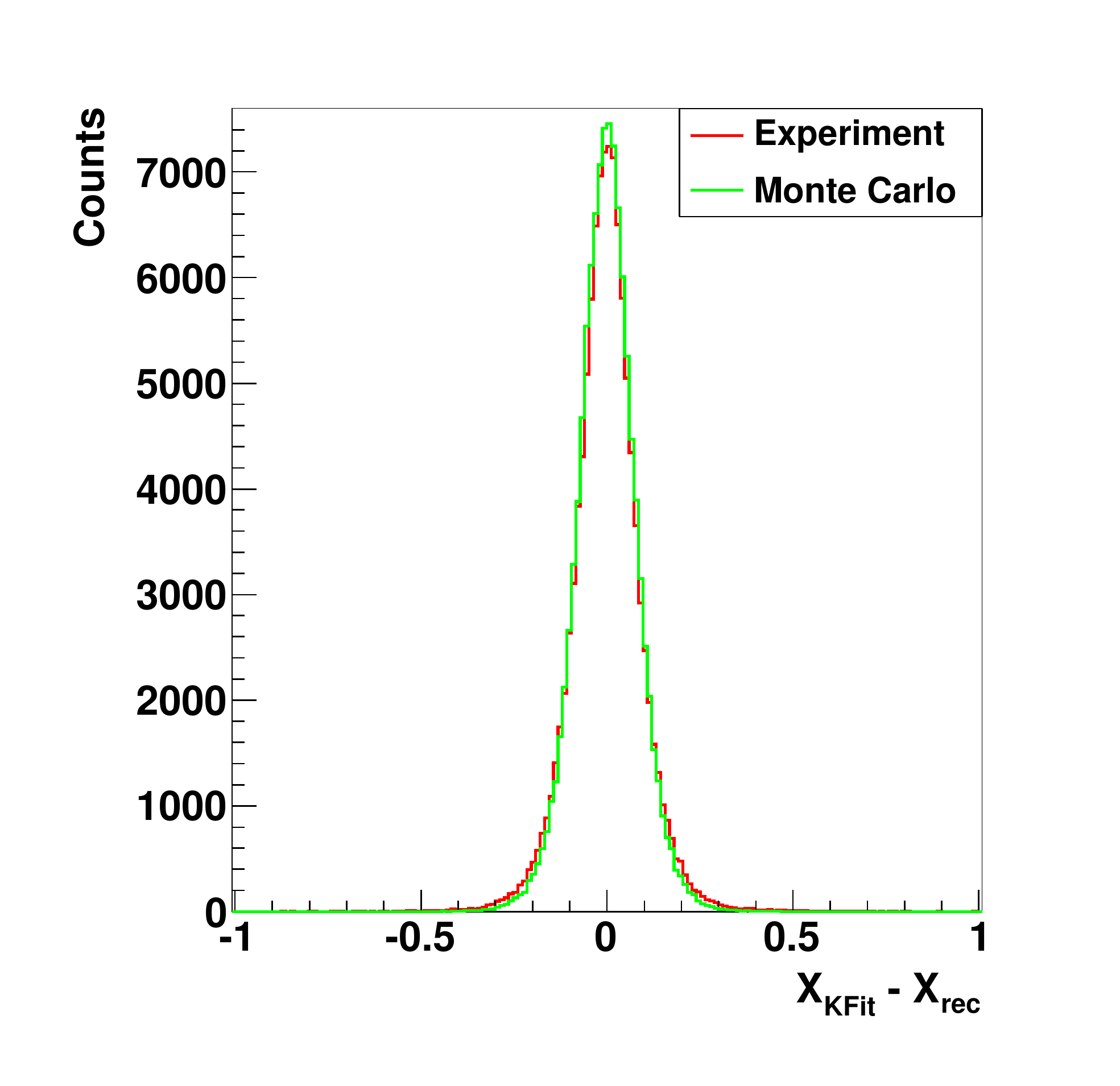,width=0.55\textwidth}}}
\hspace{.0cm}
\parbox{0.5\textwidth}{\centerline{\epsfig{file=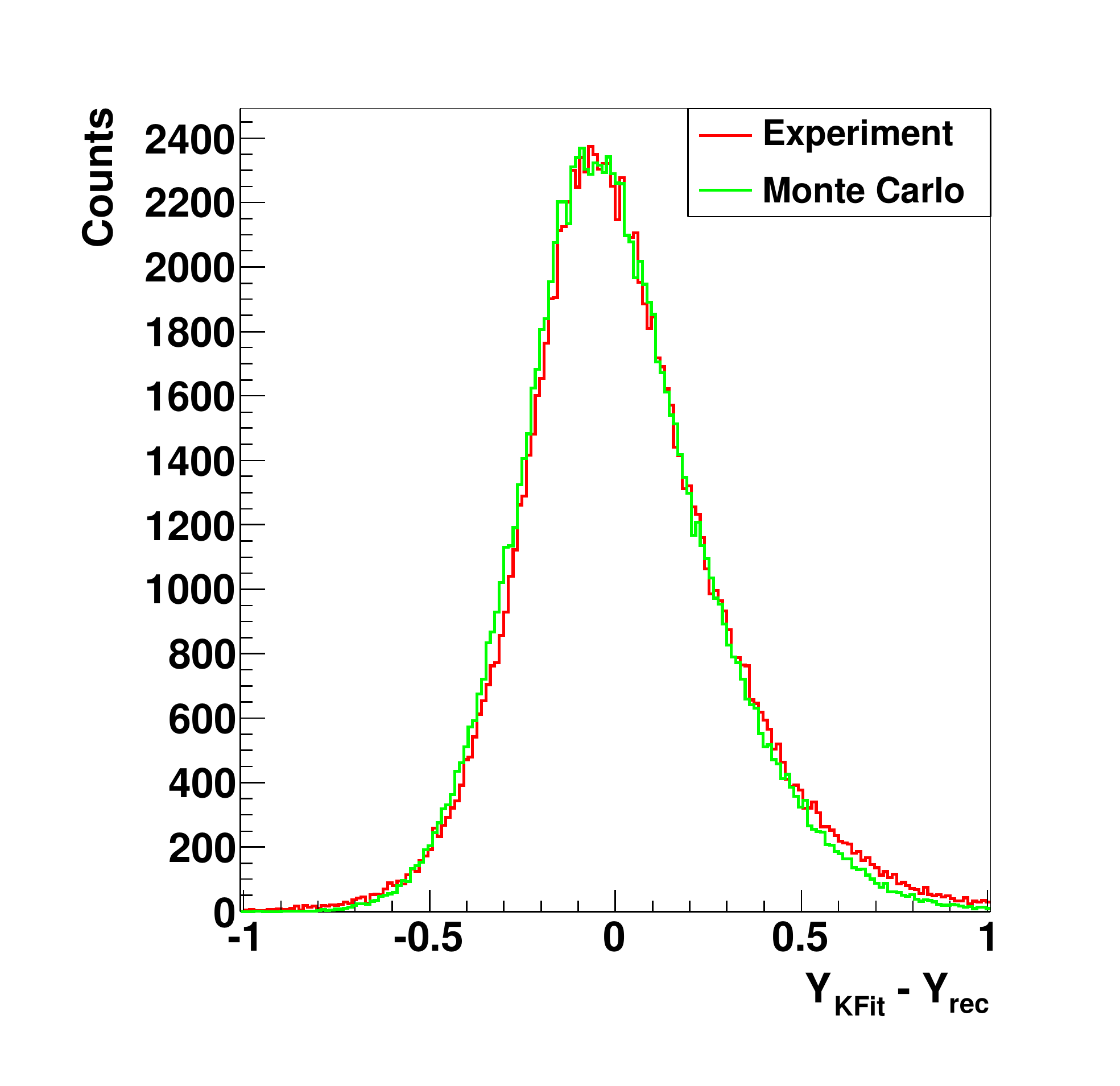,width=0.55\textwidth}}}
\caption{
Distribution of differences for {\bf{(left)}} $X$ and {\bf{(right)}} $Y$ variable between values after and before 
the kinematical fit. Red histogram stands for the experimental data and green for the Monte Carlo 
simulations. The simulated distributions were normalized to the experimental data.  
}
\label{MCRAWagree}
\end{figure}

First  to check the agreement between simulations and experiment we have determined distribution of
the difference between the values of X and Y before and after the kinematical fit for simulations and 
measured data:
\begin{equation}
\Delta X = X_{KFit} - X_{rec},
\end{equation}
\begin{equation}
\Delta Y = Y_{KFit} - Y_{rec}.
\end{equation}
The result is presented in Fig.~\ref{MCRAWagree} where the left panel shows the distribution
of $\Delta X$ and right of $\Delta Y$. The simulated data were normalized to the experimental distributions
such that integrals of both are the same.
It is clearly seen that for both observables a good agreement between simulated and measured data 
were achieved. However, one can also notice that the distribution of $\Delta Y$ variable is not symmetric and 
shifted to lower values. One of the reasons of the asymmetry in the $\Delta Y$ variable
could be high and rapidly changing values of errors at the edges of the acceptance for each type of particle as it was shown in Fig.~\ref{errparam}, and the lack of the availability in present WASA-at-COSY software of inclusion into the fit of a full covariance matrix. Observed effect influence the tiny values of searched asymmetry parameters. Therefore, for the further extraction of these parameters
in the $\eta\to\pi^+\pi^-\pi^0$ decay, we will use values of X and Y as reconstructed before the fit, which are 
independent of the simulations of errors, and their correlations~\footnote{It is worth to mention that the
kinematically fitted events for the $\eta\to\pi^+\pi^-\pi^0$ will be further used in the analysis 
of the $\eta\to\pi^0 e^+e^-$ where the three pion decay is used as a normalization channel for the
estimations of the branching ratio. In this case kinematic fit improves significantly the resolution of the 
invariant mass distribution, and and in turn it improves also the identification of the decay channel 
and allows for the more effective background suppression.}.
In order to, avoid the large uncertainties of the energy and polar angle for charged pions, at the edges of the 
acceptance, for the determination of the asymmetry parameters, we accept only pions with kinetic energy greater 
than 50~MeV and with scattering angle greater than 35$^o$. 

\begin{figure}[!h]
\hspace{.0cm}
\parbox{0.5\textwidth}{\centerline{\epsfig{file=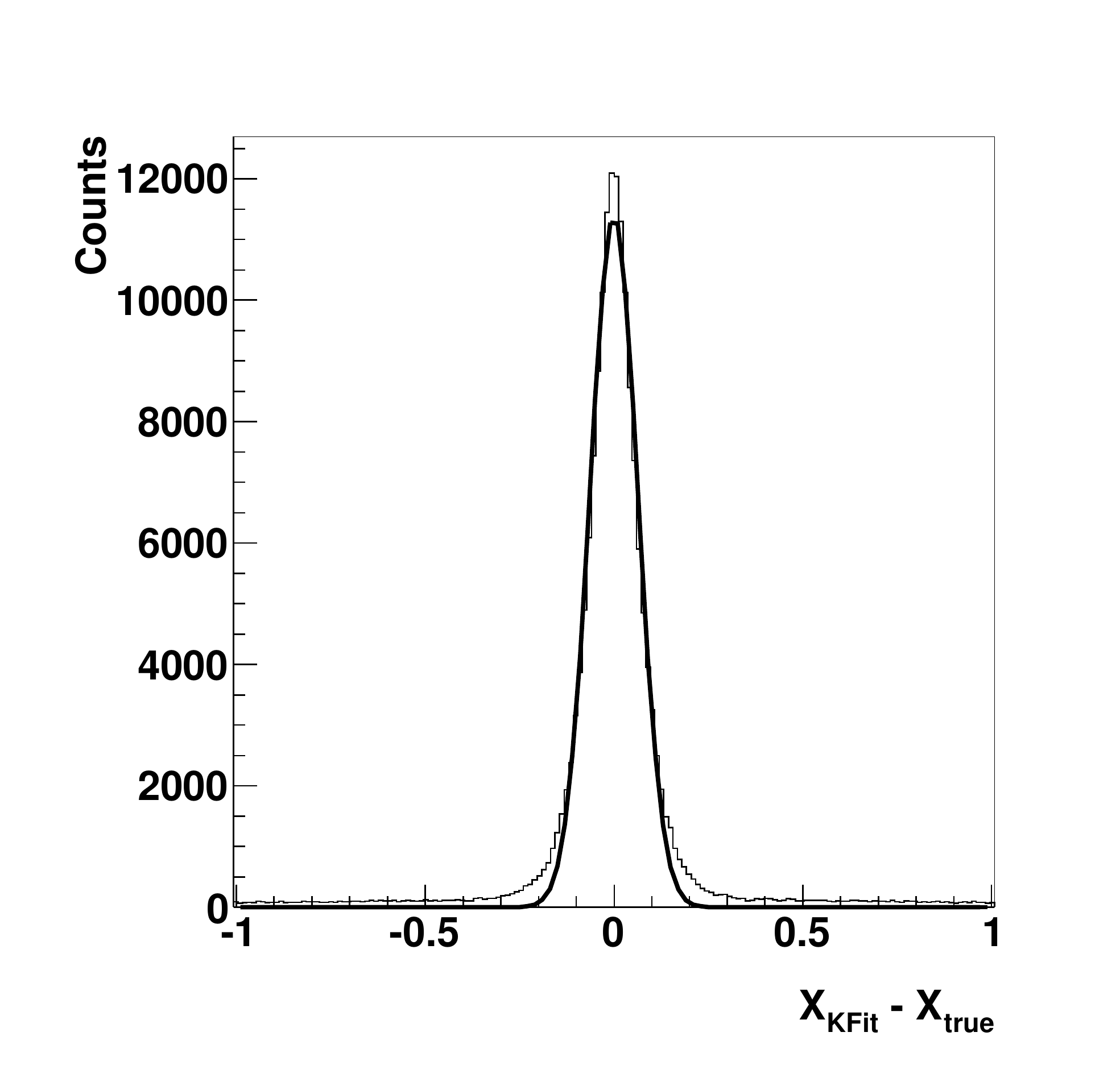,width=0.55\textwidth}}}
\hspace{.0cm}
\parbox{0.5\textwidth}{\centerline{\epsfig{file=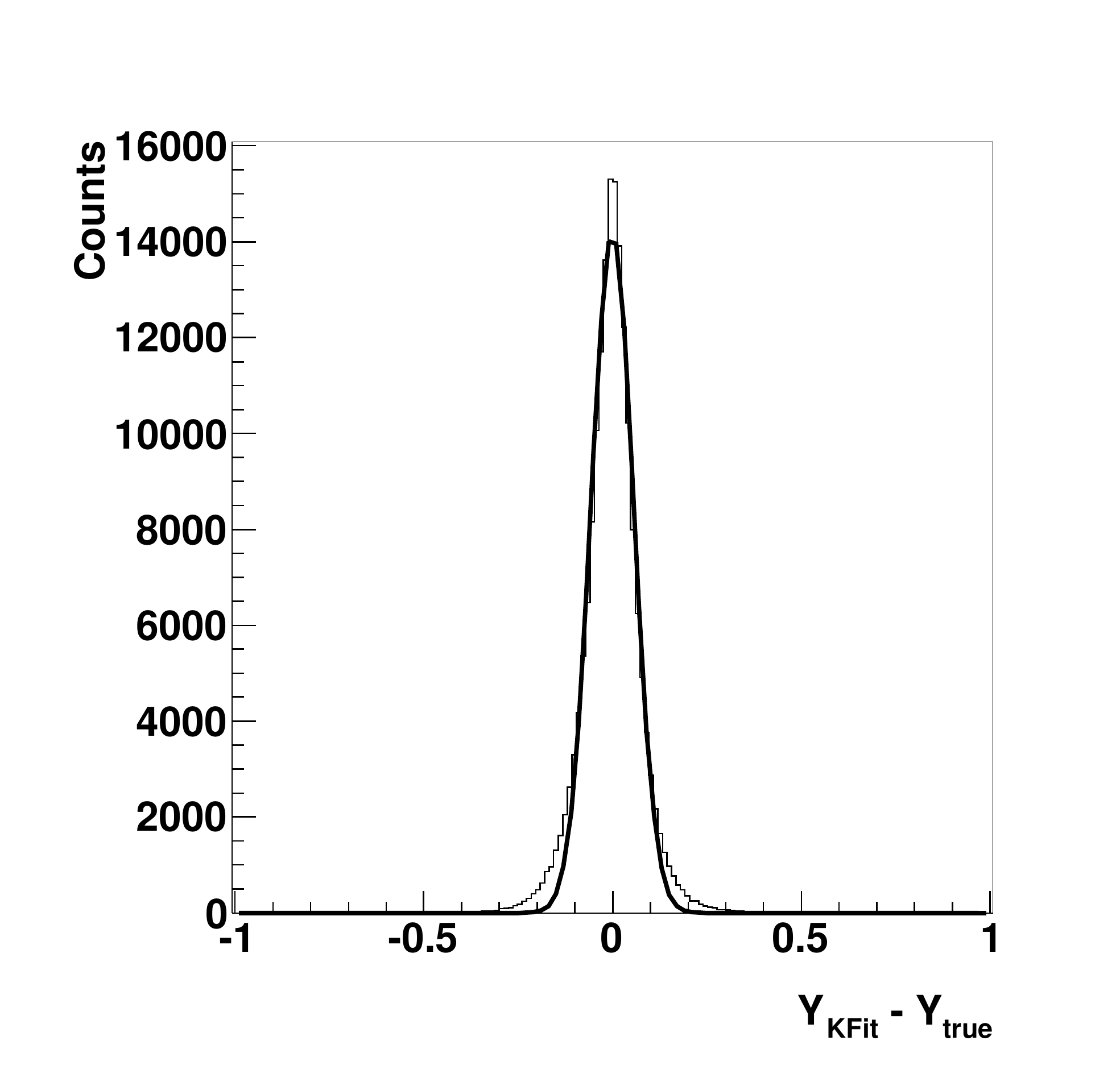,width=0.55\textwidth}}}
\caption{
The distributions of $\Delta X_{res}$ {\bf{(left)}} and $\Delta Y_{res}$ {\bf{(right)}} obtained 
from simulations. The superimposed lines indicate the result of fit with the Gaussian function.
}
\label{XYres}
\end{figure}

In order to estimate the resolution of the reconstruction of the X and Y we have taken simulated 
data, and for each event compared generated values with values reconstructed after the kinematical fitting:
\begin{equation}
\Delta X_{res} = X_{KFit} - X_{gen},
\end{equation}
\begin{equation}
\Delta Y_{res} = Y_{KFit} - Y_{gen},
\end{equation}
where the subscript $gen$ stands for the value calculated from generated 
four momentum vectors, and $KFit$ denotes the values calculated from momentum vectors reconstructed 
after the kinematical fit. The resulting distributions of $\Delta X_{res}$ and $\Delta Y_{res}$ 
were fitted with the Gauss functions. The procedure is illustrated with the spectra shown in Fig.~\ref{XYres}.
Obtained resolution for the determination of X is equal to $\sigma_X = 0.061$, and for 
Y is $\sigma_Y = 0.055$. 
Therefore, the binning of the Dalitz plot histogram was chosen to be in intervals of 0.2, which 
are roughly equal to about three standard deviations.

\section{Dalitz Plot}
\hspace{\parindent}
The final data sample used to determine the Dalitz plot contains predominantly only events from searched 
$pp\to pp\eta \to pp\pi^0\pi^+\pi^-$ process. The background which is left 
in majority originates from the direct production of three pions. 
In order to plot the Dalitz distribution this background has to be subtracted bin by bin. 

To remove the background one could use the Monte Carlo simulations, but for beam energy in range of 1.4~GeV
neither total nor differential cross sections for the three pion production are known, thus it is not possible 
to describe the mass distribution of the $\pi^+\pi^-\pi^0$ in a reliable way. 
Therefore, we will subtract the background using the method of polynomial fit outside the peak region. 
We apply the formula of a fourth order polynomial function, which reads\cite{Klaja:2009si}:
\begin{equation}
f(mm,a,b,c,d,e) = 
 a + b\cdot mm + c\cdot mm^2 + d\cdot mm^3 + e\cdot mm^4
\label{tloKlajus}
\end{equation}
where $a,b,c,d,e$ are free parameters varied during the fit.
\begin{figure}[H]
\mbox{
\hspace{.1cm}
\parbox{0.30\textwidth}{\centerline{\epsfig{file=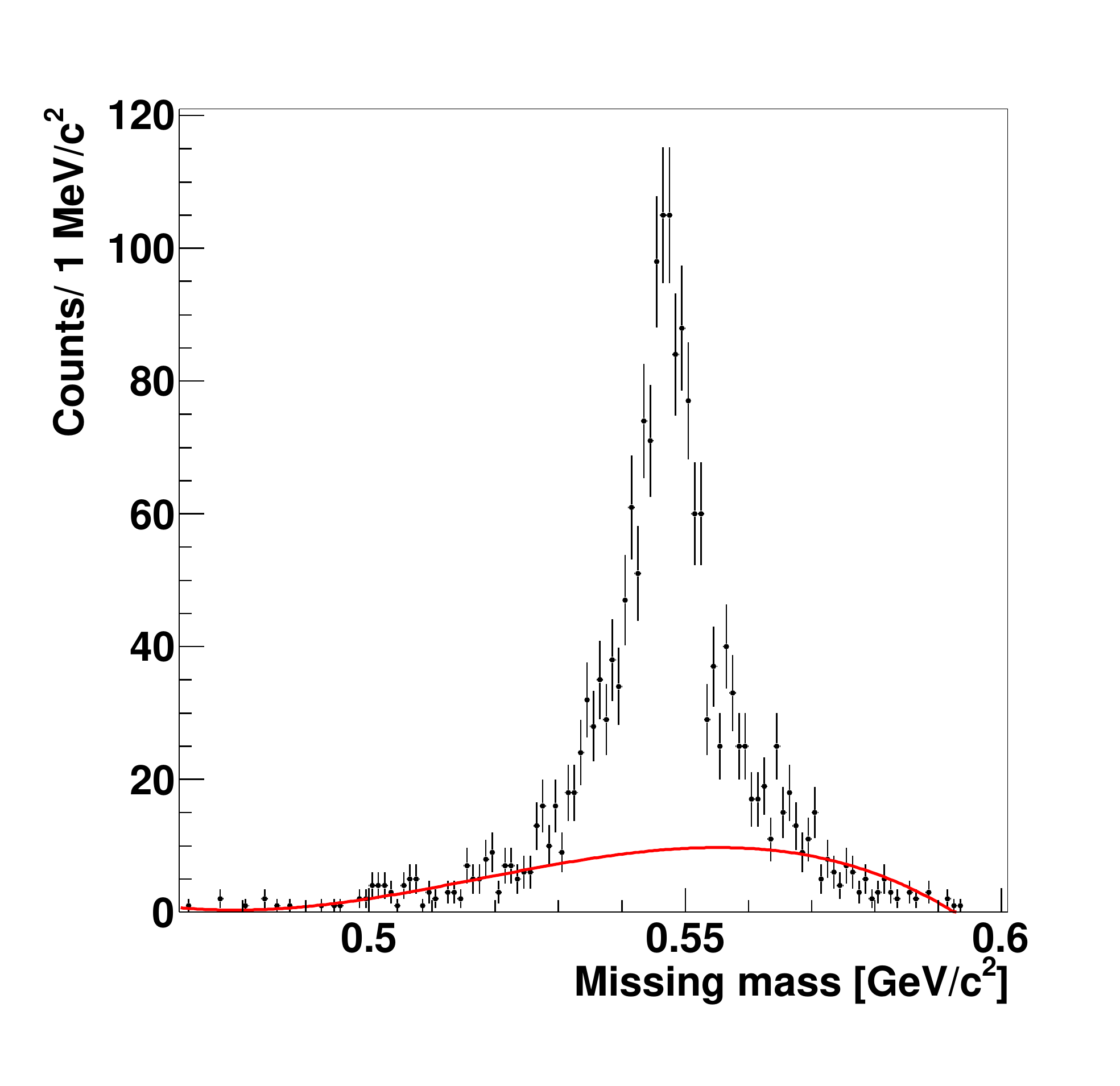,width=0.35\textwidth}}}
\hspace{.2cm}
\parbox{0.30\textwidth}{\centerline{\epsfig{file=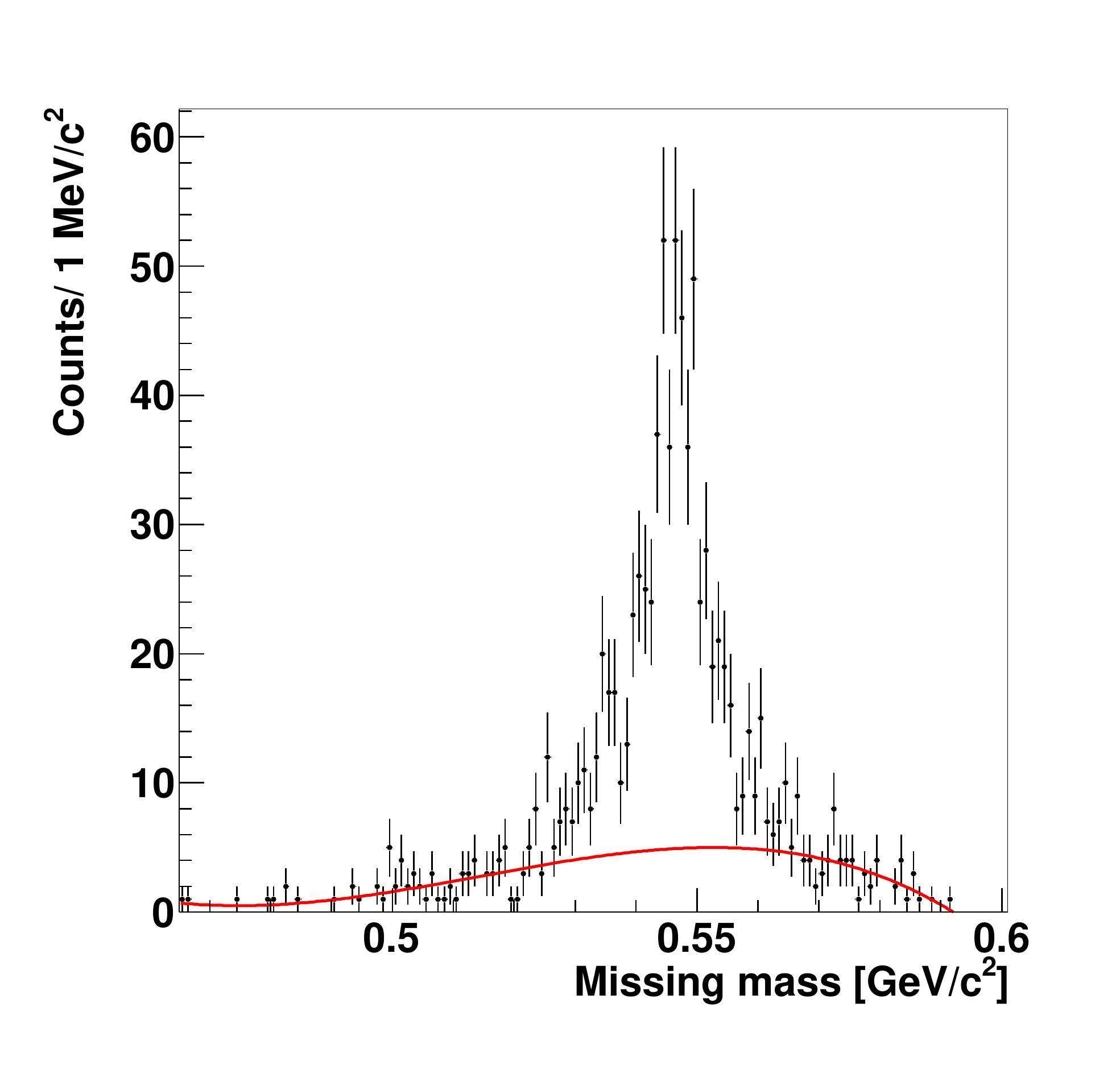,width=0.35\textwidth}}}
\hspace{.2cm}
\parbox{0.30\textwidth}{\centerline{\epsfig{file=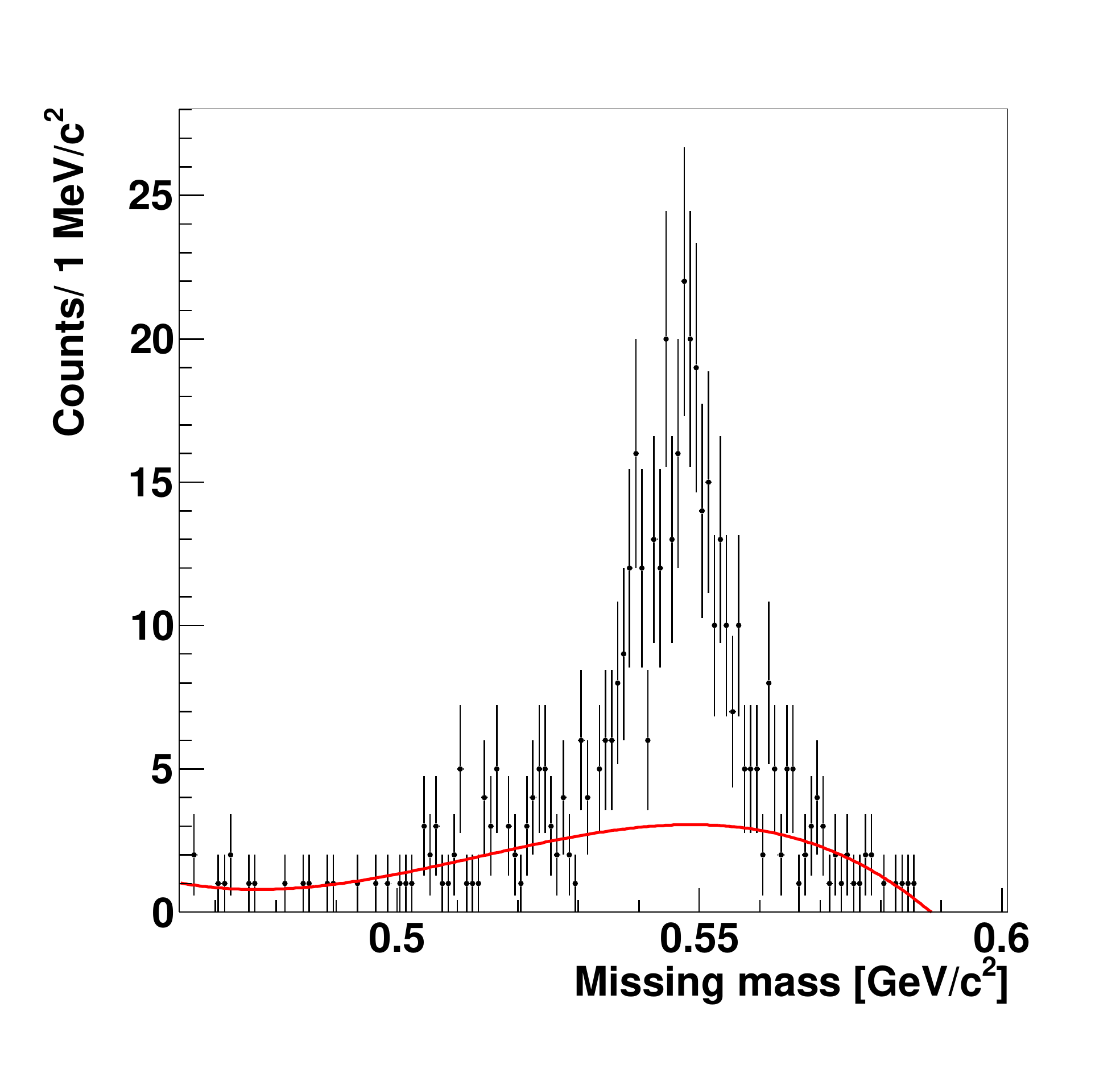,width=0.35\textwidth}}}}\\
\mbox{
\hspace{.1cm}
\parbox{0.30\textwidth}{\centerline{\epsfig{file=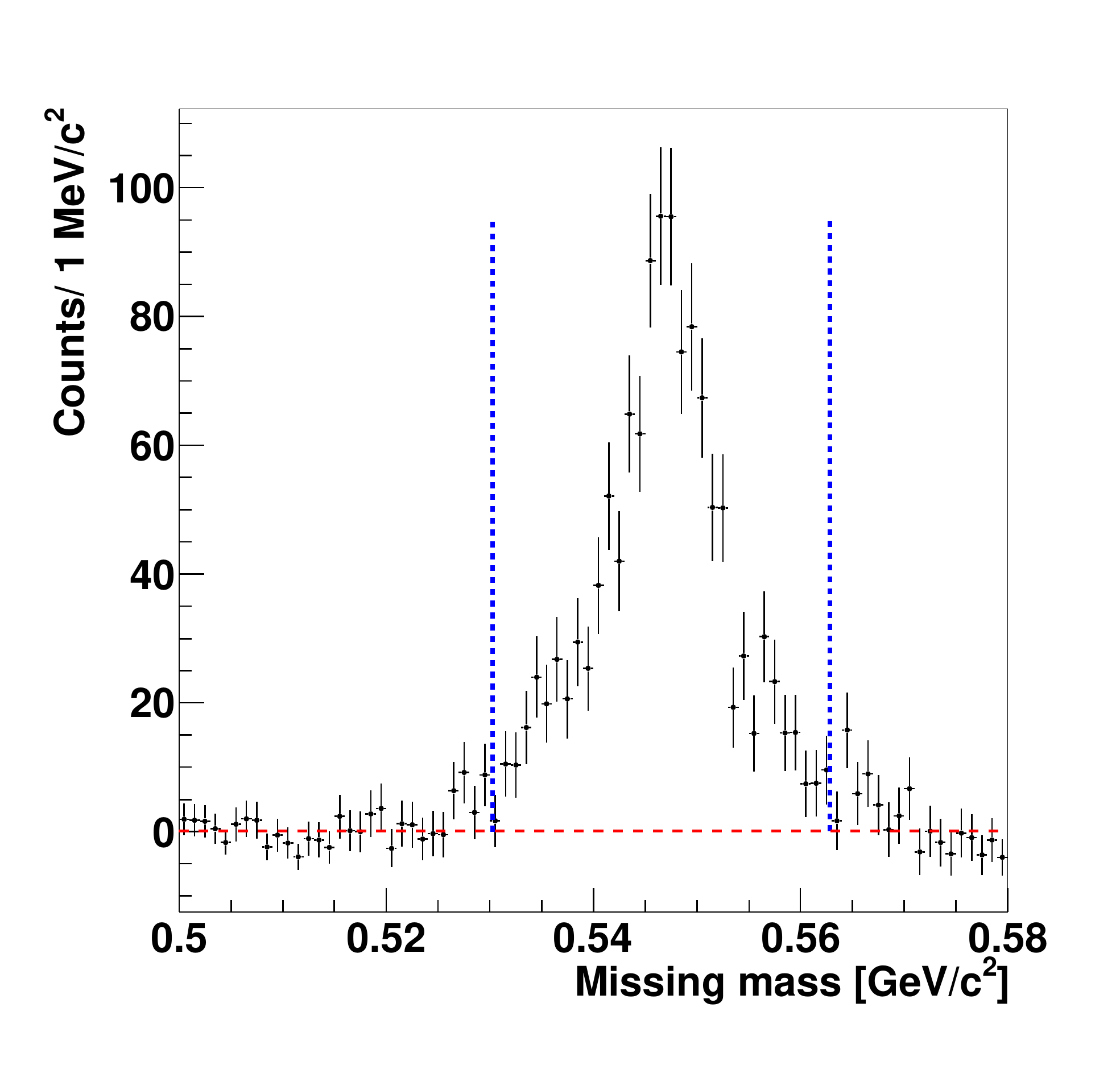,width=0.35\textwidth}}}
\hspace{.2cm}
\parbox{0.30\textwidth}{\centerline{\epsfig{file=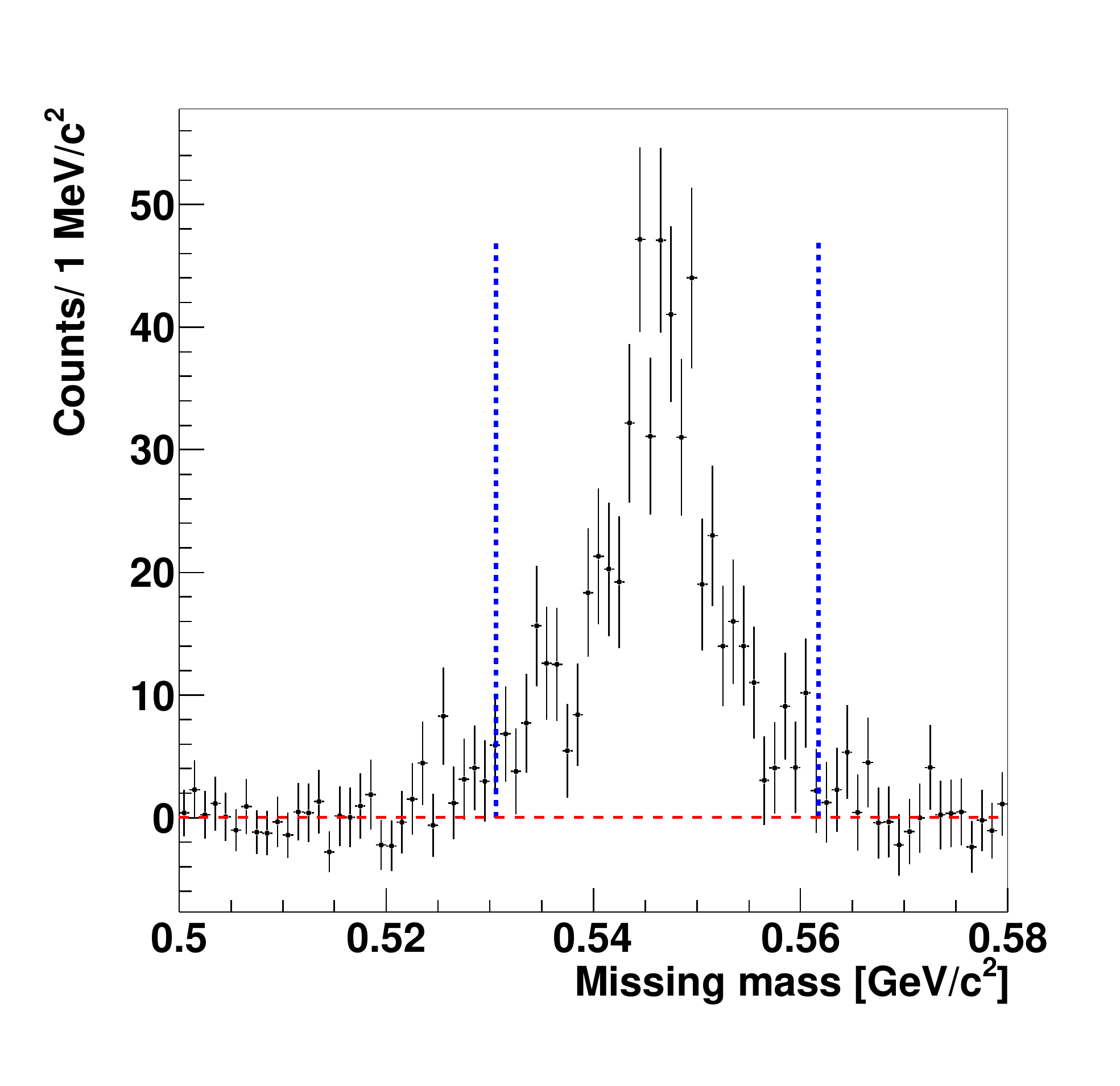,width=0.35\textwidth}}}
\hspace{.2cm}
\parbox{0.30\textwidth}{\centerline{\epsfig{file=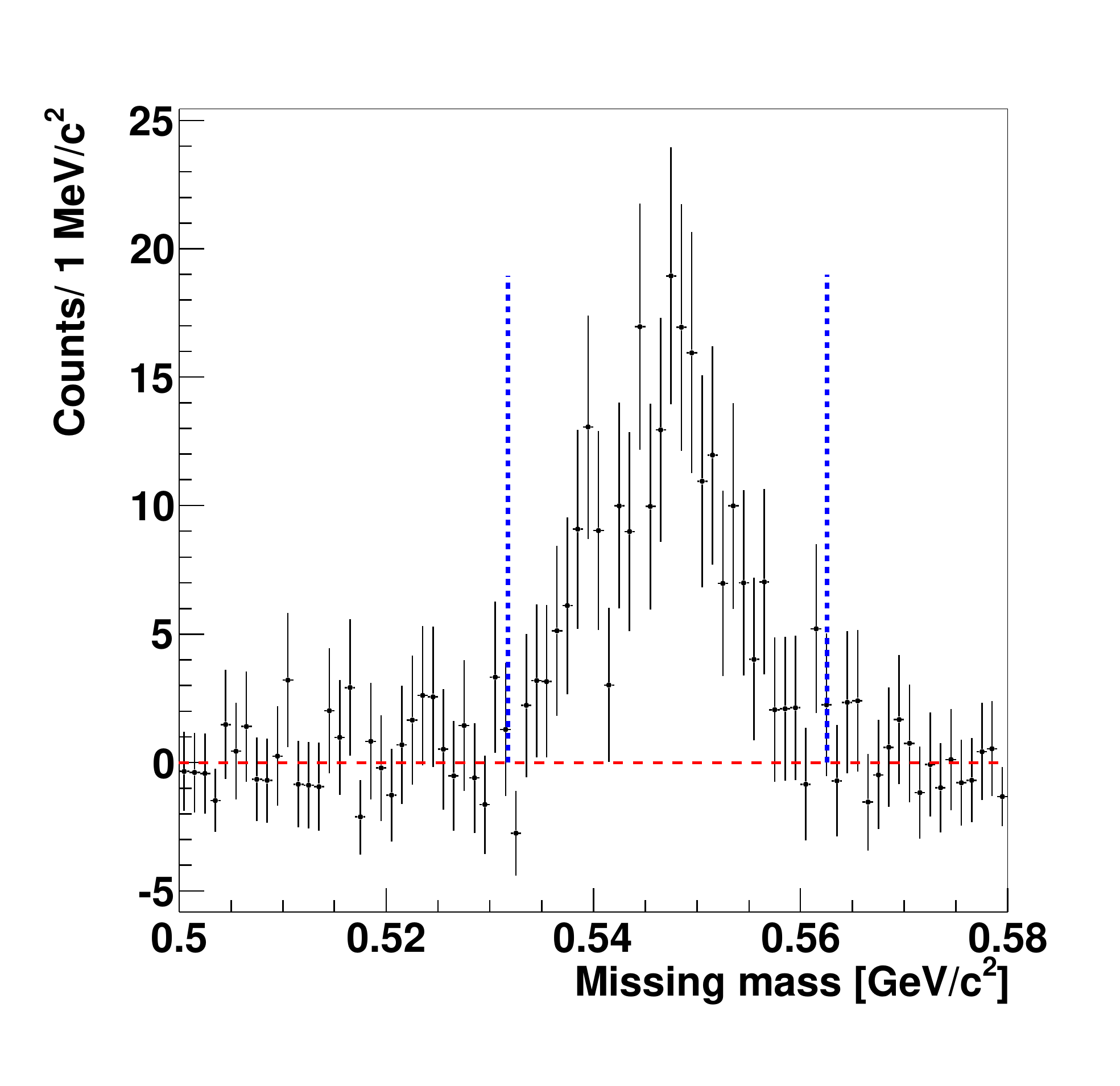,width=0.35\textwidth}}}}
\caption{
Experimental distributions of the missing mass determined for three exemplary bins of the Dalitz plot:
{\bf{(upper panel)}} before the background subtraction, and {\bf{(lower panel)}} 
after the background subtraction using a polynomial fit. The fitted function is shown as a red solid line. 
Vertical dashed lines indicates the region in which the number of signal events were counted. 
}
\label{FITsub}
\end{figure}

The upper panel of Fig~\ref{FITsub} presents exemplary missing mass for three bins of the Dalitz plot.
The superimposed red lines denote the result of the fit with the function defined in Eq.~\ref{tloKlajus}.
The polynomial was fitted to the data outside the maximum corresponding to the signal.
For each presented in Fig.~\ref{FITsub} missing mass spectrum, a very clear peak can be seen 
at the mass of the $\eta$ meson, above a continues almost flat background.  
One can also see that the fitted function is in a good agreement with the shape of the background. 
After subtraction of background we counted the number of events in the intervals marked by the dashed lines.
   
\begin{figure}[!h]
\mbox{
\hspace{.0cm}
\parbox{0.45\textwidth}{\centerline{\epsfig{file=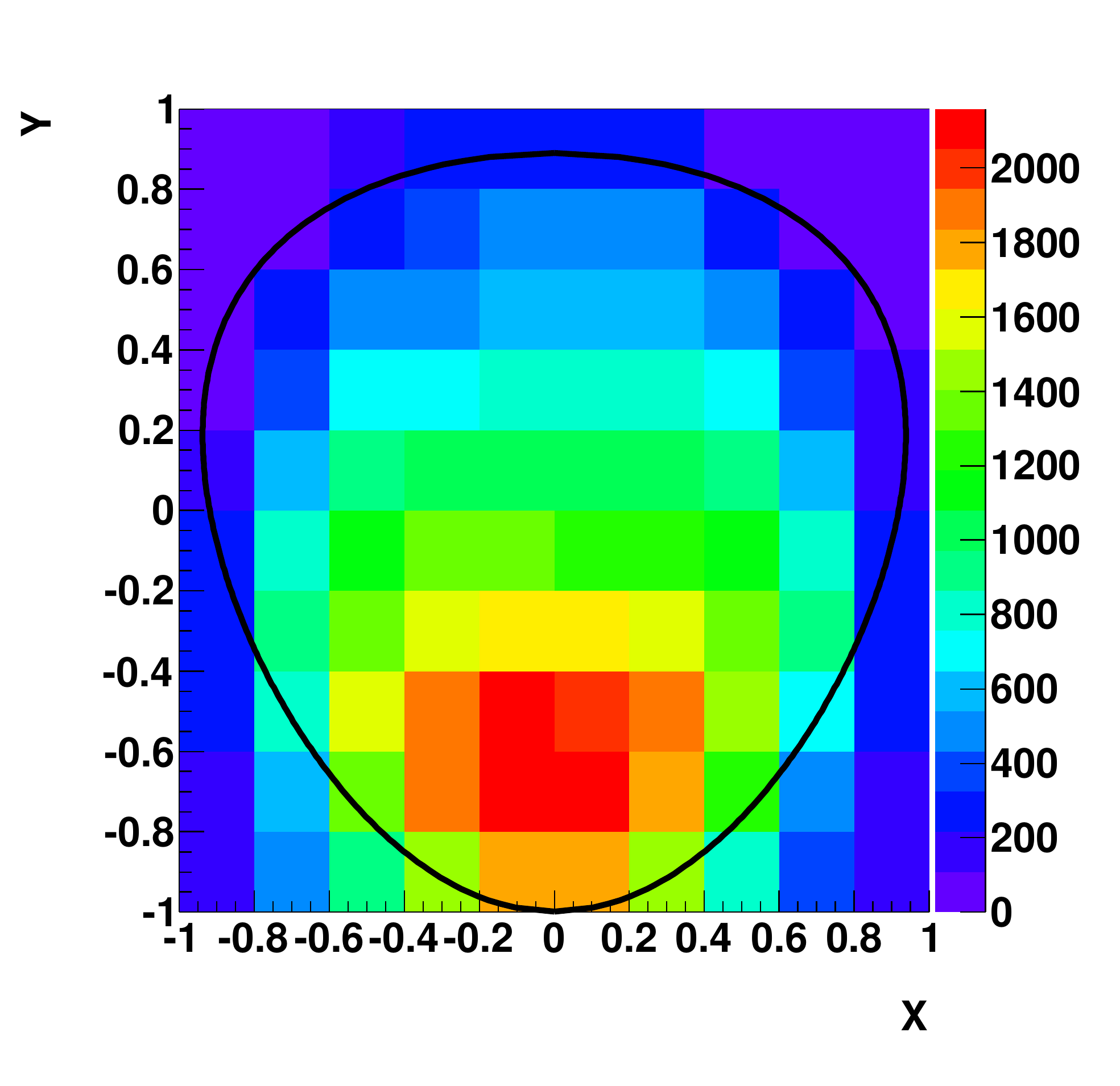,width=0.50\textwidth}}}
\hspace{.5cm}
\parbox{0.45\textwidth}{\centerline{\epsfig{file=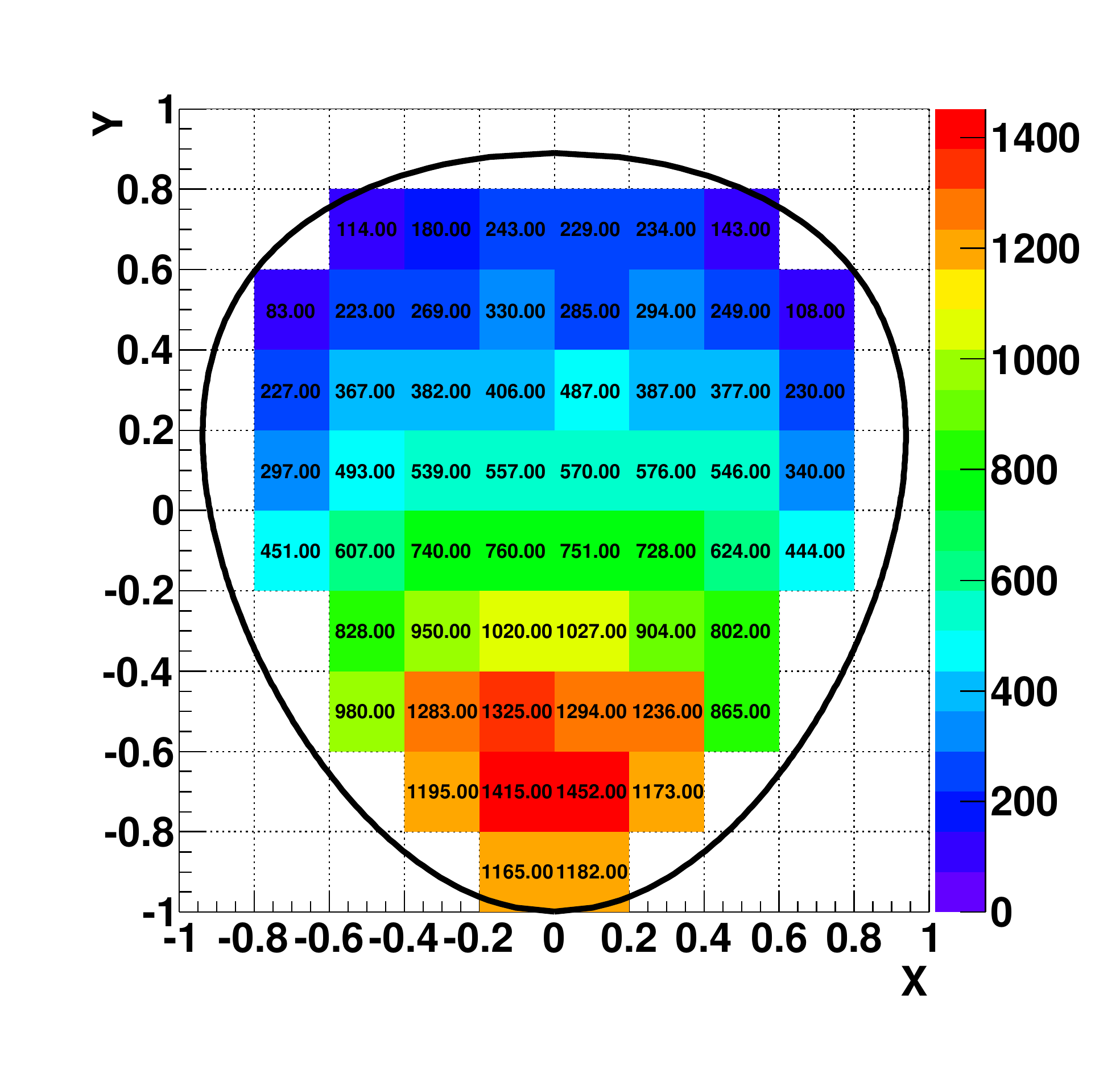,width=0.50\textwidth}}}}
\caption{
Experimental distributions of the Dalitz plot: before background subtraction {\bf{(left)}}, and after
background subtraction {\bf{(right)}}. The black solid line indicates the kinematical boundary of the 
Dalitz plot. In the right panel number of events in each interval is indicated.
}
\label{DalitzXY}
\end{figure}

Figure \ref{DalitzXY} (left) shows measured Dalitz plot for the $\eta\to\pi^+\pi^-\pi^0$ decay before background subtraction. One can see that part of the events is outside the kinematical boundaries of the Dalitz plot.
Therefore for further analysis we will consider only event in bins which centers are fully inside the kinematical boundaries. The distribution on the right shows the result obtained after the background subtraction, performed separately for each bin.

In the next step the experimental distributions have to be corrected for the detection and reconstruction
efficiency to compare to the model expectations. 
The overall efficiency was determined in Sec~\ref{Sec:eff} and it was defined as a fraction of number 
initially simulated events to the number of reconstructed events after the applied analysis.  
In order to be able to draw a conclusion as much model independent as possible the efficiency 
correction has to be determined separately for each interval of the distribution of interest.
Therefore, for the Dalitz plot the correction has been done using the two-dimensional 
efficiency distribution derived for each bin. This method enables to be independent of the model used for the 
description of the decay amplitude in the simulations.
The only dependence may occur due to the finite bin size, but anyhow this influence is only
due to the non-linearity of the distribution within the bin limits and this in our case is negligible. 
To determine the efficiency we have 
simulated a sample of events assigning a weight to each event according to the ChPT Leading Order 
predictions~\cite{Bijnens:2007pr,Bijnens:2007jk} (see Tab.~\ref{tab:parametry}). 
The relation for the efficiency is given by:
\begin{equation}
\epsilon(X,Y) = \frac{\sum_j w_{j}^{rec}(X,Y)}{\sum_i w_{i}^{gen}(X,Y)},
\label{eff}
\end{equation}
where $w_{j}^{rec}$ and $w_{i}^{gen}$ denotes the weights of the events and the summation is done over
reconstructed and generated events within given X,Y interval of the Dalitz plot. 
\begin{figure}[!t]
\mbox{
\hspace{.0cm}
\parbox{0.30\textwidth}{\centerline{\epsfig{file=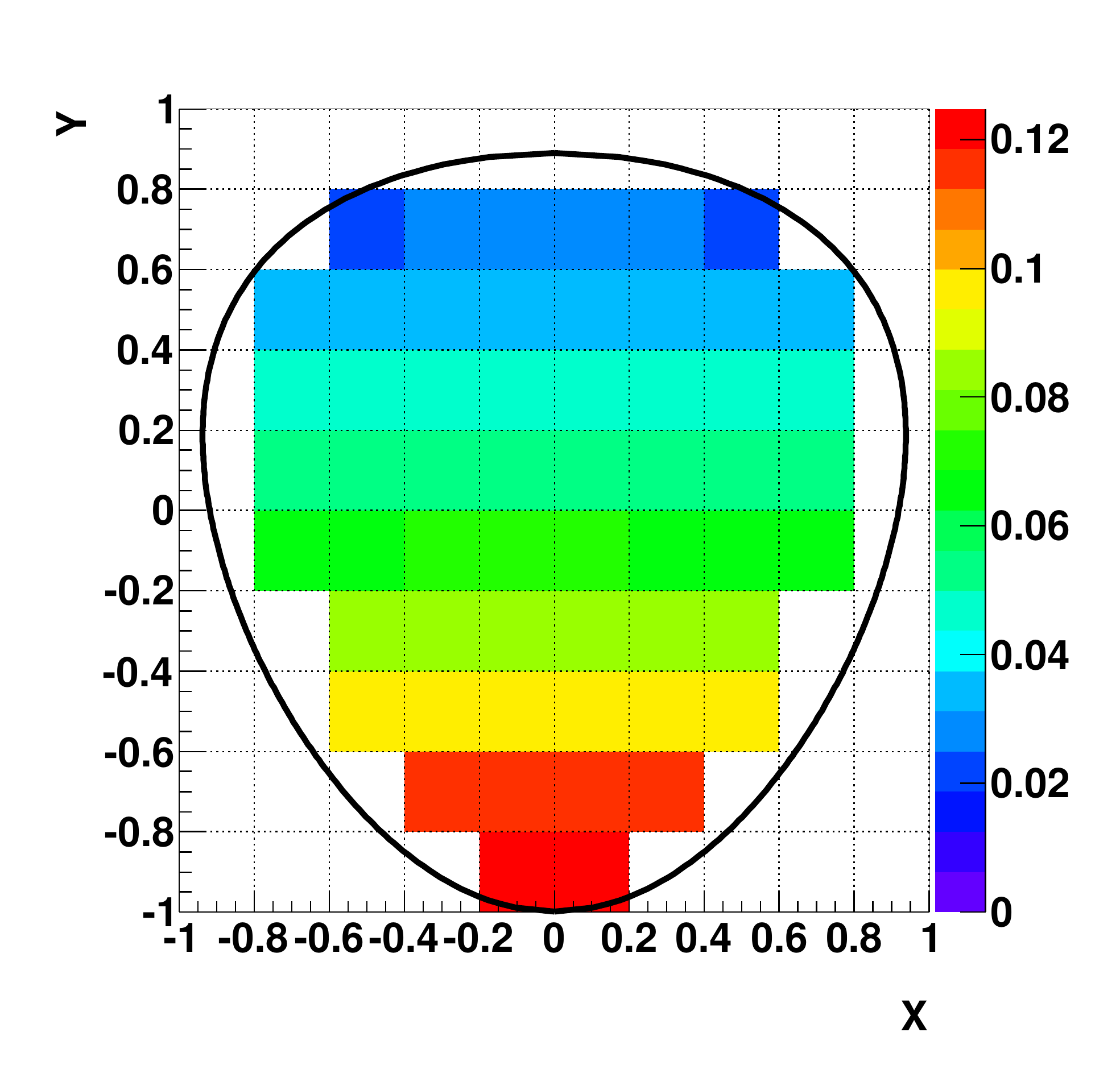,width=0.33\textwidth}}}
\hspace{.3cm}
\parbox{0.30\textwidth}{\centerline{\epsfig{file=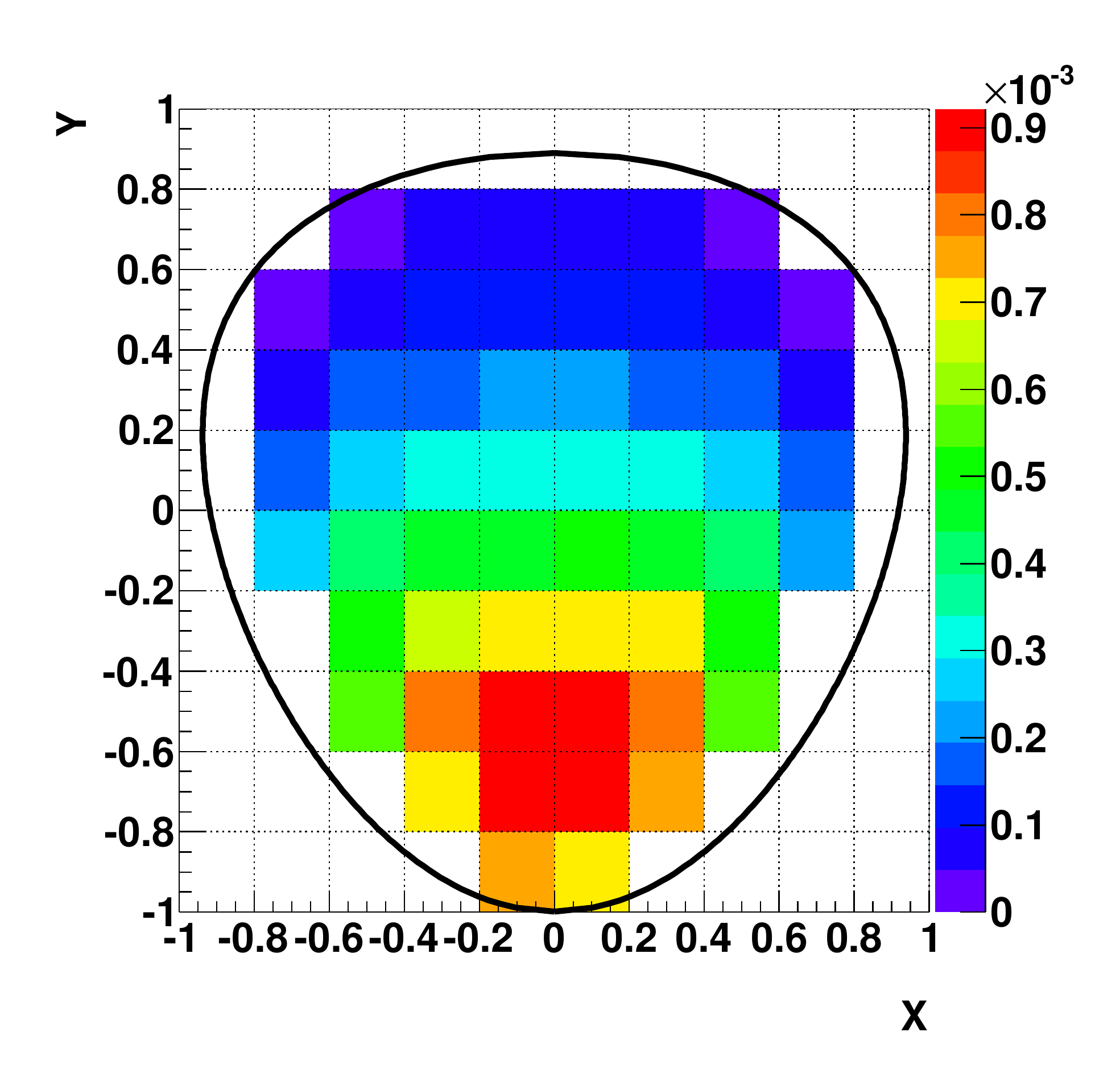,width=0.33\textwidth}}}
\hspace{.3cm}
\parbox{0.30\textwidth}{\centerline{\epsfig{file=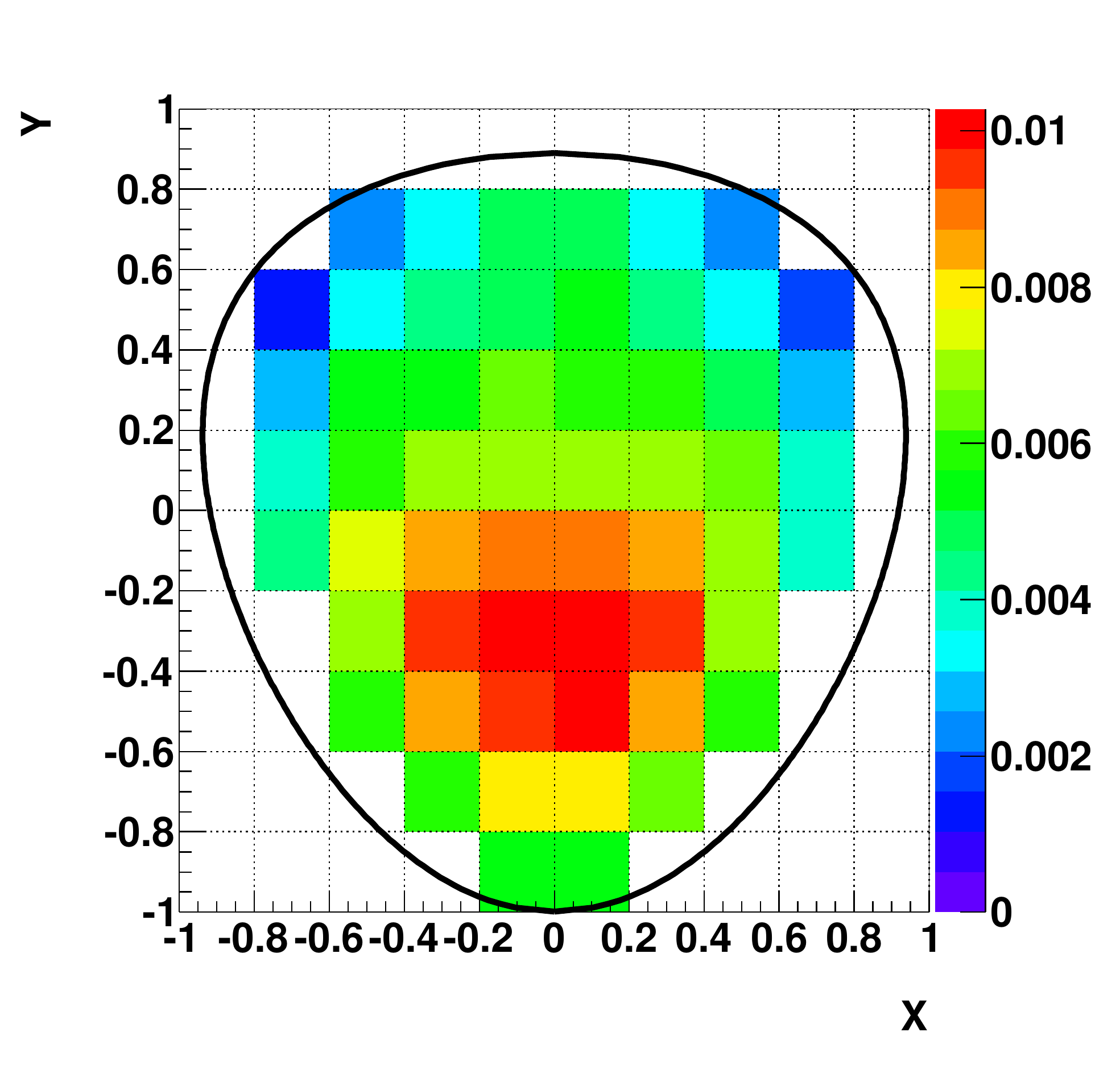,width=0.33\textwidth}}}}
\caption{
Dalitz plot distribution for: {\bf{(left)}} all generated events weighted with the ChPT Leading 
Order predictions, {\bf{(middle)}} reconstructed  events after passing whole analysis chain. 
{\bf{(right)}} The efficiency determined as a function of the Dalitz plot. 
}
\label{DXYeff}
\end{figure}

The Dalitz plot for model generated events is presented in Fig.~\ref{DXYeff} (left). 
Generated four momentum vectors of all particles were used in the Monte Carlo simulations
of the response of the detector. Next, all events were subject  
to the same analysis chain as the experimental data. The resulting spectrum for
reconstructed events is shown in Fig.~\ref{DXYeff} (middle).  
The two dimensional efficiency determined  as a function of each bin of the 
Dalitz plot is shown in the right panel of Fig.~\ref{DXYeff}. 
The maximum efficiency was found for values of Y close to -0.5 and X values close to 0.
\begin{figure}[!h]
\hspace{3.3cm}
\parbox{0.50\textwidth}{\centerline{\epsfig{file=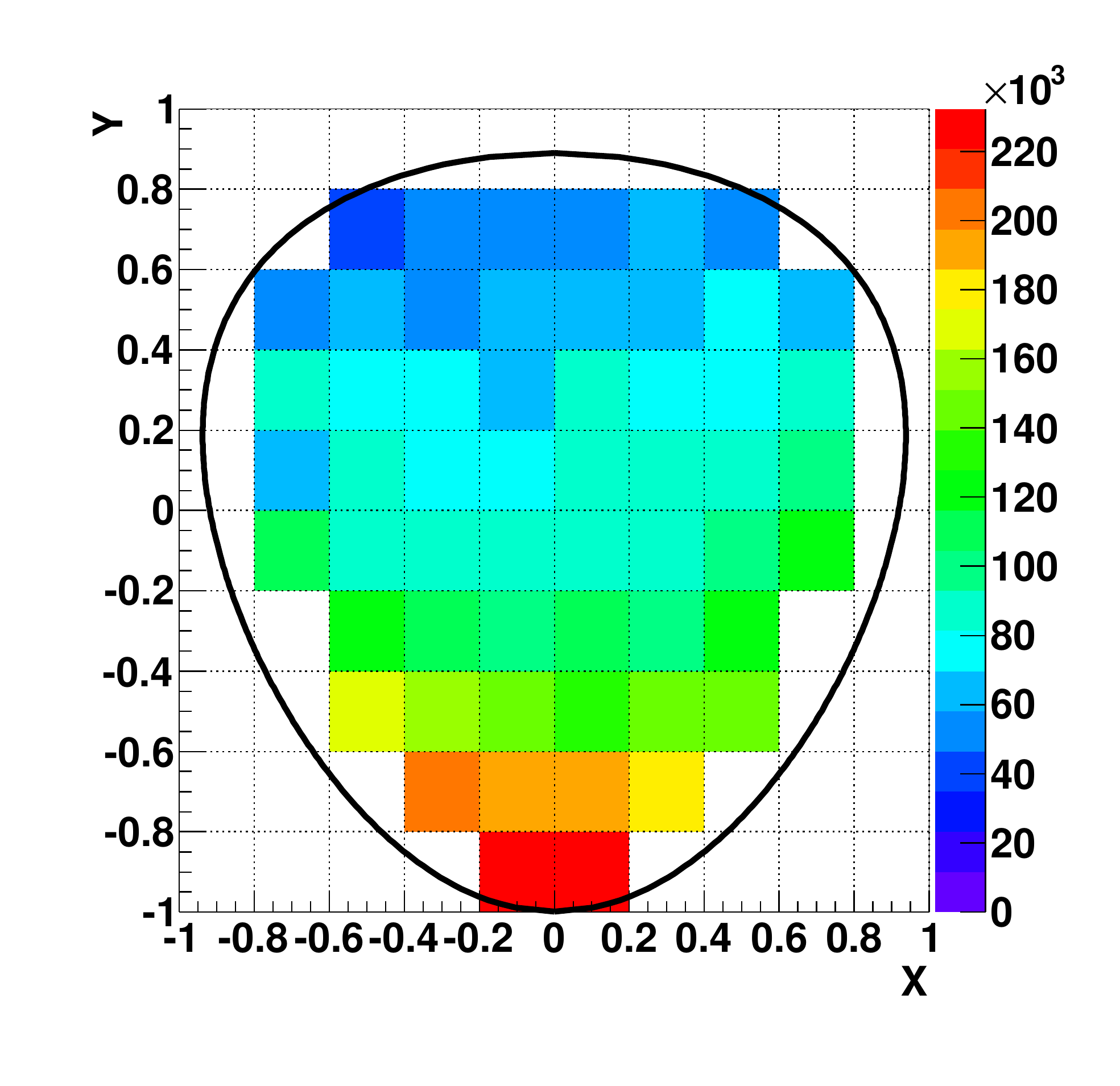,width=0.55\textwidth}}}
\caption{
Final Dalitz plot after background subtraction and efficiency correction.
}
\label{DXYbcgacc}
\end{figure}
Figure~\ref{DXYbcgacc} shows the efficiency corrected Dalitz plot for the $\eta\to\pi^+\pi^-\pi^0$ system.
The population is decreasing with increasing values of Y variable which is proportional 
to the energy of neutral pion. 

The Dalitz plot density population is expected to be described by transition amplitude 
$\vert M\vert^2$ given by Eq.~\ref{amplituda}. 
\begin{table}[!h]
\centering
\begin{tabular}{|c|c|c|c|}
\hline
dof & a & b & c\\\hline\hline
55  & -1.039 (fixed)  &  0.27 (fixed) & $0.05 \pm 0.10$\\\hline
\end{tabular}
\caption{
Results of the two dimensional fit to the Dalitz plot distribution of the 
amplitude $\vert M\vert^2 = A_0^2( 1 + a\cdot Y + b\cdot Y^2 + c\cdot X)$.
}
\label{DPfit}
\end{table}
Having the background free and acceptance corrected Dalitz plot one can fit the two dimensional distribution 
by minimizing the $\chi^2$ function:
\begin{equation}
\chi^2(A_0^2,c) = \sum_{X,Y} \frac{(N_{X,Y} - \vert M\vert^2(A_0^2,c))^2}{\sigma^2},
\end{equation}
where $N$ stands for the number of the events corresponding to $\eta\to\pi^+\pi^-\pi^0$ decay after the background
subtraction, and $\sigma^2$ is the uncertainty of the acceptance corrected number of signal events.
The bins indicated in white are not included in the fit. We fitted $\vert M\vert^2$ 
including terms up to squared X, however we have fixed $a$ and $b$ parameters as predicted by ChPT Leading Order. Obtained results for fit is given in Tab.~\ref{DPfit}. The coefficient $c$ which reflects the degree of C-invariance violation is consistent with zero. As it will be shown in the 
next section this is consistent with the values of the obtained left-right asymmetry. 

\section{Determination of asymmetries}
\hspace{\parindent}
The integrated asymmetries of the Dalitz plot are sensitive for the C violation in the 
amplitudes of a given $I$ state, therefore it is important to estimate their values. 
The definition of all three asymmetries were given in Chapter 2 by equations~\ref{ALR},~\ref{AQ},~\ref{AS}. 
The left-right asymmetry $A_{LR}$ is sensitive to the C violation averaged over all $I$ states,
however the quadrant asymmetry $A_Q$ and sextant asymmetry $A_S$ can test C invariance for specific 
$I = 1$ and $I = 0,2$ isospin states of the $\eta\to\pi^+\pi^-\pi^0$ decay, respectively. 

\begin{figure}[p]
\vspace{-0.5cm}
\mbox{
\hspace{0.3cm}
\parbox{0.45\textwidth}{\centerline{\epsfig{file=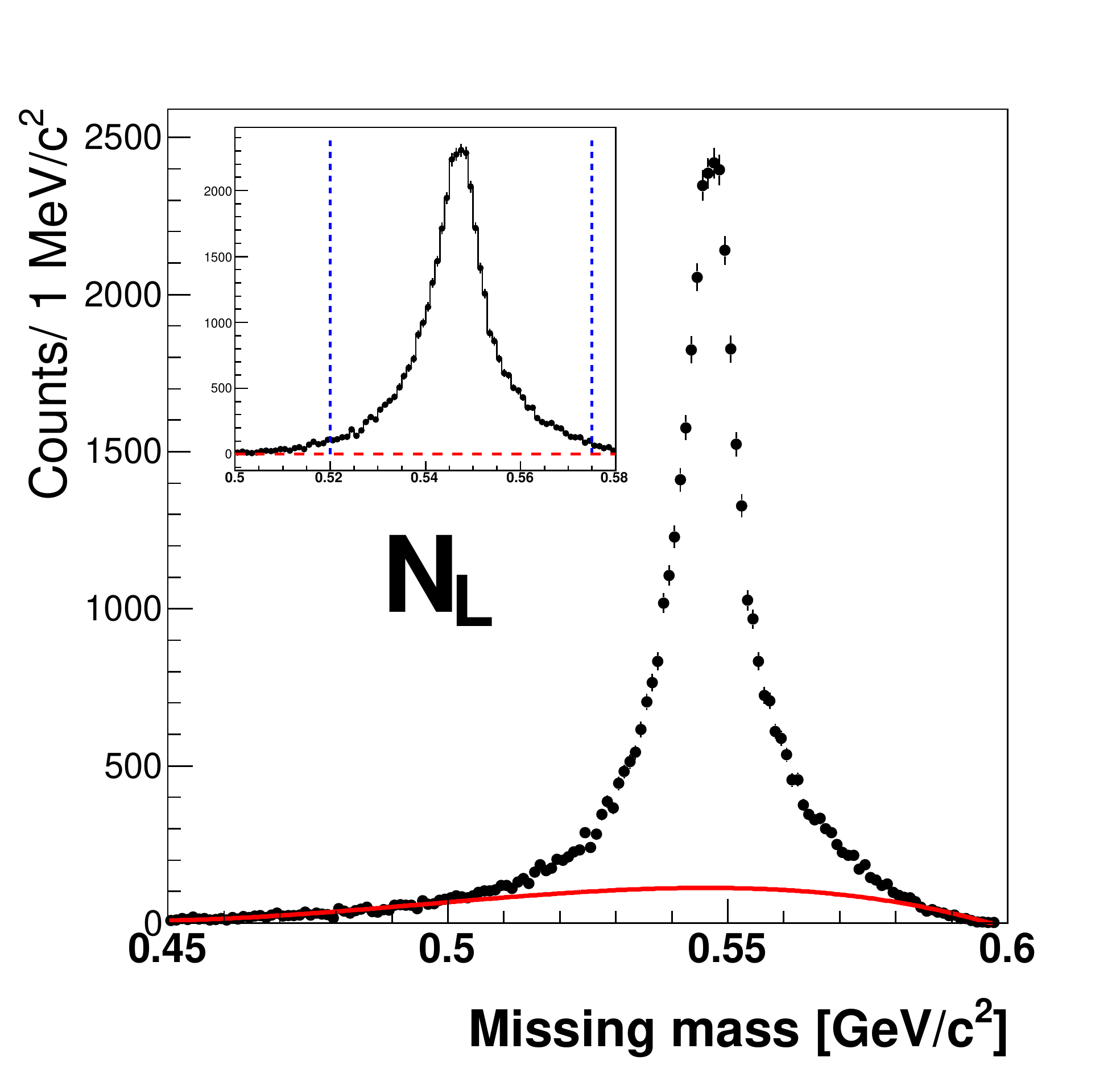,width=0.50\textwidth}}}
\hspace{0.4cm}
\parbox{0.45\textwidth}{\centerline{\epsfig{file=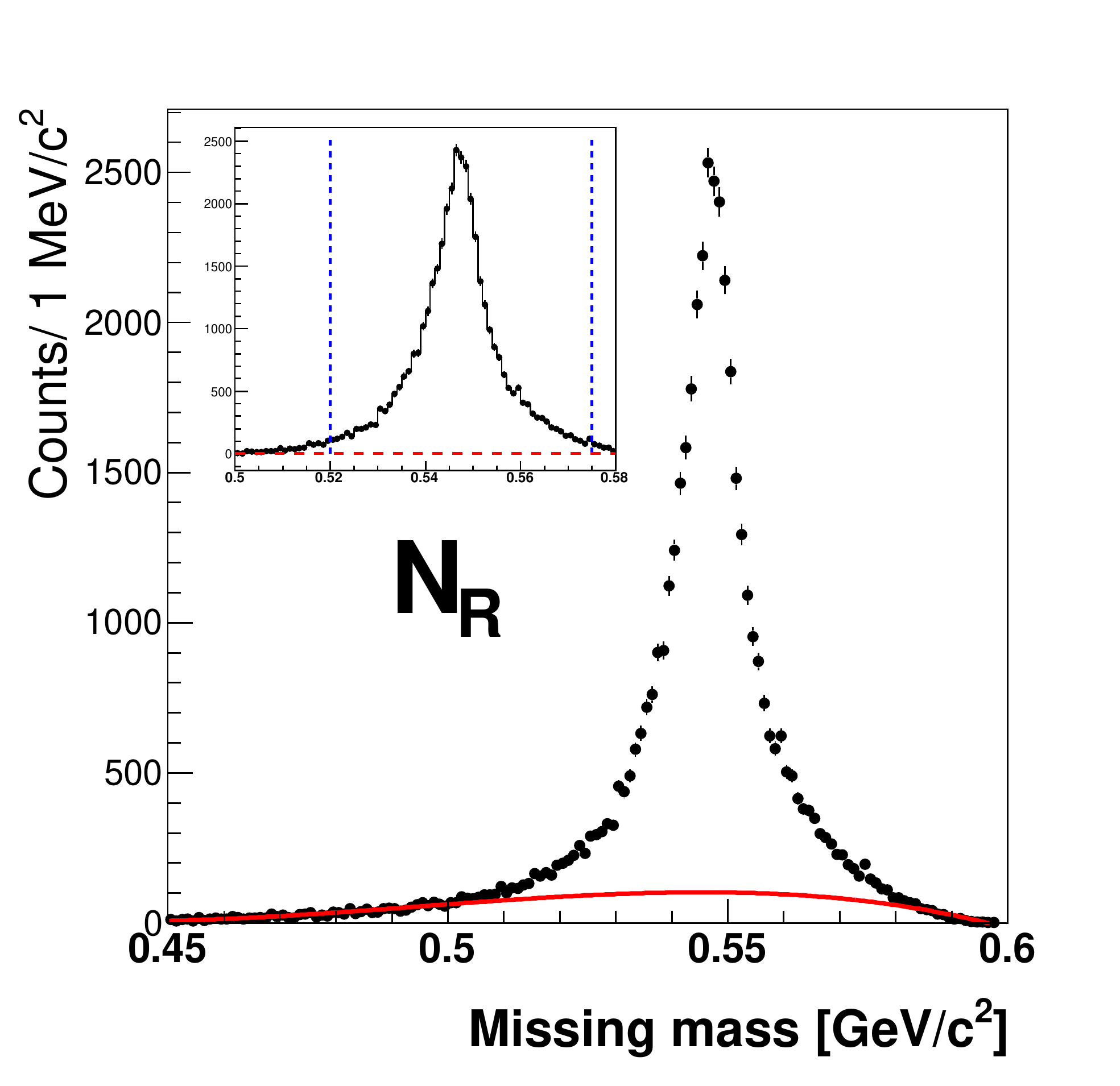,width=0.50\textwidth}}}}\\
\hspace{0.3cm}
\mbox{
\hspace{0.3cm}
\parbox{0.45\textwidth}{\centerline{\epsfig{file=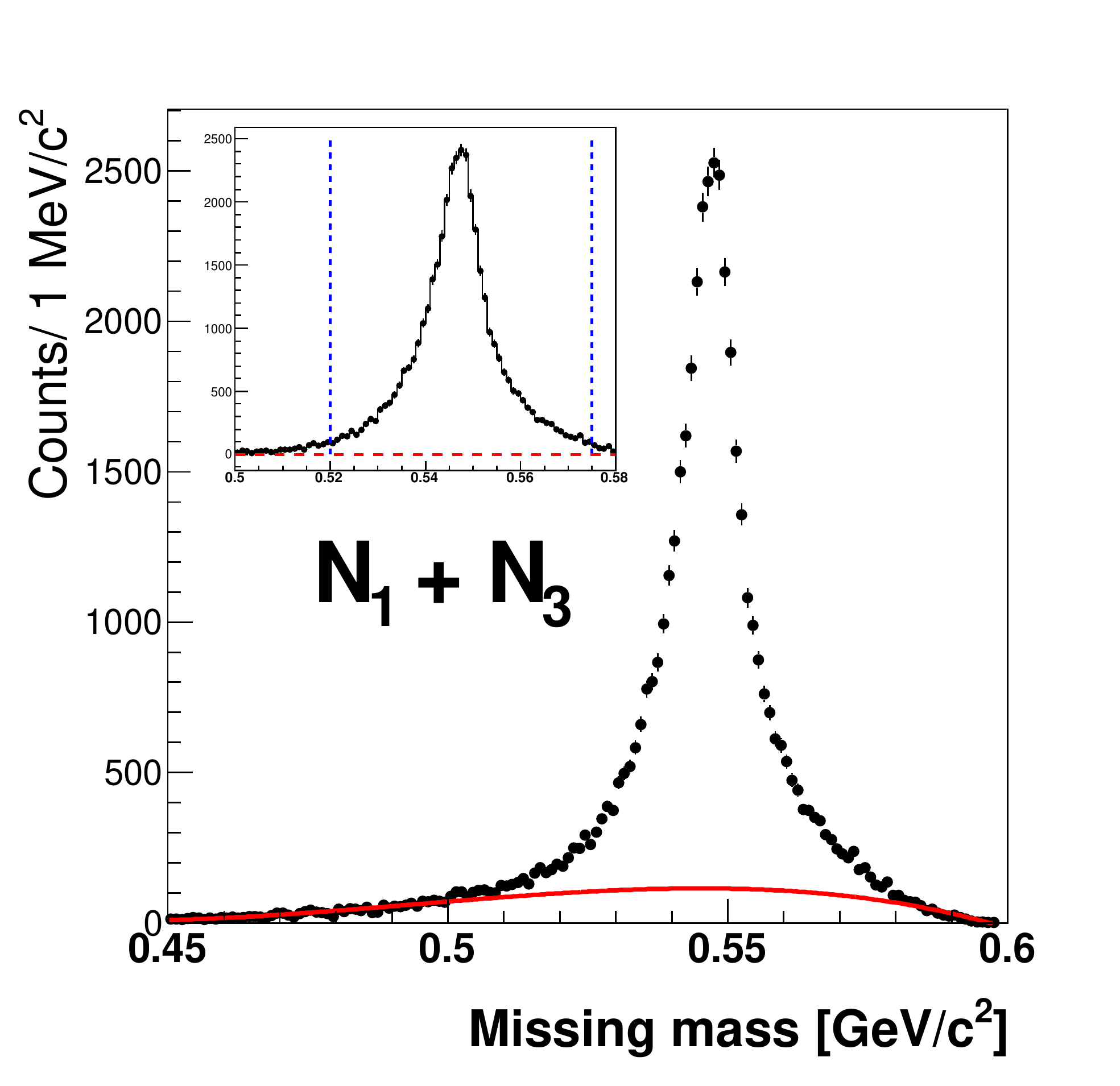,width=0.50\textwidth}}}
\hspace{0.4cm}
\parbox{0.45\textwidth}{\centerline{\epsfig{file=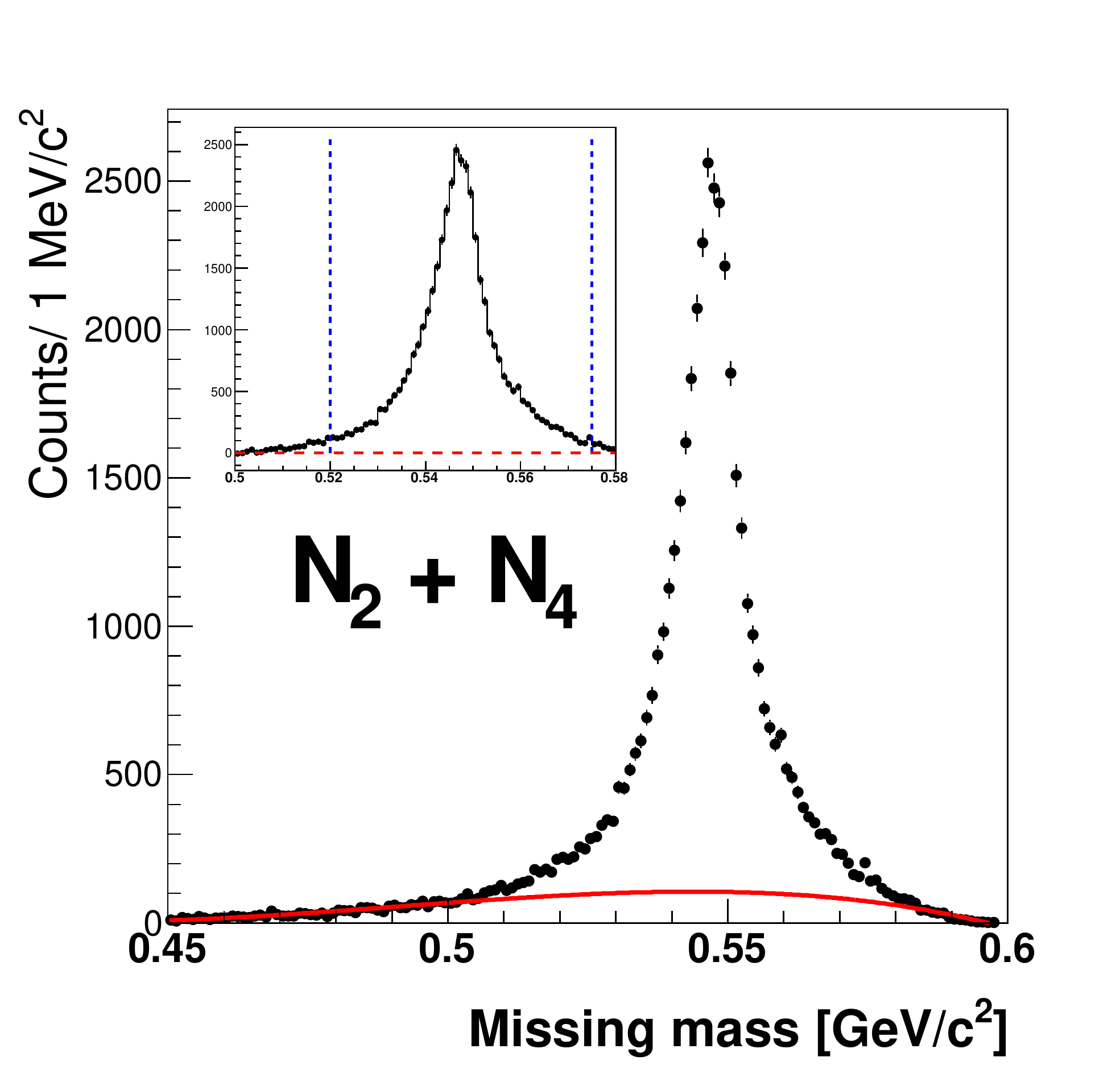,width=0.50\textwidth}}}}\\
\hspace{0.3cm}
\mbox{
\hspace{0.3cm}
\parbox{0.45\textwidth}{\centerline{\epsfig{file=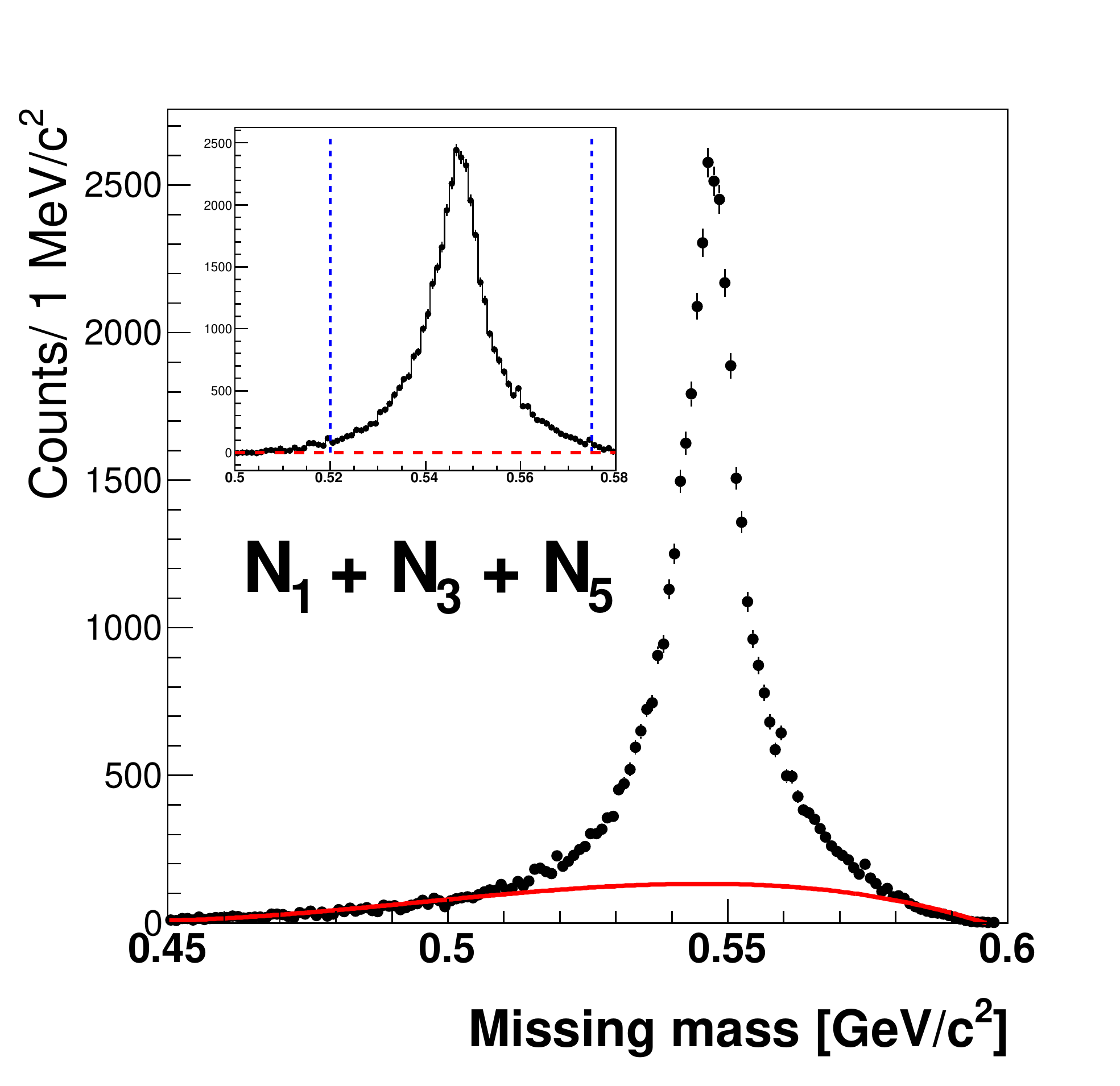,width=0.50\textwidth}}}
\hspace{0.4cm}
\parbox{0.45\textwidth}{\centerline{\epsfig{file=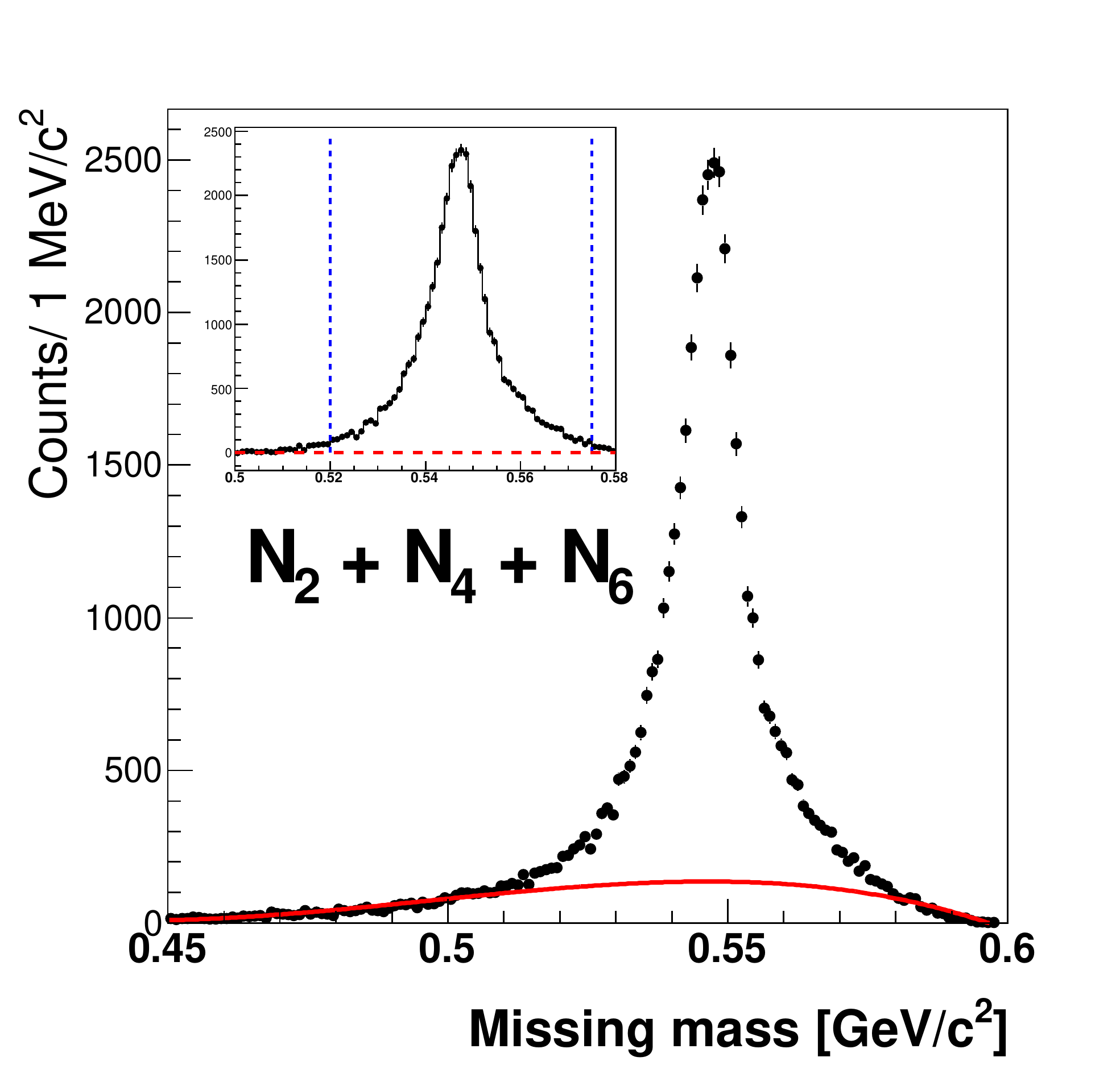,width=0.50\textwidth}}}}\\
\caption{
Missing mass distribution for the $pp\to pp\eta$ reaction determined to calculate the asymmetry parameters
of the Dalitz plot. From the top in rows: (1) left-right asymmetry, (2) quadrant asymmetry, (3) sextant 
asymmetry. The red solid line indicate the fit of the polynomial function to subtract the background. 
}
\label{ASYMimm}
\end{figure}

To determine the asymmetry parameters the Dalitz plot was divided into regions as it is shown 
in Fig.~\ref{DXY-asymetrie}. Then according to formulas~\ref{ALR},~\ref{AQ},~\ref{AS}
we summed the events separately for odd and even regions of the Dalitz plot. Next for each summed region we
reconstruct the missing mass of the $pp\to pp\eta$ reaction as it is shown in Fig.~\ref{ASYMimm}.
Furthermore, to determine the number of events corresponding to the $\eta\to\pi^+\pi^-\pi^0$ decay 
in each region we have subtracted the background using the polynomial fit method, the same which was applied 
in previous section according to the formula~\ref{tloKlajus}. Figure~\ref{ASYMimm} shows the 
corresponding missing mass distributions. 
Moreover, inside of each plot the corresponding spectrum of the missing mass after the subtraction of the background is shown. 
\begin{table}[!h]
\centering
\begin{tabular}{|c|c|}
\hline
Dalitz plot region & Efficiency $\epsilon$\\\hline\hline
 $\epsilon_{L}$    &  0.6656\%\\
 $\epsilon_{R}$    &  0.6659\%\\\hline
 $\epsilon_{13}$   &  0.6719\%\\
 $\epsilon_{24}$   &  0.6721\%\\\hline
 $\epsilon_{135}$  &  0.6731\%\\
 $\epsilon_{246}$  &  0.6709\%\\\hline
\end{tabular}
\caption{
Efficiency for the different regions of the Dalitz plot calculated from the simulations of the
$pp\to pp\eta\to pp\pi^+\pi^-\pi^0$ weighted according to the ChPT Leading order predictions. 
}
\label{tab:eff}
\end{table}
The number of $\eta\to\pi^+\pi^-\pi^0$ events were counted in the region denoted by dashed lines 
in Fig.~\ref{ASYMimm}.
The reconstruction efficiency was estimated based on the simulations of signal reaction. 
The efficiency was estimated separately for each region of 
the Dalitz plot as ratio of number of reconstructed and generated events. Obtained values of the reconstruction efficiency for different regions of the Dalitz plot are collected in Tab.~\ref{tab:eff}. 
One can see that in all regions the efficiency is almost the same and it is equal to around 0.67\%  

\begin{figure}[p]
\vspace{-0.5cm}
\mbox{
\hspace{0.3cm}
\parbox{0.45\textwidth}{\centerline{\epsfig{file=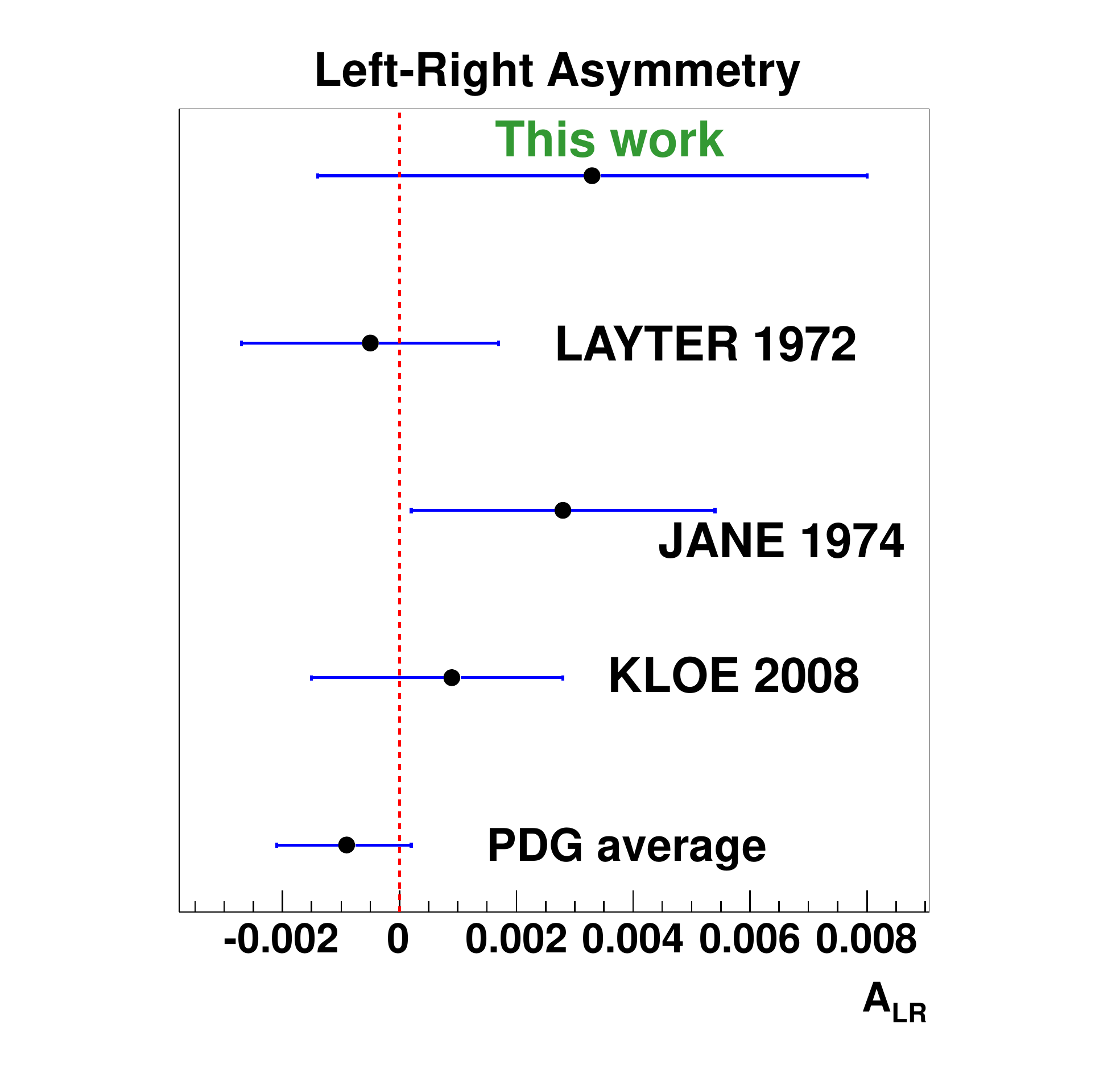,width=0.50\textwidth}}}
\hspace{0.5cm}
\parbox{0.45\textwidth}{\centerline{\epsfig{file=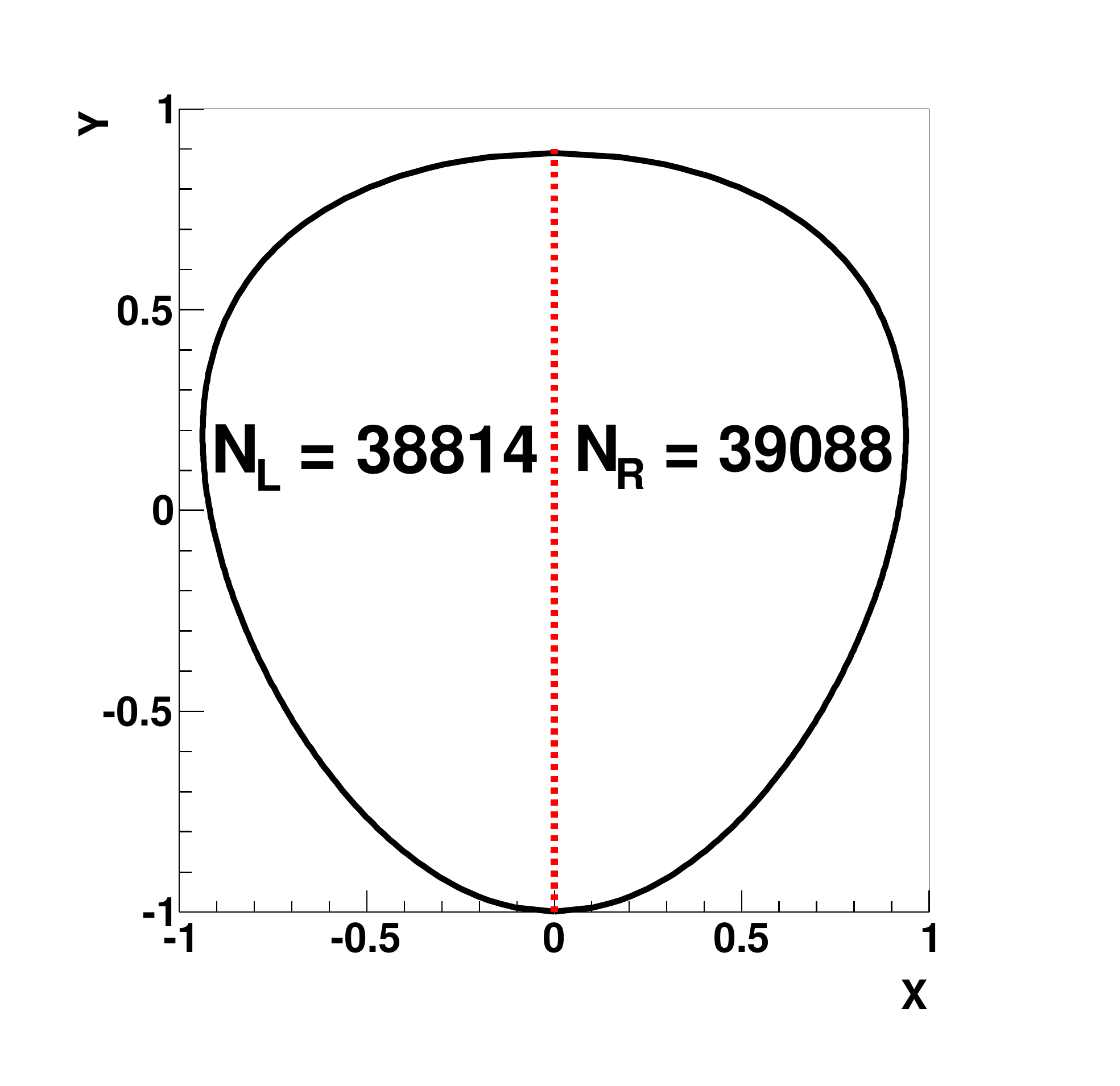,width=0.50\textwidth}}}}\\
\vspace{-0.4cm}
\mbox{
\hspace{0.3cm}
\parbox{0.45\textwidth}{\centerline{\epsfig{file=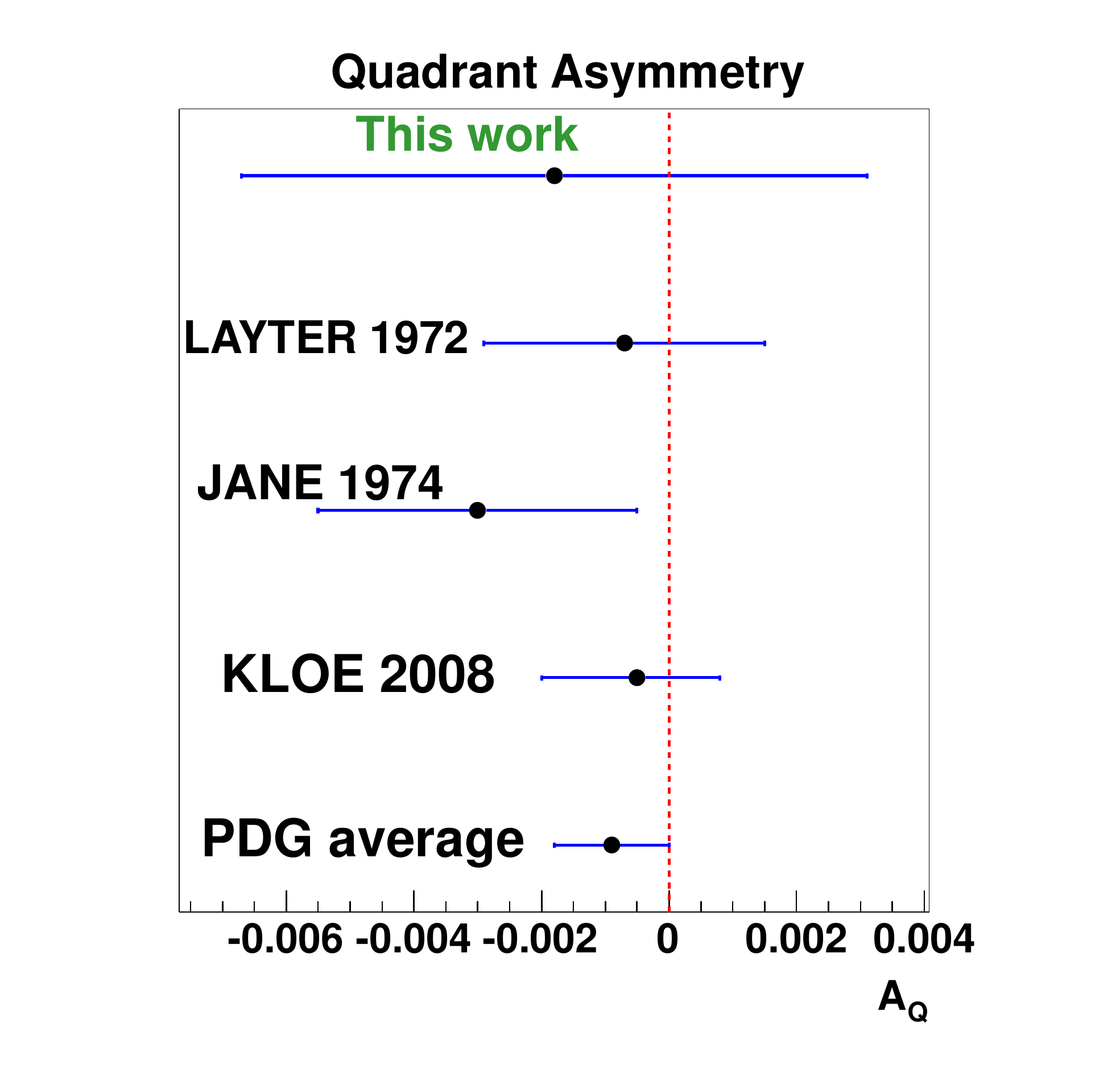,width=0.50\textwidth}}}
\hspace{0.5cm}
\parbox{0.45\textwidth}{\centerline{\epsfig{file=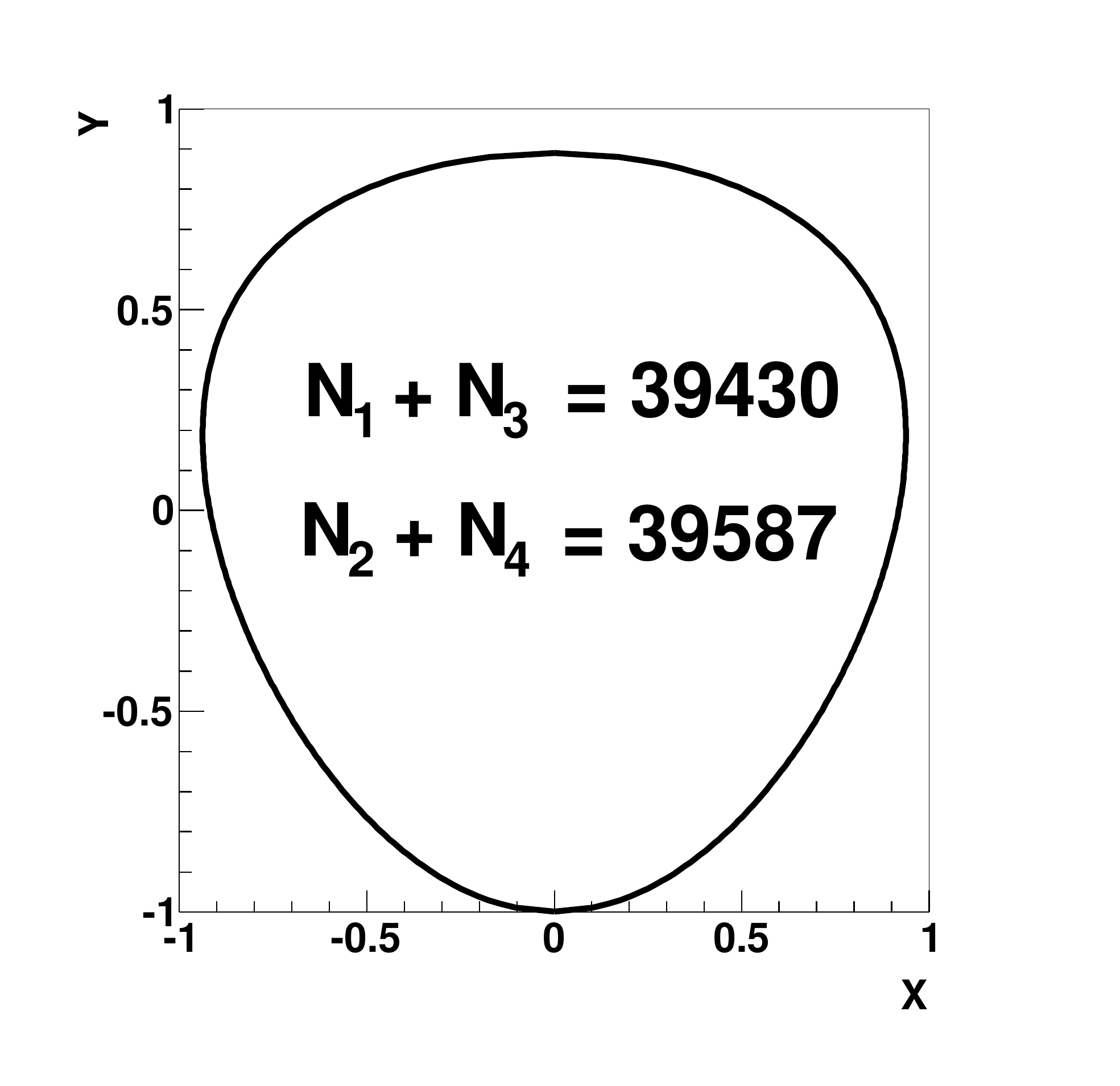,width=0.50\textwidth}}}}\\
\vspace{-0.4cm}
\mbox{
\hspace{0.3cm}
\parbox{0.45\textwidth}{\centerline{\epsfig{file=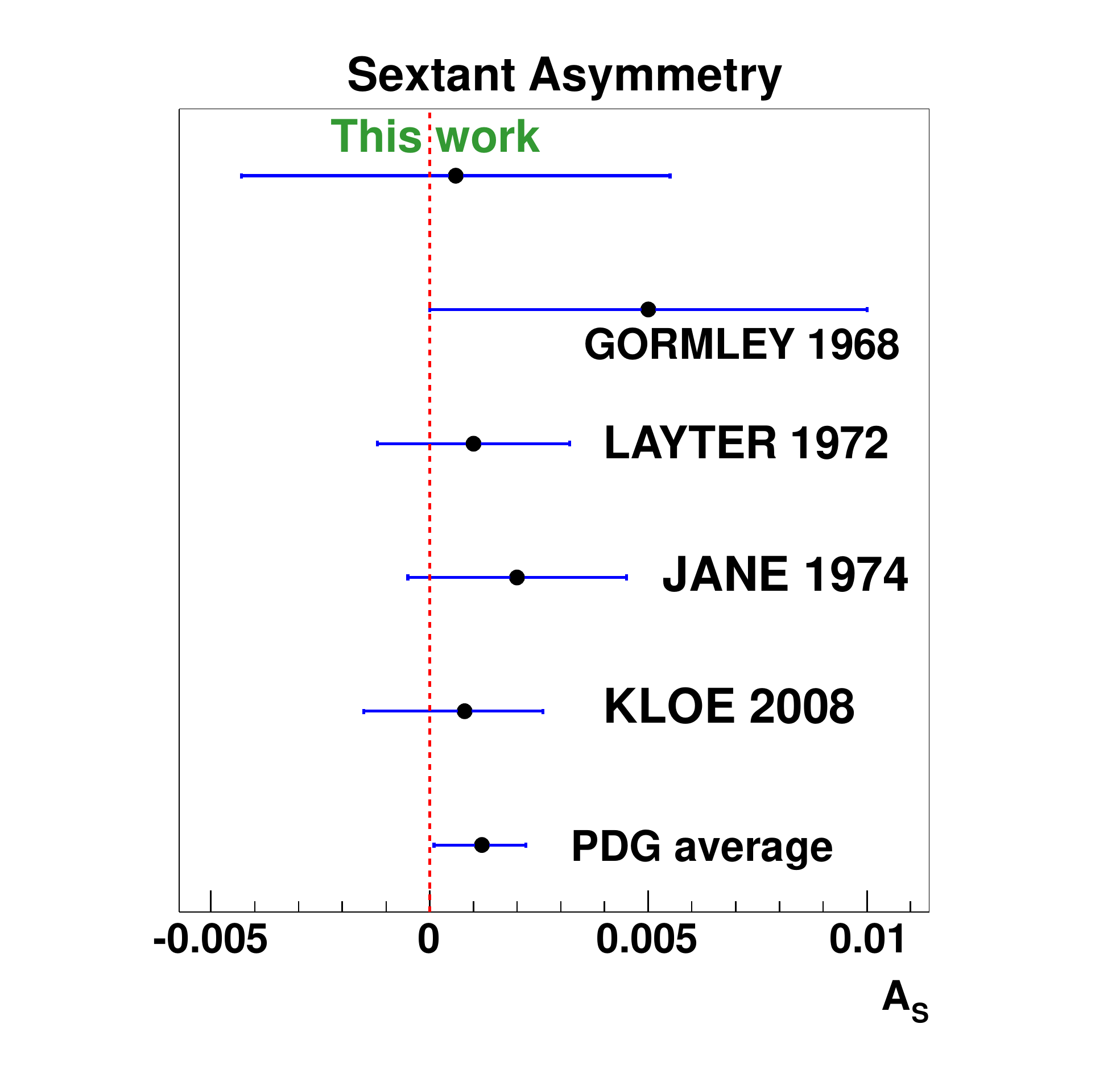,width=0.50\textwidth}}}
\hspace{0.5cm}
\parbox{0.45\textwidth}{\centerline{\epsfig{file=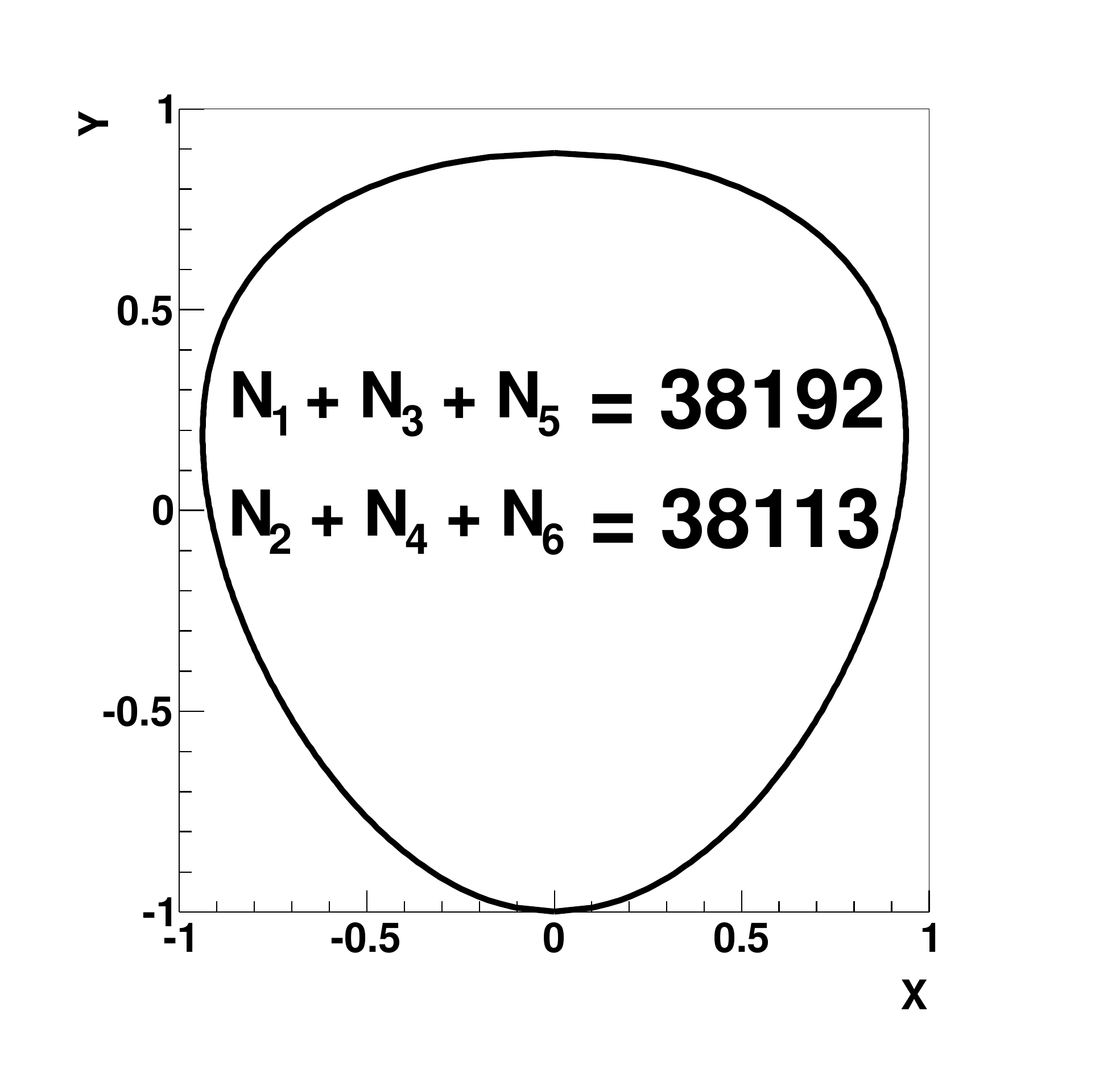,width=0.50\textwidth}}}}\\
\caption{
{\bf{(left panel)}} Comparison of values of asymmetries obtained in this work with results determined 
by previous experiments, and a value given by PDG.
{\bf{(right panel)}} Number of events reconstructed which were used for calculation of asymmetries 
according to formulas~\ref{ALR},~\ref{AQ}, and~\ref{AS}.
}
\label{CompASYM}
\end{figure}

The final results for the asymmetry parameters are:
\begin{equation}
A_{LR} =  (+ 0.33 \pm 0.38_{stat})\times 10^{-2}
\end{equation}
\begin{equation}
A_{Q}  =  (- 0.18 \pm 0.38_{stat})\times 10^{-2}
\end{equation}
\begin{equation}
A_{S}  =  (+ 0.06 \pm 0.39_{stat})\times 10^{-2}
\end{equation}
where the second number denotes the statistical errors.
The statistical error was calculated according to the formula for the error propagation which reads:
\begin{equation}
\sigma(N_{LR}) = \frac{2}{(N_{R} + N_{L})^2}\sqrt{ N_{R}^2 \cdot \sigma^2(N_{L}) + N_{L}^2 \cdot \sigma^2(N_{R})},
\label{bladALR}
\end{equation}
where the $\sigma(N_{L/R}) = \sqrt{N_{L/R} + B_{L/R}}$, the $N_{L/R}$ denotes the number of signal events, 
and $B_{L/R}$ stands for the number of background events under the $\eta$ meson peak. The evaluation method of 
the systematic error will be given in the next section. The formula given by equation~\ref{bladALR} can be 
also generalized for the calculation of the statistical uncertainties $\sigma(A_{Q})$ and $\sigma(A_{S})$.
In right panel of Fig.~\ref{CompASYM} the result of this work is compared to previously determined values of 
asymmetries. The errors shown are the statistical and systematic for all presented measurements.  

\section{Estimation of the systematic uncertainty}
\hspace{\parindent}
Beside the statistical uncertainty of the measured observables one has to investigate the systematic 
effects which may change the result. To evaluate the systematic uncertainty of established 
values of asymmetry parameters, we have used the method described in~\cite{Barlow:2002yb}, 
as the method which is recommended by the WASA-at-COSY collaboration. In this work~\cite{Barlow:2002yb}
the systematic uncertainty $\Delta_{sys}$ is calculated as a difference between the result obtained 
with cuts used in the analysis and a result obtained after changes of these cuts. Next $\Delta_{sys}$
is compared to $\sigma_{sys}$ defined as:
\begin{equation}
\sigma_{sys} = \sqrt{ \sigma^2_{changed} - \sigma^2_{original}},
\label{syst}
\end{equation}
where the $\sigma_{changed}$ denotes the statistical uncertainty of asymmetry obtained after the 
change of the cut, and $\sigma_{original}$ indicates uncertainty of the originally obtained result.
The ratio of $\frac{\Delta_{sys}}{\sigma_{sys}}$ is than used as a measure of the significance of the 
determined systematic error. If this ratio is smaller than one the systematic error due to the tested 
effect is negligible. 

Below we will discuss the applied changes of the selection criteria. Each change was done separately for 
individual cuts assuming that the sources of the systematic errors are uncorrelated. After the change of 
one condition at a time the others were kept as in the original analysis, and the whole analysis was repeated.
The obtained changes of the asymmetries ($\Delta A$) due to variation of the condition listed below are 
presented in Tab.~\ref{tab:systematyka}.   
\begin{enumerate}
\item{\bf{Identification of charged particles in SEC}}\\
The identification of charged particles is based on the correlation between energy deposit in the 
electromagnetic calorimeter and the momentum determined from the signals measured by drift chamber. The corresponding distribution
is shown in Fig.~7.4. The pair of charged particles is treated as a $\pi^+$ and $\pi^-$ if both of them 
fall into region which is below the cut lines. The cut lines are parameterized with the following function:
$\Delta E_{SEC} = a\cdot\vert\vec{p}\vert\cdot q + b$. To estimate the systematic influence caused 
by the charged pion identification in the calorimeter, the red lines shown in Fig.~7.4 separating leptons
from pions are moved closer to the pion bands by changing the parameter b. We decreased the allowed area
for pions by 5\%, and repeated the analysis. 
  
\item{\bf{Identification of charged particles in PSB}}\\
Second method of particles identification for pions is based on the energy deposited in the 
Plastic Scintillator Barrel as a function of the reconstructed momentum. The method 
is shown in Fig.~7.5. In order to estimate the systematic effect, the area marked on the figure 
by the red lines was decreased by 5\%, and analysis was repeated. 
 
\item{\bf{Suppression of the $2\pi$ background via missing mass squared for the $pp\to pp\pi^+\pi^-X$ reaction}}\\
This cut was dedicated to reduce the large amount of the two pion background originating from the 
direct production. The cut can be seen in Fig.~7.9 (left). To estimate the influence of this cut on the 
final results we repeated the analysis reducing the condition by 0.0025~GeV$^2$/c$^4$ towards the lower missing masses. 

\item{\bf{Suppression of the $2\pi$ background via missing mass for the $pp\to \pi^+\pi^-X$ reaction}}\\
In the original analysis to suppress the rest of the two pion background we have used the condition which was 
presented in Fig.~7.9 (right). To establish it influence on the obtained result we have change the 
cut by 0.05~GeV/c$^2$ toward the higher missing masses, and repeated the analysis. 

\item{\bf{Background subtraction}}\\
The direct production of pions obscures the interesting signal if one wants to look at missing mass 
distributions. In order, to subtract this background a polynomial function was fitted to the continues 
background outside the signal peak region. In order to check how the choice of the order of the polynomial
changes the results we have fitted the background with fifth-order polynomial function instead of originally 
used fourth-order.
\end{enumerate}

The results of performed systematic checks are gathered in Tab.~\ref{tab:systematyka}, where the values of the 
change of asymmetry parameters $\Delta A$ obtained after the change of individual cuts together with the estimated  $\sigma_{sys}$ for the cut are given.
\begin{table}[!h]
\centering
\begin{tabular}{c|c|c|c|c|c|c}
\hline
1 &      2    &  3   & 4    & 5 & 6  & 7\\\hline\hline
Effect &          &  A   & $\sigma$      & $Delta$ A & $\Delta\sigma_{sys}$ & $\frac{\Delta A}{\Delta\sigma_{sys}}$\\\hline\hline
Oryg.& $A_{LR}$ &  0.0033  &   0.0038      &       --     &   --   &  --    \\
values& $A_{Q}$  &  0.0018  &   0.0038      &      --     &   --   &  --    \\
     & $A_{S}$  &  0.0006  &   0.0039      &       --     &   --   &  --    \\
     & c        &  0.05    &   0.10        &       --     &   --   &  --    \\\hline\hline
 1. & $A_{LR}$ & 0.00392  &  0.00401      &  0.00062      &  0.00129   &  0.48  \\
    & $A_{Q}$  &-0.00241  &  0.00401      &  0.00070      &  0.00128   &  0.56  \\
    & $A_{S}$  & 0.00101  &  0.00399      &  0.00041      &  0.00084   &  0.49  \\
    & c        & 0.072    &  0.120        &  0.023        &  0.067     &  0.34  \\\hline
 2. & $A_{LR}$ & 0.00353  &  0.00391      &  0.00023      &  0.00093   &  0.25  \\
    & $A_{Q}$  &-0.00151  &  0.00392      &  0.00029      &  0.00094   &  0.31  \\
    & $A_{S}$  & 0.00081  &  0.00380      &  0.00021      &  0.00087   &  0.24  \\
    & c        & 0.055    &  0.110        &  0.005        &  0.04701   &  0.11  \\\hline
 3. & $A_{LR}$ & 0.00264  &  0.00361      &  0.00067      &  0.00118   &  0.56  \\
    & $A_{Q}$  &-0.00101  &  0.00362      &  0.00079      &  0.00117   &  0.67  \\
    & $A_{S}$  & 0.00110  &  0.00379      &  0.00050      &  0.00092   &  0.54  \\
    & c        & 0.049    &  0.099        &  0.001        &  0.006     &  0.14  \\\hline
 4. & $A_{LR}$ & 0.00370  &  0.00378      &  0.00040      &  0.00039   &  1.03  \\
    & $A_{Q}$  &-0.00231  &  0.00377      &  0.00051      &  0.00048   &  1.08  \\
    & $A_{S}$  & 0.00121  &  0.00386      &  0.00061      &  0.00056   &  1.10  \\
    & c        & 0.046    &  0.105        &  0.004        &  0.034     &  0.13  \\\hline
 5. & $A_{LR}$ & 0.00412  &  0.00388      &  0.00082      & 0.00078    &  1.04  \\
    & $A_{Q}$  & 0.00274  &  0.00391      &  0.00094      & 0.00092    &  1.02  \\
    & $A_{S}$  & 0.00140  &  0.00382      &  0.00080      & 0.00079    &  1.01  \\
    & c        & 0.060    &  0.116        &  0.00941      & 0.05807    &  0.16  \\\hline
\end{tabular}
\caption{Results of the studies of systematic effects. First column indicates the number of the 
investigated effect as listed in the text. Third and fourth columns shows the asymmetry value and
its statistical uncertainty as determined in this work. Fifth column indicates the estimated 
contribution to the systematic effect ($\Delta A$), and sixth column shows its uncertainty.}
\label{tab:systematyka}
\end{table}
According to the methodology presented in reference~\cite{Barlow:2002yb} we consider contributions to the 
systematic error as non-significant and hence negligible if $\frac{\Delta A}{\sigma_{sys}}$ is less than one. 
Hence, the final systematic uncertainties amount to:
\begin{equation*}
\sigma_{sys}(A_{LR}) = 0.09\cdot 10^{-2},
\end{equation*}
\begin{equation*}
\sigma_{sys}(A_{Q}) = 0.11\cdot 10^{-2},
\end{equation*}
\begin{equation*}
\sigma_{sys}(A_{S}) = 0.10\cdot 10^{-2},
\end{equation*}
and systematic error for the determination of the parameter $c$ is negligible.
For the comparison we have also estimated the systematic errors more conservatively by adding in quadrature all
contributions to the systematic error. In this case errors amount to:
\begin{equation*}
\sigma_{sys}(A_{LR}) = 0.13\cdot 10^{-2},
\end{equation*}
\begin{equation*}
\sigma_{sys}(A_{Q}) = 0.15\cdot 10^{-2},
\end{equation*}
\begin{equation*}
\sigma_{sys}(A_{S}) = 0.12\cdot 10^{-2},
\end{equation*}
and are still much lower than the statistical uncertainty.
One also has to note, that beside presented above sources of the systematic effects, 
there are still another checks which may be done in further studies of investigated reaction. 

\section{Discussion}
\hspace{\parindent}
In the first part of this thesis the charge conjugation has been tested in $\eta\to\pi^+\pi^-\pi^0$ decay.
As a main result we have obtained three asymmetries: $A_{LR}$, $A_{A}$ and $A_{S}$ of the Dalitz plot 
sensitive for the C symmetry breaking in different isospin states of the final particles. The estimated values of the asymmetries are 
listed below together with the statistical and systematic uncertainties:
\begin{equation*}
A_{LR} =  (+ 0.33 \pm 0.38_{stat} \pm 0.09_{sys})\times 10^{-2},
\end{equation*}
\begin{equation*}
A_{Q}  =  (- 0.18 \pm 0.38_{stat} \pm 0.11_{sys})\times 10^{-2},
\end{equation*}
\begin{equation*}
A_{S}  =  (+ 0.06 \pm 0.39_{stat} \pm 0.10_{sys})\times 10^{-2}.
\end{equation*}
One can see that obtained values of all studied parameters are consistent with zero and with the previous 
measurements (see Fig.\ref{CompASYM}). Therefore, we can conclude that the charge conjugation is invariant 
in strong interaction at the level of achieved accuracy.

Moreover, we have fitted the Dalitz plot density according to the phenomenological parameterization 
of the transition amplitude given by equation~\ref{amplituda}. As expected from C-invariance the $c$ 
coefficient is equal to zero within uncertainty:
\begin{equation*}
c = 0.05 \pm 0.10_{stat}.
\end{equation*}

\chapter{Extraction of $\eta\to\pi^0 e^+ e^-$ decay}
\hspace{\parindent}
In order to investigate the charge conjugation invariance in the electromagnetic interactions 
one can study the decay of the $\eta$ meson into the $\pi^0 e^+ e^-$ system. This rare process till now 
has been not observed in any experiment. This decay has a similar final state as the decay 
$\eta\to\pi^+\pi^-\pi^0$ which was described in previous chapter. Thus, the reconstruction,
preselection of tracks and proton identification for both processes are the same. These stages of the 
analysis were shown in chapters 5 and 6. Therefore, in next sections we will describe only the selection 
aiming at identification of the final state with two oppositely charged leptons $e^+$ and $e^-$ and two photons
originating from the $\pi^0$ decay.  

The search for rare decays requires a good understanding of a physical processes, detector response, reconstruction methods and background subtraction. Therefore, for the purpose of the investigation of 
the $\eta\to\pi^0 e^+ e^-$ decay we have simulated using Monte Carlo methods several processes which may 
contribute as a background to the searched decay. Table \ref{tab:simpi0ee} shows the simulated processes 
together with the number of generated events used in the simulations. The table is divided into three parts
indicating (i) signal reaction, (ii) background originating from the decay of the $\eta$ meson, and (iii) 
background from the direct multi-pion production. 
\begin{table}[!h]
\centering
\begin{tabular}{c|c|c}
\hline
No. & Reaction & Number of generated events\\\hline\hline
1 &$pp\to pp\eta\to pp e^+e^-\pi^0\to ppe^+e^-\gamma\gamma$ (signal)     &  $5\times 10^6$\\\hline
2 &$pp\to pp\eta\to pp \pi^+\pi^-\pi^0\to pp\pi^+\pi^-\gamma\gamma$      &  $20\times 10^6$\\
3 &$pp\to pp\eta\to pp \pi^+\pi^-\pi^0\to pp\pi^+\pi^- e^+e^-\gamma$     &  $1\times 10^6$\\
4 &$pp\to pp\eta\to pp\pi^+\pi^-\gamma$                                  &  $5\times 10^6$\\
5 &$pp\to pp\eta\to pp e^+ e^-\gamma$                                    &  $20\times 10^6$\\
6 &$pp\to pp\eta\to pp\pi^0\gamma\gamma\to pp\gamma\gamma\gamma\gamma$   &  $3\times 10^6$\\
7 &$pp\to pp\eta\to pp\pi^0\gamma\gamma\to pp e^+ e^-\gamma\gamma\gamma$ &  $3\times 10^6$\\
8 &$pp\to pp\eta\to pp\gamma\gamma$                                      &  $6\times 10^6$\\\hline
9 &$pp\to pp\pi^+\pi^-\pi^0\to pp\pi^+\pi^-\gamma\gamma$                 &  $20\times 10^6$\\
10 &$pp\to pp\pi^+\pi^-$                                                 &  $58\times 10^6$\\
11 &$pp\to pp\pi^0\pi^0\to pp\gamma\gamma\gamma\gamma$                   &  $18\times 10^6$\\
12 &$pp\to pp\pi^0\pi^0\to pp e^+e^-\gamma\gamma\gamma$                  &  $8\times 10^6$\\\hline
\end{tabular}
\caption{List of simulated reactions which may contribute to the background in the 
search for the $\eta\to e^+e^-\pi^0$ decay. Right column indicate the numbers of initially 
generated events.}
\label{tab:simpi0ee}
\end{table}
Reactions two and nine possess similar final state as signal process with two charge particles and 
two gamma quanta in the final state. The misidentification of the reaction 
could be due to mistaken identification of the pions as electrons. 
In the case of third reaction the splitting of signals in the calorimeter and misidentification 
of pions as electrons, and bending of electrons inside the beam pipe may be wrongly recognized as a signal. 
The fourth, fifth and tenth  
reaction can be misidentified as a signal due to bremsstrahlung effect in the calorimeter. 
Reactions six, eight and eleven have two or more photons in the final state, and these can 
cause the external conversion of gamma quanta on the beam pipe, resulting in production of electron-positron pair. In the reaction seven and twelve merging of clusters in the calorimeter may lead to wrong identification of these reactions as a signal. 

In the next section we will describe the selection criteria, established based on the simulations, 
aming at suppression of background and the identification of the signal reaction.

\section{Identification of $\pi^0$ meson and leptons $e^\pm$}
\hspace{\parindent}
The final state of $\eta\to\pi^0 e^+e^-$ decay consists from two photons originating from the decay of the 
neutral pion and two oppositely charged leptons. As a first step in identification of the  
desired decay one has to recognize all final state particles. 
The reconstruction of the $\pi^0$ meson is based on the identification of two photons in the calorimeter. 
From the analysis we accept only these events for which two or more clusters were reconstructed. 
However, part of identified clusters are due to splitting
and bremsstrahlung of the charged particles. 
\begin{figure}[b!]
\vspace{-.5cm}
\hspace{3.5cm}
\parbox{0.5\textwidth}{\centerline{\epsfig{file=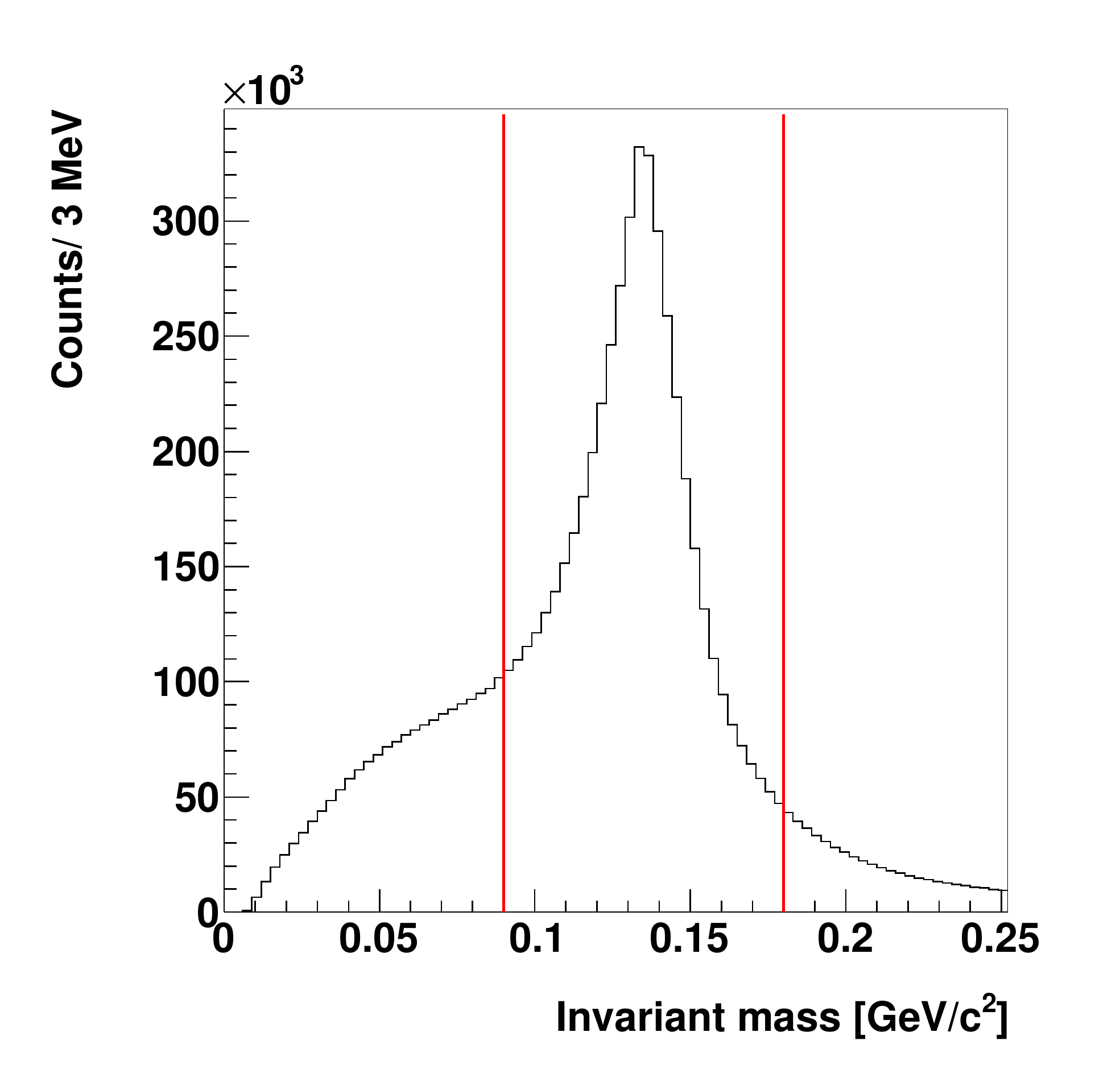,width=0.55\textwidth}}}
\caption{
Experimental invariant mass spectrum for two gamma quanta registered in the calorimeter.
In the case of three photon candidates only these pairs were taken into account for which the 
$\chi^2$ defined in eq.~\ref{chigg} was minimal. The superimposed vertical solid lines at 
90~MeV/c$^2$ and 180~MeV/c$^2$ indicate invariant mass region accepted for the further analysis.
}
\label{EtaPi0EE:imgg}
\end{figure}
Thus, from the data sample one has to choose only events which 
contain two gamma quanta originating from the decay of the $\pi^0$ meson. To this end, we will use the 
chi-square test according to the equation~\ref{chigg} and calculate for each $\gamma_i\gamma_j$ pair an
invariant mass using formula~\ref{Eqimgg}. For the further analysis we will accept only these photon candidates
for which the $\chi_{ij}^2$ is minimal. To suppress splitting effects we restricted the opening angle between two clusters to be greater than $20^o$. The distribution of the invariant mass of accepted photon pairs is shown 
in Fig.~\ref{EtaPi0EE:imgg}. Furthermore, we have restricted the 
values of the invariant mass of two photons to be in the range from 90~MeV/c$^2$ to 180~MeV/c$^2$. This condition is indicated as a red solid line in Fig.~\ref{EtaPi0EE:imgg}.   

\begin{figure}[!h]
\hspace{0.1cm}
\parbox{0.45\textwidth}{\centerline{\epsfig{file=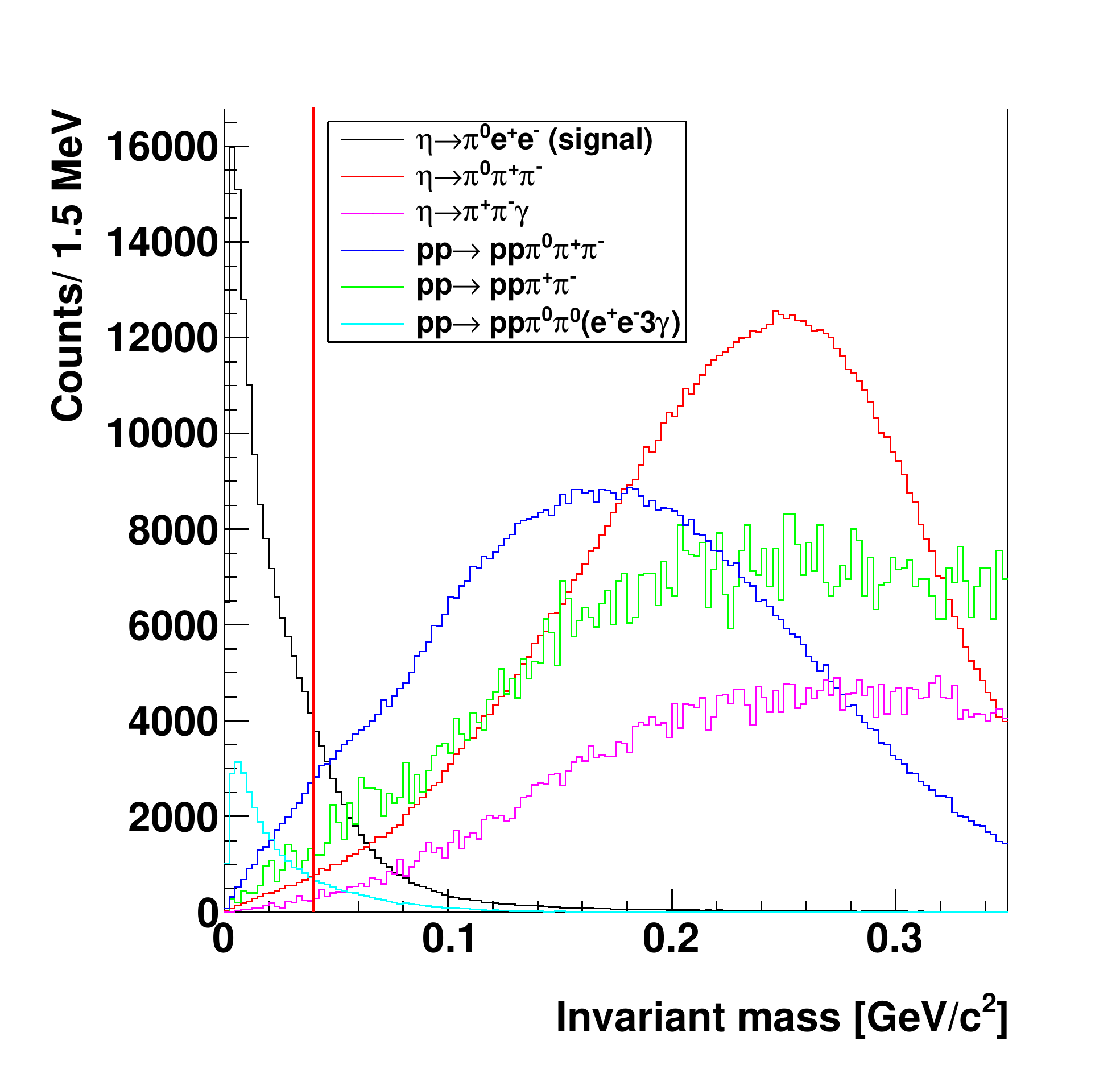,width=0.50\textwidth}}}
\hspace{0.6cm}
\parbox{0.45\textwidth}{\centerline{\epsfig{file=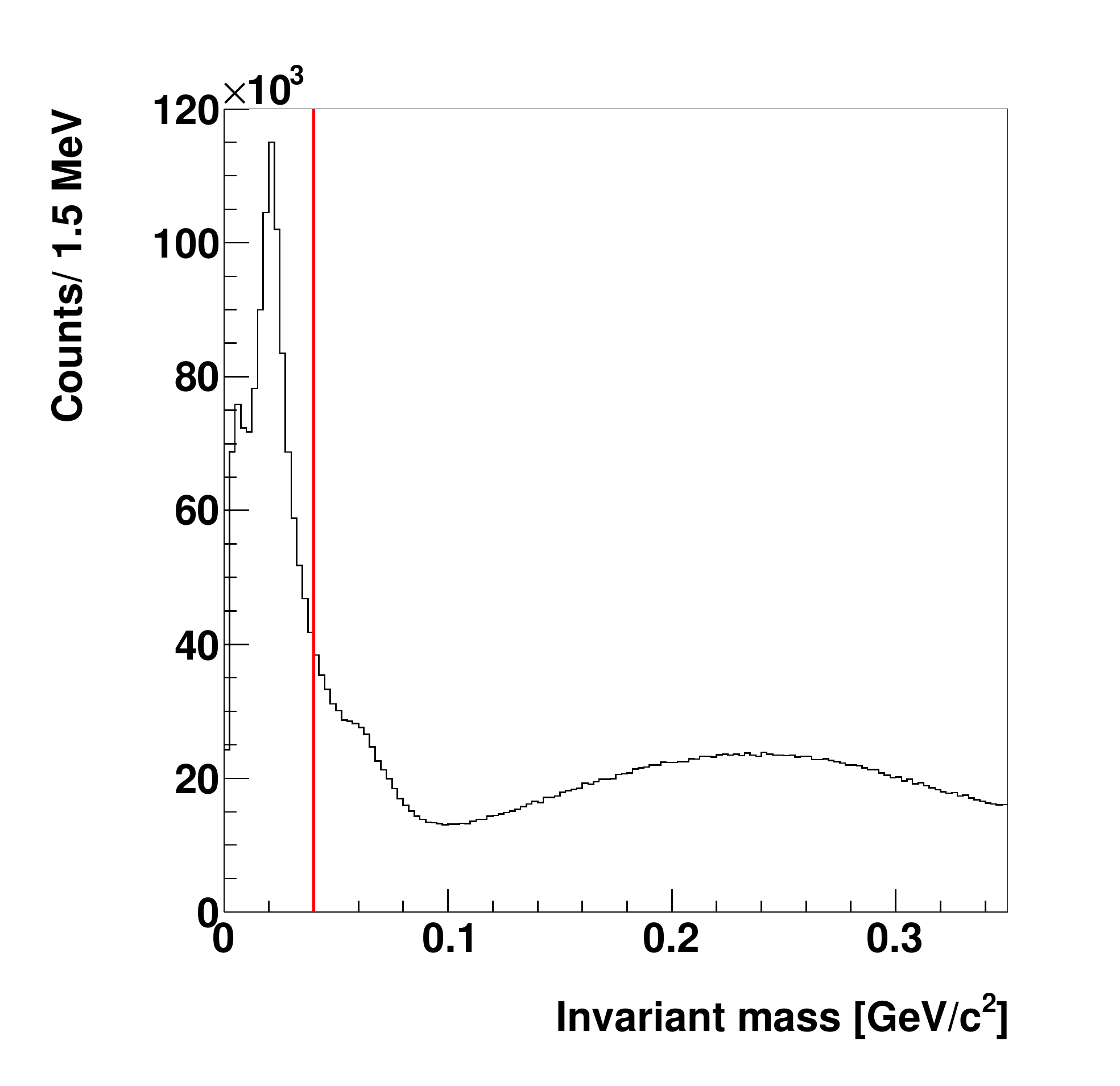,width=0.50\textwidth}}}
\caption{
Invariant mass distributions of two oppositely charged particles registered in central detector.
{\bf{(left)}} Distribution of simulated events for $\eta\to\pi^0 e^+ e^-$ signal decay (solid black line) 
and background reactions with pions as indicated inside the figure. {\bf{(right)}} The experimental 
distribution of invariant mass. The superimposed vertical solid line indicates the cut used for 
accepting only events with invariant masses smaller than 40~MeV/c$^2$.
}
\label{EtaPi0EE:imll}
\end{figure}

Before the identification of electrons and positrons one has to deal with a large background coming from the
other charged particles, mostly pions originating from direct production reactions and the $\eta$ meson decays. 
In order to reduce this background we will use the invariant mass distribution for two oppositely charged
particles detected in central detector. In case of $e^+e^-$ pair the invariant mass due to small mass of the electron and positron, the fact that they originate from the virtual photon, and the form factor mass dependence, the spectrum should be peaked near zero~\cite{Hodana:2012phd,Stepaniak:2002ad}. 
While for the pions the distribution should be shifted
towards much higher invariant masses. The situation is illustrated in Fig.~\ref{EtaPi0EE:imll} where on the 
left panel simulations are shown for signal ($\eta\to\pi^0 e^+ e^-$) and background reactions. It is worth
to notice that in Fig.~\ref{EtaPi0EE:imll} we present only channels with pions in the final state, as this cut are used to suppress 
multi-pion background. In the right panel of Fig.~\ref{EtaPi0EE:imll} the corresponding experimental
distribution of the invariant mass is presented. 
To suppress the pion background we accept only these events for which the invariant 
mass of two charge particles is smaller than 40~MeV/c$^2$. The applied condition is indicated as a solid red line
in Fig.~\ref{EtaPi0EE:imll}. It reduces the pion background almost by the 99\%, decreasing the efficiency 
of the signal reconstruction by 24\% only. 

\begin{figure}[!h]
\hspace{3.4cm}
\parbox{0.5\textwidth}{\centerline{\epsfig{file=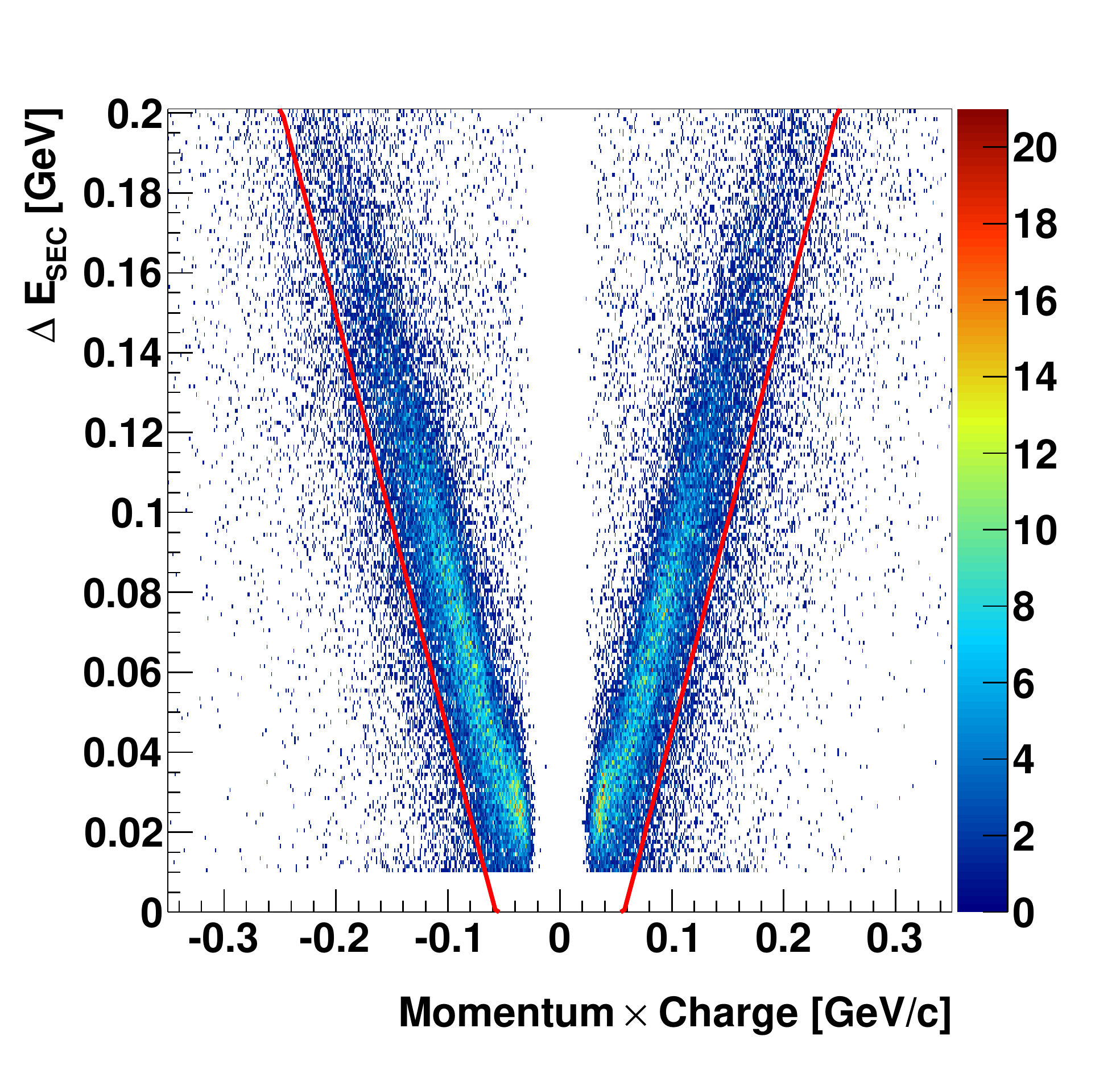,width=0.55\textwidth}}}
\caption{
Experimental $\Delta E - \vert\vec{p}\vert$ spectrum used for electron and positron identification.
The $\Delta E$ corresponds to energy deposited in the electromagnetic calorimeter and the 
$\vert\vec{p}\vert$ was reconstructed based on the bending of the trajectory in the magnetic field.
The superimposed lines indicate a cut used to select candidates for electrons and positrons. 
}
\label{EtaPi0EE:dExp}
\end{figure}

After reduction of the pion background one can process with the identification of electrons and 
positrons emerged into the electromagnetic calorimeter using the $\Delta E - \vert\vec{p}\vert$ method 
(the detailed description of this method was given in Sec.\ref{sec:pipi}). Charged particles scattered into 
the calorimeter deposit their total kinetic energy which at given momentum is larger for electrons than pions.
This situation can be observed in the left panel of Fig.\ref{MCpidCDC} where deposited energy in 
the calorimeter by charged particles as a function of their absolute momentum for simulated 
$\eta\to\pi^+\pi^-\pi^0(\pi^0\to e^+ e^-\gamma)$ decay is shown. The four clearly separated bands can be seen.
The experimental distribution after applying previous cuts is  shown in Fig.~\ref{EtaPi0EE:dExp}. In this figure
only bands coming from electrons and positrons are clearly visible since pions are strongly reduced 
due to application of cut on the invariant mass of charged particles described previously 
(see Fig.~\ref{EtaPi0EE:imll}). 
However, still a small fraction of pions can be seen in the regions bellow the $e^+$ and $e^-$ bands.
Therefore, in order to improve the selection of electrons and positrons, we imposed an identification condition given by an 
inequality: $\Delta E_{SEC} > 1.05\cdot\vert\vec{p}\vert\cdot q - 0.06$, which is indicated as a solid 
red line in Fig.~\ref{EtaPi0EE:dExp}.  

\section{Suppression of split-off and conversion background}\label{sec:conversion}
\begin{figure}[!h]
\hspace{0.1cm}
\parbox{0.45\textwidth}{\centerline{\epsfig{file=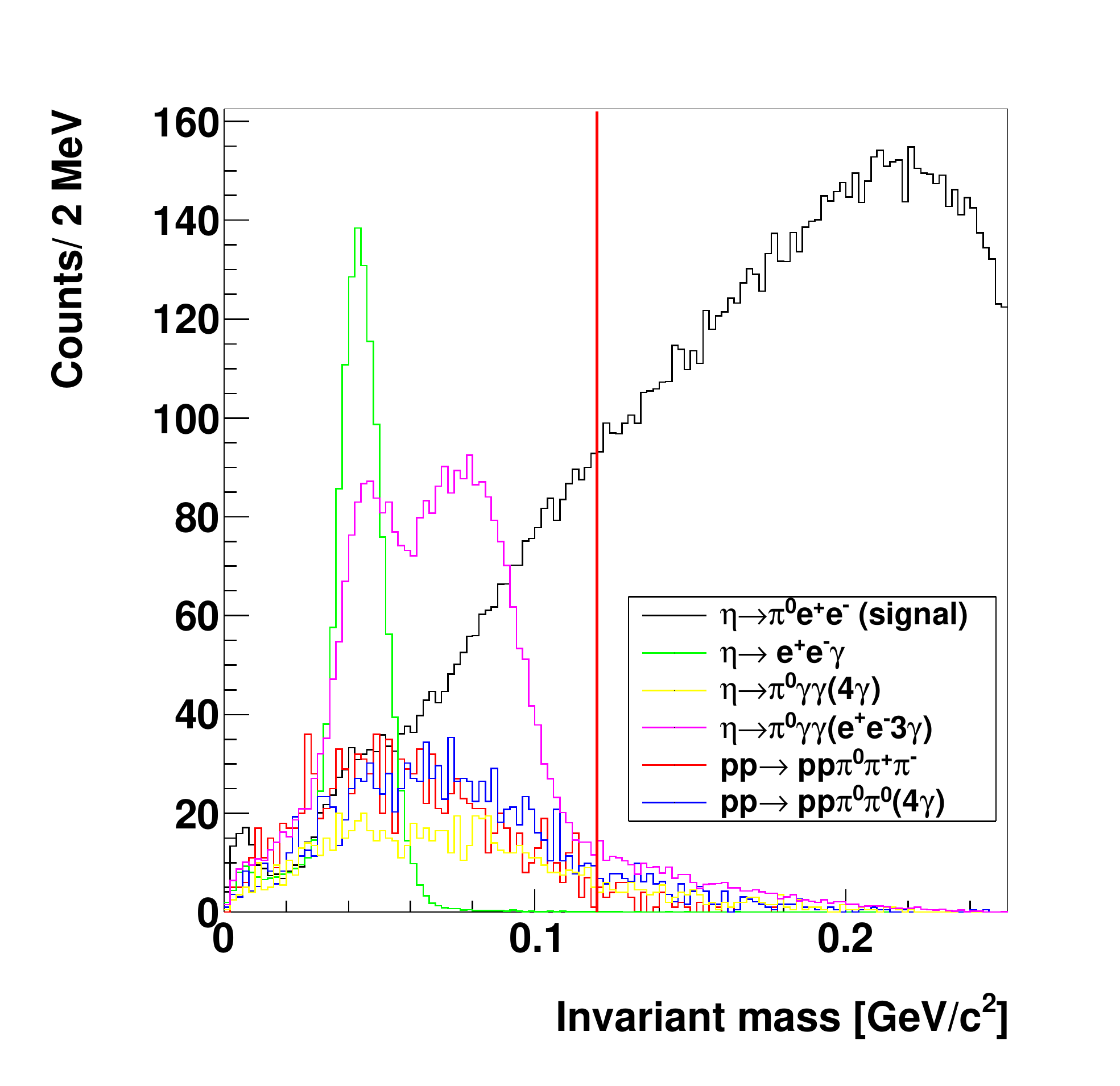,width=0.50\textwidth}}}
\hspace{0.6cm}
\parbox{0.45\textwidth}{\centerline{\epsfig{file=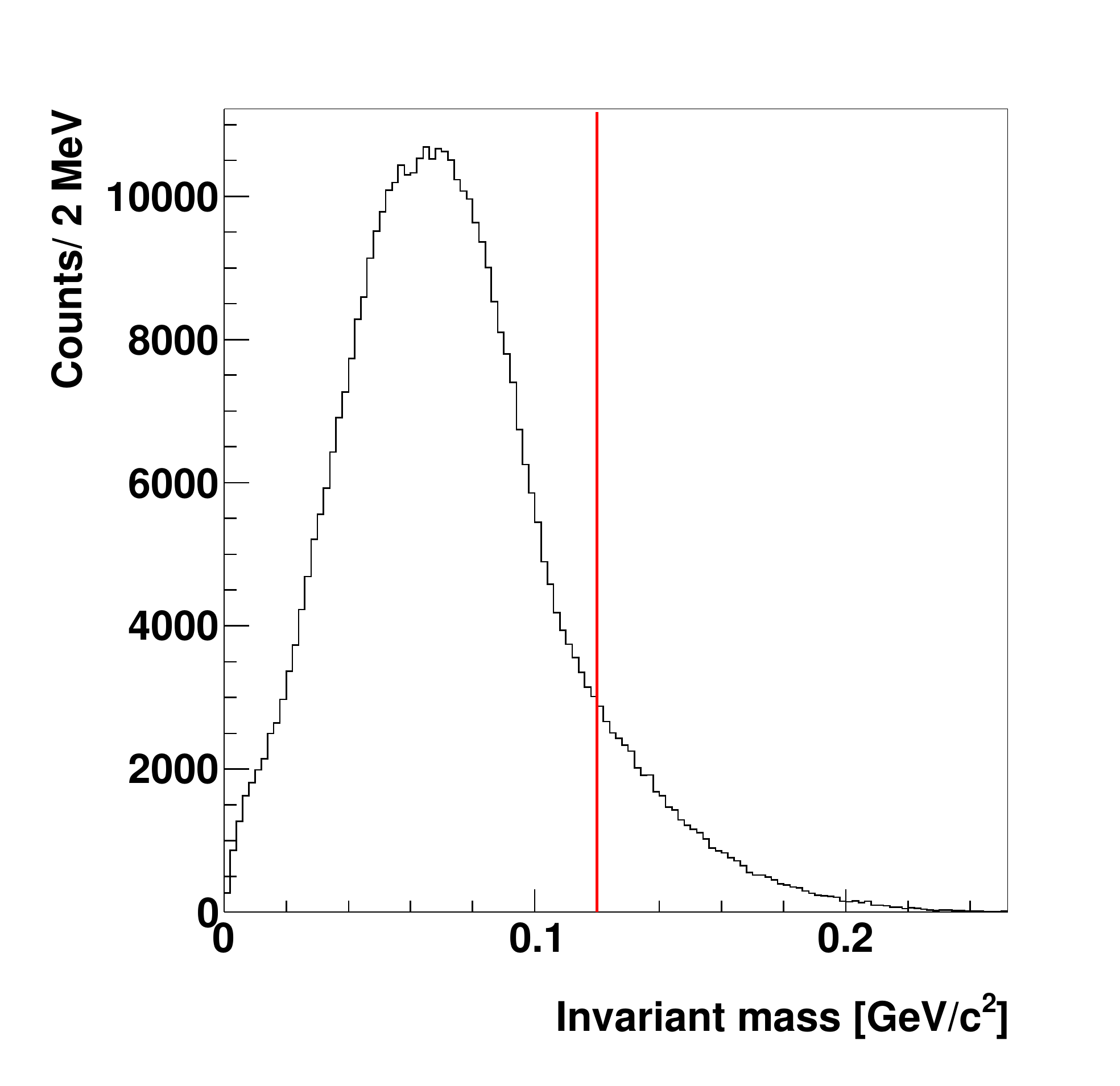,width=0.50\textwidth}}}
\caption{
The smallest invariant mass of pairs of charged and neutral particle candidates: {\bf{(left)}}  
obtained from the 
simulation, and {\bf{(right)}} from the experimental sample. The solid red line denotes the cut used 
in order to suppress events with the bremsstrahlung effect. 
}
\label{EtaPi0EE:imCDC_DCN}
\end{figure}

\hspace{\parindent}
Several reactions can obscure signal of searched decay due to split-off effect caused by the 
{\it{Bremsstrahlung}} of a charge particles in the calorimeter.  The bremsstrahlung results in emission 
of an additional photon when a charge particle is passing close to the nuclei. In such a case 
the resulting photon emerges with a small opening angle to the parent particle and with small kinetic energy.
Therefore, for example in case of the decay $\eta\to e^+e^-\gamma$ one additional gamma quantum arising 
from the bremsstrahlung can imitate the signal channel. In order to suppress this type of background 
we will restrict the smallest invariant mass of a charged and neutral particle pairs to the values larger 
than 120~MeV/c$^2$. The smallest invariant mass was taken as a lowest value of the mass calculated for all the
combinations of pairs of charged and neutral particles candidates.
In case of the signal reaction ($\eta\to\pi^0 e^+ e^-\to\gamma\gamma  e^+ e^-$) an average relative momentum between 
$\gamma$ quanta and leptons are much larger than it is in the case of wrongly identified splitted clusters. 
The corresponding spectrum of invariant mass obtained from 
simulations for signal and several background channels is shown in Fig.~\ref{EtaPi0EE:imCDC_DCN} (left).
Results of simulations are shown for these reactions which may be misidentified as signal due to the 
splitting effects in electromagnetic calorimeter. 
One can see that in case of the split-off reaction the invariant mass is peaked close to zero due to 
small opening angle and small energy of photon relative to the charged particle. For the signal decay the 
invariant mass distribution is much broader and shifted towards higher invariant masses.
The experimental distribution of the minimal invariant mass of a charged and neutral particles is
presented in Fig.~\ref{EtaPi0EE:imCDC_DCN} (right). To suppress events with bremsstrahlung photons we accept
only events for which the invariant mass is larger than 120~MeV/c$^2$. 

Another important effect which has to be studied when considering rare decays of the $\eta$ meson containing
electrons is the external conversion of photons in the detector material
(see also Sec.~\ref{sec:pipi}). The $\gamma$ quanta emerged from the interaction point before reaching the 
electromagnetic calorimeter have to pass through a Beryllium beam pipe with a thickness of 1.2~mm. 
However, when a photon passes through the material it can interact with the nuclei, and convert 
into $e^+e^-$ pair, which in specific cases can obscure the signal from the $\eta\to\pi^0 e^+e^-$ reaction.
\begin{figure}[t]
\hspace{0.7cm}
\parbox{0.31\textwidth}{\centerline{\epsfig{file=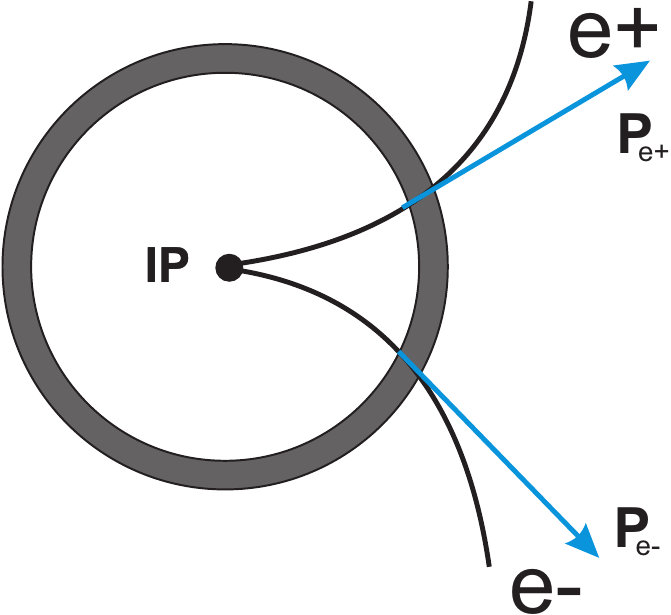,width=0.36\textwidth}}}
\hspace{1.5cm}
\parbox{0.45\textwidth}{\centerline{\epsfig{file=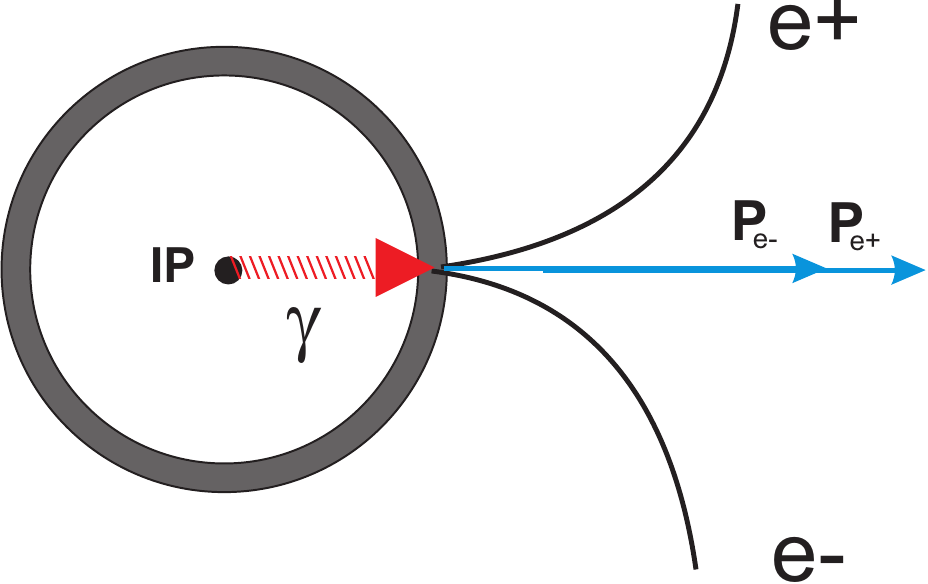,width=0.50\textwidth}}}
\caption{
{\bf{(left)}} Illustration of a primary vertex originating e.g from the $\eta\to e^+e^-\gamma$ decay occurring
in the interaction point (IP).
{\bf{(right)}} Diagram of an external conversion of a photon in the material of a beam pipe resulting 
in production of an electron-positron pair.
}
\label{EtaPi0EE:konwersja}
\end{figure} 
\begin{figure}[!h]
\hspace{3.1cm}
\parbox{0.5\textwidth}{\centerline{\epsfig{file=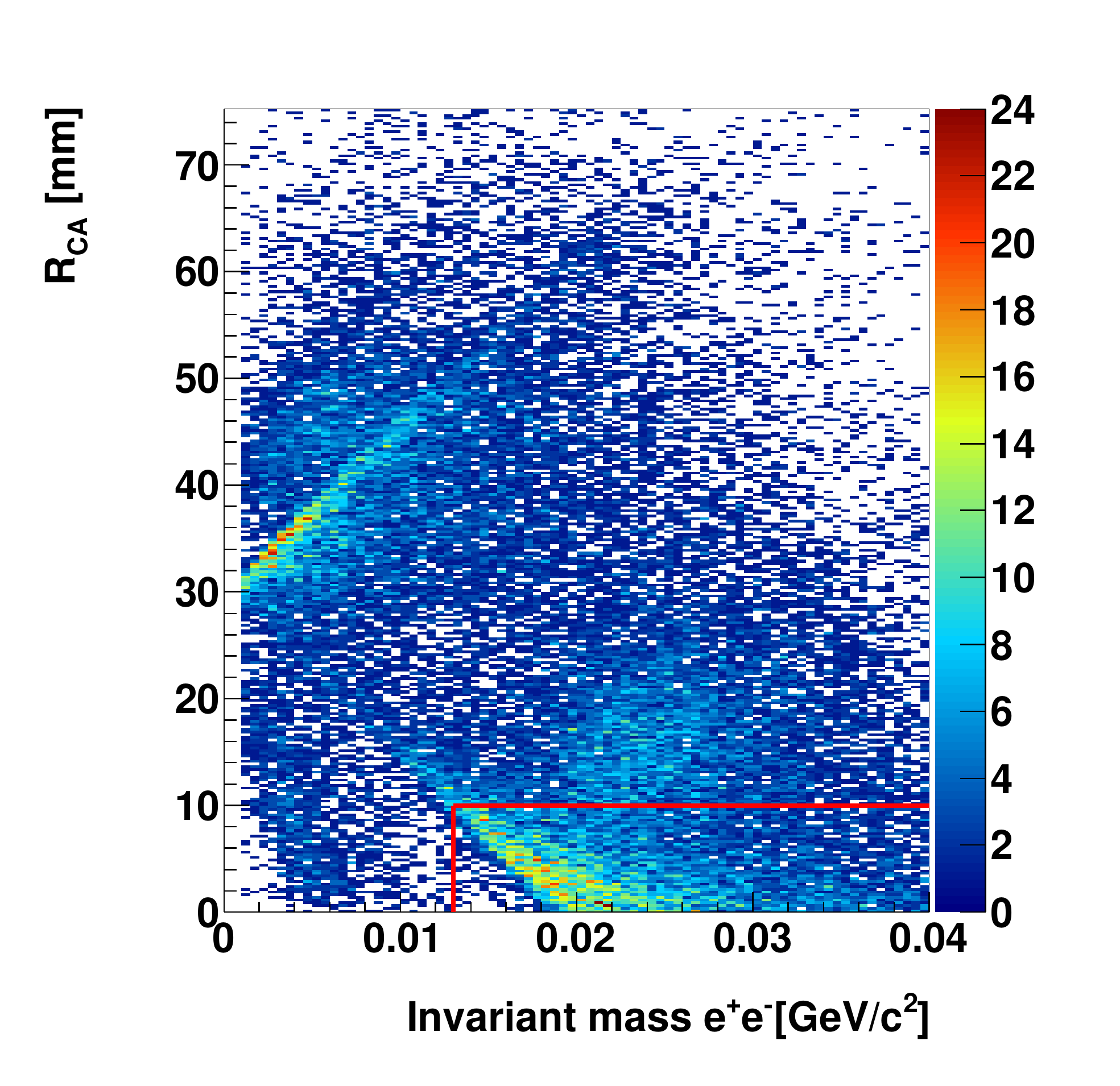,width=0.55\textwidth}}}
\caption{
Experimental distribution of the distance $R_{CA}$ between the interaction point and reconstructed $e^+e^-$
vertex as a function of invariant mass of lepton pair calculated assuming that this pair originates from the 
beam pipe. The red solid line indicate the cut applied in order to suppress the external conversion.
}
\label{EtaPi0EE:RCA}
\end{figure}
In case of electron and positron emerging from the interaction point (IP) their momentum vectors calculated under the assumption that they originate from the beam pipe, will have ''wrong'' directions due 
to the curvature of the tracks in the magnetic field (see Fig.~\ref{EtaPi0EE:konwersja} (left)). 
As a result, the invariant mass calculated based on these vectors will be larger than the mass of 
$e^+e^-$ pair emitted from the beam pipe region. In contrary, for the case of $e^+e^-$ pair originating 
form the 
$\gamma$ conversion at the beam pipe the momentum vectors at the beam pipe will be almost parallel to each 
other (see Fig.~\ref{EtaPi0EE:konwersja} (right)), thus resulting invariant mass will be relatively small. 
Therefore, to suppress the background from pair creation one can use the correlation of the distance between center of
the interaction region and the point of the closest approach of two helices reconstructed ($R_{CA}$) and the 
invariant mass of a lepton pair calculated under the assumption that they were created in the beam pipe.
The corresponding experimental distribution of the $R_{CA}$ as a function of the invariant mass 
of the $e^+e^-$ pair is presented in Fig.~\ref{EtaPi0EE:RCA}. One can see two regions populated on the plot. 
One which is located at small values of invariant mass and radius above 30~mm, and second around invariant
mass values of 20~MeV/c$^2$ and radius below 10~mm. First mentioned region corresponds to the external conversion
of $\gamma$ quanta and, the second region to the reactions where $\eta$ meson decayed in the interaction region (IP) into channel with  
$e^+e^-$ pair in the final state. However, the conversion region is not so enhanced as it was seen in 
Sec.~\ref{sec:pipi}. This is because on this stage of the analysis we have already strongly suppressed events with photons which might cause the conversion by restriction on the minimal invariant mass on charged particle and neutral particle candidate. To suppress the conversion background we applied a cut which is indicated as a solid red line in Fig.~\ref{EtaPi0EE:RCA}. The condition results in rejection of nearly 100\% of all conversion events.

\section{Further background reduction}
\hspace{\parindent}
At this stage of the analysis one has complete reaction chain identified: 
$pp\to pp\eta\to e^+e^-\pi^{0}\to e^+e^-\gamma\gamma$ with all the particles in the final state. 
However, to extract events corresponding to the interesting decay one has to further suppress 
background originating from several reactions. For that purpose one can use the missing mass of two protons 
and the invariant mass of decay products. In this section we will introduce further selection criteria 
aiming at the background suppression. 

\begin{figure}[t]
\hspace{0.1cm}
\parbox{0.45\textwidth}{\centerline{\epsfig{file=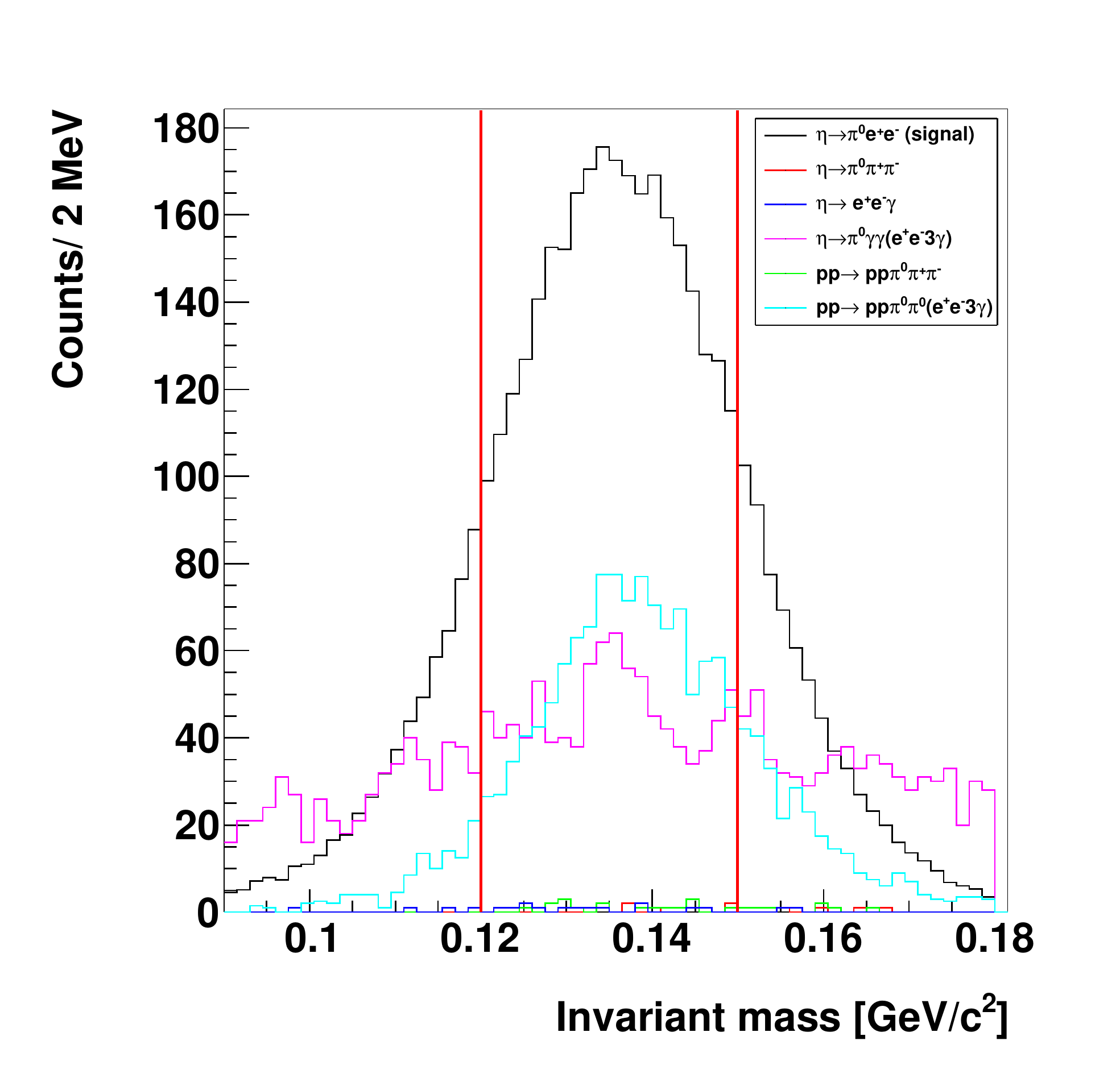,width=0.50\textwidth}}}
\hspace{0.6cm}
\parbox{0.45\textwidth}{\centerline{\epsfig{file=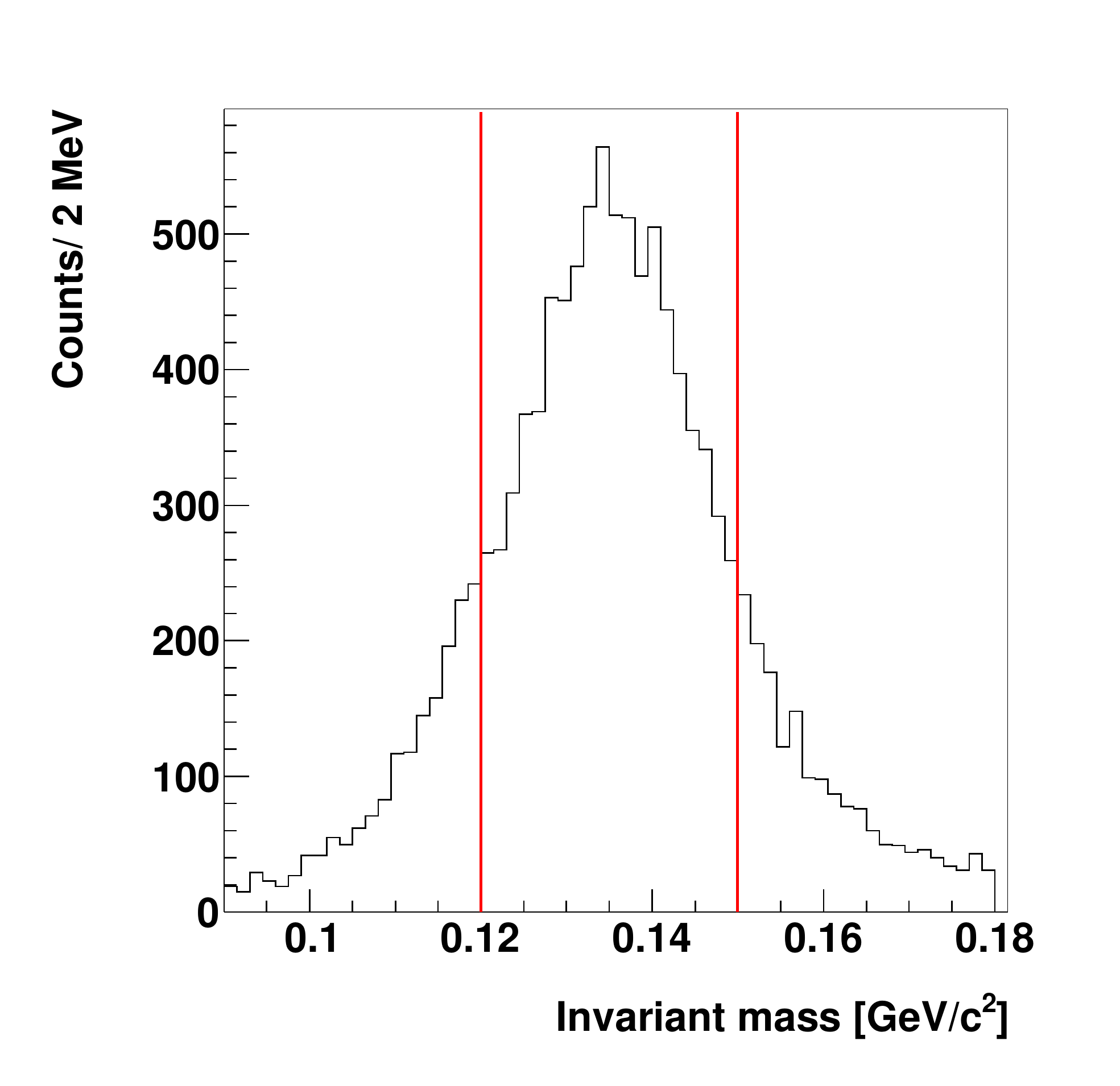,width=0.50\textwidth}}}
\caption{
Invariant mass of two gamma quanta for: {\bf{(left)}} simulations, and {\bf{(right)}} measured data.
The red solid line indicate region of invariant masses  from 120~MeV/c$^2$ to 150~MeV/c$^2$ used to select  
candidates for the signal reaction.
}
\label{EtaPi0EE:IMpi0A}
\end{figure}

First we will strongly restrict invariant mass of two photons identified as originating from the 
neutral pion. This condition will allow to reduce the remaining background from the misidentified 
reactions like: $\eta\to\pi^0\gamma\gamma\to e^+e^-\gamma\gamma\gamma$ with additional gamma quanta.
Figure \ref{EtaPi0EE:IMpi0A} (left) presents the invariant mass of two registered photons for simulated 
background and signal processes. To limit this background we accept only masses in the region from 
120~MeV/c$^2$ to 150~MeV/c$^2$ which corresponds to the FWHM of the 
simulated signal distribution. The spectrum of the invariant mass from experimental data can be seen in 
Fig.~\ref{EtaPi0EE:IMpi0A} (right). The solid red lines indicate applied cut.

\begin{figure}[!h]
\hspace{0.1cm}
\parbox{0.45\textwidth}{\centerline{\epsfig{file=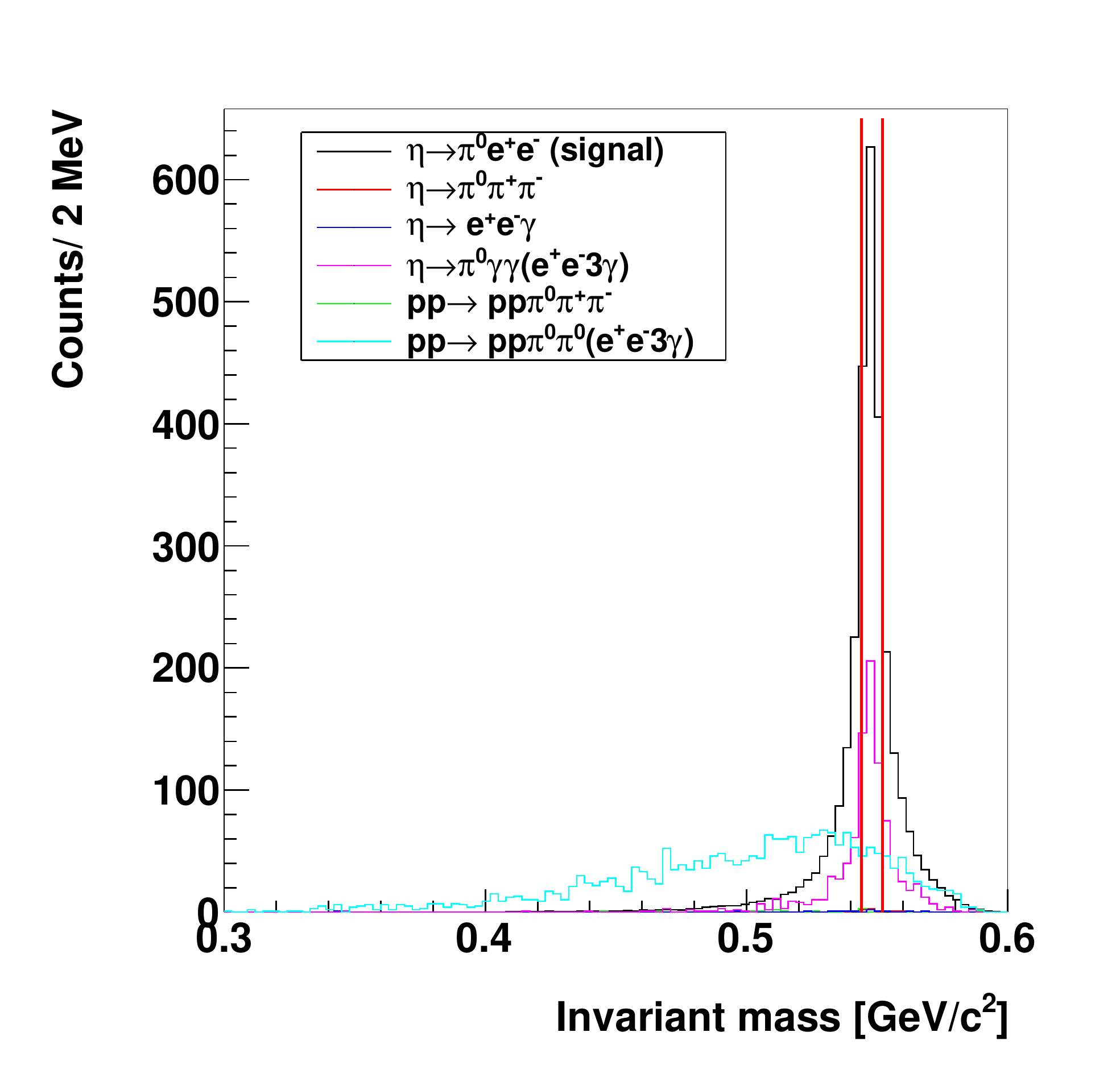,width=0.50\textwidth}}}
\hspace{0.6cm}
\parbox{0.45\textwidth}{\centerline{\epsfig{file=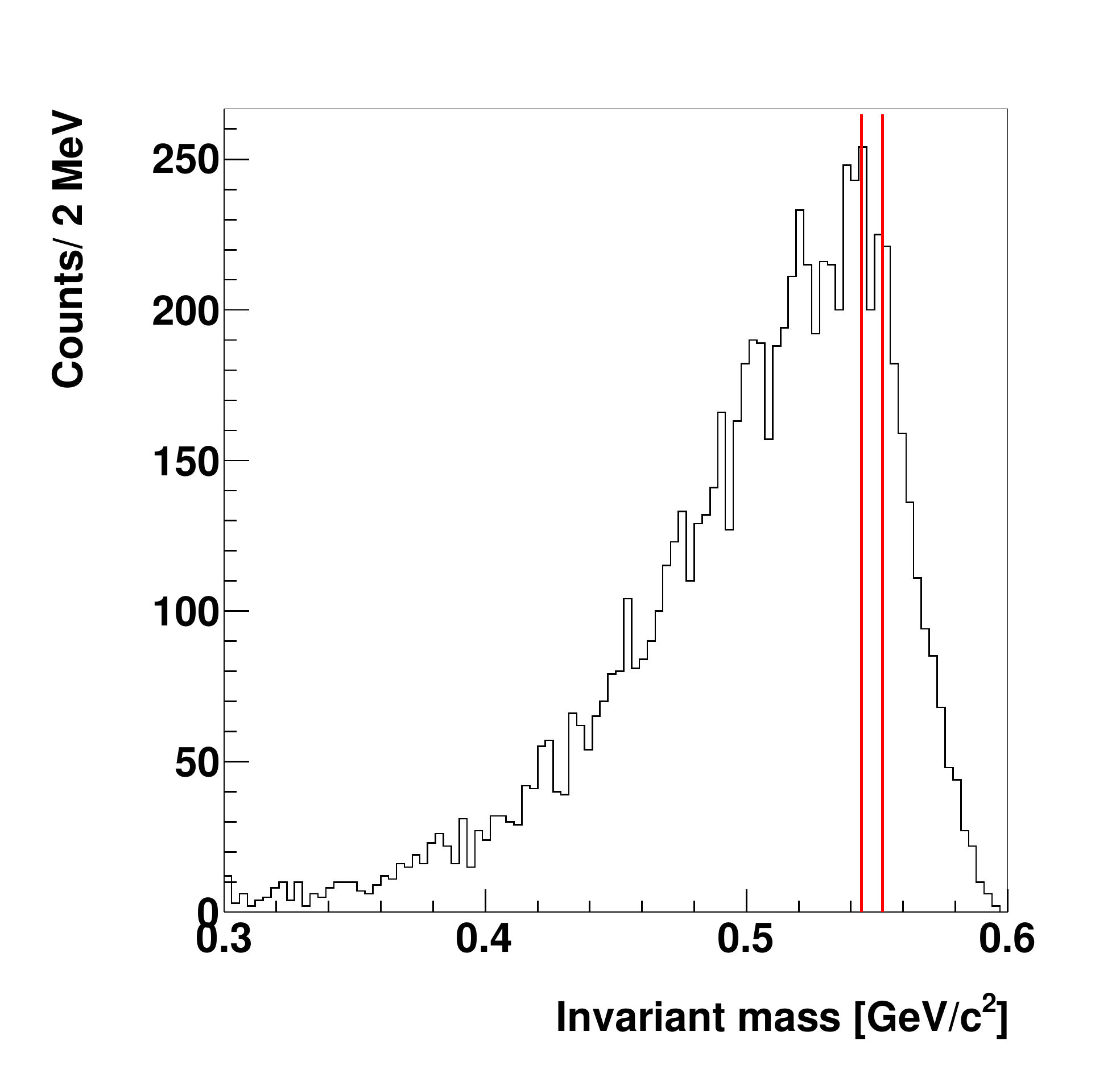,width=0.50\textwidth}}}
\caption{
Distribution of the missing mass of two protons: {\bf{(left)}} from simulations, and {\bf{(right)}} 
measured data. The red solid line indicates selected region close to the $\eta$ mass in the range 
from 544~MeV/c$^2$ to 552~MeV/c$^2$.  
}
\label{EtaPi0EE:MMpp}
\end{figure}

The missing mass distribution of two protons for events remaining after all previous cuts is 
still contaminated with background originating from the $\eta$ meson decay into pions and also direct 
two and three pion production (see Fig.~\ref{EtaPi0EE:MMpp}). As the branching ratio is unknown for searched decay and also the total cross
section for the direct three pion production remains unmeasured at the investigated beam energy it 
is not possible to estimate the absolute contribution from background. However, to remove as much background 
as possible we will restrict missing mass of two protons only to the narrow 
range around the $\eta$ meson mass from 544~MeV/c$^2$ to 552~MeV/c$^2$ (this range corresponds to the value of the 
FWHM of the resolution of missing mass distribution). The events which fulfill this condition will 
be considered further for the analysis.
The simulated missing mass of two protons for background reactions and the signal is presented in 
Fig.~\ref{EtaPi0EE:MMpp} (left). The analogous distribution for the experimental data is presented 
in Fig.~\ref{EtaPi0EE:MMpp} (right). 
\begin{figure}[!h]
\hspace{0.1cm}
\parbox{0.45\textwidth}{\centerline{\epsfig{file=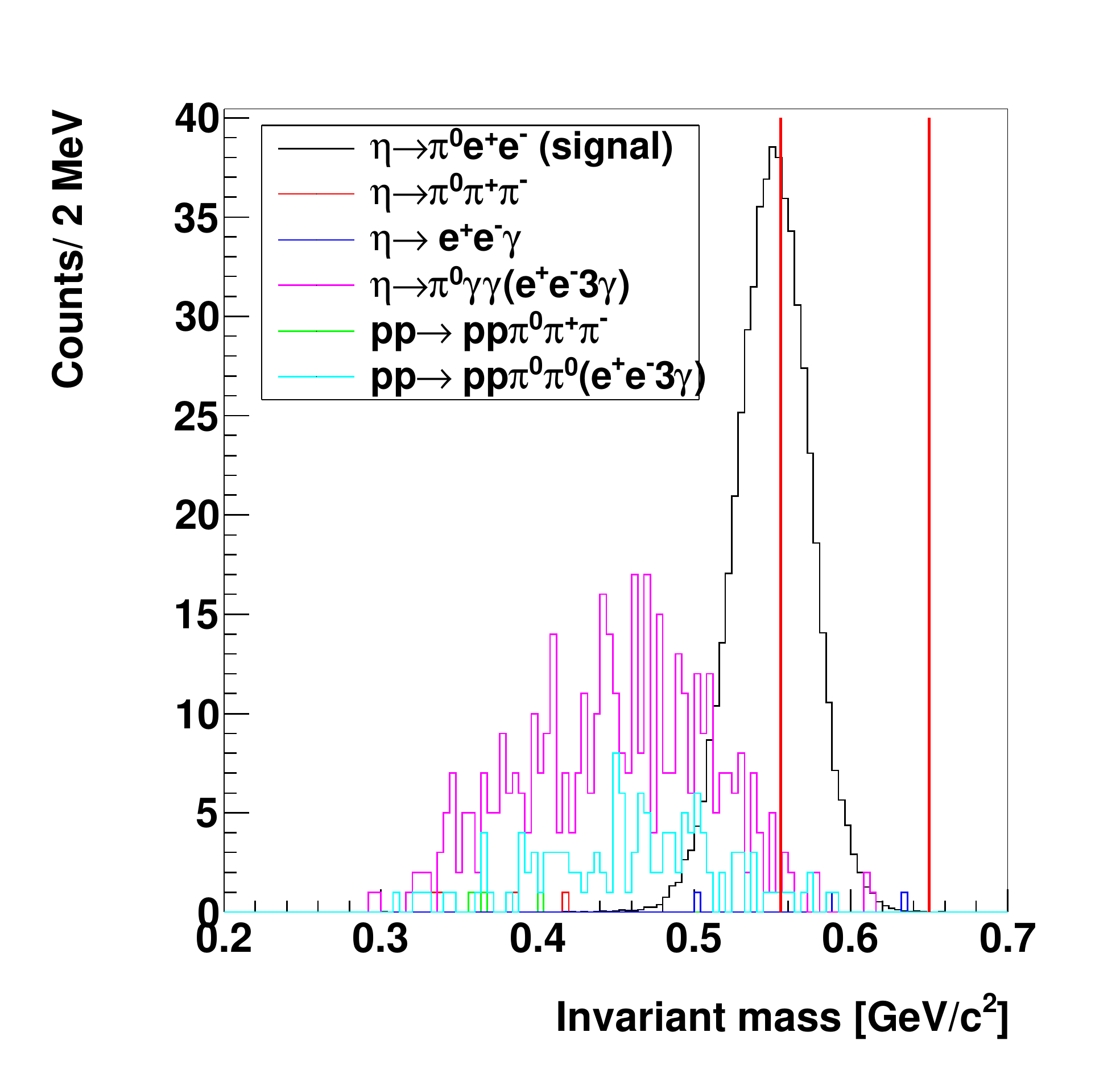,width=0.50\textwidth}}}
\hspace{0.6cm}
\parbox{0.45\textwidth}{\centerline{\epsfig{file=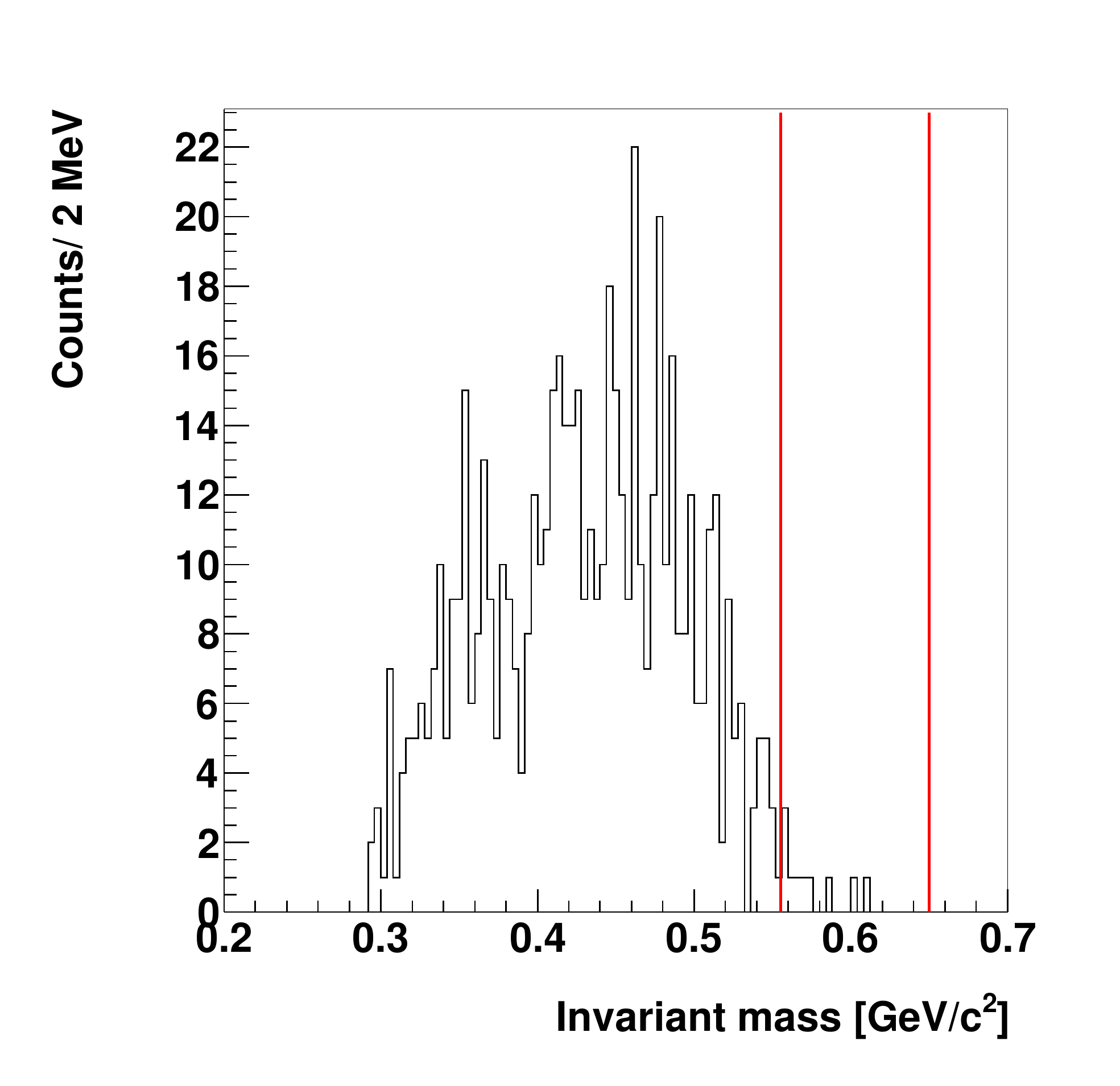,width=0.50\textwidth}}}
\caption{
Invariant mass of particles identified as $\pi^0 e^+e^-$ from: {\bf{(left)}} simulations, 
and {\bf{(right)}} measured data. The red solid line indicates the region of 
invariant masses from 555~MeV/c$^2$ to 650~MeV/c$^2$ used to select candidates for the signal reaction.
}
\label{EtaPi0EE:IMeepi0}
\end{figure} 

Finally, one can plot the invariant mass of particles in the final state: $\pi^0 e^+ e^-$,
and select the range corresponding the signal reaction. Figure~\ref{EtaPi0EE:IMeepi0} (left)
presents simulated signal and background reactions remained after all previous selection conditions. 
One can see that the background reactions populate the lower invariant masses, and the signal is 
pronounced around the mass of the $\eta$ meson. Therefore, to suppress the background we select invariant
mass region from 555~MeV/c$^2$ to 650~MeV/c$^2$. This reduces the background nearly by 99\%, decreasing the
signal only by 53\%. Later on this selection condition will be further optimized.
The experimental distribution of corresponding invariant mass is shown in Fig.~\ref{EtaPi0EE:IMeepi0} (right).
The summary of all applied selection criteria with the number of experimental events remaining after the cuts 
and with the selection efficiency of the signal obtained from the simulation are shown in 
Tab.~\ref{tab:EtaPi0EE:EFF}. 
\begin{table}[t]
\centering
\begin{tabular}{c|l|c|c}
\hline
Cut  & Cut description & $\epsilon_{S}$  & Data \\\hline\hline
 PRESEL            & Preselection of two protons in FD, two        & 13\%    & 7485153  \\
                   & oppositely charged particles in CD and        &         &          \\
                   & two or more neutral clusters in SEC.          &         &          \\\hline
 $IM(\gamma\gamma)$& Selection of two photons originating from     & 90\%    & 4952520  \\
                   & the decay of $\pi^0$ meson with invariant     &         &          \\
                   & mass in the range from 90 - 180 MeV/c$^2$     &         &          \\\hline
 $IM(l^+l^-)$      & Restriction on invariant mass of two particles& 76\%    & 1154086  \\
                   & measured in MDC to be less than 40 MeV/c$^2$  &         &          \\\hline
 SEC               & Particle identification in the                & 56\%    & 432813   \\
                   & Scintillating Electromagnetic Calorimeter.    &         &          \\\hline
 Split-off         & Suppression of events with the bremsstrahlung & 79\%    & 44668    \\
                   & effect for charged particles.                 &         &          \\\hline
 Conversion BP     & Suppression of external conversion of photons & 43\%    & 11617    \\
                   & on the beam pipe.                             &         &          \\\hline
 $IM(\gamma\gamma)$& Additional restriction on mass of $\pi^0$     & 69\%    & 8129     \\
                   & meson from 120 - 150 MeV/c$^2$.               &         &          \\\hline
 MM(pp)            & Missing mass of two protons with the          & 47\%    & 583      \\
                   & restriction of 544 - 552 MeV/c$^2$.           &         &          \\\hline
 IM($e^+e^-\pi^0$) & Invariant mass of final state particles with  & 47\%    & 10       \\
                   & restriction from 555 - 650 MeV/c$^2$.         &         &          \\\hline\hline
  {\bf{TOTAL}}     & {\bf{Total efficiency for signal selection}}  & {\bf{0.26\%}}  & {\bf{10}}       \\
                   & {\bf{and final number of selected events.}}   &         &          \\\hline\hline
\end{tabular}
\caption{The list of conditions used for selecting the $\eta\to e^+e^-\pi^0$ decay. Third column include the efficiency reconstruction for the signal determined from the simulations. Fourth column indicates the 
number of experimental events left after application of subsequent cuts.}
\label{tab:EtaPi0EE:EFF}
\end{table}

After application of all above described conditions to the experimental data number of event candidates 
for $\eta\to\pi^0 e^+ e^-$ decay is equal to $N^{exp} = 10 \pm 3_{stat}$. However, the estimated number 
of $N^{exp}$ may account not only for events corresponding to the searched decay $\eta\to\pi^0 e^+ e^-$
but it may also be due to the misidentification of the background reactions.
In order to estimate the number of background events we have performed the same analysis of the simulated 
samples of background reactions (see Tab.~\ref{tab:background}).
\begin{table}[!h]
\centering
\begin{tabular}{c|c|c|c|c}
\hline
No. & Reaction & Number of & Number of   & Number of\\
    &          & generated & reconstru-  & expected\\
    &          & events    & cted events & events\\\hline\hline
1 &$pp\to pp\eta\to pp\pi^0 e^+e^-\to ppe^+e^-\gamma\gamma$ (signal)     &  $5\times 10^6$  & 9208 & ?\\\hline
2 &$pp\to pp\eta\to pp \pi^+\pi^-\pi^0\to pp\pi^+\pi^-\gamma\gamma$      &  $20\times 10^6$ & 0    & 0\\
3 &$pp\to pp\eta\to pp \pi^+\pi^-\pi^0\to pp\pi^+\pi^- e^+e^-\gamma$     &  $1\times 10^6$  & 0    & 0\\
4 &$pp\to pp\eta\to pp\pi^+\pi^-\gamma$                                  &  $5\times 10^6$  & 0    & 0\\
5 &$pp\to pp\eta\to pp e^+ e^-\gamma$                                    &  $20\times 10^6$ & 2    & 0\\
6 &$pp\to pp\eta\to pp\pi^0\gamma\gamma\to pp\gamma\gamma\gamma\gamma$   &  $3\times 10^6$  & 0    & 0\\
7 &$pp\to pp\eta\to pp\pi^0\gamma\gamma\to pp e^+ e^-\gamma\gamma\gamma$ &  $3\times 10^6$  & 13   & 0\\
8 &$pp\to pp\eta\to pp\gamma\gamma$                                      &  $6\times 10^6$  & 0    & 0\\\hline
9 &$pp\to pp\pi^+\pi^-\pi^0\to pp\pi^+\pi^-\gamma\gamma$                 &  $20\times 10^6$ & 0    & 0\\
10 &$pp\to pp\pi^+\pi^-$                                                 &  $58\times 10^6$ & 0    & 0\\
11 &$pp\to pp\pi^0\pi^0\to pp\gamma\gamma\gamma\gamma$                   &  $18\times 10^6$ & 0    & 0\\
12 &$pp\to pp\pi^0\pi^0\to pp e^+e^-\gamma\gamma\gamma$                  &  $8\times 10^6$  & 7    & 13\\\hline
\end{tabular}
\caption{List of simulated reactions. Third column includes the number of initially generated events,
and fourth column indicates the number of events left after all selection conditions. 
Last column indicates number of events expected to fulfill all selection criteria after taking into account a cross section and branching ratio.}
\label{tab:background}
\end{table}

After application of selection criteria to the simulated background reactions, and  
after taking into account the cross section and branching ratios values the only background channel left is 
$pp\to pp \pi^0\pi^0\to pp e^+e^-\gamma\gamma\gamma$ reaction which amounts to $N_{B} = 13 \pm 4_{stat}$.
Therefore the number of reconstructed events in the measured data is consistent statistically with the estimated 
amount of background. Thus we state that in the measured data sample of $pp\to pp\eta$ reaction we 
did not observe a signal from the searched $\eta\to\pi^0 e^+e^-$ decay. In the next chapter we will estimate 
the upper limit for the branching ratio of investigated decay channel.  

\chapter{Results for the $\eta\to\pi^0 e^+e^-$ decay}
\hspace{\parindent}
In this thesis we aim at establishing the unknown branching ratio for the $\eta\to\pi^0 e^+ e^-$ decay
which might violate charge conjugation invariance. However, as it was shown in previous section as a  
result of the conducted analysis of an experimental data for the $pp\to pp\eta$ reaction we did not observe
a statistically significant signal over the background. The final number of candidates events for the 
$\eta\to\pi^0 e^+ e^-$ process selected from experimental events is equal to 
$N^{exp} = 10 \pm 3_{stat}$, and the expected number of background events obtained from the 
simulations is equal to $N_{B} = 13 \pm 4_{stat}$. In such case, where the expected number of signal events is equal to zero, it is impossible to calculate the value of the branching ratio, and only an upper limit 
can be estimated. 

In order to calculate the upper limit for the branching ratio one has to know the 
number of expected: (i) signal, (ii) background, (iii) normalization channel events, (iv) selection
efficiency, and (v) the branching ratio for the normalization channel.  
In case of this thesis for the normalization we choose the $\eta\to\pi^+\pi^-\pi^0$ decay. 
This channel has the same topology of the final state, as the signal reaction, thus the systematic effects 
should affect both processes nearly in the same way. Therefore, this will allow to cancel out many of the systematic uncertainties.  

\section{The upper limit for the branching ratio $BR(\eta\to\pi^0 e^+e^-)$}
\hspace{\parindent}
The number of the signal events can be given by the following formula:
\begin{equation}
N_{S} = L \cdot \sigma_{\eta}\cdot BR_{\eta\to\pi^{0}e^+e^-} \cdot \epsilon_{S},
\label{eqsignal}
\end{equation}
and for the normalization channel:
\begin{equation}
N_{norm} = L \cdot \sigma_{\eta}\cdot BR_{\eta\to\pi^+\pi^-\pi^{0}} \cdot \epsilon_{norm},
\label{eqnorm}
\end{equation}
where $L$ denotes the integrated luminosity, the $\sigma_{\eta}$ is the total cross section for production of the $\eta$ meson, 
$N_{S}$ and $N_{norm}$ stands for the number of events for the signal and normalization channel, 
$\epsilon_{S}$ and $\epsilon_{norm}$ denotes the detection efficiency for the signal and 
normalization reaction, respectively. 
Therefore, the formula for the investigated branching ratio obtained from the two above given equations
will take following form:
\begin{equation}
BR_{\eta\to\pi^{0} e^+e^-} = \frac{N_{S} \cdot BR_{\eta\to\pi^+\pi^-\pi^{0}}  \cdot \epsilon_{norm}}{N_{norm} \cdot \epsilon_{S}}.
\label{brbr}
\end{equation}
The branching ratio for the normalization channel 
$\eta\to\pi^+\pi^-\pi^0$ is equal to $BR(\eta\to\pi^+\pi^-\pi^0) = (22.74 \pm 0.34)\%$~\cite{Nakamura:2010zzi}.
The signal for the searched decay $\eta\to\pi^0 e^+e^-$ was not observed in the investigated data sample, and therefore the equation (\ref{brbr}) will take form of an inequality:
\begin{equation}
BR_{\eta\to\pi^{0} e^+e^-}^{UL} < \frac{N_{S}^{UL} \cdot BR_{\eta\to\pi^+\pi^-\pi^{0}}  \cdot \epsilon_{norm}}{N_{norm} \cdot \epsilon_{S}},
\label{brUL}
\end{equation}
where the $N_{S}^{UL}$ denotes the number of expected signal events at a given confidence level. 

However, to test and optimize the last selection condition on the invariant mass of the $\pi^0 e^+e^-$ 
we have varied this cut and for each case we counted the number of events for measured and 
simulated data.
The result of these tests are presented in Tab.~\ref{tab:CutTests}, together with  efficiency calculated for the signal. The cut discussed in Section~9.3 is shown as no.~2 in Tab.~\ref{tab:CutTests}.
\begin{table}[!h]
\centering
\begin{tabular}{c|c|c|c|c|c}
\hline
No.&IM($\pi^0e^+e^-$)  & $N^{exp}$ & $\epsilon_{S}$ & $N^{expected}_{B}$ & $\rho=\frac{\epsilon_{S}}{N^{expected}_{B}}$\\\hline\hline
1.& 550 - 650 MeV/c$^2$      & 12        &    0.28\%      & 17      & 1.64\\\hline
2.& 555 - 650 MeV/c$^2$      & 10        &    0.26\%      & 13      & 1.96\\\hline
3.& 560 - 650 MeV/c$^2$      & 7         &    0.20\%      & 11      & 1.82\\\hline
4.& 580 - 650 MeV/c$^2$      & 3         &    0.05\%      & 4       & 1.25\\\hline
\end{tabular}
\caption{Table showing the number of events obtained from the measured data and from simulations of the background channel $pp\to pp \pi^0\pi^0\to pp e^+e^-\gamma\gamma\gamma$ for different values of the cut on the invariant mass of the $\pi^0 e^+e^-$ system. }
\label{tab:CutTests}
\end{table}
One can see that by narrowing the condition on the invariant mass the number of
events is decreasing both for experimental data and simulated background. Also the efficiency of the signal 
reconstruction decreases. As an criterion for choosing the best cut range we defined the significance parameter as:
\begin{equation}
\rho=\frac{\epsilon_{S}}{N^{expected}_{B}}.
\end{equation}
The $\rho$ parameter is increasing with the higher efficiency of the signal reconstruction and 
decreasing with number of background events. Thus as an optimum cut we have chosen cut number 2 listed 
in Tab.~\ref{tab:CutTests} for which the $\rho$ coefficient was the highest.  

After the cut optimization one can estimate the upper limit for the searched branching ratio.
The overall detection and reconstruction efficiencies for the signal and normalization channels were determined based on the simulations and they are equal to: $\epsilon_{norm} = 0.89\%$ and $\epsilon_{S} = 0.26\%$, respectively. From the Tab.~\ref{tab:CutTests} one can see that we have measured 
10 events in experimental data and expect based on the simulations 13 background events. Thus, the value of the upper limit of the expected signal is equal to $N_{S}^{UL} = 3.95$~\cite{Feldman:1997qc}, at the confidence level of 90\%. The final result for the upper limit of the branching ratio equals to:
\begin{equation}
BR_{\eta\to\pi^{0} e^+e^-}^{UL} < \frac{ 3.95 \cdot 0.2274 \cdot 0.89\%}{ 82725 \cdot 0.26\%},
\end{equation}
\begin{equation}
BR_{\eta\to\pi^{0} e^+e^-}^{UL} <  3.7 \cdot 10^{-5}~~~~~~\text{(CL~=~90\%)}.
\end{equation}
Discussion of obtained result will follow in the next section.

\section{Discussion}
\begin{figure}[!h]
\hspace{3.1cm}
\parbox{0.55\textwidth}{\centerline{\epsfig{file=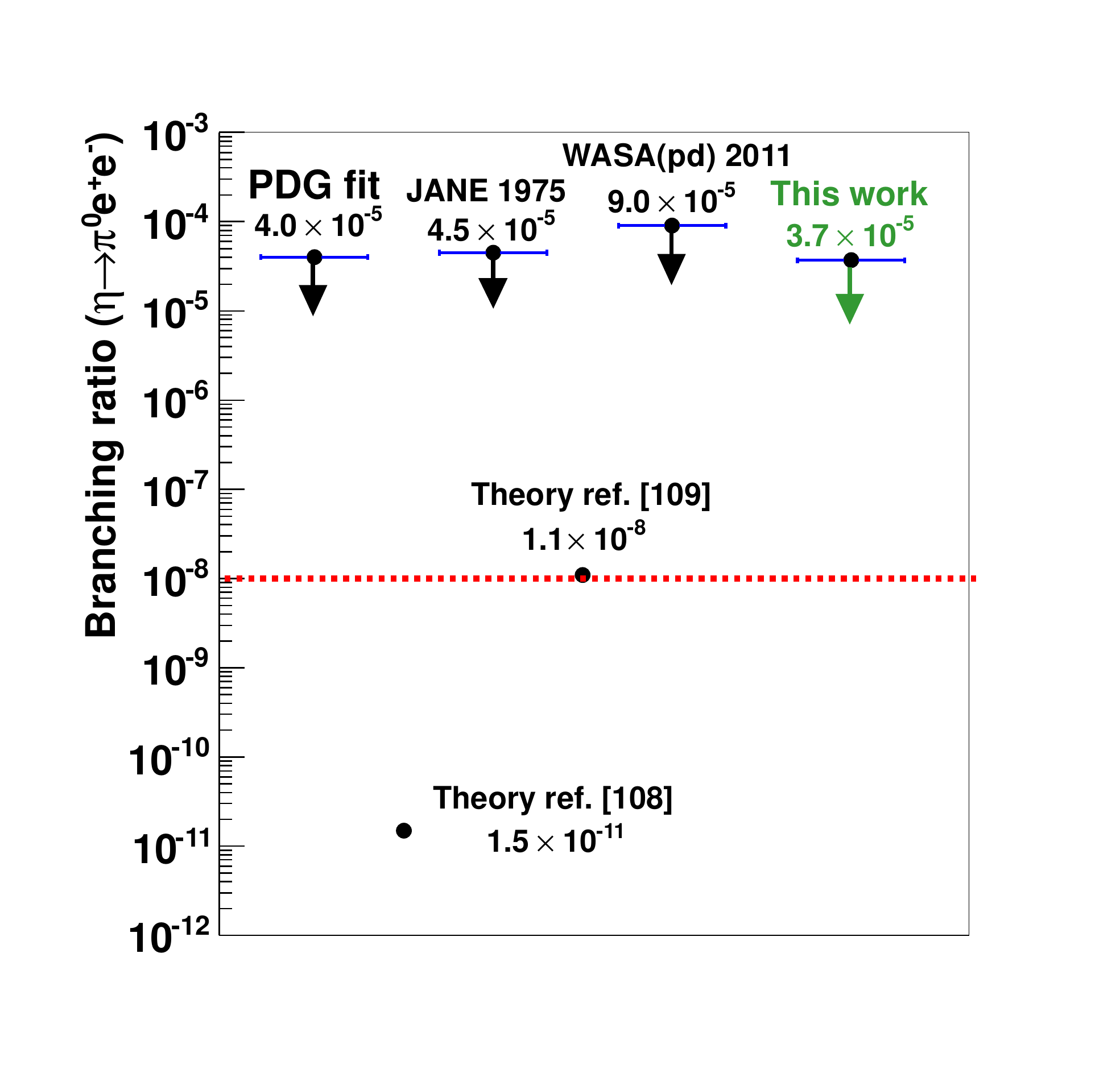,width=0.60\textwidth}}}
\caption{
Comparison of obtained value of the upper limit for the branching ratio of the decay $\eta\to\pi^0 e^+ e^-$
with previous measurements and the PDG fitted value. The red dashed line indicates the limit of the 
Standard Model predictions. 
}
\label{EtaPi0EE:BRlimit}
\end{figure}

\hspace{\parindent}
Presented analysis for search of the $\eta\to\pi^0 e^+ e^-$ decay is the first from the WASA-at-COSY data
where the $\eta$ meson was produced in proton-proton collisions.  
Obtained result is smaller than presently known upper limit given 
by the PDG group~\cite{Nakamura:2010zzi} based on previous results~\cite{Jane:1975nt}.
This result, constitutes a next step in the search for rare decay of the $\eta$ meson by means of the WASA-at-COSY detector. Previously this decay was investigated by WASA-at-COSY in $pd\to^{3}He\eta$ reaction where 
obtained value of the upper limit for the branching ratio was 
$BR(\eta\to\pi^0 e^+e^-) < 9 \cdot 10^{-5}$~\cite{Winnemoller:2011phd}. The result obtained in this work 
is compared to previous results and the theoretical prediction in Fig.~\ref{EtaPi0EE:BRlimit}.

In previous chapter, we have shown that analysis conducted in the framework of this thesis revealed no  
observe signal from the $\eta\to\pi^0 e^+ e^-$ decay. The experimentally obtained number of signal 
candidate events was the same as the number of estimated background events, with respect to the statistical uncertainty. Therefore, we determined an upper limit for the branching ratio for this decay.
In case of small statistical signal there are many approaches which can be 
used for the estimation of the upper limits~\cite{Silarski:2012Phd}. However, as a result of 
this thesis we will take the upper limit estimated in previous section
according to the prescription given by Feldman and Cousins~\cite{Feldman:1997qc}, assuming the Poisson 
distribution of the variables where the limit estimations on the physical quantities are based on the likelihood ratios.

\chapter{Summary and perspectives}
\hspace{\parindent}
The main motivation of this dissertation was the test of the charge conjugation C in the strong 
and electromagnetic interactions. For this purpose we studied the $\eta\to\pi^+\pi^-\pi^0$ 
and $\eta\to\pi^0 e^+e^-$ decays, where the $\eta$ meson was produced via $pp\to pp\eta$ reaction 
near the kinematical threshold. One of the goals was to determine the three asymmetries in the Dalitz
plot population ($A_{LR}$, $A_{Q}$, $A_{S}$) for the strong isospin violating decay $\eta\to\pi^+\pi^-\pi^0$ 
in order to learn about the C invariance in different isospin states of the $\pi^+\pi^-\pi^0$ system. The second 
goal was to study the branching ratio for the $\eta\to\pi^0 e^+e^-$ decay which can break the C 
symmetry in electromagnetic interactions, when occurring by forbidden first order transition.

The measurement described in this thesis was performed in the Research Center J\"{u}lich in Germany 
by means of the WASA-at-COSY detector installed at the Cooler Synchrotron COSY.
The initial $pp\to pp\eta$ reaction was induced by the collision of a proton beam with the momentum 
of 2.14~GeV/c on a proton pellet target. Nucleons emerged from the interaction have been registered in the 
WASA forward detector, whereas the decay products of the $\eta$ meson were detected in the central detector.
For the identification of the $\eta$ meson the missing and invariant mass techniques were applied. The 
background originating from the direct two pion production and other $\eta$ meson decays has been reduced to 
negligible level by applying the momentum and energy conservation lows, and by 
using the missing and invariant mass distributions, as well as by performing a kinematic fit method. 
For the $\eta\to\pi^+\pi^-\pi^0$ decay the remaining physical background originating from the direct production 
of three pions via $pp\to pp\pi^+\pi^-\pi^0$ was subtracted for each studied phase space interval separately.

The three asymmetry parameters for the $\eta\to\pi^+\pi^-\pi^0$ decay, sensitive to the violation of the charge 
conjugation invariance were determined for the first time from the WASA-at-COSY data where the $\eta$ meson was  produced in proton-proton reaction. The final sample contained about $10^{5}$ events which enabled to estimate the following values of the left-right,
quadrant, and sextant asymmetry parameters:
\begin{equation*}
A_{LR} = 0.0033 \pm 0.0038_{stat} \pm 0.0009_{sys},
\end{equation*} 
\begin{equation*}
A_{Q} = - 0.0018\pm 0.0038_{stat} \pm 0.0011_{sys},
\end{equation*} 
\begin{equation*}
A_{S} =  0.0006 \pm 0.0039_{stat} \pm 0.0010_{sys}.
\end{equation*} 
Established values of the asymmetry parameters are consistent with
zero within the range of the statistical and systematic uncertainty, which allows to conclude 
that the charge conjugation symmetry C is conserved in strong interactions on the level of the achieved accuracy. 
Obtained result is also in agreement with previously measured 
values~\cite{Layter:1973ti,Layter:1973ti,Ambrosino:2008ht} and the average of the Particle 
Data Group~\cite{Ambrosino:2008ht}.

Furthermore the Dalitz plot distribution was analyzed. 
The Dalitz plot was represented in the X and Y variables proportional to the difference of the $\pi^+$ and 
$\pi^-$ energies, and $\pi^0$ energy, respectively. The distribution was corrected for each bin for the 
geometrical acceptance of the WASA-at-COSY apparatus as well as, the reconstruction efficiency and 
was determined free from the background of the multi-pion production.
The obtained density distribution was fitted with the phenomenological model. 
As a result of the fit we have obtained coefficients $c = 0.05\pm 0.10$
which is standing in the odd powers of X variable, and it is sensitive for C violation. 
Obtained result is consistent with zero within estimated uncertainty.

In second part of the thesis the investigation of the $\eta\to\pi^0 e^+e^-$ decay is presented. 
Based on analyzed experimental data $N^{exp} = 10 \pm 3_{stat}$ event candidates 
have been identified from the two weeks of data taking. 
The background originating from the other $\eta$ meson decays was suppressed to a negligible level. 
The only remaining background which was left in the signal region is originating from the $pp\to pp\pi^0\pi^0\to pp e^+e^-\gamma\gamma\gamma$ reaction. Next, we have estimated that the expected number of background events 
in the signal region should be equal to $N_{B} = 13 \pm 4_{stat}$. Based on given number of measured events, number of obtained background, and the estimated uncertainty we conclude that no signal was observed and therefore 
we have established the upper limit for the branching ratio of the $\eta\to\pi^0 e^+e^-$ decay: 
\begin{equation*}
BR( \eta\to\pi^0 e^+e^-) < 3.7 \cdot 10^{-5}~~~~{(\text{90\% C.L.)}}
\end{equation*}
Obtained result is slightly better than presently know value of this branching ratio from the 
fit of the PDG group~\cite{Nakamura:2010zzi}.

Results presented in this thesis were based on the data sample of about $5\cdot 10^7$ $\eta$ mesons. 
As a perspective for further improvement of the results on both decays: $\eta\to\pi^+\pi^-\pi^0$ and  
$\eta\to\pi^0 e^+e^-$, it would be interesting to use full statistic sample, since the WASA-at-COSY 
currently has collected around $10^{9}$ $\eta$ mesons in proton-proton collisions, which is 
one of the world's largest data sample for the $\eta$ meson. 
Therefore, the investigations presented in this thesis will be continued. 
Such statistics should enable to lower the statistical uncertainties 
for the determination of the asymmetry parameters which were investigated in this thesis by a factor of ten.  
For the  $\eta\to\pi^0 e^+e^-$ decay it would be possible to lower the sensitivity of the branching ratio
determination about an order of magnitude. 

\newpage
\pagestyle{fancy}
\fancyhead{}
\fancyfoot{}
\renewcommand{\headrulewidth}{0.8pt}

\fancyhead[RO]{\textbf{\sffamily{{{\thepage}}~}}}
\fancyhead[RE]{\bf\footnotesize{\nouppercase{ }}}
\fancyhead[LE]{\textbf{\sffamily{{{\thepage}}}}}
\fancyhead[LO]{\bf\footnotesize{{\nouppercase{ }}}}

\advance\headheight by 5.3mm
\advance\headsep by 0mm

\newpage
\thispagestyle{empty}
\begin{center}\end{center}
\begin{center}\end{center}\begin{center}\end{center}
\begin{center}\end{center}\begin{center}\end{center}
\begin{center}\end{center}\begin{center}\end{center}
\begin{center}\end{center}\begin{center}\end{center}
\begin{center}\end{center}
\begin{center}\end{center}
\begin{center}\end{center}
\begin{center}\end{center}
\begin{center}\end{center}
\begin{center}\end{center}
\begin{center}\end{center}
\begin{center}\end{center}
\begin{center}\end{center}
\begin{center}\end{center}
\begin{flushright}
{\bf{"Anyone who has never made a mistake\\
has never tried anything new."}}\\
(Albert Einstein)
\end{flushright}

\newpage
\thispagestyle{empty}
\begin{center}\end{center}

\newpage
\thispagestyle{empty}

\begin{center}{\LARGE{\bf Acknowledgments}}\end{center}
\vspace{1.cm}
\hspace{\parindent}
It would not have been possible to write this dissertation without 
the support and help of the many people around me during the years of Ph.D. studies, 
and therefore I would like to take the opportunity to thank everyone in this short note.

First and foremost, my utmost gratitude is due to Prof. dr. hab. {\bf{Pawe\l{} Moskal}} who has 
introduced me to a beautiful and interesting world of particle physics, and 
for last six years was a guide in my journey always encouraging and motivating me in daily work.
I would like to thank you Pawe\l{} for your infinite patience even in difficult times. 
I am also indebted to you for reading and correcting each new version of this thesis,
and for solving many problems which were always waiting "around the corner".

I am also greatly indebted to Prof. dr hab. {\bf{Walter Oelert}} who has introduced me to 
extraordinary anti-world of particle physics, where with his support, help and guidance I had 
a great opportunity get to know methods of producing and trapping antihydrogen atoms. 
I also wish to thank you Walter for the hospitality during my stays in Geneva, and for 
allowing me to use you car in CERN always when I needed it. 

I would like to thank Dr {\bf{Dieter Grzonka}} who was always a great support when I was visiting J\"{u}lich 
and Geneva, and for teaching me many things especially during our work in CERN. Also I would like to thank you Dieter for many interesting discussions.

I am grateful to Dr hab. {\bf{Andrzej Kup\'{s}\'{c}}} for his patience in answering my stupid questions and
for many valuable suggestions and ideas.

Many thanks are due to Dr {\bf{Joanna Klaja}}, Dr {\bf{Pawe\l{} Klaja}} and {\bf{Wojciech Klaja}} 
for theirs great friendship, and for always keeping me feel like home during my stays in J\"{u}lich, for 
the morning coffees, and many interesting discussions about physics and life. Also I wish to thank 
you Joanna and Pawe\l{} for reading and correcting the part of this thesis.

I would like to thank Master of Science {\bf{Wojciech Krzy\.{z}yk}} for his great friendship during last 
four years, for keeping me motivated and for many discussions. I also like to thank you for our many trips 
to distant parts of the world. 

I wish to thank Master of Science {\bf{Jaros\l{}aw J. Zdebik}} for his friendship, support, and 
a lot of discussions. Also I would like to thank you Jaros\l{}aw for keeping me encouraged during the Ph.D
studies, and for all the positive things which happened during the last six years.   

I am greatly indebted to Master of Science {\bf{Lech Sk\'{o}rski}} for his friendship and for many hours
of discussion (mostly "off-topic"), and for keeping a pleasant atmosphere during the daily work in 
rooms 244 and 212. 

I would like to express great gratitude to my roommates in room 244 Dr {\bf{Izabela Ciepa\l{}}} and Dr  {\bf{Aleksandra Wro\'{n}ska}}, who are always keeping a nice atmosphere during the days of work, and for 
many interesting discussions which we had. Also I would like to thank you very much Iza for very careful
reading and correcting the first part of my thesis, and for many valuable suggestions. 

I am very grateful to Prof. dr hab. {\bf{Bogus\l{}aw Kamys}} for letting me work in the Nuclear Physics
Division of the Jagiellonian University. 

Many thanks are due to Prof. dr hab. {\bf{James Ritman}} for support of my activities in the Nuclear Physics 
Institute at the Research Centre J\"{u}lich, and for financial support of my stays in J\"{u}lich.

I wish to thank: 
Mgr {\bf{Tomasz Bednarski}}, 
Dr {\bf{Eryk Czerwi\'{n}ski}}, 
Dr {\bf{Bartosz G\l{}owacz}}, 
Dr {\bf{Ma\l{}gorzata Hodana}}, 
Dr {\bf{Wojciech Krzemie\'{n}}}, 
Mgr {\bf{Szymon Nied\'{z}wiecki}},
Mgr {\bf{Iryna Ozerianska}},
Mgr {\bf{Andrzej Pyszniak}},
Mgr {\bf{Izabela Pytko}}, 
Dr {\bf{Dagmara Rozp\k{e}dzik}},
Mgr {\bf{Magdalena Skurzok}}, and
Mgr {\bf{Micha\l{} Silarski}}
for a nice atmosphere during the daily work in Krak\'{o}w and J\"{u}lich.

I am also grateful to all the people from the WASA-at-COSY Collaboration,
and the Colleagues from the ATRAP group, for a very nice cooperation.  

I am very grateful to my parents Barbara and Jerzy, my sister Sylwia, and my grandmother Brabara,
for their support during the whole time.

\newpage




\end{document}